\documentclass[twoside,11pt]{report}
\usepackage{amsthm,url,latexsym,amssymb,amsfonts,array}
\usepackage[centertags,intlimits]{amsmath}
\usepackage[normalsize,hang,bf]{caption}
\usepackage{graphicx}
\usepackage{xr-hyper}
\usepackage{hyperref}
\usepackage{makeidx}
\usepackage{epsfig} 

\usepackage{color}

\usepackage{nicefrac}
\usepackage{multirow}
\usepackage{arydshln}

\def\bearl{\begin{array}{l}}
\def\bearll{\begin{array}{ll}}
\def\bearlll{\begin{array}{lll}}
\def\eear{\end{array}}
\def\beqe{\begin{equation}}
\def\eeqe{\end{equation}}

\def\be{\begin{equation}}
\def\ee{\end{equation}}
\def\beq{\begin{eqnarray}}
\def\eeq{\end{eqnarray}}
\newcommand{\CC}{\mathbb C}

\newcommand{\R}{\mathbb R}
\newcommand{\Z}{\mathbb Z}

\newcommand{\h}{\,{\bf H}}
\renewcommand{\t}{\,{\bf T}}
\newcommand{\e}{\,{\bf E}}
\newcommand{\f}{\,{\bf F}}
\newcommand{\s}{\,{\bf S}}

\newcommand{\x}{\,{\bf X}}
\newcommand{\y}{\,{\bf Y}}
\newcommand{\lb}{\,{\bf L}}
 \newcommand{\SL}{\mbox{SL}(2,\mathbb{R})}
 \newcommand{\ls}{{\frak{sl}(2,\mathbb{R})}{}}
\def\lambdabar{{\mathchar'26\mkern-12mu\lambda}}

\def\stm{\overset{M}{\star}}
\newcommand{\p}{\partial}
\newcommand{\pb}{\bar{\partial}}
\newcommand{\di}{d}

 \def\ts#1#2{\ \mbox {\kern0.1ex\raise1.6ex\hbox{\tiny
 #1}\kern-0.45em\mbox{$\star$}\kern-0.45em\lower1.2ex\hbox{\tiny
 #2}}}

\def\st#1#2{\ \mbox {\kern0.1ex\raise1.6ex\hbox{\tiny
#1}\kern-0.45em\mbox{$\star$}\kern-0.45em\lower1.2ex\hbox{\tiny
#2}}\;}

\def\a#1#2{\alpha_{\bf #1\,\bf #2}}
\def\al#1#2{\alpha_{\bf #1\,\bf #2}}
\def\un#1#2{\nu_{\bf #1\,\bf #2}}
\def\n#1#2{n_{\bf #1}-\exp[-2\,\alpha_{\bf #1\,\bf #2}]\,n_{\bf #2}}
\def\bbf#1{{\bf #1}}

\newcommand{\ddt}{\frac{d}{dt}}
\newcommand{\ddto}{{\left.\frac{d}{dt}\right|_0}}
\newcommand{\fin}{\end{document}}
\def\a#1#2{\alpha_{\bf #1\,\bf #2}}
\def\un#1#2{\nu_{\bf #1\,\bf #2}}
\def\n#1#2{n_{\bf #1}-\exp[-2\,\alpha_{\bf #1\,\bf #2}]\,n_{\bf #2}}

\def\lambdabar{{\mathchar'26\mkern-12mu\lambda}}
\def\rpartial{\mathrel{\partial\kern-.75em\raise1.75ex\hbox{$\rightarrow$}}}
\def\lpartial{\mathrel{\partial\kern-.75em\raise1.75ex\hbox{$\leftarrow$}}}
\def\b#1{{\bf #1}}

\newcommand{\EBTZ}{\widetilde{\rm BTZ}}
\newcommand{\ca}{{\cal A}{}}
\newcommand{\cn}{{\cal N}{}}
\newcommand{\AN}{\ca\,\cn}

\def\soun{\mathfrak{so}(1,n)}

\def\sod{\mathfrak{so}(2)}
\def\son{\mathfrak{so}(n)}
\newcommand{\dpt}[3]{#1\,:\,#2\to #3}

\def\beq{\begin{eqnarray}}
\def\eeq{\end{eqnarray}}

\theoremstyle{plain}

\theoremstyle{definition}

\theoremstyle{definition}

\theoremstyle{definition}

\def\a{\alpha}
\def\b{\beta}
\def\k{\kappa}
\def\l{\lambda}

\def\m{\mu}
\def\n{\nu}

\def\o{\omega}

\def\x{\xi}
\def\z{\zeta}

\def\ca{{\cal A}}
\def\cb{{\cal B}}

\def\cd{{\cal D}}
\def\ce{{\cal E}}
\def\cf{{\cal F}}

\def\ck{{\cal K}}

\def\cn{{\cal N}}

\def\cp{{\cal P}}

\def\cs{{\cal S}}

\newcommand{\PP}{\mathcal P}
\newcommand{\Zx}{\mathcal Z}

\newcommand{\footpourmoi}[1]{}

\newcommand{\footpourtoi}[1]{}

\newcommand\hsp[1] {\mbox{\hspace{#1 em}}}
\def\lefthook      {{\vrule height5pt width0.4pt depth0pt}}
\def\leftrighthookfill{$\mathsurround=0pt \mathord\lefthook
                   \hrulefill\mathord\righthook$}
\def\righthook     {{\vrule height5pt width0.4pt depth0pt}}

\newcommand\Cont[1]{\hsp{.4}\vtop{\ialign{##\crcr$\hfil\displaystyle
{\hsp{-.4}#1}\hfil\hsp{-.4}$\crcr\noalign{\kern-1.9pt\nointerlineskip\vskip2pt}
\leftrighthookfill\crcr}}\hsp{.4}}

\newcommand\normord[1] {\,\raisebox{.033em}{\large\bf:}#1
                   \raisebox{.033em}{\large\bf:}\,}

\newcommand\norm[1] {\,|| #1 || \,}
\newcommand\ind[1] {\emph{#1}\index{#1}}

\newcommand\abs[1] {\,| #1 | \,}

\def \a{{\alpha}}
\def \be{{\beta}}

\def \S{{\cal S}}

\def \ex{{\rm e}}

\def \ra{{\, \rightarrow \,}}
\def \lra{{\longrightarrow}}

\def \eqdef{{\,\, \overset{\triangle}{=} \,\,}}

\def \p{{\partial}}
\def \pb{{\overline{\partial}}}

\def \zb{{\overline{z}}}
\def \wb{{\overline{w}}}
\def \Jb{{\overline{J}}}
\def \Tb{{\overline{T}}}
\def \C{{\mathbb C}}
\def \R{{\mathbb R}}
\def \Z{{\mathbb Z}}
\def \abar{{\bar{\alpha}}}
\newcommand{\re}[1]{(\ref{#1})}

\def\hb{{\bar{h}}}
\def\cd{{\cal D}}
\def\ex{{\mathrm{e}}}

\def\nn{{\nonumber}}

\def\g{{\gamma}}
\def\d{{\delta}}
\def\cL{{\cal{L}}}
\def\cH{{\cal{H}}}
\def\o{{\omega}}

\def\si{{\sigma}}
\def\Si{{\Sigma}}
\def\bz{{\bar{z}}}
\def\bo{{\bar{w}}}

\makeindex

\begin{document}



\begin{titlepage}
\thispagestyle{empty}
\begin{center}

\begin{figure}
\includegraphics*[scale=0.13]{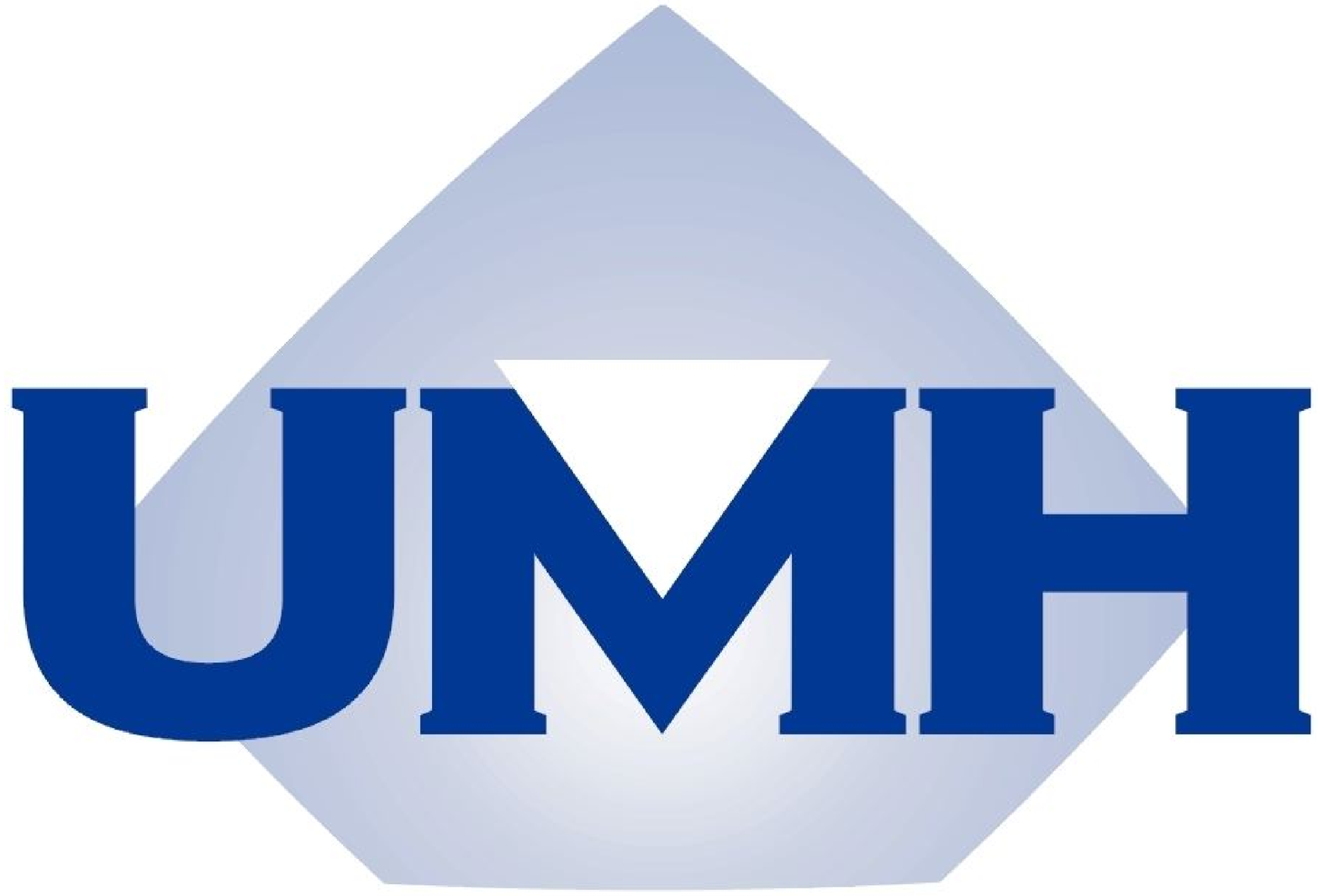} \hfill
\includegraphics*[scale=0.3]{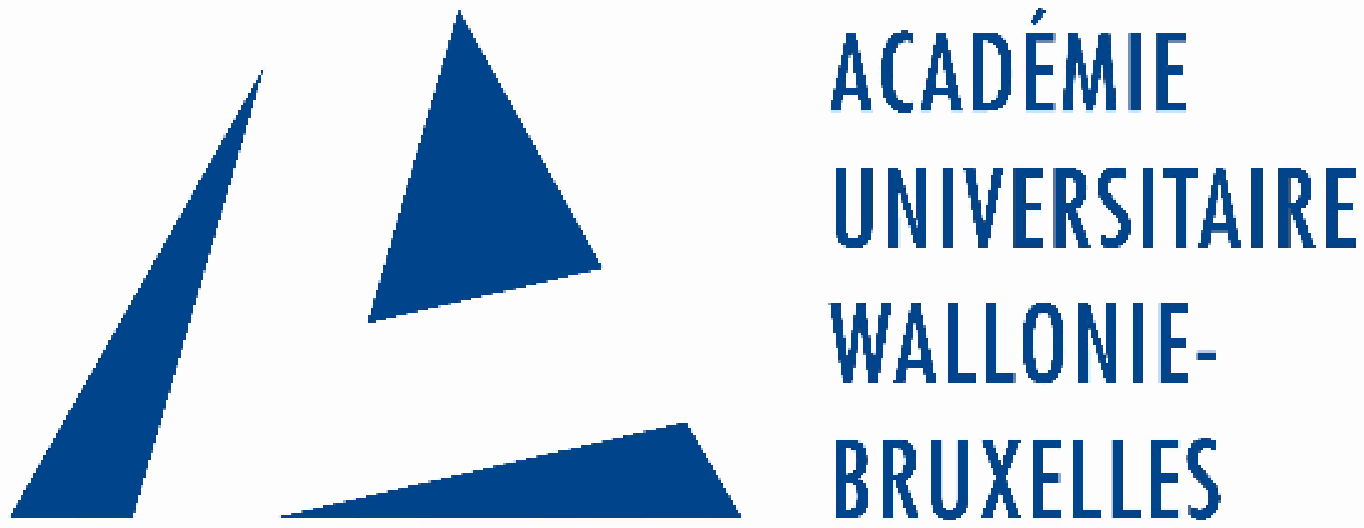}
\end{figure}
{\small Universit\'e de Mons-Hainaut \\
Facult\'e des Sciences \\
Service de M\'ecanique et Gravitation\\}

\vspace{4.5cm}

{\huge \bf
Deformations of anti-de Sitter black holes}

\vspace{2.75cm}

\vspace{1.5cm}
{\it St\'ephane Detournay }\\
\vspace{.2cm}
{\small Boursier F.R.I.A.} \\

\vspace{7.5cm}
\end{center}

{\small Directeur de th\`ese: Philippe Spindel} \hspace{4.5cm}
{\small Ann\'ee acad\'emique 2005--2006}

\end{titlepage}



\cleardoublepage
\pagestyle{empty}
\tableofcontents\thispagestyle{empty}

\cleardoublepage
\newpage
\section*{Acknowledgments}
\thispagestyle{empty}

\emph{Ma premi\`ere rencontre avec Philippe Spindel fut assez d\'econcertante. Disposant d'un bagage quelque peu inorthodoxe, j'avais la ferme intention d'entamer une th\`ese en physique th\'eorique. Philippe Spindel accepta de me recevoir en me fixant rendez-vous \`a son bureau. Comme je lui parlais de mon envie d'entamer une th\`ese sur les trous noirs, il passa au tableau, y griffonna quelques signes et me demanda s'ils me rappelaient quelque chose. A ce moment-l\`a, je pensais que c'en \'etait fait de moi, car jamais je n'avais vu de tels symboles (il me semble apr\`es coup qu'il s'agissait des \'equations de structure de Cartan), et que je n'avais plus qu'\`a tourner les talons. Pourtant, il me donna ma chance. Il m'int\'egra au Service de M\'ecanique et Gravitation de l'UMH, o\`u je pus prendre au vol un projet sur lequel il travaillait avec Claude Gabriel et Musongela Lubo, et il me donna l'occasion de l'accompagner \`a une conf\'erence \`a Tours, \`a laquelle il se rendait avec Robert Brout. Un an plus tard, j'entamais sous sa direction la th\`ese que voil\`a, \`a l'aide d'une bourse du Fonds pour la Recherche dans l'Industrie et l'Agricuture. Au cours de celle-ci, j'ai eu l'occasion d'\'etudier des sujets passionnants, en rapport avec mon travail de recherche. Dire qu'il a contribu\'e \`a ma formation est un euph\'emisme: en toutes circonstances (parfois cocasses), il s'est montr\'e disponible pour r\'epondre \`a mes interrogations, m'orienter, me conseiller, parfois bien au-del\`a de la physique. J'irai m\^eme jusqu'\`a dire que les trajets en voiture Bruxelles-Mons ``\`a la fine pointe de l'aube", malgr\'e le fait qu'une certaine fraction de ceux-ci commen\c{c}aient par un froid d\^u \`a mon manque de ponctualit\'e pathologique, vont me manquer. Pour ceci, et pour tout ce que je n'arrive pas \`a exprimer dans ces quelques mots, je voulais lui dire merci. }

\emph{Pierre Bieliavsky a \'egalement jou\'e un r\^ole important pour moi au cours de ces ann\'ees de th\`ese. 
Son enthousiasme vous ferait aimer l'analyse fonctionnelle, lorsqu'il vous explique que c'est beau comme une fugue de Bach.
Les nombreuses discussions que nous avons eues, qu'elles aient eu lieu \`a l'Universit\'e, dans un restaurant vietnamien, une baraque \`a frites ou
au t\'el\'ephone, et souvent seulement interrompues par des \'el\'ements ext\'erieurs d\'echa\^in\'es (fermeture du restaurant, d\'echarge de batterie du portable,...) ont \'et\'e, et sont toujours pour moi une source \'enorme de satisfaction et de d\'ecouvertes.}

\emph{Un \'etudiant en th\`ese serait bien isol\'e s'il ne pouvait pas partager ses angoisses, ses inqui\'etudes, ses joies et ses projets avec d'autres doctorants et jeunes chercheurs qui se trouvent dans la m\^eme situation que lui. Merci pour cela \`a la joyeuse bande de jeunes physiciens (et ing\'enieurs physiciens): Claire No\"{e}l, Sandrine Cnockaert, Gerald Goldstein, Pierre Capel, Marc Theeten, Geoffrey Comp\`ere, Nazim Bouatta, Paola Aliani, Sophie de Buyl, Fabien Buisseret, Vincent Mathieu, Nicolas Boulanger, Serge Leclercq, Laura Lopez-Honorez, Benoit Roland, Elizabete Rodriguez, ainsi qu'aux gais lurons des Weinberg Lectures, Jarah Evslin, Daniel Persson, Stanislav Kuperstein, Carlo Maccaferri, Mauricio Leston, I hope I'll find guys as crazy as you in postdoc! Merci \'egalement \`a Denis Haumont pour ces  interminables discussions dans les vestiaires du centre sportif de la For\^et de Soignes sur la pr\'ecarit\'e du statut de chercheur, et sur les coordonn\'ees de Pl\"ucker! Merci \`a la Modave team, Nicolas Boulanger, Xavier Bekaert, Vincent Bouchard, Sandrine Cnockaert, Vincent Mathieu, Alex Wijns, Stijn Nevens, Nazim Bouatta et Geoffrey Comp\`ere pour tous ces bons moments et ceux \`a venir.}

\emph{Merci \`a tous les physiciens th\'eoriciens et (g\'eographiquement) assimil\'es de l'UMH: Jean Nuyts, Yves Brihaye, Francis Michel, Claude Semay, Serge Leclercq, Philippe Grosjean, Martine Dumont, Marie-Anne Carlier, Fabien Buisseret, Vincent Mathieu,... , et une mention particuli\`ere pour Nicolas ``Old Fellow" Boulanger!}

\emph{Merci \`a Alex Sevrin, Marc Henneaux, Marios Petropoulos, Pierre Bieliavsky, Jean Nuyts, Yves Brihaye et Claude Semay d'avoir accept\'e de faire partie de mon jury de th\`ese.}

\emph{Pour avoir grandement facilit\'e mes collaborations avec l'ULB, je tenais \`a exprimer ma gratitude \`a Jean-Marie Fr\`ere, Serge Massar et Marianne Rooman; mes remerciements amicaux vont \'egalement \`a tous les membres du Service de Physique Th\'eorique de l'ULB pour leur accueil, dans un ordre inspir\'e de celui utilis\'e pour ma conclusion:  Xavier Calmet, Laura Lopez-Honorez, Michel Tytgat, Jean-Marie Fr\`ere, Fran\c{c}ois Englert, Robert Brout, Josep Oliver, Emmanuel Nezri, Marianne Rooman, Fu-Sin Ling, Michael Bebronne, Gilles Vertongen, Isabelle Reynders, Serge Massar, Paola Aliani, Peter Tyniakov, et tous ceux que j'oublie!}

\emph{Merci \`a Paola Aliani pour les conseils en Anglais et les cours d'Italien, et d'avoir accept\'e mon invasion du salon ces derniers mois...}

\emph{J'exprime mes remerciements les plus sinc\`eres \`a ma famille, qui n'a jamais cess\'e de me soutenir. En particulier merci \`a Papa, dont les premiers cours de thermodynamique et de math\'ematiques ne sont probablement pas \'etrangers \`a ce qui m'a men\'e au pr\'esent travail et pour m'avoir donn\'e le go\^ut de me surpasser (ainsi qu'un ordinateur portable!); merci \`a Maman pour les apports vitaux d'ice-tea, et pour me donner cette impression que quoi qu'il arrive, je pourrai toujours compter sur elle!}

\emph{Merci \`a Oph\'elie pour une remarque cruciale au bas de la page S11 de mes notes manuscrites. }

\emph{Merci \`a Maman et Bernard d'abord, puis M. et Mme. Foriers ensuite, d'avoir endur\'e et support\'e les bruits r\'ep\'etitifs de mes pas au-dessus de leurs t\^etes \`a des heures souvent incongrues.}

\emph{Je voulais remercier toutes les personnes avec qui j'ai eu l'occasion de collaborer aux cours de ces ann\'ees, qui sont un peu comme une seconde ``famille": Philippe Spindel, le ``papa", pas toujours commode mais toujours de bon conseil, dont on pourrait dire qu'il a ``une main de velours dans un gant de fer"; Marianne Rooman, la ``maman", dont l'aide a \'et\'e pr\'ecieuse aux premiers stades de ce travail, et dont on esp\`ere qu'elle reviendra bient\^ot au foyer ``BTZ"; Pierre Bieliavsky, le ``grand fr\`ere" super fort sur les traces duquel on aimerait marcher; Marios Petropoulos, l'``oncle" bienveillant et sage; Laurent Claessens, le cousin alter-mondialiste et blagueur; Domenico Orlando, le cousin aventureux; Michel Herquet, le petit fr\`ere avec lequel on a travers\'e les premi\`eres gal\`eres. Merci \'egalement \`a Geoffrey Comp\`ere, Laurent Houart, Glenn Barnich, Francois Englert, Jan Troost, Gaston Giribet, Marc Henneaux, Yuji Satoh, Riccardo Argurio, Daniel Sternheimer, Domenico Orlando, Xavier Calmet, Nicolas Boulanger pour des interactions tr\`es enrichissantes \`a diff\'erents niveaux. Je remercie \'egalement Dimitri Pourbaix, Claude Gabriel et Musongela Lubo avec qui j'ai eu la chance de collaborer avant d'entamer cette th\`ese. 
En esp\'erant que la famille s'agrandisse encore!}

\emph{Pour leur soutien dans mes d\'emarches post-doctorales, je remercie Marc Henneaux, Paul Windey, Marcel Arnould, Arlette Noels, Philippe Spindel, Marios Petropoulos et Pierre Bieliavsky.}

\emph{Je remercie le Fonds pour la Recherche dans l'Industrie et l'Agriculture de m'avoir octroy\'e la bourse ayant financ\'e le pr\'esent travail. Pour m'avoir donn\'e l'opportunit\'e d'assister \`a diverses Ecoles et conf\'erences, qui ont grandement contribu\'e \`a ma formation et \`a mes relations scientifiques,
 je voudrais \'egalement exprimer ma reconnaissance \`a l'Universit\'e de Mons-Hainaut, Philippe Spindel, Pierre Bieliavsky, Marianne Rooman, ainsi qu'au Fonds National de la Recherche Scientifique.}

\emph{Merci \`a tous ceux qui m'ont \'ecout\'e d'une oreille, tant\^ot curieuse, tant\^ot amus\'ee, tant\^ot perplexe, lorsque je tentais de les convaincre de l'importance primordiale des trous noirs \`a 2 dimensions spatiales. Pour ces \'episodes et tous ceux que je ne citerai pas ici,
 je voulais remercier Lo\"ic, Caroline, Alexandre, Laetitia, Denis, Carole, Marco, Anthony, Caroline, J-F, Fred, B\'en\'e, Dominique, Clara, Denis, les irr\'eductibles du Parc, le Bouchon, l'orchestre de l'ULB et les camarades de l'ARWSL 
 .}

 \emph{J'exprime ma reconnaissance et mon admiration \`a quelques personnes qui ont eu une influence d\'ecisive sur mes orientations: mes parents, Roger Delchambre, Melinda Van Durme-Buyse, Yves De Rop et Philippe Spindel.}  

\emph{Merci \`a Sophie de Buyl pour sa contribution \`a toutes les \'etapes de ce travail, du soutien psychologique \`a la relecture, en passant par l'encodage des notes que j'ai eu la mauvaise id\'ee de r\'ediger \`a la main dans un premier temps.}
\thispagestyle{empty}

\emph{Je remercie chaleureusement tous ceux qui ont contribu\'e \`a la relecture de cette th\`ese, en tout ou en partie: Philippe Spindel, Pierre Bieliavsky, Laurent Claessens, et tout particuli\`erement Christiane Schomblond, Sophie de Buyl, Claude Ballantine et Jo\"elle Claes pour le dernier rush! Un grand merci \`a Christoph Sieg pour m'avoir aid\'e \`a surmonter les obstacles qui ont tent\'e de m'emp\^echer de mettre cette th\`ese sur ArXiv!}

\emph{Enfin, merci \`a celle dont les talents d'agent immobilier, de physicien,  de photographe, de banquier, de chanteuse, de danseuse, de dactylographe, de relaxeur, d'assistant psychologique, de n\'egociateur n'en finissent pas de m'\'etonner et de m'\'emerveiller ...}

\emph{``M\^eme si tu pars loin..."  }
\thispagestyle{empty}

\cleardoublepage \chapter*{Introduction}
\setcounter{page}{1}
\pagestyle{myheadings}
\addcontentsline{toc}{chapter}{\numberline{}Introduction}

Black holes
 are among the most intriguing and fascinating objects of the physics landscape. The idea of the existence of bodies so massive that not even light could escape was put forward by the English geologist John Michell in 1784, and later promoted by the French mathematician Pierre-Simon Laplace in the first and second editions of his book {\itshape Exposition du syst\`eme du Monde} (1795). They noted that, according to Newtonian gravity and Newton's corpuscular theory of light,
``{\itshape un astre lumineux de m\^eme densit\'e que la terre, et dont le diam\`etre serait deux cents cinquante fois plus grand que celui du soleil , ne laisserait en vertu de son attraction , parvenir aucun de ses rayons jusqu'\`a nous}" (see e.g. \cite{Laplace}, cited in \cite{MisnerThorneWheeler:1973}). 

Nevertheless, it is in the light of the modern description of gravitation, provided by Einstein's General Relativity, that the concept of black holes takes all its meaning.
Only a few months after the publication of General Relativity, in 1916, the German physicist and astronomer
Karl Schwarzschild produced the first exact solution to the general gravitational equations \cite{Schwarzschild:1916uq}, subsequently recognized as describing the gravitational field outside a spherical, non-rotating mass $M$ \footpourmoi{CORRECT?}.
Although this was not completely realized at that time,
 this solution, now called the \ind{Schwarzschild solution}, turns out to describe a \ind{black hole} (a name invented by Wheeler in 1968),
when the whole mass $M$ is confined in a sphere of radius less than $r_H = \frac{2 G M}{c^2} \sim 3\frac{M}{M_\odot} \mbox{km}$ (where $c$ is the speed of light, $G$ the Newton's constant and $M_\odot$ the mass of the sun). The \ind{Schwarzschild radius} $r_H$ is now known to be the radius of the \ind{event horizon} of a non-rotating black hole. The presence of an event horizon can be stated as defining a black hole.
It can be described as a (hyper-) surface within which all light-like paths, and hence all
trajectories of infalling observers, necessarily end on the \ind{black hole singularity}, conventionally located at $r=0$. At this point, characterized mathematically by the fact that some scalar invariants diverge in general (\ind{curvature singularity}), an observer would feel infinite tidal forces. General relativity not only predicts the existence of black holes, but also gives a precise account of how they can form. In 1939, Oppenheimer, Snyder and Volkoff studied the gravitational collapse of a homogeneous sphere of pressure-free fluid using general relativity, and discovered that if the body develops a core heavier than three solar masses, the pressure of neutron degenerate matter cannot support against gravitational collapse, and nothing prevents contraction until a ``point" of zero volume and infinite density. Hawking and Penrose showed, in the sixties, that the appearance of singularities does not rely on some simplifying assumptions, but is a generic feature of collapsing objects.
By now, indirect astronomical observations strongly point toward the existence of stellar ($4-15 M_{\odot}$), intermediate, or supermassive ($10^5-10^{10} M_{\odot}$) black holes, including one at the center of our own Milky Way.

Besides their observational interest, black holes are at the center of some of the most long-standing challenges of contemporary theoretical physics.
The presence of the black hole singularities are intimately related to short distance physics in the quantum domain, while the current day description of gravitation is essentially classical. Therefore they are believed to signal the breakdown of general relativity in certain regimes, which a quantum theory of gravity should be able to cure by letting new phenomena come into play.

Black holes have 
aroused an increasing theoretical interest after the work by Bardeen, Carter and Hawking\cite{Bardeen:1973gs} which produced the
so called four laws of black hole mechanics\footpourmoi{les connaitre! Voir TheseLemos}, giving strong indications that black holes could be seen as thermodynamic objects. In particular, the second law states that the area of the black hole horizon cannot decrease during any physical process. This prompted Bekenstein to suggest that every black hole has an entropy, associated with the area of its horizon \cite{Bekenstein:1972tm}. This idea was first considered very suspiciously, because it implies that a black hole is hot and radiates. However, it rapidly received support, thanks to a result by Hawking who showed, using a semi-classical treatment in which the gravitational field of the black hole is treated classically but the matter is treated quantum mechanically, that black holes do radiate, and that the frequency spectrum is characteristic of a blackbody \cite{Hawking:1974sw}. Just as the macroscopic thermodynamical properties of perfect gases hinted at their microscopic atomistic structure, the classical thermodynamical properties of black holes suggest the existence of quantized micro-states, whose dynamics would account for the macroscopic production of entropy.  
A theory of quantum gravity would be expected to meet the challenge of \emph{explaining the nature of these micro-states}, thereby solving the \ind{black hole entropy problem}.

Hawking's result that black holes emit a thermal radiation stands at the origin of another thorny problem, called the \ind{black hole information paradox} \cite{Hawking:1976ra} (see also for example \cite{Page:1993up,Einhorn:2005bi,Peet:2000hn}), which can be summarized as follows.
Since the Hawking temperature of a black hole is inversely proportional to its mass, it is quantum mechanically unstable because, when losing mass, 
its temperature grows and the particle emission rate grows\footpourmoi{et pour un TN general??}.
Therefore, the black hole can ultimately \emph{evaporate}\index{black hole evaporation}.
Then comes the paradox: in the case of gravitational collapse, one may suppose that matter starts in a pure state, collapses to form a black hole, which eventually evaporates leaving the universe in a mixed thermal state. However, unitary evolution in quantum mechanics implies that pure states evolve to pure states, so that somewhere in this process, there must be a breakdown of unitarity. The question is then whether this process is like the burning of a book, in which there is no doubt that the information is encoded in the radiation even though it would be impossibly difficult to recover, or if the information is really lost. A third possibility would be that the information is neither lost nor present in the radiation, but is stored in a remnant of the evaporating black hole.  An analogous statement of the problem relies on the so-called \ind{no-hair theorem}\footpourmoi{tient seulement a $3+1$ dimensions?}, according to which all black hole solutions of the Einstein-Maxwell equations can be completely characterized by only very few externally observable parameters: mass $M$,
 charge $Q$, and angular momentum $J$. The fact that the thermal Hawking radiation depends only upon these parameters entails a loss of information, since two distinct objects bearing the same values of $M$, $J$ and $Q$ would give rise to the same Hawking radiation according to an observer outside the horizon.

Faced with these questions, a rich variety of original approaches have been proposed toward a formulation of a quantum theory of gravity (see e.g. \cite{Rovelli:1997qj,Carlip2,Kiefer:2005uk} for reviews). We will briefly recall some of them here, focusing on the aspects relevant to the present work.

It has often been a good strategy to tackle 
 field theory via models with a reduced number of degrees of freedom; general relativity is no exception to this. In particular, general relativity in (2+1) dimensions \cite{Carlip3,Carlip4,Carlip:1998uc} has received a lot of attention and been revealed as a useful arena in which to explore the foundations of classical and quantum gravity,
 as settled by the seminal work of Deser, Jackiw and 't Hooft \cite{DT,Deser:1988qn,'tHooft:1986gp}\footpourmoi{resume: voir TheseLemos p29}, Achucarro and Townsend \cite{Achucarro:1987vz}, and Witten \cite{Witten:1988hc,Witten:1989sx}. An essential feature is that (2+1)-dimensional gravity has no local degrees of freedom.
  In this case, the Riemann curvature tensor is algebraically determined by the Ricci tensor. As a consequence, a vacuum solution with a vanishing cosmological constant is necessarily flat, and no gravitational waves can exist. This,
  and the fact that general relativity in (2+1) dimensions has no Newtonian limit (i.e. static point sources feel no gravitational force)\footpourmoi{voir TheseLemos}, first seemed to indicate that the model is too physically unrealistic to give much insight into real gravitating systems in (3+1) dimensions. It therefore came as a surprise when Ba$\tilde{\mbox{n}}$ados, Teitelboim and Zanelli showed that (2+1)-dimensional gravity with a negative cosmological constant $\Lambda$ admitted black hole solutions \cite{BTZ}, whose properties were subsequently studied in detail in \cite{BHTZ}. These black holes, referred to as \ind{BTZ black holes}, enjoy remarkable properties which brought them to the forefront of research on quantum aspects of black hole physics, with applications well beyond the scope of (2+1)-dimensional gravity, especially in string theory as we will discuss later on. Following the same reasoning as above, any vacuum solution of the Einstein's equations in (2+1) dimensions with a negative cosmological constant has constant curvature, i.e. it is locally anti-de Sitter. It has furthermore been shown that BTZ black holes can be obtained from the three-dimensional anti-de Sitter space ($AdS_3$) by performing discrete identifications along orbits of well-chosen Killing vectors. The BTZ black holes display most of the features of their (3+1)-dimensional counterparts: they have an event horizon and, in the rotating case, an inner horizon; they appear as the final state of gravitational collapse \cite{Ross:1992ba} and their thermodynamic properties can be investigated using standard techniques \cite{Carlip1,Carlip4}\footpourmoi{Hawking rad??}. They differ however from usual black hole solutions in one important respect: they have \emph{no curvature singularity} at the origin, which was to be expected since they are locally $AdS_3$.

A compelling argument in favor of the microscopic origin of black hole entropy
has been given in an elegant way in the framework of (2+1)-dimensional gravity \cite{Strominger:1997eq,Birmingham:1998jt} (for a review, see also e.g. \cite{Carlip4}). The first step rests on a remarkable observation by Brown and Henneaux \cite{BrownHenneaux}, who investigated the asymptotic symmetries of (2+1)-dimensional asymptotically anti-de Sitter spaces. They found that the generators of diffeomorphisms that preserve the asymptotic structure of the $AdS$ metric (the ``asymptotic Killing vectors'') form two copies of commuting Virasoro algebras, without central charge.  Since the boundary of an asymptotically $AdS$ space is a cylinder, it is not really surprising that the asymptotic diffeomorphisms are related to the diffeomorphisms of the cylinder\footpourmoi{symetries asymptotiques de $AdS_3$ = groupe conforme du bord, c-a-d du cylindre..?}. What came as somewhat unexpected was that, when these symmetries are realized as canonical transformations, the original Virasoro algebras acquire a central extension, with central charge given by $c = 3l/2G$, where $\Lambda = -1/l^2$.  Subsequently, Strominger \cite{Strominger:1997eq} and Birmingham, Sachs and Sen \cite{Birmingham:1998jt} independently pointed out that this could be used to compute the asymptotic density of states. Their argument appealed to the so-called Cardy formula \cite{Cardy:1986ie,Bloete:1986qm}, expressing that the asymptotic density of states in a two-dimensional conformal field theory (whose symmetry algebra is a Virasoro algebra) is fixed by a few features of the symmetry algebra, independent of any details of the dynamics. They showed that the entropy computed from the Cardy formula agrees precisely with the Bekenstein-Hawking entropy, given in terms of the surface (actually, the perimeter) of the BTZ black hole horizon. This impressive result unfortunately does not give any clue about the nature of the underlying quantum mechanical degrees of freedom, and seems to indicate that the latter are located at infinity rather than at the horizon. It can
 be seen as a concrete realization of the \ind{holographic principle}, as proposed by 't Hooft \cite{'tHooft:1993gx} and improved by Susskind \cite{Susskind:1994vu}, according to which the combination of quantum mechanics and gravity requires the world to be an image of data that can be stored on a lower-dimensional projection, much like a holographic image\footnote{This principle already implicitly shows through Bekenstein's proposal that information should be associated with the surface of the horizon rather than with a volume.}\footpourmoi{A comprendre!!}\footpourmoi{Lire Susskind: The world as an hologram}.

 


A more ambitious approach to quantum gravity is provided by \ind{string theory}, which is, at present, by far the most investigated research direction\footnote{For this section, standard references are the textbooks \cite{GSW1,GSW2,Pol1,Pol2}.}. 
String theory offers not only a framework for a unitary theory of quantum gravity, but also represents an attempt to unify all the known forces in nature. The basic idea of string theory is to replace the basic point-like constituents of the Universe with one-dimensional extended objects of extremely small scale, called strings. These can vibrate in different modes, each mode appearing as a different particle; their interactions are represented by letting them split and combine.
 The idea that string theory might describe a quantum theory of gravity actually goes back to the beginning of the seventies, where it was noticed that the string spectrum contains a particle that could be identified with the graviton, the hypothetical elementary particle that transmits the force of gravity in the framework of quantum field theory. Between 1984 and 1986, it was realized that more sophisticated string theories could describe all elementary particles and their interactions, so that string theory was promoted to the status of most promising idea to unify theories of physics.
The principles 
 of string theory naturally lead to
the introduction of new physical concepts, such as supersymmetry (a symmetry between bosons and fermions) and extra spatial dimensions, whose existence is to be tested in the near future, either in accelerator experiments or by future cosmological and astrophysical data\footpourmoi{Comment, en pratique??}. When taking supersymmetry into account, one ends up with five consistent \ind{superstring theories}, interconnected by a web of dualities. Strong evidence has been provided in the nineties that all these theories might be different limits of one unique theory, named \ind{M-theory}.

At present, the approach to string theory is essentially perturbative. The only accessible quantities are scattering amplitudes computed in a given vacuum of the theory, i.e. around a given space-time geometry, which does not result from dynamical degrees of freedom of the theory. That is, no complete string \emph{field} theory exists. However, the consistency of the theory puts strong constraints on the allowed geometry on which it makes sense to study the propagation of a single string.
In string theory, an acceptable background should be interpretable as a coherent state of string excitations. Then comes a kind of magic: the  constraints imposed on the gravitational field are nothing other than Einstein's equations! More precisely, general relativity appears as a low-energy effective theory of string theory (i.e. to first order in the \ind{inverse string tension} $\a'$, the only free parameter of the theory).
When including the other massless modes of the string spectrum as well as supersymmetry, the resulting low energy effective theories are the \ind{supergravity theories}. A given background will be called an \ind{exact string background} if it satisfies the appropriate constraints to \emph{all orders in $\a'$}, i.e. beyond the supergravity approximation.
These are very strong restrictions, and only a few exact string backgrounds are known. Besides the flat case, an important class of such backgrounds is provided by the Wess-Zumino-Witten (WZW) models and their gauged versions \cite{Novikov:1982ei,Witten:1983ar,Polyakov:1983tt,KZ,GepWitt, DiFr}. Remarkably enough, some of them turn out to describe black holes, which is a windfall for extracting information on their properties in the context of string theory. The first example has been given by Witten \cite{Witten:1991yr}, Dijkgraaf, Verlinde and Verlinde \cite{DVV} and others, through the so-called two-dimensional black hole.
Another example is the BTZ black hole already introduced above and which was recognized to be part of an exact string background soon after its discovery\cite{HoroWelch}. Later on, it has become ubiquitous when discussing black hole physics in the context of string theory. Unfortunately, it is not known whether realistic, (3+1)-dimensional black hole solutions of general relativity can be lifted to exact string backgrounds, nor if a string theoretic description is attainable. Even if they are very interesting in their own right, WZW models and their gauged versions represent very special points in the moduli space of possible string theory vacua, in particular the more symmetric ones. They do, however, enjoy the additional interesting property that they allow for \ind{exact marginal deformations} \cite{Chaudhuri:1989qb}. This means that it is possible to deform these models, while keeping exact backgrounds, allowing a wide variety of less symmetric solutions to be reached and linking already known theories.  

Another key feature of string theory arose in the mid-nineties, when Polchinski discovered that the theory requires the inclusion of higher-dimensional objects, called \ind{D-branes}. These were initially seen as hyper-surfaces of different dimensionalities (one speaks of a Dp-brane when it has $p$ space-like dimensions) on which open strings could end, ``D" standing then for ``Dirichlet" boundary conditions. The crucial observation of Polchinski is that these D-branes are truly dynamical, non-perturbative\footpourmoi{comprendre pourquoi: leur masse exprimee en termes de $g_S$...} objects of the theory. On one hand, they can emit and absorb closed strings (one of their modes is the graviton, other modes include gauge bosons), and therefore have mass and charge.
On the other hand, the open strings (whose modes include the photon for example) attached to a D-brane give rise to gauge theories living on it. This can be summarized by saying that D-branes are gravitational sources, on which a gauge theory lives. 

A definite theoretical success of string theory
is related to the black hole entropy problem. Strominger and Vafa showed \cite{Strominger:1996sh} that some black holes might be seen as bound states of D-branes, and they gave a first-principle calculation of their microscopic entropy (for reviews, see \cite{Peet:2000hn,David:2002wn,Mathur:2005ai}). This derivation was originally restricted to non-rotating, extremal, supersymmetric, charged black holes. The charges actually serve as tags that help identify their microscopic constituents in string theory, while supersymmetry and extremality (combined into the \emph{BPS} condition) ensure that the counting of states at weak coupling can safely be extrapolated to strong coupling.\footpourmoi{Precisement!!}\footpourmoi{pour un TN extremal, l'horizon est en $r=0$? Mais alors son aire est nulle, non? Voir Mathur}\footpourmoi{trou noirs comme etats lies de branes: comprendre le lien entre vision supergravite et vision cordes-comptage d'etats-. Ou est le trou noir? Ou sont les branes? Voir Mathur: comment voit-on le passage strong-weak coupling??} Variants of the original computation can however be applied to more general black holes (see \cite{Emparan:2006it} and references therein), but the simplest and more familiar solutions, Schwarzschild for example, are still out of reach (see however \cite{Argurio:1998xm,Sfetsos:1997xs}). The relevance of the aforementioned BTZ solution is also apparent through the fact that almost all black holes for which the entropy can be computed in string theory have a near-horizon geometry containing the BTZ solution (see \cite{Peet:2000hn})\footpourmoi{En quoi cette near-horizon limit est-elle interessante?}.
Another breakthrough with far-reaching consequences in string theory is the $AdS/CFT$-correspondence, proposed by Maldacena, expressing that string 
theory on $AdS_5 \times S^5$ is equivalent to a supersymmetric N=4 Yang-Mills gauge theory defined on the 4-dimensional boundary of $AdS_5$ \cite{Maldacena:1998re,Aharony:1999ti,Witten:1998qj} (see also \cite{Petersen:1999zh}). More generally, it suggests a duality between a string theory, i.e. a theory including gravity, defined on a space, and a quantum field theory \emph{without} gravity defined on the conformal boundary of this space, whose dimension is lower by at least one. This correspondence, a glimpse of which could already be foreseen in Brown and Henneaux's work, can be viewed as yet another illustration of the holographic principle, and has led to new insights into the black hole information paradox \cite{Peet:2000hn,Bigatti:1999dp} as well as into the nature of black hole and cosmological singularities \cite{Craps:2006yb}\footpourmoi{A VOIR Absolument!}.

In spite of its intrinsic elegance and successes, it is fair to say that string theory, as a physical theory, is still at an embryonic stage, and suffers from some theoretical difficulties (see e.g. \cite{Kiefer:2005uk,Rovelli:1997qj,Rovelli:2003wd}). Therefore, even if the position of string theory is clearly dominant in the (very) high energy physics landscape, alternative approaches have flourished and made their own way (for a review, see e.g. \cite{Rovelli:1997qj,Carlip2}), sometimes even revealing deep connections with string theory. This is the case, in particular, for \ind{noncommutative geometry}. The notion of noncommutativity is deeply ingrained in us, in particular as the central mathematical concept expressing uncertainty in quantum mechanics, where it applies to any pair of conjugate variables, such as position and momentum. 
A natural way, albeit maybe too naive, of quantizing gravity, in the sense of quantizing the structure of space-time, might be to imagine that position measurements might fail to commute, and describe this using the noncommutativity of the coordinates. This idea dates back to Snyder \cite{Snyder1}, who realized that by starting with a conventional field theory, and interpreting the fields as depending on noncommuting coordinates, one can follow the usual development of perturbative quantum field theory with surprisingly few changes, to define a large class of \ind{noncommutative field theories}\footpourmoi{Invariance de Lorentz? v. CarlipProgressp47, voir surtout Douglas-Nekrasov!!}. A motivation for postulating noncommutativity could rely on the inherent ``uncertainty principle" in quantum gravity which prevents one from measuring positions to better accuracies than the Planck length. As noticed by DeWitt, the momentum and energy required to make such a measurement would then itself modify the geometry at these scales. One might wonder if these effects could be modeled by noncommuting coordinates. The resulting noncommutative quantum field theories 
have been analyzed in some detail and exhibited some peculiar features, including non-locality and the coupling of high- and low-energy
degrees of freedom (``UV/IR mixing") (for a review, see \cite{Douglas:2001ba}). Recently, noncommutative theories of gravity, generalizing the Einstein-Hilbert action to a space with noncommuting coordinates, have also been proposed \cite{Calmet:2005qm,Aschieri:2005zs,Alvarez-Gaume:2006bn}. 
One may also mention the approach of Kempf, Mangano and Mann: they implement a minimal length uncertainty, keeping the position operators commuting but modifying the commutation relation between position and momentum by the addition of an extra momentum-dependence \cite{Kempf:1994su}. This simple modification has been shown \cite{Brout:1998ei} to be able to cure the trans-planckian problem of black hole physics, lying in the fact that the original derivation of Hawking radiation involved field modes with frequencies near the black hole horizon arbitrarily high, in particular, higher than the inverse Planck time, which a priori raised doubts about its validity. 

Noncommutative geometry is a branch of mathematics introduced 30 years ago by A. Connes \cite{Connes:1994yd}, and which experienced a dramatic development over the last years due to its importance in mathematics and in many physical applications.
In this framework, the concept of a \emph{noncommutative manifold}\index{noncommutative manifold} can be settled through algebraic instead of geometric tools.
That this is possible rests upon a celebrated result, by Gel'fand, that knowledge of a topological space is essentially equivalent to knowledge
of the algebra of continuous functions on that space. Riemannian geometry, in particular the notion of distance, can be captured
algebraically as well, via the introduction of an object called the \ind{spectral triple} (for a pedagogical account, see \cite{Martinetti:2003ig,martinetti-2001-})\footpourmoi{A LIRE!!+ voir le modele standard!}. It consists, in addition to an algebra, of the Hilbert space formed by all the spinor fields on a riemannian (spin) manifold and of the Dirac operator associated with its metric. Connes not only showed that the Riemannian manifold can be reconstructed from the spectral triple, but also that the Yang-Mills
functional on the manifold can be obtained from it in a purely operatorial way. A noncommutative riemannian manifold is the idea of describing a manifold by a spectral triple in which the algebra is noncommutative. 


A suitable framework for such noncommutative algebras is that of \ind{deformation quantization}, introduced originally by Bayen, Flato, Fronsdal, Lichnerowicz and Sternheimer in the late seventies \cite{BFFLS}, as an alternative method for quantizing a classical system. The traditional method for quantizing a system induces a radical change in the nature of the physical observables, from functions defined on the classical phase space to operators acting on a Hilbert space. Deformation quantization, instead, suggests reading the (noncommutative and associative) composition of operators at the level of functions. This way, the quantization amounts to ``deform" the commutative algebra of functions on the phase space into a noncommutative algebra by replacing the (commutative) pointwise multiplication of functions by a one-parameter (usually Planck's constant $\hbar$) noncommutative operation, called \ind{star product}, in a prescribed way. Namely, a star product must define an associative operation (because so is the composition of operators) and exhibit the right ``classical limit", as $\hbar$ tends to zero. The simplest example of such quantization is that of a free particle in $\R^n$ (``Weyl's quantization") leading to the \ind{Moyal-Weyl product}, which deforms the algebra of function on the symplectic phase space $\R^{2n}$. Originally, star products were expressed as formal power series in the deformation parameter, leading to \ind{formal deformation quantization}, in which one does not worry about the convergence of the series. The notion of \ind{strict deformation quantization} was later introduced by Rieffel \cite{Rieffel1}, in which the product of two functions is again a function, in some suitable functional framework. The Moyal-Weyl 
product is one example of such a strict deformation quantization.
The notion of deformation quantization has been extended to general Poisson manifolds, and a long-standing open question has been to know whether or not any Poisson manifold could be quantized. The question was answered positively by Kontsevich\cite{Kontsevich:1997vb} in a 1997 pre-print, where a recipe to construct explicit formulae in terms of formal power series was given. The question for strict deformation quantization is far more involved. As pointed out by Connes and Marcolli \cite{connes-2006-}, the notion of a strict deformation quantization should be regarded as a
notion of integrability for formal solutions, similarly to the case of formal and actual solutions of ordinary differential equations.
In particular, a lot is known about \emph{formal} deformation quantizations, but when one tries to pass to \emph{strict} deformation quantizations,  
there are cases where existence fails, and others where uniqueness fails\footpourmoi{Voir ConnesMarcolli!}. Beyond the flat case, the most notable examples were introduced by Rieffel for manifolds admitting an action of an abelian group \cite{Rieffel2}, and by Bieliavsky, Bonneau, Massar and Maeda \cite{PierreStrict,BielBonneau,BielMassar,BielMassar2} for manifolds admitting an action of solvable groups. 

One reason why physicists' interest in noncommutative geometry got rekindled over the last few years comes from the the deep and unexpected connections it has revealed with string theory (see \cite{SeibWitt}, and references therein). This relationship can be expressed as follows. In perturbation theory, D-branes' dynamics is described by the excitations of the open string ending on them. When considering a flat D-brane in Minkowski space-time, carrying a magnetic field (called B-field), one observes upon quantization that the open string coordinates living in the D-brane no longer commute, due to the presence of the B-field: the D-brane is deformed into a noncommutative manifold, the pointwise multiplication of functions on it being deformed into the Moyal-Weyl one. Moreover, in a special limit of the theory, where the massive open string modes decouple, one observes that the dynamics of the massless fields is completely governed by the noncommutative product. Namely, the low-energy effective theory is given by a noncommutative version of Yang-Mills theory\footpourmoi{details: these Sieg, these Larsson, Seiberg-Witten}. Beyond the flat case, for curved D-branes in curved backgrounds, very little is known, because on one hand string theory is much less understood, and on the other hand, because it is not clear what could replace the Moyal-Weyl product in the presence of curvature (see \cite{Szabo:2006wx})\footpourmoi{lire ce truc!!}\footpourmoi{lien avec les cordes: refs 15 et 76 de Rovelli, tore non-commutatif, etc?? Voir aussi etudiants Weinstein pour le tore non-commutatif. Voir Connes-Marcolli pour relation avec theorie des cordes!}.




\newpage
{\bf This thesis} 

The central objects of the present work will be the BTZ black holes, from which we will depart from time to time. We are mainly interested in \emph{deforming} these black holes, in the two aforementioned senses: deformation quantization and exact marginal deformation. The first will explore how some geometrical properties of BTZ black holes allow them to be embedded in noncommutative geometry, in obtaining noncommutative versions of these spaces, and will suggest how the relationship to string theory could manifest itself in this non-trivial situation. The second will allow us to construct, starting from the BTZ solution, a variety of new exact string theory backgrounds, encompassing already known solutions. We will show that some BTZ space-times support exact D-brane configurations, ``winding around" the black hole, crossing horizons and singularities, that may be connected to similar-looking D-branes in the two-dimensional black hole by means of marginal deformations. We will also pay some attention to higher dimensional generalizations of the BTZ black holes, pointing out a geometric argument that allows us to characterize them in a simple way. 
We will allow for a small digression, coming out as a natural continuation of
the deformation quantization of BTZ spaces. That is, we will address the question of the existence of invariant deformations of the hyperbolic plane, and discuss their underlying geometry.


{\bf Plan}

We start, in the Prelude, with a brief description of the BTZ black holes and how they are constructed out of $AdS_3$ space, by performing discrete identifications. We follow the original papers \cite{BTZ,BHTZ}, to which we refer for further details.

The first part of this thesis is dedicated to emphasizing on remarkable geometrical properties of these black holes that naturally suggest a route for formulating a ``noncommutative'' version of these spaces. In the first chapter, we start by following the steps of Bieliavsky, Rooman and Spindel \cite{BRS}, and recall how Lie group/algebra techniques can be used to obtain a global dynamical description of the non-rotating massive BTZ black hole. This latter rests upon a foliation of $AdS_3$, identified with the group manifold of $\SL$, in two-dimensional twisted conjugacy classes, stable under the identifications leading to the black hole. We complement their analysis by determining the black hole horizons using group-theoretical arguments \cite{BHAdSl}. We show that both black hole singularities and horizons are closely connected to solvable subgroups of $\SL$, which completely encode the black hole structure \cite{Keio}. Finally, we use the analogy between the leaves of the foliation of the non-rotating BTZ black hole and Misner space to define a maximally extended non-rotating BTZ space-time \cite{BDRS}. Although causally badly behaved, due to the presence of closed time-like curves, this extension enjoys an interesting property: each two-dimensional leaf therein admits an action of a solvable subgroup of $\SL$, denoted $AN$. This will be important in the sequel, since this property will allow us to deform the algebra of functions on these leaves, which will, moreover, soon be recognized as D1-branes in the BTZ background. In the last part, we turn to the analysis of the generic rotating BTZ black holes, for which an adapted foliation and a global description is also possible, though quite different from the non-rotating case. Again, the $AN$ subgroup is found to act naturally on the leaves of the foliation and to be related to the black hole's inner and outer horizons \cite{BDHRS}. In Chapter 2, we deal with higher-dimensional generalizations of the BTZ construction. It was indeed realized, soon after the discovery of the BTZ black holes, that a similar construction could, in some cases, be performed in higher-dimensional $AdS$ spaces to obtain new black hole solutions. Exploiting the fact that anti-de Sitter spaces are homogeneous, and can furthermore be endowed with a symmetric space structure,  we point out that, as in the three-dimensional case, the black hole singularities are tied to solvable groups. More precisely, we show that closed orbits of solvable subgroups of the isometry group  define the location of black hole singularities, in the same sense as in the BTZ case. We also use simple symmetric space arguments that allow us to prove the existence of horizons \cite{BHAdSl}.
The third chapter is devoted to the deformation quantization of some of the spaces referred to above. After a general introduction to the context, we concentrate on \ind{strict WKB quantization}, where the deformed product is internal on some functional space, and is completely determined by a three-point kernel.  To construct such products, we will make use of so-called \emph{universal deformation formulae}\index{universal deformation formula}.
Given a manifold on which a Lie group acts, and given a deformed product with appropriate invariance properties on that Lie group, such formulae give us a recipe to construct a star product on the manifold itself. This makes clear the importance of the appearance of solvable subgroups' actions on the BTZ black holes' geometry. 
Indeed, by constructing an invariant deformed product on the $AN$-group manifold, which is a tractable problem by virtue of the group structure,
we can automatically induce a deformed product on each leaf of the BTZ foliation on which this group acts, and hence induce a deformed product on the whole BTZ space. We therefore construct a family of $AN$-invariant products on the $AN$-group manifold, and then induce them onto the whole maximally extended non-rotating massive BTZ space \cite{BDRS}. 
This construction suggested a connection with another problem, namely the existence of \ind{quantum Riemann surfaces}, where the adjective ``quantum" in the mathematical literature stands for ``noncommutative". It is based on the following two facts. First, from the \ind{uniformization theorem}, ``almost" all Riemann surfaces can be obtained as a quotient of the hyperbolic plane (see Sect.~\ref{SectHyp} for a more precise statement). Second, as a homogeneous (and symmetric) space, the hyperbolic plane is identified with the $AN$-group manifold. However, these $AN$-invariant products mentioned above are not totally adapted, because one would expect a deformation of the hyperbolic plane to be $\SL$-invariant, and not only $AN$-invariant. To solve this problem, we show that it is possible to ``twist" $AN$-invariant products so to as to ensure the $\SL$-invariance. In particular, the invariant kernel we are looking for is obtained by solving a second order hyperbolic differential equation. We make a link with a proposition by Weinstein on the geometrical interpretation of the \emph{phase} of WKB kernels in the quantization of hermitean symmetric spaces, whose hyperbolic plane is one of the simplest examples. We prove 
that any $\SL$-invariant product on the hyperbolic plane has to be of the form we obtained \cite{PlanHyp}. 

The second part of this work deals with string theoretical aspects of black holes. It starts, in Chapter \ref{ChapStrings}, with a comprehensive introduction to string theory,  
with the aim to show how exact string backgrounds are associated with two-dimensional conformal field theories (CFTs) on the string worldsheet\footnote{The \ind{worldsheet} is the surface swept out by a string during its evolution, generalizing the concept of particle's worldline}. It will also draw attention to D-branes, and to their low energy dynamics. Chapter \ref{ChapWZW} presents the main features of the Wess-Zumino-Witten (WZW) models, which allow us to describe string propagation on group manifolds. It turns out that any Lie group manifold, supplemented with an appropriate three-form field strength, can be promoted to an exact string theory background. This is the reason why the BTZ black holes, which are quotients of $AdS_3$, or equivalently of $\SL$, are exact conformal backgrounds, described by an orbifold of the $\SL$ WZW model. We discuss a class of D-brane configurations in the WZW models, called \emph{symmetric D-branes}\index{symmetric D-branes}. In the geometric regime, these are represented by (twisted) conjugacy classes of the group manifold. From this, we observe that,
 using the previous geometric analysis of the non-rotating BTZ black holes, every two-dimensional leaf of the foliation, on which we deformed the algebra of functions, is indeed a symmetric WZW D1-brane. We pursue the overview of general features of WZW models by presenting their gauged versions, and then turn to their marginal deformations. Marginal deformations allow us to reach more general and less symmetric backgrounds, and allow us to relate various models. For example, \ind{symmetric deformations} relate a WZW model to one of its gauged versions, while \ind{asymmetric deformations} relate strings moving on a group manifold to strings moving on the corresponding geometric coset, through a continuous line of deformations. These connections also apply for the D-brane configurations of the corresponding models.  We indeed observe that symmetric D1-brane configurations in BTZ spaces survive along the whole symmetric deformation line, and yield, at the endpoint, known D0-brane configurations in the two-dimensional black hole background, described by a $\SL/U(1)$ gauged WZW model.
A further point of interest concerns marginal deformations of $AdS_3$. Starting from the fact that black hole solutions can be obtained by performing identifications in $AdS_3$ along orbits of Killing vectors, we can then go on to ask what kind of solutions can be obtained by performing identifications in a \emph{marginal deformation} of $AdS_3$?
We analyze the different kinds of backgrounds that can be obtained using symmetric, asymmetric and combined deformations (double deformation). We show that identifications in asymmetric deformations do not lead to viable black hole solutions (there are naked singularities), while the double deformation  (with or without identifications) leads to a generalization of the black string solution found in \cite{Horne:1991gn}. We analyze the causal structure of such solution and compute the associated charges.

A fair proportion of this thesis consists of (hopefully pedagogical) introductions to various subjects. Chapter \ref{ChapStrings} presents some basics of string theory, that should be sufficient for our purpose, and the first part of Chapter \ref{ChapWZW} deals with some rudiments of WZW models. To be  self-contained, these should be supplemented by an introduction to conformal field theory (see \cite{CoursModave} for a lot of useful references) and another to affine Lie algebras (a good account which gets straight to the point can be found in \cite{DiFr}). The Appendices introduce basic notions of symmetric spaces as well as of symplectic and Poisson geometries. A prerequisite to Lie group theory and differential geometry is assumed, and can be found for example in \cite{Faraut} and in \cite{GeoSpindel}, where a good account about the geometry of Lie group manifolds is also given.

This thesis results from collaborations and discussions with 
P. Bieliavsky, L. Claessens, M. Herquet, D. Orlando, M. Petropoulos, M. Rooman, F. Rouvi\`ere and Ph. Spindel, and is based on the following articles, preprints and yet-to-be-finished work:
\begin{itemize}

\item {\itshape ``Global geometry of the 2+1 rotating black
hole"}, P. Bieliavsky, S. Detournay, M. Herquet, M. Rooman, Ph.
Spindel, Physics Letters B 570 (2003) 231, hep-th/0306293 \cite{BDHRS}

\item {\itshape ``Star products on extended massive non-rotating
BTZ black holes"}, P. Bieliavsky, S. Detournay, M. Rooman, Ph.
Spindel, J. High Energy Phys. 06 (2004) 031, hep-th/0403257 \cite{BDRS}

\item {\itshape ``Noncommutative locally Anti-de Sitter black
holes"}, P. Bieliavsky, S.Detournay, M. Rooman, Ph. Spindel, in
"Noncommutative geometry and Physics", edited by Y. Maeda, N. Tose, N. Miyazaki, S. Watamura
and D. Sternheimer, World Scientific, math.QA/0507157 \cite{Keio}

\item {\itshape ``Three-dimensional black holes from deformed
anti-de Sitter"}, S.Detournay, D. Orlando, M. Petropoulos, Ph.
Spindel, J. High Energy Phys. 07(2005)072, hep-th/0504231 \cite{3DBH}

\item {\itshape ``Solvable symmetric black holes in anti-de Sitter spaces"}, L. Claessens, S. Detournay, pre-print :
math.DG/0510442, accepted for publication in Journal of Geometry and Physics \cite{BHAdSl}

\item {\itshape "Non formal quantizations of symmetric surfaces"},
P. Bieliavsky, S. Detournay, F. Rouvi\`ere, Ph. Spindel, in preparation \cite{PlanHyp}

\end{itemize}

\cleardoublepage \chapter*{Prelude : the BTZ black holes}
\addcontentsline{toc}{chapter}{\numberline{}Prelude : the BTZ black holes}





We will take, as our starting point,
 the vacuum Einstein's equations in (2+1)-dimensions, in the presence of a negative cosmological constant $\Lambda=-1/l^2$: 
\begin{equation}\label{VacuumE}
 R_{\m \n} -\frac{1}{2} g_{\m\n} R = -\Lambda g_{\m\n},
\end{equation}
where $g_{\m\n}$ denotes the space-time metric, $R_{\m\n}$ and $R$ its associated Ricci tensor and Ricci scalar respectively (the gravitational constant $G$ will be set to $1/8$ throughout the text). An important property arising in (2+1)-dimensions is that the Riemann curvature tensor is \emph{completely} determined by the Ricci tensor (see e.g. \cite{Carlip1}, eq. (2.1)), as appears from a simple counting argument (since the number of independent components of the Riemann tensor - $\frac{d^2 (d^2 -1)}{12}$ - equals that of the Ricci tensor -$\frac{d(d+1)}{2}$-, when $d=3$). The full Riemann tensor is determined as $R_{\m \n \lambda \rho} = \Lambda (g_{\m \lambda} g_{\n \rho} - g_{\n \lambda} g_{\m \rho})$, and describes, by definition, a solution of constant negative curvature.

An important class of solutions to \re{VacuumE} is called \ind{anti-de Sitter spaces} ($AdS$). In $d$ dimensions, $AdS_d$ is the maximally symmetric space (i.e. it has $\frac{d(d+1)}{2}$ Killing vectors), and as a consequence has constant negative curvature. We will return to the general description of anti-de Sitter spaces in $d$ dimensions in Sect. \ref{AdSL}. For the time being, we will briefly discuss the three-dimensional case, $AdS_3$, since the BTZ black holes (which we will soon define as black hole solutions to \re{VacuumE}) will appear to be intimately tied to this space. The reason for this is already apparent: since in three dimensions, any solution to \re{VacuumE} has constant negative curvature, it has to be {\itshape locally} $AdS_3$\footpourmoi{solution a courbure cstte unique? c-a-d, si Riemann satisfait cbure cstte, alors g est totalement fixee. Semble evident}, from which it can differ only by global properties. Any solution could thus be globally obtained by gluing patches of $AdS_3$. $AdS_3$ space can be represented as the hyperboloid
\begin{equation}\label{hyperb}
  u^2 + t^2 - x^2 - y^2 = -1/\Lambda \quad,
  \end{equation}
  embedded in the four dimensional flat space with metric $ds^2 = -du^2 - dt^2 + dx^2 +
  dy^2$. 
A useful system of coordinates covering the whole of the $AdS_3$ manifold
(up to trivial polar singularities) may be introduced by setting
\beq \label{CoordGlobAdS3}
 u= l \cosh\xi\,\cos\lambda\qquad ,& \hspace{10mm} &
t= l \cosh\xi\,\sin\lambda \quad ,\nonumber\\
x=l \sinh\xi\,\cos\varphi \qquad ,& \hspace{10mm} & y=l
\sinh\xi\,\sin\varphi \quad, \eeq
in which the invariant\footpourmoi{comme round metric sur la sphere} metric on $AdS_3$ reads
\begin{equation}
\label{metricAdS3} ds^2= l^2( -\cosh^2\xi\, d\lambda^2 + d\xi^2 +
\sinh^2\xi\,d\varphi^2)\qquad,
\end{equation}
with $l^2 = -\frac{1}{\Lambda}$. Note that, as appears in \re{CoordGlobAdS3}, the time coordinate $\lambda$ of $AdS_3$ is periodic, $\lambda \in [0,2\pi[$. This is cured by "unwrapping" $\lambda$, by letting it vary in $\R$. The space we get is the universal covering of $AdS_3$, with topology $\R^3$. This is the space we will be considering in the following, which we simply refer to as $AdS_3$. The further change of radial
coordinate\footpourmoi{pourquoi pas $r \in[-\pi/2,\pi/2[$?} :
\begin{equation}
\tanh\xi=\sin r \hspace{10mm} r \in[0,\pi/2[ \qquad ,
\end{equation}
yields a conformal embedding of $AdS_3$ in the Einstein static space $E_3$ with topology $\R \times \S^2$\cite{HE}\footpourmoi{Clarifier $E_3$ versus cylindre plein + attention: representation conforme ne preserve pas la topologie...}:
\begin{equation}\label{he} ds^2=\frac{l^2}{\cos^2 r }\left( -d\lambda^2+d r
^2+\sin^2 r \,d \varphi ^2\right) \qquad .
\end{equation}
We will use this representation extensively in the following.
  
As we anticipated, a remarkable class of solutions of equations \re{VacuumE} in (2+1)-dimensions is provided by the so-called \ind{BTZ black holes}. These were discovered by Ba$\tilde{\mbox{n}}$ados, Henneaux, Teitelboim and Zanelli \cite{BTZ,BHTZ}, who found stationary and axisymmetric solutions to \re{VacuumE} in the form
\beq\label{SolBTZGen}
 ds^2 = -(N^\perp)^2 dt^2 + (N^\perp)^{-2} dr^2 + r^2 (d\phi + N^\phi dt)^2 \quad , \quad 0 \leq \phi < 2 \pi, 0\leq r < \infty,  t\in \R, 
\eeq
with 
\beq
  N^\perp =\left( -M + \frac{r^2}{l^2} + \frac{J^2}{4 r^2}\right)^{1/2} \quad , \quad N^\phi = -\frac{J}{2 r^2} \quad , \quad (|J|\leq M l).  
\eeq

This solution shares most features of the Kerr black hole solution in (3+1) dimensions: in particular, it has an inner and an outer horizon, an ergosphere, and its Penrose diagram has the overall structure of a black hole (see \cite{BHTZ, Carlip1}, also see Chap. \ref{Chap-BTZ}). Also, the BTZ solution arises naturally as the final state of gravitational collapse in (2+1)-dimensions \cite{Ross:1992ba} \footpourmoi{Carlip1,p10:pq gravite (2+1) n'a pas de limite newtonienne : voir TheseLemosBlackHoles!}(see also \cite{Cruz:1994ar,Lubo:1998ue} for the investigation of properties of 
in (2+1)-dimensional stars).
The constant parameters $M$ and $J$, which determine the asymptotic behavior of the solution, can be shown to correspond to the standard mass and angular momentum, associated with asymptotic time translations and rotations (see e.g. \cite{Carlip1} and references therein).  

The BTZ black holes differ however from the Kerr (and Schwarzschild, for $J=0$) solutions in some important respects. First, they are asymptotically anti-de Sitter, rather than asymptotically flat, which is reflected in the asymptotic region of the Penrose diagrams. Then, as may appear peculiar at first sight, \emph{there is no curvature singularity at $r=0$}. At second sight, this was of course expected, since BTZ black holes are locally $AdS_3$, as we discussed before. This led the authors of \cite{BTZ,BHTZ} to the idea that it was likely that the BTZ black holes could be seen as a \emph{quotient} of $AdS_3$. This indeed appeared to be the case: they showed that the solutions \re{SolBTZGen} could be obtained by performing discrete identifications in $AdS_3$ along orbits of well-chosen Killing vectors \footnote{For Euclidian signature, there exists a theorem by W.P. Thurston stating that any geodesically complete space of negative constant curvature is a quotient of the three-dimensional hyperbolic space, or Euclidian $AdS_3$, denoted $H_3$, by a discrete subgroup of its isometry group $O(3,1)$. For BTZ black holes, this was not obvious (see discussion [4] in \cite{BHTZ}).}. We may sketch the construction as follows. From \re{hyperb}, the isometry group of $AdS_3$, denoted by $Iso(AdS_3)$, is the group $O(2,2)$. The generators of $Iso(AdS_3)$, the Killing vectors, thus form a $so(2,2)$ Lie algebra\footnote{This is similar to the $so(3)$ Lie algebra formed by the Killing vectors of the round metric on the two-sphere $S^2$.}. Each Killing vector $\Xi$ generates a one-parameter subgroup of isometries of $AdS_3$:
\beq
 P \ra {\rm e}^{t \Xi}P  \quad , \quad P\in AdS_3.
\eeq 
These one-parameter subgroups have been first classified in \cite{BHTZ}, where it was shown that the inequivalent one-parameter subgroups\footnote{Two one-parameter subgroups $\Gamma$ and $\Gamma'$ were called equivalent when they are conjugate in $SO(2,2)=Iso(M)$. In this case, the quotients $M/\Gamma$ and $M/\Gamma'$ are isometric, the isometry being induced from that of $G$ which conjugates $\Gamma$ and $\Gamma'$ \cite{Figueroa, Figueroa-O'Farrill:2001nx}.} fall into 7 distinct categories, which were denoted by ${\bf I}_a$, ${\bf I}_b$,${\bf I}_c$,${\bf II}_a$,${\bf II}_b$,${\bf III}^+$,${\bf III}^-$. We will return to this classification in the next chapter. The different BTZ black holes can be obtained by identifying points in $AdS_3$ as
\beq
 P \sim {\rm e}^{2 \pi n \Xi}P  \quad , \quad P\in AdS_3 \quad , \quad n\in \Z,
\eeq 
with $\Xi$ chosen in the appropriate category: $\Xi \in {\bf I}_b$  for the \emph{generic} black hole ($0 \leq J < M |l|$), $\Xi \in {\bf II}_a$  for the \emph{extremal} black hole ($0 < J = M |l|$), and  $\Xi \in {\bf III}^+$  for the \emph{vacuum} black hole ($J = M =0$). Of course, regions of $AdS_3$ where $\Xi$ is time-like ($||\Xi||^2 < 0$) have to excluded from $AdS_3$ before the quotient is taken, in order to avoid closed time-like curves in the resulting space-time. The surface $||\Xi||^2=0$ appears as a singularity in the causal structure, since continuing beyond it would produce closed time-like curves. For this reason, it may be regarded as a true singularity in the quotient space. With this definition, the only incomplete geodesics are those that hit the singularity, just as in the (3+1) black hole.
Let us give the explicit form of the identifications in the generic case.
The identification vector reads in this case, in the coordinates \re{hyperb}\footpourmoi{voir BHTZ A.11 pour le signe devant $r_-$!},
 \beq \label{Killutxy}
  \Xi = \frac{r_+}{l} (y \p_t + t \p_y) + \frac{r_-}{l} (x \p_u + u \p_x),
 \eeq
 where $r_-$ and $r_+$ denote the inner and outer horizons of the metric \re{SolBTZGen}:
\beq\label{rplusmoins}
 r_\pm = l \left[ \frac{M}{2} \left( 1 \pm \sqrt{1- (\frac{J}{M l})^2} \right) \right]^{1/2} ,
 \eeq
 corresponding to the roots of $N^\perp = 0$\footpourmoi{Revoir pourquoi c'est donne par ca!}. 
After having excluded the regions of $AdS_3$ where $||\Xi||^2 < 0$, the remaining ones, which we will often call \emph{safe regions}\index{safe region}, here given by $ t^2 - y^2 > \frac{- r_-^2 l^2}{r_+^2 - r_-^2}$, may be covered by an infinite set of coordinate patches of three types: type $I$ (where $r_+^2 <||\Xi||^2 < +\infty$), type $II$ ($r_-^2 <||\Xi||^2 < r_+^2$), and type $III$ ($0 <||\Xi||^2 < r_-^2$). One may then choose three contiguous regions of type $I$, $II$ and $III$ respectively, and introduce a $(t,r,\phi)$ parametrization as follows:
\begin{itemize}
\item Region I : $r > r_+$
\beq
 t= \phantom{-}\sqrt{A(r)} \cosh \tilde{\phi}(t,\phi) \quad &,& \quad y= \phantom{-}\sqrt{A(r)} \sinh \tilde{\phi}(t,\phi) \nn \\
 x= \phantom{-}\sqrt{B(r)} \cosh \tilde{t}(t,\phi) \quad &,& \quad u= \phantom{-}\sqrt{B(r)} \sinh \tilde{t}(t,\phi)
 \eeq
 
 \item Region II : $r_- < r < r_+$
\beq
 t= \phantom{-}\sqrt{A(r)}\phantom{-} \cosh \tilde{\phi}(t,\phi) \quad &,& \quad y= \phantom{-}\sqrt{A(r)}\phantom{-} \sinh \tilde{\phi}(t,\phi) \nn \\
 x= -\sqrt{-B(r)} \sinh \tilde{t}(t,\phi) \quad &,& \quad u= -\sqrt{-B(r)} \cosh \tilde{t}(t,\phi)
 \eeq
 
 \item Region III : $0 < r < r_-$
\beq
 t= \phantom{-}\sqrt{-A(r)} \sinh \tilde{\phi}(t,\phi) \quad &,& \quad y= \phantom{-}\sqrt{-A(r)} \cosh \tilde{\phi}(t,\phi) \nn \\
 x= -\sqrt{-B(r)} \sinh \tilde{t}(t,\phi) \quad &,& \quad u= -\sqrt{-B(r)} \cosh \tilde{t}(t,\phi)
 \eeq
  \end{itemize}
 with 
 \beq
 A(r)= l^2 \left( \frac{r^2 - r_-^2}{r_+^2 -r_-^2} \right) \quad &,& \quad B(r) = l^2 \left( \frac{r^2 - r_+^2}{r_+^2 -r_-^2} \right) \nn \\
 \tilde{t} = (1/l) (r_+ t/l - r_-) \phi \quad  &,& \quad \tilde{\phi} = (1/l) (-r_- t/l - r_+ \phi).
 \eeq

 These regions correspond to the three different regions of the Penrose diagram: $0 < r < r_-$, $r_- < r < r_+$ and $r_+ < r < + \infty$. It can be checked that the $AdS_3$ metric in these coordinates reads precisely  as \re{SolBTZGen}, \emph{but} with $-\infty < \phi < +\infty$. Since the Killing vector \re{Killutxy}, expressed in the coordinates $(t,r,\phi)$ is nothing other than 
 \beq
  \Xi = \p_\phi \quad ,
  \eeq
  we conclude that we get the black hole background exactly, since the identifications amount to
  \beq
   \phi \sim \phi + 2 n \pi  \quad , \quad n \in \Z.
  \eeq

   \cleardoublepage \part{Geometry and deformations}
 
 \cleardoublepage \chapter{Lie group description of BTZ black holes}\label{Chap-BTZ}

In this chapter, we will see why and how one can use Lie-algebraic techniques to analyze the structure of some BTZ black holes. We will focus on the generic BTZ black holes, and show that it is possible to construct global coordinate systems on safe regions in $AdS_3$, which are furthermore well-adapted to the identifications leading to them. These will allow us to get a nice picture of the black hole geometry \cite{BRS,BDHRS}. We will be led to use somewhat different constructions, depending on whether we deal with the rotating or non-rotating black hole. We will then use group-theoretical arguments to determine and characterize the horizons and singularities \cite{Keio}. We will also define the notion of maximally extended non-rotating BTZ black holes and emphasize on interesting properties of this space \cite{BDRS}. An outcome of special interest for our future purposes will be to show that the black hole structure is closely tied to solvable subgroups of $\SL$.



\section{$AdS_3$ as a Lie group manifold}
\subsection{$AdS_3$ and $\SL$}

A remarkable property of $AdS_3$ space is that it can be identified with a Lie group manifold. Indeed, consider the simple Lie group $\SL$ of two by two real matrices with unit determinant, parametrized as 
\begin{equation}\label{AdSl}
 \SL = \{g = \left(\begin{array}{cc} u+x & y+t\\ y-t & u-x
         \end{array}
         \right)| x,y,u,t\in\mathbb{R},\ \det g =1 \} \qquad .
         \end{equation}
It follows from \re{hyperb} (where in the following, we set $\Lambda = -1$) that points in $AdS_3$ are in one-to-one correspondence with elements of $\SL$ (which has also a manifold structure, by definition of a Lie group). $AdS_3$ and $\SL$ can further be identified as pseudo-Riemannian spaces, when we endow $G = \SL$ with its bi-invariant \ind{Killing metric} $\beta_g : T_g G \times T_g G \ra \R$. It is defined at the identity $e$ through the Killing form $B$ of its Lie algebra
$\mathfrak{g}=sl(2,\R)$, identified with the tangent space at the identity: 
\beq \label{MetriqueKill}
 \beta_e (X,Y) = B(X,Y) \equiv \frac{1}{8}\mbox{Tr}(ad_X . ad_Y)= \frac{1}{2}\mbox{Tr}(X.Y) \quad, X,Y \in
 sl(2,\R) \quad ,
\eeq

where the last equality holds because we are dealing with a Lie
algebra of traceless matrices.
The Lie algebra
 \begin{equation}
         \label{matsl} sl(2,\R)=\{\left(\begin{array}{cc} z^H & z^E \\ z^F &
         -z^H
         \end{array}\right):= z^H\h+z^E\e+z^F\f\} \qquad ,
         \end{equation} is expressed in terms of the generators $\{\h , \e, \f
         \}$ satisfying the commutation relations:
         \begin{equation}
         \label{comrel} [\h ,\e]=2\e \quad , \quad [\h,\f]=-2\f \quad ,
         \quad [\e,\f]=\h \quad .
         \end{equation}

Parametrizing $g \in G$ as (see App.\ref{antids})
\begin{equation}
g = \mathrm{e}^{ {\frac{\lambda +\varphi}{2}} \t}
\mathrm{e}^{\xi {\bf S}} \mathrm{e}^{-{\frac{\lambda-\varphi}{2}}
\t} \quad , \quad \t = \e - \f, \;   {\bf S} = \e + \f,
\end{equation}
which provides good global coordinates for $\SL$ when
$\lambda \in [0,2\pi[$, $\xi\in [0,\infty[$, and $\varphi\in
[0,2\pi[$, we obtain that the \ind{bi-invariant metric}
\footnote{Let $G$ be a Lie group manifold and $\mathfrak{g}$ its Lie algebra, with generators $\{T_a\}$, $a=1,\cdots,dim G$, and Killing form $B$. We denote by $\{\theta^a\}$ and $\{\sigma^a\}$, $a=1,\cdots,dim G$ the left and right-invariant one-forms on $G$, dual to the left and right-invariant vector fields, see \re{InvVF},\re{InvVF2}. The left-invariant one-forms may be used to define a left-invariant metric $g_L$ on $G$ as $g_L = g_{ab} \theta^a \otimes \theta^b$. It turns out that that for $g_{ab}= B(T_a,T_b) \equiv \beta_{ab}$, this metric is also right-invariant (see \cite{GeoSpindel}). This stems from the fact that $g_L$ will also be right-invariant iff $R^*_h g_L = g_L$, or infinitesimally ${\cal L}_{l_a} g_L = 0$, where $R_h$ denotes the right-translations on $G$, which are generated by the left-invariant vector fields $\{l_a\}$. This condition can be brought to $g_{ij} C^i_{kc} + g_{kl} C^l_{jc} = 0$, which can be seen to be trivially satisfied by $\beta$, using the property of the Killing form $B(ad_X Y,Z) = B(X, ad_Y Z)$, see \cite{Barut,Faraut}. If $G$ is a simple group, it admits only one (up to a scale factor) \ind{bi-invariant metric}, called the \ind{Killing metric}. From \re{MetriqueKill}, its components can be determined in terms of the structure constants of $\mathfrak{g}$, defined by $[T_a,T_b] = ad_{T_a} T_b =  C_{ab}^c T_c$. The matrix $(C_a)^c_b$ is the representative of $T_a$ in the adjoint representation of $\mathfrak{g}$, the indices $c$ and $b$ referring to lines and columns respectively. Thus $B(T_a, T_b) =\frac{1}{8} \mbox{Tr} (ad_{T_a} ad_{T_b}) = \frac{1}{8} \mbox{Tr} ((C_a)^c_d (C_b)^d_e)= \frac{1}{8} \mbox{Tr} ( (C_a.C_b)^d_e ) = \frac{1}{8}  (C_a.C_b)^e_e =  \frac{1}{8} C^e_{ad} C^d_{be} \equiv \beta_{ab}$. Using the fact that $g^{-1} dg = \theta^a T_a$, the bi-invariant Killing metric can then easily be determined by $\beta = \mbox{Tr} (g^{-1} dg \otimes g^{-1} dg)$.}
precisely coincides with \re{metricAdS3}. Again, the time coordinate $\lambda$ has to be unwrapped, which is tantamount to taking the universal covering of $\SL$, $\widetilde{\SL}$. In the sequel, we will always assume we are dealing with $\widetilde{\SL}$, which we denote $\SL$ to lighten the notation.


For later purpose, we also define the one-parameter subgroups of
$\SL$ :
 \begin{equation} A=\exp(\R\,\h) ,\quad N=\exp(\R\,\e) ,\quad \bar{N}=\exp(\R\,\f)
         ,\quad K=\exp(\R\,\t)\  \qquad ,
\end{equation}
which are the building blocks of its Iwasawa decomposition (see
\cite{Barut,Helgason}, and also Sect. \ref{SectD2}):
\begin{equation} \label{Iwasawa}
 K \times A \times N \rightarrow \SL : (k,a,n)
\rightarrow k.a.n \,\, \mbox{or} \, \, a.n.k \, ,
  \,k\in K, \, a\in A, \, n\in N \quad,
 \end{equation}
the mapping being an analytic diffeomorphism\footnote{For this particular case, a direct computation shows that this application is indeed a $C^\infty$ bijective map.} of the product
manifold $K \times A \times N $ onto $\SL$.
We also mention important subgroups of $\SL$: 
\beq\label{Solvables}
 AN = \exp(\R\,\h).\exp(\R\,\e) \quad \mbox{and} \quad A\bar{N} = \exp(\R\,\h).\exp(\R\,\f),
\eeq
which are \emph{solvable subgroups}\index{solvable subgroup}, sometimes called \ind{minimal parabolic subgroups} or \ind{Iwasawa subgroups} (see also Sect. \ref{SectD2}). They will play an important role in the structure of the BTZ black holes.


\subsection{Isometry group of $AdS_3$}

We have just seen that $AdS_3$ is identified with the group $\SL$ endowed with its Killing metric. Let us now return to its isometry group, $Iso(AdS_3) = O(2,2)$ (see Prelude). Its elements can be visualized as 4 by 4 matrices acting on vectors $(u\; t \; x \; y)^T \in \R^{2,2}$.  This group is constituted by four connected parts, whose identity component is $SO(2,2)$. The other components can be reached using a parity (${\cal P} : x\rightarrow -x$) and/or time reversal (${\cal T}:t\rightarrow -t$) transformation.
From the bi-invariance of the Killing metric on $AdS_3$, $O(2,2)$ is locally isomorphic to
 $\SL \times \SL$ (they share the same Lie algebra) , the action being given by
 \beq\label{actisom}
  (\SL \times  \SL) \times  \SL \rightarrow  \SL :
  ((g_L,g_R),z) \rightarrow g_L \, z \, g_R^{-1} \quad.
  \eeq
This action corresponds to the component of $Iso(AdS_3)$
connected to the identity transformation (because $\SL$ is
connected). The full action of $Iso(AdS_3)$ (including the
components not connected to the identity) can be obtained by
considering the parity and time reversal transformations
\begin{eqnarray}
{\cal P}(g)&=&\left
(\begin{array}{cc}u-x&y+t\\y-t&u+x\end{array}\right)\quad,\label{revx}\\
    {\cal T}(g)&=&\left
(\begin{array}{cc}u+x&y-t\\y+t&u-x\end{array}\right)\quad.\label{revt}
\end{eqnarray}

There exists a Lie algebra isomorphism between $sl(2,\R) \times
sl(2,\R)$ and $iso(AdS_3) \equiv Lie(Iso(AdS_3)) $. It is given by
 \beq\label{Iso}
 \Phi : sl(2,\R) \times sl(2,\R) \rightarrow iso(AdS_3) : (X,Y)
 \rightarrow  \overline{X} - \underline{Y} \quad ,
\eeq

where $\overline{X}$ (resp. $\underline{Y}$) denotes the
right-invariant (resp. left-invariant) vector field on $\SL$
associated to the element $X$ (resp. $Y$) of its Lie algebra, that
is \beq \label{InvVF}
 \overline{X}_g &=& \frac{d}{dt}_{|0} L_{\exp t X} \, g =
 \frac{d}{dt}_{|0} \exp (t X) \, g = X g \quad, \\ \label{InvVF2}
\underline{Y}_g &=& \frac{d}{dt}_{|0} R_{\exp t X} \, g =
 \frac{d}{dt}_{|0}  g\, \exp (t Y) = g Y \quad,
\eeq where $L_g$ and $R_g$ denote the left and right translations
on the group manifold. The two last equalities hold because $\SL$
is a matrix group. Sometimes the notations $\overline{X}_g = (r_X)_g$ and $\underline{Y}_g = (l_Y)_g$ are also used.

Thus, to the generator $(X,Y) \in sl(2,\R)\times sl(2,\R)$, we
associate the one-parameter subgroup of (the connected part of)
$Iso(AdS_3)$
\begin{equation}
 \Psi_t(g)= \exp(tX) \, g \, \exp(-tY) \quad,\quad g\in \SL, t\in
 \R \quad.
 \end{equation}

As a consequence, the classification
 of the one-parameter subgroups of $SO(2,2)$ of \cite{BHTZ} now amounts to the classification, up to conjugation, of the one-parameter
subgroups of $\SL \times \SL$. This has been achieved in
\cite{BRS}. Two subgroups $(G_L,G_R)$ and $(G'_L,G'_R)$  are
said to be conjugated if
\begin{enumerate}
{\item[(i)]
they can be related by combinations of
transformations $\cal P$ and $\cal T$;}

{\item[(ii)] there exist $g_L$ and $g_R$ $\in \SL$ such that
$G_i = g_i \, G'_i\, g_i^{-1}$, for $i=L,R$.}
\end{enumerate}

Note that (ii) corresponds to a conjugation by an
element of $SO(2,2)$, while (i) corresponds to conjugation by an
element not connected to the identity. Let us start with the equivalence relation (ii).
Two subgroups $(\exp tX,\exp tY)$ and $(\exp tX',\exp tY')$ are said conjugated by an element of $SO(2,2)$
 if there exist $g_L, g_R \in \SL$ such that
 \beq
   && g_L \exp t X' g_L^{-1} = \exp
t X \, , \quad g_R \exp t
Y' g_R^{-1} = \exp t Y \\
  &\Leftrightarrow& Ad(g_L) \exp t X' = \exp t X \, , \quad Ad(g_R) \exp t Y' = \exp t
 Y \\
&\Leftrightarrow& \exp t Ad(g_L)X' = \exp t X \, , \quad \exp t
Ad(g_R)Y' = \exp t Y \\
&\Leftrightarrow& X=Ad(g_L)X' \, , \quad Ad(g_R)Y' = Y\quad.
\eeq
We used the same notation for the adjoint action of $G$ on $G$ and on $\mathfrak{g}$.
Thus, $(X,Y) \sim (X',Y')$ implies that $X$ and $X'$ belong to the
same adjoint orbit in $\mathfrak{g}$ (and so do $Y$ and $Y'$). These orbits in $\ls \simeq \R^{1,2}$ are of six different types: $Ad(G)(a \h)$ is a one-sheet hyperboloid, $Ad(G)(\pm b^2 \t)$ are the upper or lower sheet of a two-sheet hyperboloid, $Ad(G)(\e)$ is a future-directed cone, $Ad(G)(\f)$ is a past-directed cone and $Ad(G)(0)$ is the origin of $\ls$. The constants $a$ and $b$ are related to the radii of the orbit. For example, writing $Ad(g)(a H) = z^H \h + z^S \s + z^T \t$, with $g\in \SL$, we find that $(z^H)^2 + (z^S)^2 - (z^T)^2 = a^2$, which is represented in $\ls$ as a one-sheet hyperboloid of radius $a$.

The Killing vectors of
$AdS_3$ (i.e. the generators of the one-parameter subgroups of
$\SL \times \SL$) can consequently all be brought into the following form
 \beq
 (X, Y) \quad , \quad X,Y = \{a \h, \pm b^2 \t,\e,\f\} \quad , \quad a,b \in \R_0.
 \eeq

The equivalence under the ${\cal P}$ and ${\cal T}$
transformations further reduces the number inequivalent
one-parameter subgroups. For example, one may check that ${\cal T}\left( {\rm e}^{t \h} {\cal T}(g)\right) = g {\rm e}^{t \h}$ and ${\cal T}\left({\cal T}(g) {\rm e}^{t \e}\right) = {\rm e}^{t \f} g$, for all $g\in \SL$, expressing that the subgroups $({\rm e}^{t \h},{\rm e}^{t \e})$ and $({\rm e}^{t \f},{\rm e}^{t \h})$ are conjugated under a ${\cal T}$ transformation.

Proceeding this way, one arrives at the complete classification\footnote{We correct by the way a misprint in \cite{BRS}, where the subgroups $III^+$ and $III^-$ had been switched.} of the inequivalent one-parameter subgroups of $\SL \times \SL$ \cite{BRS} :
\begin{eqnarray} \label{Class}
I_a&&(a \t\, , \,   b \h)\sim (b \h\, , \,   a \t)\qquad
\; (a > 0,\ b \geq 0)\qquad ,\nonumber\\
I_b&&(a\h\,,\, b\h)\sim ( b \h\, , \,  a \h)\qquad
(a > 0,\ b\geq 0) \qquad ,\nonumber\\
I_c&&(a \t\, , \,  b \t)\sim (b \t\, , \, a \t)\qquad \
(a>0, b\neq 0)\; \qquad ,\nonumber\\
II_a&&(a \h\, , \,\e)\sim (\e\, , \, a \h)\sim (\f\, , \,
a \h)\sim (a \h,\f)\  (a\ge 0) \ , \nonumber\\
II_b&&(a \t\, , \,\e)\sim (\e\, , \, a \t)\sim (\f\, , \,
-a \t)\sim
(-a \t\, , \,\f)\  (a\ge 0) \ , \nonumber\\
III^+&&(\e\, , \,\e)\sim(\f\, , \,\f)\qquad,\nonumber\\
III^-&&(\e\, , \,-\e)\sim(\f\, , \,-\f)\qquad,
\end{eqnarray}
where the 7 different subgroups are named according to the original classification of \cite{BHTZ}.

From equations \re{Killutxy} and \re{rplusmoins}, we know that the norm of the chosen identification Killing vector $\Xi$ will be related to the mass $M$ and angular momentum $J$ of the black hole solution. In the present language, we have $\Xi = (X_L,X_R)$, and the relation can be translated into \cite{BRS}
\beq
M(\Xi)&=& \frac 1 2 \left ( \|X_L\|^2+\|X_R\|^2 \right )\qquad,\label{mass}\\
J(\Xi)&=&\frac 1 2  \left ( \|X_L\|^2-\|X_R\|^2 \right
)\qquad,\label{angm} \eeq
 where $\|X\|^2=B (X,X)$\footpourmoi{Lien precis entre cette definition et celle de BHTZ?}.



 

\section{Non-rotating massive BTZ black hole}\label{NRBTZ}

\begin{figure}[ht]
\begin{center}
\includegraphics*[scale=0.65]{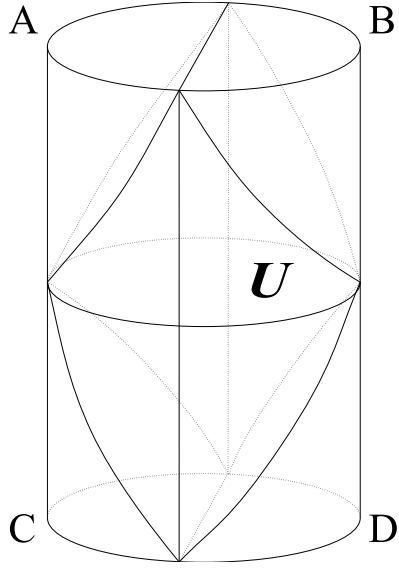}

  \caption{A finite-time section of the conformal representation of $AdS_3$. In this figure, the time $\lambda$ flows vertically, while the coordinates $r$ and $\varphi$ can be seen as radial and angular coordinates respectively (see Prelude). The light-like surfaces correspond to the singularities ${\cal S}$, which demarcate a connected, diamond-shaped, safe region ${\cal U}$.\label{Cylindre1}}
\end{center}
\end{figure}

From \re{mass} and \re{angm}, as well as from the Prelude and the classification \re{Class}, the identification vector in the non-rotating massive case is
\beq \label{KillNonRot}
 \Xi = a(\h,\h) \quad ,
\eeq
where the constant $a$ is related to the mass of the black hole through $a = \sqrt{M}$. In the coordinates \re{AdSl}, it is expressed as 
\beq \label{KillNonRot2}
  \Xi = a(y \p_t + t \p_y) .
\eeq
see e.g. \re{Killutxy} and \re{rplusmoins}.
 We define the \ind{BHTZ subgroup} 
as the one-parameter subgroup of $\SL \times \SL$ generated by the
Killing vector \re{KillNonRot}:
  \beq \label{BHTZ}
 \psi_t (z) = \exp(t a \h) \, z \, \exp(-t a \h) \, , \quad z\in
 \SL.
 \eeq
A non-rotating massive BTZ black hole is then obtained as the
quotient $\SL / \psi_\Z$ by an isometric action of $\Z$, that is,
by performing the following identifications:\footpourmoi{Je ne suis pas sûr des facteurs de $\pi$! Mais coincide avec StarP et BRS.}
\beq \label{Identif}
 z \sim \exp(n a \pi \h) \, z \, \exp(-n a \pi \h) \quad,\quad n\in \Z \quad.
\eeq
As we already stressed, in order to avoid closed time-like curves
in the quotient space, the regions of $AdS_3$ where the orbits of
the BHTZ action (\ref{BHTZ}) are time-like, or equivalently where
$\beta_z (\Xi,\Xi) < 0$, must be excluded. We will henceforth
consider a \ind{safe region}, i.e. an open and connected domain ${\cal U}\subseteq AdS_3$
where $\beta_z (\Xi,\Xi) > 0$, $\forall z\in {\cal U}$ (see Fig.~\ref{Cylindre1}).
 The black hole's singularities ${\cal S}$ will be defined as the
 surfaces where the identifications becomes light-like :
 \beq
 {\cal S} =\{ z \in AdS_3 \,\, | \,\,  \beta_z(\Xi,\Xi) = 0\} \quad.
\eeq

\subsection{Twisted Iwasawa decomposition and global description}\label{Sect:TwistedIwasawa} 
One would now like to find a parameterization of the region ${\cal U}$ in which the BTZ identifications are easy to implement. We closely follow \cite{BRS}. To this end, we will first find a foliation of $AdS_3$ whose two-dimensional leaves are stable under the BHTZ action, and are all isomorphic. Thereby, we will be brought to a two-dimensional problem, which will be easier to handle, and we could restrict our attention to a single leaf (say, the one through the identity).

Let us first introduce an external automorphism of $\ls$, which can be chosen
as
\begin{equation}
\label{defsig} \sigma(\h)=\h,\ \sigma(\e)= -\e ,\ \sigma(\f) = -\f
\qquad .
\end{equation}
Viewing $\SL$ as a subgroup of ${GL(2,\mathbb{R})}$, one may
express the automorphism $\sigma$ as:
\begin{equation}
\sigma=Ad(\h) \qquad \mbox{where} \qquad \h=\left(
\begin{array}{cc}
1 & 0\\
0 & -1
\end{array}
\right)
  \in GL(2,\mathbb{R}) \qquad .
\end{equation}
The corresponding external automorphism of $\SL$ (which we again
denote by $\sigma$) is $\sigma (g) = \h \,g \h$ (note that $\h =
\h^{-1}$). We then consider the following \ind{twisted action} of $\SL \subset SO(2,2)$ on itself :
 \beq \label{TwistedAction}
\tau : \SL \times \SL \rightarrow \SL: (g,z) \rightarrow \tau_g(z) = g \, z\,
 \sigma(g^{-1}) \quad.
\eeq The fundamental vector field associated to this action is
given by : \beq X^*_z =  \frac{d}{dt}_{|t=0} \tau_{\exp(-tX)}(z) =
-X z + z \sigma(X)
 \quad,
 \eeq
 by noting that $\sigma(\exp (tX))= \exp (t \, \sigma(X))$, with
 a slight abuse of notation (we use the same notation for the Lie group and Lie algebra automorphisms).
We note that the BHTZ action $\Z \times \SL$ ( eq.(\ref{BHTZ}))
can equivalently be written as \beq \label{BHTZTwisted}
 \psi_t (z) = \tau_{\exp (a\,t\,\h)} (z) \quad .
 \eeq
Notice that this observation is true \emph{only} in the non-rotating case (see \re{angm}). This is what leads us to consider the rotating case separately.
The orbit of a point $z$ under this action is 
\beq \label{orbit}
 {\cal O}_z =  \{\, \tau_g (z) \quad, \quad
 g\in\SL\} \quad
 \eeq
and is usually referred to as a \ind{twisted conjugacy class}. This particular structure will become important in Chapter 6, especially in Sect. \ref{DBrBTZ}, where we will discuss D-branes in this background.
 Using the matrix representation of (\ref{AdSl}), we find that
 ${\cal O}_z$ is constituted by the elements $g\in\SL$ such that
 $x(g) = x(z)=cst$. We see from (\ref{hyperb}) that the orbits
 are two-dimensional one-sheet hyperboloids in flat
 four-dimensional space $\R^{2,2}$. We will not use this representation,
 but adopt a more convenient one.

\begin{figure}[ht]
\begin{center}
\includegraphics*[scale=0.6]{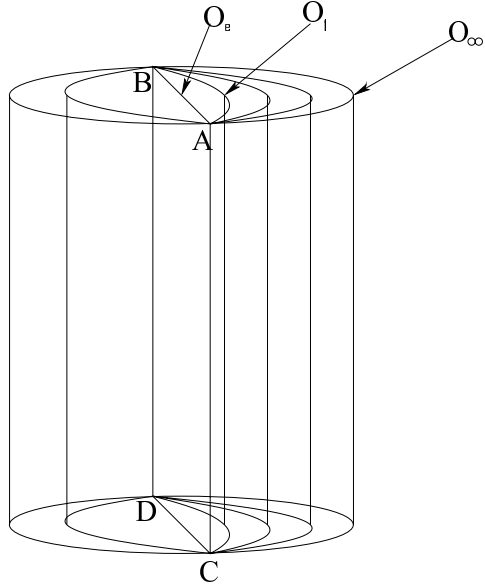}
  \caption{Foliation of $AdS_3$ in twisted conjugacy classes, as expressed by \re{phibis}. As a $G$-homogeneous space, each leaf of the foliation is isomorphic to $G/A = AdS_2$.\label{foliationCorfu2}}
\end{center}
\end{figure}

 The global description of the black hole relies on the following
 proposition. Consider the application :
 \beq \label{twistedI}
 \phi : K \times A \times N \rightarrow \SL : (k,a,n) \rightarrow
 \phi(k,a,n) = \tau_{kn}(a) \quad.
 \eeq
It may be checked that this application is a global diffeomorphism\footpourmoi{Preuve: $\phi_*$ isomorphisme local, donc $\phi$ isomorphisme. On verifie la surjectivite et l'image inverse du neutre est $(e,e,e)$.PRECISEMENT!!}, which we refer to as \ind{twisted Iwasawa decomposition}.
Furthermore, we observe, by letting $g=k n a' \in \SL$ that
 \beq
  \tau_g (a) = \phi(k,a,n) \quad \forall a \in A.
   \eeq
  Therefore, the orbit of $a \in A$ under the twisted bi-action is
given by 
\beq
 {\cal O}_a = \phi(a,N,K) \cong NK = G/A \quad ,
\eeq
where the last equality holds because of the Iwasawa decomposition
of $G=\SL$ (\ref{Iwasawa}). The application can be rewritten as
\beq \label{phibis}
 \phi : A \times G/A \rightarrow G :
(a,[g])\rightarrow g \, a \, \sigma(g^{-1}) \quad ,
 \eeq

where $[g]  = \{ gA,\, g \in G\}\in G/A.$ This is exactly the
decomposition we needed to reduce the geometrical description of
$AdS_3$ and the black hole from three to two dimensions. Indeed,
it provides us with a foliation (a trivial fibration) of $AdS_3$
over $A \simeq \R$ whose leaves are two-dimensional orbits of the
action (\ref{TwistedAction}). This foliation is represented in Fig.~\ref{foliationCorfu2}. With $a \in A$ parametrized as $a = {\rm e}^{\rho \h}$, $\rho \in \R$, a leaf will indistinctly be denoted by ${\cal O}_a$ or ${\cal O}_\rho$. 

Each of the leaves is stable under
the BHTZ subgroup, that is, $ z \in {\cal O}_a \Rightarrow \psi_t
(z) \in {\cal O}_a ,\, a\in A\,,z\in \SL$. Moreover, the BHTZ
action is fiberwise\footnote{i.e., we may consider the action of
the BHTZ subgroup "leaf by leaf".} , which means that

\beq \label{fiberwise}
 \tau_h (\phi(a,[g])) = \phi(a,h.[g]) = \phi(a,[hg]) \quad.
 \eeq

This, with the fact that the orbits are all diffeomorphic,
allows us to restrict ourselves to the study of the leaf through
the identity and focus on the space $G/A$\footnote{The stabilizer
of $e$ for the twisted bi-action is $Stab_G(e)=A$, $G$ acting on
${\cal O}_e$. Therefore, ${\cal O}_e$ is homeomorphic to $G/A$.
This is another way of seeing the orbits as homogeneous spaces.
Last, to go to a general orbit, we use the fact that all the
orbits are diffeomorphic.}. This space can be realized as the
adjoint orbit of $\h$ in $\ls$. Indeed, consider the application
\beq
 \Upsilon : G/A \rightarrow Ad(G)\h : [g] \rightarrow \Upsilon([g]) =
 Ad(g)\h\quad.
\eeq
Because $\Upsilon([e])=\Upsilon([-e])=\h$, $G/A$ is a $\Z_2$
covering of $Ad(G)\h$. It is well defined because $Ad(a)\h = \h \,
\forall a\in A$.

With $A\cong\R$ and using the application $\Upsilon$, $\phi$ can
further be rewritten as
\beq \label{diffeoGl2} \phi : \R \times Ad(G)\h \rightarrow G :
(\rho,Ad(kn)\h) \rightarrow \phi(\rho,Ad(kn)\h) =
\tau_{kn}(\exp(\rho\h)) \quad. \eeq

 Restricting $\phi$ to the leaf at identity ${\cal O}_e$ (for
which $\rho=0$) , we define the application
\beq
 \label{iota} \iota : Ad(G)\h \rightarrow {\cal O}_e : Ad(kn)\h
\rightarrow \tau_{kn} (e) \quad,
 \eeq
  and consider the adjoint
orbit of $\h$, which corresponds to a one-sheet hyperboloid in
$\ls$.

\begin{figure}[ht]
\begin{center}
\includegraphics*[scale=0.5]{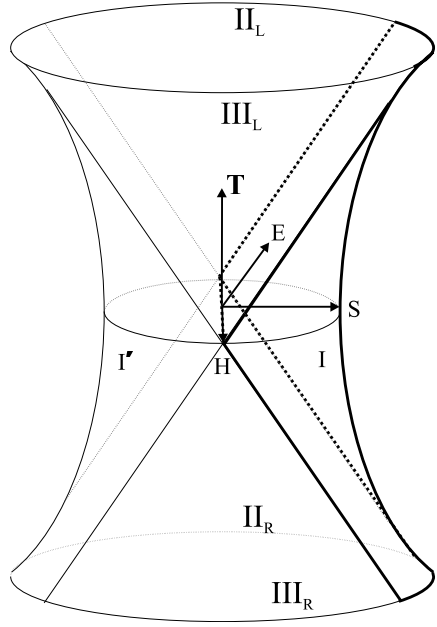}
  \caption{ An adjoint orbit of $\h$ under $\SL$,
identified to a leaf of the foliation. The regions $I$ and $I'$ correspond
to the regions where the BHTZ orbits are space-like. They correspond to the intersection of the safe region ${\cal U}$ in $AdS_3$ with a leaf of the foliation, see Figs.\ref{Cylindre1} and \ref{foliationCorfu2}. Also represented are the regions $II_L$, $II_R$, $III_L$ and $III_R$ where the BHTZ orbits are time-like.\label{OrbiteThese}}
\end{center}
\end{figure}

It is useful to visualize how the BHTZ subgroup \re{BHTZ} acts within this picture. Let us consider the orbit of a point $X=Ad(kn)\h \in Ad(G)\h$, 
 corresponding
to a point $x=\tau_{kn}(e) \in {\cal O}_e$,
 under the BHTZ
subgroup.
The orbit of $x$ under the BHTZ subgroup is \beq \psi_{n}(x) =
\tau_{\exp(n \h a)}(x) = \tau_{\exp(n \h a)kn}(e) \quad, \eeq
corresponding to the curve $Ad(\exp (n \h a)) X$ on $Ad(G)\h$. One
then checks,
 using the $Ad$-invariance of the Killing form, that
\beq
 \beta_e (\h , Ad(\exp(n \h a)X - X) = 0 \quad.
  \eeq
The orbit of $X$ is thus obtained as the intersection of the orbit with the plane perpendicular to the
$\h$-axis, passing through $X$. The planar sections given by the intersections of the orbit with the planes perpendicular to $\h$ passing through $\h$ and $-\h$ divide the orbit into six connected components, labelled by $I$, $I'$, $II_L$, $II_R$, $III_L$ and $III_R$ (see Fig.~\ref{OrbiteThese}). Using \re{phibis}, \re{iota} and the fact that $||\Xi||^2 = t^2-y^2$ (in terms of the parameterization \re{AdSl}), we may identify therein the regions where the orbits of the BHTZ subgroup are space-like, i.e. we may determine the intersection of a safe region with the orbit. 
 They are given by 
 \beq \label{BonDomaine2}
 \{ X = x^{\h} \h + x^{\e} \e + x^{\f} \f \in Ad(G)\h \quad  | \quad -1< x^{\h} < 1
 \} \quad.
 \eeq
These safe regions are represented in Fig.~\ref{OrbiteThese}, where they are labelled by $I$ and $I'$. In the following, we will restrict on a single connected safe region, say $I$.




Now, it is not difficult to check that this safe region can be parameterized as\footnote{Note that a
parametrization of the whole hyperboloid would be given by
 $Ad(\exp(\beta/2 \t)\exp(\gamma/2
\h)) \s $, leading to a global metric on $AdS_3$.}
\beq
 X = Ad\left(\exp(\frac{\theta}{2}\h) \, \exp(-\frac{\tau}{2}
 \t)\right)\, \h \quad , \, 0<\tau<\pi \, ,\,  -\infty<\theta<+\infty
 .
\eeq
In this parameterization, the orbits of the BHTZ action are the
 $\tau=cst$ lines. Using (\ref{iota}), any point $x\in {\cal O}_e$
belonging to ${\cal U}$ can be parameterized by the two
coordinates $(\theta,\tau)$ via
\beq \label{CoordOe}
 z(\theta,\tau) = \tau_{\exp(\frac{\theta}{2})\h \,
 \exp(-\frac{\tau}{2})
 \t} (e)
 \eeq

 Note that $e$ and $-e$ belong to the boundary of ${\cal U}$.
 Finally, we obtain a global (up to polar singularities) parameterization of points in ${\cal
 U}$ with the help of (\ref{phibis}), as
\beq \label{CoordGlob} z(\rho,\theta,\tau) =
\tau_{\exp(\frac{\theta}{2}\h) \,
 \exp(-\frac{\tau}{2}
 \t)}(\exp(\rho\h)) \quad ,
 \eeq
 or explicitly
\begin{equation}
  \label{gReg2}
  z(\rho,\theta_,\tau) =
  \begin{pmatrix}
    \cos\tau \cosh\rho + \sinh \rho &   e^\theta \cosh \rho \sin \tau \\
    -e^{-\theta} \cosh \rho \sin \tau &  \cos\tau \cosh\rho - \sinh \rho
  \end{pmatrix}
 \quad.
\end{equation}
The action of the BHTZ subgroup reads in these coordinates
\beq (\tau , \rho, \theta) \rightarrow (\tau, \rho, \theta + 2 n \pi a)
\quad .
 \eeq

Eq. (\ref{gReg2}) allows to derive a global expression for the
metric of the black hole :
\beq \label{Metric}
 ds^2= d\rho^2 + \cosh^2\rho(-d\tau^2 +
\sin^2\tau d\theta^2) \quad, \quad -\infty < \rho < +\infty,\,
0<\tau<\pi,\, 0<\theta<2 \pi a \quad.
 \eeq
 \footpourmoi{la periodicite de $\theta$ n'est pas $2\pi$?}



\begin{figure}[ht]
\begin{center}
\includegraphics*[scale=0.5]{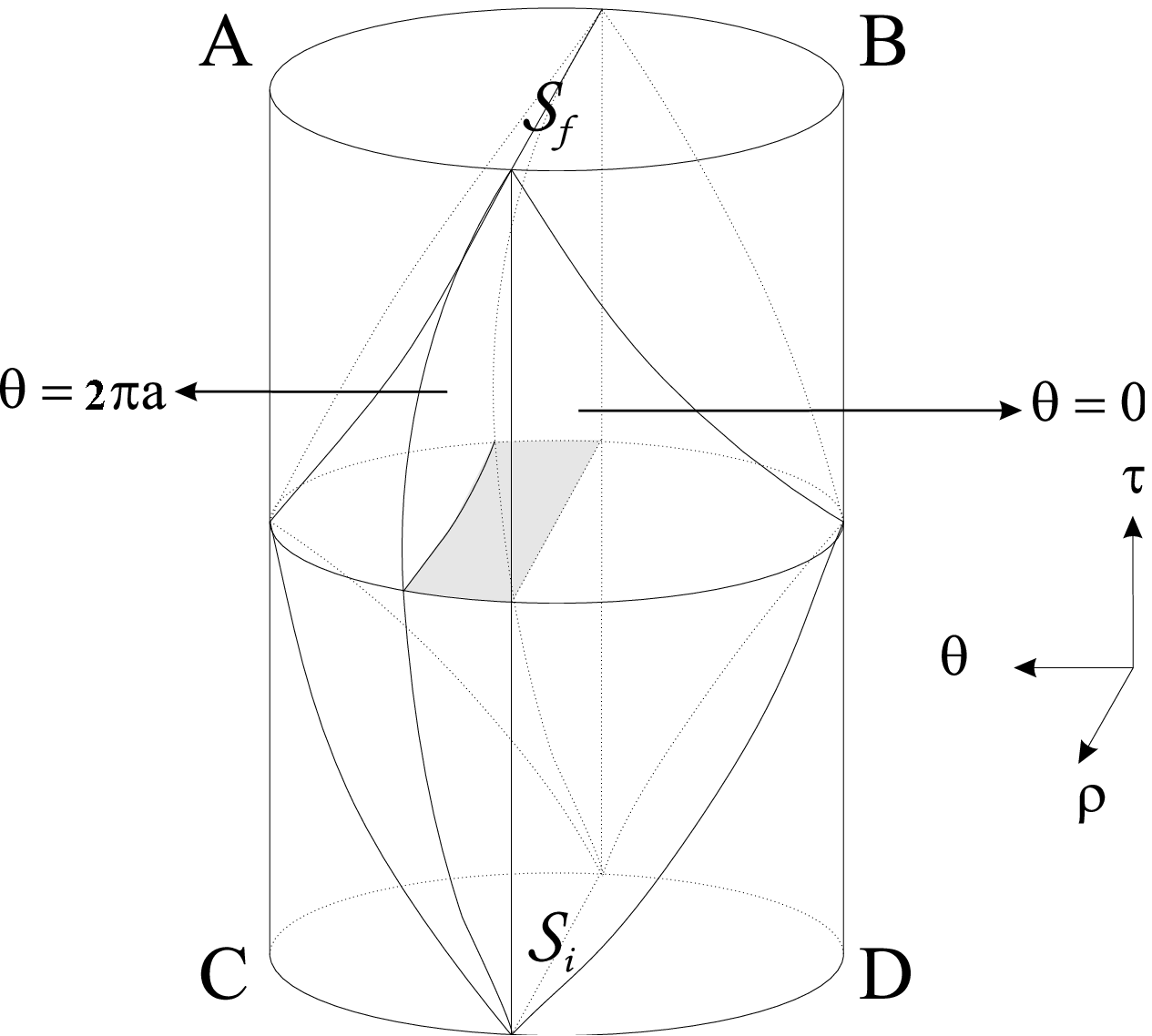}
  \caption{See text.\label{BTZCylThese}}
\end{center}
\end{figure}

Using this global parameterization, we can represent the dynamical evolution of the
black hole, from its initial to its final singularity (denoted
${\cal S}_i$ and ${\cal S}_f$). The shaded region is a constant-time section of a fundamental domain of
the BHTZ action, (see Fig.~\ref{BTZCylThese}).



\begin{figure}[ht]
\begin{center}
\includegraphics*[scale=0.55]{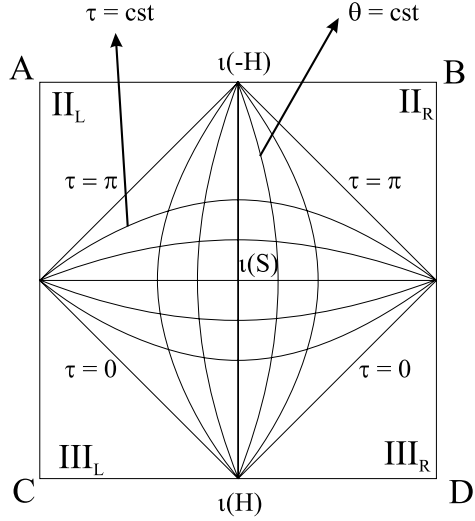}
  \caption{$\rho=0$ section of the preceding figure.\label{feuillerho}}
\end{center}
\end{figure}

From this three-dimensional picture, we can also go to a two-dimensional one. Fig.(\ref{feuillerho}) is a
section of Fig.(\ref{Cylindre1}) by the $\rho=0$ surface. We
represented the coordinate lines $\theta=cst.$ and $\tau=cst.$
(the latter corresponding to orbits of the BHTZ subgroup). On the
straight lines, the identifications become light-like. Note that,
after identifications, the manifold structure is destroyed at
$e=\iota(\h)$ and $-e=\iota(-\h)$, which are fixed points of the
BHTZ action (see sect.$5.8$ of \cite{HE}, and \cite{BHTZ}).
Everywhere else in $I$, the action of the BHTZ subgroup is
properly discontinuous. We will return to this point in Sect. \ref{SubSec-ExtendedBTZ}. 

\newpage
\subsection{Characterization of the horizons and singularities}\label{Sect-HorNon}
In this section, we would like to stress the fact that the non-rotating black hole structure is closely related to the solvable subgroups $AN$ and $A\bar{N}$ of $\SL$. In particular, we will make the two following observations \cite{Keio}:
\begin{enumerate}
\item[(i)]  the non-rotating BTZ black hole singularities are
given by a union of solvable subgroups of $G=\SL$:
\begin{equation}
{\cal S} = Z(G) A N \cup Z(G) A \overline{N},
 \end{equation}
 where $Z(G)=\{e,-e\}$ denotes the center of $\SL$;
 \item[(ii)] the non-rotating BTZ black hole horizons\footpourmoi{j'ai vire le $Z(G)$} correspond
to a union of lateral classes of solvable subgroups of $G=\SL$:
\begin{equation}
 {\cal H} = 
  A N J \cup 
   A \overline{N} J \quad,
 \end{equation}
 where $J=\exp(\frac{3\pi}{2}\t) \in K$ satisfies $J^2 = e$.
\end{enumerate}

The first remark is immediate. Indeed, from \re{KillNonRot2}, the singularities\footpourmoi{pourrait-on prendre les singularites BTZ "ailleurs" sur les surfaces $t^2 - y^2=0$, comme pour l'extension? D'ailleurs, les singularites de BTZ ICI sont donnees par $y=t=0$! Mais dans ce cas les singularites sont donnees par $A$, et pas par $AN$!} are given by 
\beq
 {\cal S} \equiv t^2 -y^2 = 0
 \eeq
Because general elements of $AN$ are of the form $\begin{pmatrix}
    e^a &   n e^a \\
    0 &  e^{-a}
  \end{pmatrix}$, while those of $A\bar{N}$ are $\begin{pmatrix}
    e^a &   0 \\
    m e^a &  e^{-a}
  \end{pmatrix}$, with $a,n,m\in \R$, the conclusion follows. We emphasize the fact that the singularities mentioned in (i) are the loci where the identification vector $\Xi$ (see \re{KillNonRot}) becomes light-like, represented by the light-like surfaces in Fig.~\ref{Cylindre1}. The black hole singularities $\tau = 0, \pi$ of the preceding section (see \re{Metric} and Fig.~\ref{BTZCylThese}) are given, on the other hand, by the intersection of the surfaces $t^2 - y^2=0$ with a given fundamental domain of the action of the BHTZ subgroup. 
  
 We now turn to the black hole horizons. To determine them, we will use two
 methods adapted to the group-theoretical nature of the space under consideration. The first one has been introduced in \cite{Hyvrier}, while the second one has been used in \cite{BDHRS}.

Consider a light ray starting from a point $x$ in $AdS_3$
(supposed to lie in a safe region).
 It is expressed as\footpourmoi{the $-$ sign is just a choice of orientation for the time}
\begin{equation}
 l^k_x (u) = \mbox{exp}(-u Ad(k)\e).x \quad ,
 \quad u \in \R \quad ,
 \end{equation}
for a given $k \in K$. The future and past light cones at $x$ are
then given by
\begin{equation}
{\cal C}_x^\pm = \{ l^k_x (u)\}_{\{k \in K, u\in \R^\pm\}} \quad .
\end{equation}
The point $x$ will be said to lie in the \ind{interior region of
the black hole}, denoted by ${\cal M}^{\mbox{int},+}$, if
all the future-pointed light rays issued from $x$ necessarily fall
into the black hole singularity, that is
\begin{equation}\label{interior}
 x \in {\cal M}^{\mbox{int},+} \Leftrightarrow \forall k \in K, \exists u \in
 \R^+  \mbox{s.t.}  \|\underline{\h} - \overline{\h}\|^2_{l^k_x (u)}=0 \quad .
 \end{equation}
The \emph{future horizon} ${\cal H^+}$ is defined as the boundary
of ${\cal M}^{\mbox{int},+}$.

The equation to study, using the bi-invariance of the
Killing metric and the Ad-invariance of the Killing form, reduces to
\begin{equation}\label{EqHoriz}
B(\h,\h) - B(\h,Ad(\rm{e}^{-u Ad(k)\e}) Ad(x) \h) = 0 \quad.
\end{equation}

One would like to show that the domain ${\cal M}^{\mbox{int},+}$ is $A$-bi-invariant.

Equation \re{EqHoriz} determining this region is clearly invariant under $x \ra x.a \, , \, a\in
A$. Under $x \ra a.x$, one uses the cyclicity of the trace to
bring the second term to 
\beq
B(\h,Ad(Ad(a^{-1})\rm{e}^{-u Ad(k)\e})
Ad(x) \h).
\eeq
 But $Ad(a^{-1})\rm{e}^{-u Ad(k)\e} = \rm{e}^{-\tilde{u}
Ad(\tilde{k})\e}$, with $\tilde{u} = u (\ex^{-2a} \cos^2\theta +
\ex^{2a} \sin^2 \theta)$ and $\cot t = \ex^{-2a} \cot \theta$,
where $k = \ex^{\theta \t}$ and $\tilde{k}=\ex^{t \t}$. The net
result is thus simply a relabelling of the parameters (note that
$u$ and $\tilde{u}$ have the same sign!). This could also be seen
by writing a light ray as $l^k_x (u) = x.\mbox{exp}(u Ad(k)\e)$,
we obtain an expression invariant under $x \ra a.x \, , \, a\in
A$. Thus ${\cal M}^{\mbox{int},+}$ is A bi-invariant.

Let us now consider a light ray starting from a safe region in
$AdS_3$. A point $x \in {\cal U}$ (see preceding section) can be
parameterized as
\begin{equation}
 x(\rho,\theta,\tau) =
\tau_{\exp(\frac{\theta}{2}\h) \,
 \exp(-\frac{\tau}{2}
 \t)}(\exp(\rho\h)) \quad .
 \ee

Because of the A bi-invariance, we may restrict our study to
\begin{eqnarray}
 x &=& \rm{e}^{-\tau/2 \t} \rm{e}^{\rho \h} \sigma(\ex^{\tau/2 \t}) \\
 &=& \rm{e}^{-\tau/2 \t} \rm{e}^{\rho \h} \rm{e}^{-\tau/2 \t}
 \quad.
 \end{eqnarray}

The equation to study finally reduces to
\begin{equation}\label{EqHoriz2}
B(\h,\h) - B(\h,Ad(\rm{e}^{-u Ad(k)\e}) Ad(\ex^{-\tau/2 \t}
\ex^{\rho \h} \ex^{-\tau/2 \t}) \h) = 0 \quad , \quad \tau \in \,
]0,\pi[ \,\, , \,\, \rho \in \R .
\end{equation}

Consider points in $Ad(G)\h$ corresponding to ${\cal B}:= Ad(\ex^{-\tau/2 \t} \ex^{\rho \h} \ex^{-\tau/2 \t}) \h$, $\tau
\in \, ]0,\pi[ \,\, , \,\, \rho \in \R$. We may again visualize this region in the hyperboloid \re{OrbiteThese}, which is now used as an auxiliary space, and no longer as a single leaf (because $\rho$ varies!).

First note that $Ad(\ex^{\rho \h} \ex^{-\tau/2 \t}) \h$ would precisely correspond to
the safe region $I$ on the hyperboloid. Thus, ${\cal B}$ is the region swept out by the domain $I$ when rotating it
counterclockwise around the $\t$-axis with an angle $\pi$,
corresponding to regions $I \cup II_L \cup II_R \cup I'$.

The domain ${\cal B}$ can be decomposed into three regions :
\begin{equation}
 {\cal B} = {\cal B}_1 \cup {\cal B}_2 \cup {\cal B}_3 \quad ,
 \end{equation}
with
\begin{eqnarray}\label{B123}
{\cal B}_1 &=& Ad(A)Ad(\ex^{-\beta/2 \t})\h \quad, \, \beta \in \, ]0,2\pi[ \quad,\\
{\cal B}_2 &=& Ad(A)Ad(\ex^{t \s})\h \quad, \, t\in \R \quad,\\
{\cal B}_3 &=& Ad(A)(-\h \pm \e) \,\, \mbox{or} \,\, Ad(A)(-\h \pm \f)  \quad.\\
\end{eqnarray}

Before going further, let us exploit the $A$-bi-invariance of ${\cal M}^{\mbox{int},+}$, and see how it is reflected when working on the hyperboloid.  
Let $x\in G$ parameterized as $x= k n \ex^{\rho \h} \sigma((kn)^{-1})$,
 corresponding to a point $(\rho,Ad(kn)\h)$ on the adjoint orbit $Ad(G)\h$.
Then, to a point 
\begin{equation}
 x' = Ad(\ex^{\alpha \h})x \in G \quad
 \end{equation}
will correspond a point $(\rho,Ad(\ex^{\alpha \h})Ad(kn)\h)$ on the same adjoint orbit.
Indeed, we have 
\begin{eqnarray}
Ad(\ex^{\alpha \h})Ad(kn)\h &=& Ad(k'n'\ex^{\alpha' \h})\h \\
     &=& Ad(k'n')\h \quad.
     \end{eqnarray}
This point corresponds to a point $x'=  k' n' \ex^{\rho \h}
\sigma((k'n')^{-1}) \in G$, where $k'n'\ex^{\alpha' \h} :=
\ex^{\alpha \h}k n$.     
     
Thanks to the A-invariance and taking the preceding remark into
account, we may forget about the $Ad(A)$ in the equations \re{B123}.
We are thus led to analyze the existence of solutions of
\re{EqHoriz2} with $X \in {\cal B}$ of the form $X_1 =
Ad(\ex^{-\beta/2 \t})\h$, $X_2 =Ad(\ex^{t \s})\h$ and $X_3 =-\h
\pm \e \, , \, -\h \pm \f$.

Consider the first case. With $ Ad(\ex^{-\tau/2 \t} \ex^{\rho \h}
\ex^{-\tau/2 \t}) \h$ of the form $Ad(\ex^{-\beta/2 \t})\h$,
\re{EqHoriz2} becomes the following equation :
\begin{equation}\label{EqU}
 \frac{1}{4} u^2 (\cos \beta - \cos(\beta + 4\theta)) + u \sin
 \beta  + 2 \sin^2 \beta = 0 \quad.
 \end{equation}
We are looking for the values of $\beta$ for which this equation
admits a solution for $u>0$, for all $\theta \in \, [0,\pi]$\footpourmoi{(and
not $[0,2\pi]$, because there is a double covering!)}. By
considering the particular case $\theta = 0$, we find $u=-\tan
\frac{\beta}{2}$, thus the allowed values of $\beta$ have to lie
in the range $]\pi,2\pi[$. Let us look at the constraints imposed by
other values of $\theta$. We denote the two
roots of \re{EqU} by $u_1$ and $u_2$ . We have
\begin{eqnarray}
 u_1 . u_2 &=& \frac{4 \sin^2 \beta/2}{\sin 2\theta \sin(\beta +
 2 \theta)} \label{PrRac} \\
 u_1 + u_2 &=& -\frac{2 \sin \beta}{\sin 2\theta \sin(\beta +
 2\theta)}
 \end{eqnarray}
First note that $\forall \beta \in ]0,2\pi[$, $\sin 2\theta
\sin(\beta + 2\theta)$ may be positive or negative as $\theta$
varies in the range $[0,\pi]$. If $\sin \beta >0$, then there two
positive roots when $\sin 2\theta \sin(\beta + 2\theta)>0$, and
one positive and one negative when $\sin 2\theta \sin(\beta +
2\theta)<0$. Thus there always exists a positive solution for $u$,
for any $\theta$. If $\sin \beta < 0$, there are two negative
roots when $\sin 2\theta \sin(\beta + 2\theta)>0$.

Consequently, the interior region will correspond to points $X_1 =
Ad(\ex^{-\beta/2 \t})\h \, \, , \, \, \beta \in \,]\pi,2\pi[ $ on the
adjoint orbit.

For the second case, $X_2 =Ad(\ex^{t \s})\h$, the equation we get
is
\begin{equation}
 \frac{1}{4} u^2 (\cosh 2t - \cos 4\theta \cosh 2t + 2 \sin
 2\theta \sinh 2t) + u \cos 2\theta \sinh 2t + (1 - \cosh 2t) = 0
 \quad.
 \end{equation}
 By considering two special cases, it is easy to see that this
 equation does not admit a positive solution in $u$ for all
 $\theta$. Indeed, for $\theta = \pi/2$, one finds $u=-\tanh t$,
 while for $\theta = 0$, one gets $u=\tanh t$. Thus there is no
 $t\ne 0$ satisfying both conditions. The last case yields no
 positive solution for all $\theta$ either.

 As a conclusion we find that
 \begin{equation}
  x \in {\cal M}^{\mbox{int},+} \Leftrightarrow Ad(x)\h =
  Ad(A)Ad(\ex^{-\beta/2 \t})\h \quad , \quad \mbox{with} \,\,
  \beta \in \, ]\pi,2\pi[ \quad.
  \end{equation}
  The boundaries of the corresponding region in $Ad(G)\h$ are given by $-\h -
  r^2 \e$ and $-\h - r^2 \f$, $r\in \R$, or
  \begin{equation}
  Ad(N^-)(-\h) \cup  Ad(\bar{N}^+)(-\h) \quad ,
   \end{equation}
   with $N^- = \{\ex^{t\e}\}_{t\leq 0}$ and $\bar{N}^+ = \{\ex^{t\f}\}_{t\geq
   0}$.
The horizons can be deduced as
\begin{equation}
 x \in {\cal H^+} \Leftrightarrow Ad(x)\h = Ad(N^-)(-\h) \,\,
 \mbox{or} \,\, Ad(x)\h = Ad(\bar{N}^+)(-\h) \quad.
 \end{equation}
Because of the A-invariance, we may write $x=\tau_{\ex^{-\tau/2
\t}}(\ex^{\rho \h})$ and look for the relation between $\tau$ and
$\rho$ such that
\begin{equation}
 Ad(\tau_{\ex^{-\tau/2
\t}}(\ex^{\rho \h}))\h = Ad(N^+)Ad(\ex^{\pi/2 \t}) \h \; \mbox{or} \; Ad(\tau_{\ex^{-\tau/2
\t}}(\ex^{\rho \h}))\h = Ad(\bar{N}^-)Ad(\ex^{\pi/2 \t}) \h \quad .
\end{equation}
This amounts to impose that
\begin{equation}\label{CondHor}
 \left(\ex^{-\tau/2 \t} \ex^{\rho\h} \ex^{-\tau/2 \t}\right)^{-1} \,
 (\ex^{t^2 \e} \ex^{\pi/2 \t}) \in A \cup Z(G) \quad ,
 \end{equation}
or
\begin{equation}\label{CondHor2}
 \left(\ex^{-\tau/2 \t} \ex^{\rho\h} \ex^{-\tau/2 \t}\right)^{-1} \,
 (\ex^{-t^2 \f} \ex^{\pi/2 \t}) \in A \cup Z(G) \quad ,
 \end{equation}
giving, for $\tau \in ]\pi/2,\pi[$, $\cos\tau = -\tanh \rho$, $\rho>0$ and $\cos\tau = \tanh \rho$, $\rho<0$.



The domain ${\cal M}^{\mbox{int},-}$ is of course defined as
\begin{equation}
 x \in {\cal M}^{\mbox{int},-} \Leftrightarrow \forall k \in K, \exists u \in
 \R^-  \mbox{s.t.}  \|\underline{\h} - \overline{\h}\|^2_{l^k_x (u)}=0 \quad .
 \end{equation}
The \emph{past horizon} ${\cal H^-}$ is defined as the boundary of
${\cal M}^{\mbox{int},-}$. By proceeding the same way, we find
that
\begin{equation}
  x \in {\cal M}^{\mbox{int},-} \Leftrightarrow Ad(x)\h =
  Ad(A)Ad(\ex^{-\beta/2 \t})\h \quad , \quad \mbox{with} \,\,
  \beta \in \, ]0,\pi[ \quad,
  \end{equation}
and
\begin{equation}
 x \in {\cal H^-} \Leftrightarrow Ad(x)\h = Ad(N^+)(-\h) \,\,
 \mbox{or} \,\, Ad(x)\h = Ad(\bar{N}^-)(-\h) \quad ,
 \end{equation}
or in coordinates : $\tau\ \in \, ]0,\pi[$, $\cos \tau =
\tanh \rho$ for $\rho > 0$ and $\cos \tau =-\tanh \rho$ for $\rho<0$.


Let us end by looking at the structure of the horizons on the
group manifold $G$. First note that, for $\cos \tau = \pm \tanh
\rho$,
\begin{equation}
 \left(\ex^{-\tau/2 \t} \ex^{\rho\h} \ex^{-\tau/2 \t}\right)^{-1} \,
 (\ex^{t^2 \e} \ex^{\pi/2 \t}) = -{\bf e} \quad .
 \end{equation}
Thus,
\begin{eqnarray}
\tau_{\exp(\frac{\theta}{2}\h) \, \exp(-\frac{\tau}{2}
 \t)}(\exp(\rho\h)) &=& Ad(\exp(\frac{\theta}{2}\h)) (\ex^{-\tau/2 \t}\ex^{\rho \h}\ex^{-\tau/2 \t}) \\
  &=&  Ad(\exp(\frac{\theta}{2} \h)) (\ex^{t \e} \ex^{\pi/2 \t} (-{\bf e})) \\
  &=& \ex^{\theta/2 \h} \ex^{t \e} \ex^{3 \pi/2 \t} \ex^{-\theta/2 \h}
 \ex^{-3 \pi/2 \t} \ex^{3 \pi/2 \t} \\
 &=& \ex^{\theta/2 \h} \ex^{t \e} \ex^{\theta/2 \h} \ex^{3 \pi/2
 \t}\quad ,
 \end{eqnarray}
where the second line holds because of \re{CondHor}.

 But $\{ \ex^{\theta/2 \h} \ex^{t \e} \ex^{\theta/2 \h} \}_{\theta
 \in \R , t\in \R} = A N A = A N$, and thus the horizons (future
 and past) are given by
 \begin{equation}\label{HorizonsGroup}
   {\cal H} =  {\cal H^+} \cup  {\cal H^-} =  A N k_{3\pi/2} \cup    A \bar{N} k_{3\pi/2}\quad ,
  \end{equation}
  with $k_{3\pi/2} = \ex^{3\pi/2 \t}$.
  We also have
  \begin{equation}
  {\cal H^+} = AN^+ k_{3\pi/2} \cup A\bar{N}^- k_{3\pi/2} \quad
  \end{equation}
  and
  \begin{equation}
  {\cal H^-} = AN^- k_{3\pi/2} \cup A\bar{N}^+ k_{3\pi/2} \quad
  \end{equation}

We now remark that we can take a shortcut to determine the horizons. Eq. \re{EqHoriz2} gives, in general, a
second order equation in $u$, $A u^2 + B u + C =0$.  A point will
belong to the horizons if this equation admits a solution ``for
infinite u", or stated more precisely, if there exists a direction
$k$ such that the solutions for $u$ are rejected to infinity. This
is equivalent to the vanishing of the coefficients of $u^2$ and
$u$ in Eq. \re{EqHoriz2}, which read explicitly
\begin{equation}
A = \frac{1}{2} \sin 2\phi (\cos(2\phi+\tau)\cosh 2\rho \sin\tau +
\cos\tau \sin(2\phi + \tau) -\sin\tau\sinh2\rho) \quad ,
\end{equation}
and
\begin{equation}
B = 2 \cosh\rho \sin \tau(\cos \tau \cosh \rho + \cos 2\phi \sinh
\rho) \quad ,
\end{equation}
for $k= \ex^{\phi \t}$.
 In this way, it is possible
to find directions $k$ such that the coefficients vanish, provided
\begin{equation}
 \cos \tau = \pm \tanh \rho \quad ,
 \end{equation}
giving again the equations of the horizons\footpourmoi{Voir
7/08/05-15, NotebookMM8/08/05}.

The black hole horizons (restricted to the intersection of \re{HorizonsGroup} or $\cos \tau = \pm \tanh \rho$ with a fundamental domain of the action of the BHTZ subgroup) are drawn in Fig.~\ref{BTZHor}, yielding a three-dimensional Penrose diagram.

\begin{figure}[ht]
\begin{center}
\includegraphics*[scale=0.6]{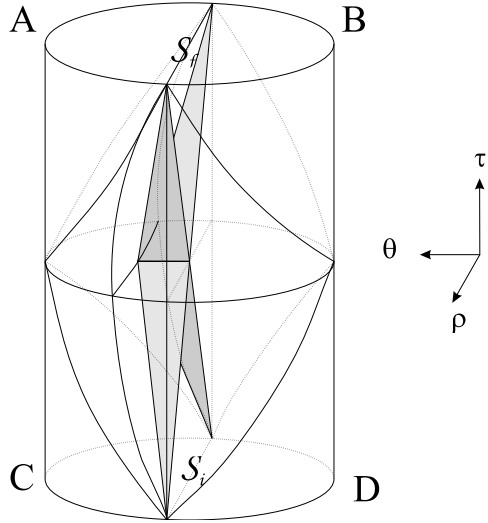}
  \caption{Representation of the black hole's horizons. These are light-like surfaces separating two regions,
one of which can be causally connected to the space-like infinity
(the exterior region), and another one causally disconnected from
it (the interior region).\label{BTZHor}}
\end{center}
\end{figure}


A useful two-dimensional picture is provided by a section by the $\theta = 0$ surface. One
then gets a usual two-dimensional Penrose diagram, where the
singularities and horizons are drawn, as displayed in
Fig.~\ref{Penrose2D}. It clearly displays the black hole causal structure, similar to that of a Schwarzschild black hole, the only difference being observed in the asymptotic region, which is $AdS$ rather than flat.
 
\begin{figure}[ht]
\begin{center}
\includegraphics*[scale=0.6]{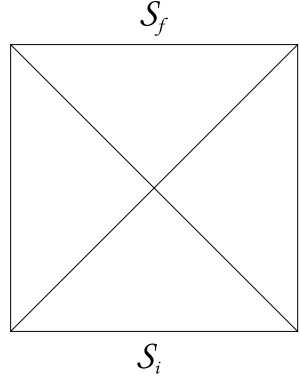}
  \caption{A $\theta=0$ section of
Fig.\ref{BTZHor}, yielding a Penrose diagram of the black hole.
The oblique lines represent the black hole's horizons. In this picture, each point represents a circle.\label{Penrose2D}}
\end{center}
\end{figure}

 


 
\subsection{Extended non-rotating massive BTZ black hole}\label{SubSec-ExtendedBTZ} 

In the original paper \cite{BHTZ}, the question was addressed to find out whether the smoothness of anti-de Sitter space as a Hausdorff manifold was preserved by the identifications. This question has been tackled in detail in Appendix B of \cite{BHTZ}. One can understand the situation by means of a simpler example, which shares all the features of the question at hand here. It is provided by the so-called \ind{Misner space} \cite{HE}. Let us consider the two-dimensional Minkowski space. The isometries of this space leaving the origin fixed form the one-dimensional Lorentz group, whose orbits are hyperbolae (see Fig.~\ref{MisnerSpaceFig}). 
\begin{figure}[ht]
\begin{center}
\includegraphics*[scale=0.7]{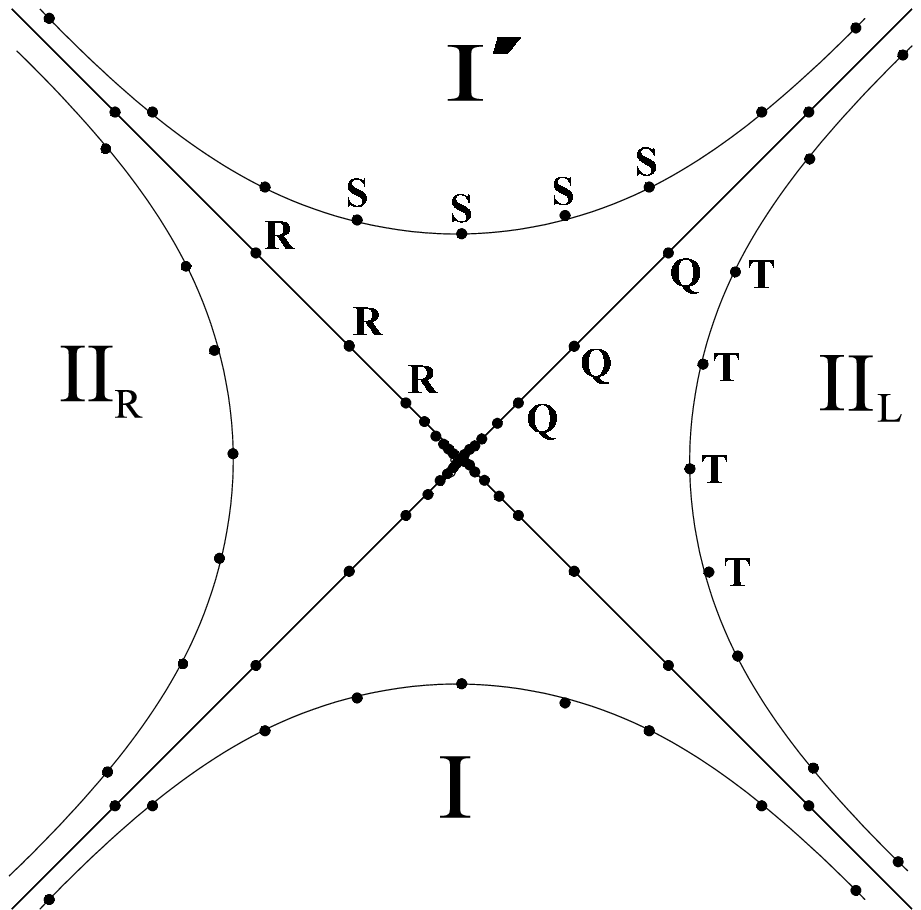}
  \caption{Two-dimensional Minkowski space $ds^2 = -dt^2 + dx^2$. Under the discrete subgroup $G$ of the Lorentz group, points $S$, $Q$, $R$ and $T$ are equivalent.\label{MisnerSpaceFig}}
\end{center}
\end{figure}
The \ind{Misner space} is defined as the quotient of the region $I'$ under the action of a discrete subgroup $G$ of the Lorentz group. It is obtained by identifying points in $I'$ according to
\beq
  (t,x) \sim (t \cosh n \pi + x \sinh n \pi, x \cosh n \pi + t \sinh n \pi) \quad , \quad n \in \Z. 
\eeq
With $\bar{t} = \frac{1}{4} (t^2 - x^2)$ and $\psi = 2 \,\mbox{arctanh}(x/t)$, the Misner space is described by a metric
\beq
 ds^2 = - \bar{t}^{-1} d\bar{t}^2  + \bar{t} \, d\psi^2 \quad , \quad \bar{t} \in \R, \psi \in [0,2\pi].
\eeq
The action of the Lorentz group in each of the regions $I'$, $I$, $II_L$ or $II_R$ is \ind{properly discontinuous}\cite{HE}. The action of a group $G$ on a manifold $M$ is said to be properly discontinuous if:
\begin{enumerate}
\item[1.] each point $Q\in M$ has a neighborhood ${\cal O}$ such that $g.{\cal O} \cap {\cal O} = \emptyset$ for each $g \in G$ which is not the identity element, and
\item[2.] if $Q,R \in M$ are such that there is no $g\in G$ with $g.Q=R$, then there are neighborhoods ${\cal O}$ and ${\cal O}'$ of $Q$ and $R$ respectively such that there is no $h \in G$ with $h.{\cal O} \cap {\cal O}' \ne \emptyset$.
\end{enumerate}
Condition 1 implies that the quotient $M/H$ is a manifold, and condition 2 implies that it is a \ind{Hausdorff manifold}\footnote{A topological space $X$ is called a  \ind{Hausdorff space} if any two distinct points of $X$ can be separated by neighborhoods.}. Therefore the Misner space is a Hausdorff space. The action of $G$ is also properly discontinuous in any two adjacent regions: $I \cup II_R$, $II_R \cup I'$, $\cdots$. However, the action of the group on the region $I'\cup II_L \cup II_R$ satisfies condition 1 but condition 2 is not satisfied for points $Q$ on the boundary between $I'$ and $II_R$ and points $R$ on the boundary between $I'$ and $II_L$. Therefore the quotient $(I' \cup II_L \cup II_R)/G$ is not Hausdorff although it is still a manifold.

The situation is quite similar for the non-rotating BTZ black hole, where we have roughly ``twice" a Misner space, the fixed points of the BHTZ action being $\iota(\h)$ and $\iota(-\h)$, see Figs.~\ref{feuillerho} and \ref{EBTZ}. Therefore, if identifications are performed in the whole $AdS_3$ space, the Hausdorff manifold structure is destroyed at these points, corresponding to the singularities (and, actually, both conditions 1 and 2 are violated at these fixed points).
 
 \begin{figure}[ht]
\begin{center}
\includegraphics*[scale=0.9]{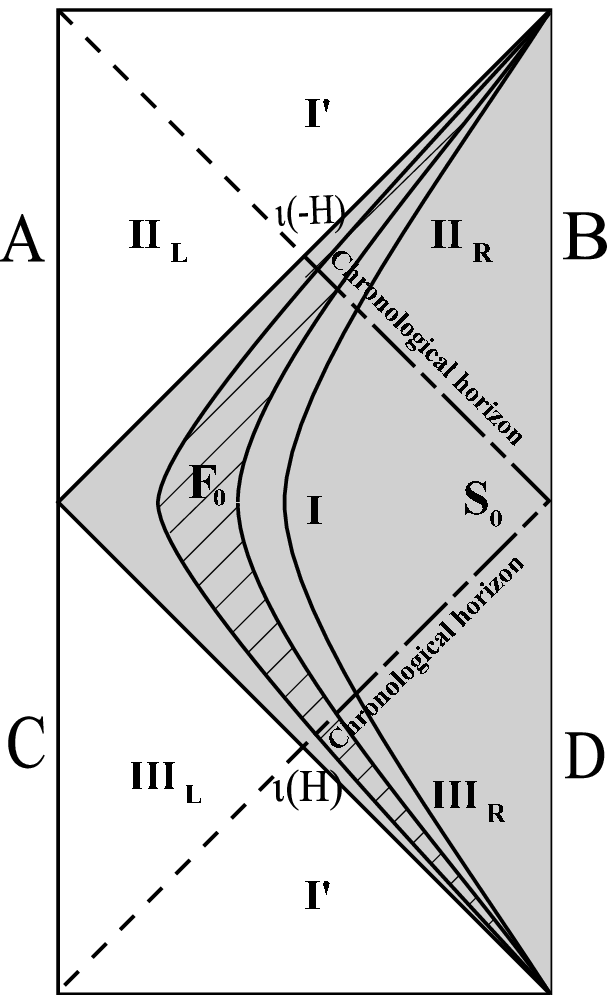}
  \caption{Penrose diagram of a $\rho=\rho_0$ section in
$\SL$. This diagram can also be seen as a
Penrose diagram of $\SL$, each point representing a line
parametrized by $\rho$, running from $-\infty$ to $+\infty$. The
region $I$ provides, after identification by
the BHTZ subgroup, the usual non-rotating BTZ
black hole space-time, bounded by the chronological horizons
$\tau=0$ and $\tau=\pi$. The maximally extended region ${\bf S_0}$
(in grey) admits an action of $\ca \cn$, and represents the
intersection of the domain $\widetilde{{\cal U}}$, defined in \re{grosU}, with a
$\rho=\rho_0$ section. This maximal extension goes beyond the
chronological singularities, where the identifications yielding the
black hole become light-like. The dashed region ${\bf F_0}$
represents a fundamental domain of the action of the BHTZ
identification subgroup on ${\bf S_0}$, i.e. a $\rho=\rho_0$
section of the extended $\EBTZ$ black hole.\label{EBTZ}}
\end{center}
\end{figure}
 
Nevertheless, there is no such problem if one restricts the identifications to a single safe region, say $I$, which are analogous to the identifications leading to the Misner space from the part $I'$ of two-dimensional Minkowski space, see Fig.~\ref{MisnerSpaceFig}. Along the same lines, one may define a \ind{maximally extended BTZ space-time} as a maximal extension of a safe region, preserving the Hausdorff manifold structure of the quotient. From Fig.~\ref{EBTZ}, we observe that there are two such inequivalent extensions: $I \cup II_L \cup III_R$ and $I \cup II_R \cup III_R$. Of course, after performing the identifications, a maximally extended BTZ space-time will exhibit closed time-like curves. Despite this, the latter extension displays an interesting property : in this case, each leaf in
the extension admits an action of the two-parameter subgroup $\AN$
of $\SL$. Indeed, going back to the hyperboloid, one may convince oneself that the region $I \cup II_R \cup III_R$
in $Ad(G)\h$ (see Fig.~\ref{OrbiteExtThese}) can be parameterized
as

\beq \label{hyperbExt} X= Ad\left( \exp(\frac{\phi}{2}\h)\exp(w\e)
\right) \s \quad ,\quad -\infty<\phi<+\infty,\, -\infty<w<+\infty
\,, \eeq

where $\s = Ad\left(\exp (-\frac{\pi}{4})\t)\right) \h$. Points in this maximally extended domain (denoted $\widetilde{\cal U} \subset AdS_3$) are thus represented by
\beq \label{grosU}
 x(\rho,\phi,w) = \tau_{r \exp (-\frac{\pi}{4}\t)}\left(\exp(\rho
 \h)\right) \quad, \quad r=\exp(\frac{\phi}{2}\h)\exp(w\e) \in \AN, 
\eeq
or explicitly 
\begin{equation}
  \label{gExt2}
  x(\rho,\phi_,w) =
  \begin{pmatrix}
    w \cosh\rho + \sinh \rho &   e^\phi \cosh \rho \left( w^2 - 1 \right)\\
    e^{-\phi} \cosh \rho  &  w \cosh\rho - \sinh \rho
  \end{pmatrix} \quad,
\end{equation}
with the corresponding metric
\begin{equation}
  \label{metricExt2}
  d s^2  = d \rho^2 + \cosh^2 \rho \left( d \phi^2 - \left( w d \phi +
      d w \right)^2\right) \quad.
\end{equation}
 Restricting $\phi$ to $[0,\, 2\pi\, \sqrt{M}]$ in (\ref{metricExt2}), we obtain a maximally extended spinless BTZ black
 hole, denoted hereafter by $\EBTZ$.

\begin{figure}[ht]
\begin{center}
\includegraphics*[scale=0.5]{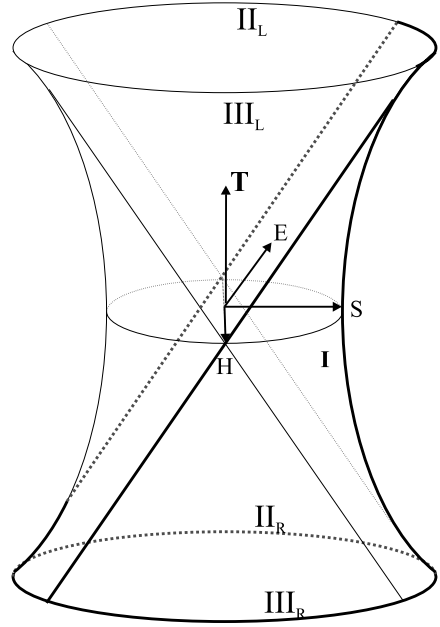}
  \caption{The intersection of a maximally extended domain with a leaf, represented as a hyperboloid (see text)  .\label{OrbiteExtThese}}
\end{center}
\end{figure}

The black hole singularities are located at $w=\pm 1$, while the horizons
correspond to the surfaces
\begin{equation}
  w = \pm \tanh \rho \quad.
\end{equation}
These coordinates are related to \re{gReg2} (on the intersection of the domains they cover) by
\begin{equation}
  \label{changecoord}
  w = \cos \tau \quad \text{and} \quad e^{\phi} = \frac{e^{\theta}}{\sin \tau} \, .
\end{equation}

It is clear from \re{grosU} that each leaf admits a left action of $\AN$:
\beq
 \tau : AN\, \times {\cal O}_\rho \ra {\cal O}_\rho : (r',x) \ra \tau_{r'}(x) = \tau_{r' r \exp (-\frac{\pi}{4}\t)}\left(\exp(\rho
 \h)\right), 
 \eeq
 as well as a right action
 \beq
 \psi : AN\, \times {\cal O}_\rho \ra {\cal O}_\rho : (r',x) \ra \psi_{r'}(x) = \tau_{r r' \exp (-\frac{\pi}{4}\t)}\left(\exp(\rho
 \h)\right).
 \eeq

The importance of this observation will become clear when we will study deformations of these BTZ space (see Sect. \ref{SectBTZDef}).




\section{Generic rotating BTZ black hole} 
\footpourmoi{{\bf QUESTIONS} : Keio p3: l'action ne s'etend a une action de $G$ par isometries?. Transversal? }
 
 In the rotating case, the identification Killing vector reads
 \beq \label{XiRot}
  \Xi = (L_+ \h , -L_- \h), 
 \eeq
or in the coordinates \re{AdSl}
\beq 
  \Xi = (L_+ + L_-) (t\p_y + y \p_t)  + (L_+ - L_-) (x \p_u + u \p_x), 
\eeq
with $L_+ > L_-$, see \re{angm}. From \re{Killutxy}, we thus have the correspondence $L_+ + L_- = r_+$ and $L_+ - L_- = r_-$, as well as
$L_\pm^2 = M \pm J$.
The identifications to perform are thus
\begin{equation}\label{BTZRot}
 z \sim \exp (2 \pi \, m \, L_+ \, \h)\, z \,
\exp(2 \pi \,  m \, L_- \, \h) \quad , \quad m \in \Z.
\end{equation}
The fact that $L_+ \ne L_-$ in the rotating case prevents us to use a foliation similar to that of the non-rotating case. Indeed, in the latter case, we could express the action of the BHTZ subgroup as an action of $A \subset \SL \subset SO(2,2)$ on $\SL$ (see \re{BHTZTwisted}), while in the rotating case this is no longer possible since the left and right subgroups are now necessarily different. 

 \subsection{Modified Iwasawa decomposition and global description} 
In the rotating case, a foliation adapted to the identifications \re{BTZRot} can be obtained by introducing a \ind{modified Iwasawa
decomposition} defining a global diffeomorphism on $\SL$\cite{BDHRS}. It reads
\begin{equation}\label{MICS}
\Phi : A \times N \times K \lra G : (a,n,k) \ra a^{L_+} \, n  \, k \, a^{L_-}
 \qquad  .
\end{equation}
This application defines a global diffeomorphism\footpourmoi{Le montrer?? Voir GBTZ-10. 1. $\phi_* (\p_a \wedge \p_n \wedge \p_k) \ne 0 \; \forall (a,n,k) \Leftrightarrow |L_-/L_+| < 1$. le montrer? Implique que $\phi$ est une submersion. 2. Montrer que $\phi$ est surjective. Ceci signifi alors que $\phi$ est revetement (voir Def. Gutt): Pq?? 3. L'image inverse du neutre est $(e,e,e)$ OU $\pi_1 (A\times N \times K) = \pi_1 (G) = e$ implique que $\phi$ est un diffeo. Pq??} iff $|L_-/L_+| < 1$. This amounts to consider $J>0$ (see \re{angm}). For $J<0$, one would have to use the decomposition $z= a^{L_-} \, n \, k \, a^{L_+}$; we hereafter assume $J>0$. Notice that the non-rotating case is excluded.

The BTZ-adapted map (\ref{MICS}) defines a global coordinate
system $(\tau, u, \phi)$ on $AdS_3$ through the application (also denoted $\Phi$)~:
\begin{equation}
 \Phi : \R^3 \rightarrow \SL : (\tau ,u,\phi) \mapsto
 \exp (L_+ \phi \h) \exp (u \e) \exp (\tau \t) \exp (L_- \phi \h) \qquad ,
\end{equation}
referred to hereafter as the \ind{Modified Iwasawa Coordinate System} (MICS).
In these coordinates, the $AdS_3$ metric reads as~:
\begin{eqnarray} \label{metric}
&&ds^2 = - d\tau^2 - du d\tau -  2 u L_+ d\tau d\phi -
L_- \sin(2\tau) du d\phi \nonumber \\
&& \qquad \qquad \qquad +2 \left[M + L_+ L_- \left( \cos(2 \tau) -
u \sin (2 \tau) \right) \right] d\phi^2 \qquad .
\end{eqnarray}
The identification (\ref{BTZRot}), yielding BTZ black holes, is
simply~:
\begin{equation} \label{btzperiod}
 (\tau, u, \phi) \mapsto (\tau, u, \phi + 2 \pi m) \qquad , \qquad
m \in \Z \qquad .
\end{equation}
Hence, when restricting the $\phi$-coordinate to $0 \le \phi < 2
\pi$, eq.(\ref{metric}) becomes a global expression for the metric
for the rotating BTZ black hole.

The decomposition (\ref{MICS}) naturally induces a foliation of
$\SL$ whose 2-dimensional leaves ${\cal F}_\tau$, corresponding to
constant $\tau$ sections, are  stable with respect to the action
of the BTZ subgroup. These leaves are obtained as follows. We
introduce the action $\nu$ of the subgroup $AN\,$ on $\SL$ as~:
\begin{equation} \label{action}
\nu : AN\, \times \SL \rightarrow \SL : (an,z) \mapsto
\nu_{an}(z)=a^{L_+} \, n  \, z \, a^{L_-} \qquad .
\end{equation}
The orbits of this action are obtained by acting with $\nu$ on a
given element of the subgroup $K$ transverse to the leaves\footpourmoi{comment se representent ces feuilles dans le cylindre?? Que se passe-t-il geometriquement quand $|L_+|=|L_-|$, c-a-d quand on retourne au cas sans rotation?? Ou sont les domaines fondamentaux + regions safe?? Voir Fig Michel}:
\begin{equation} \label{fiber}
{\cal F}_\tau  = \{ \nu_{a n} (\exp (\tau \t) ) \}\quad ,
\quad \forall \tau \in \R \quad.
\end{equation}
In particular, the subgroup $AN\,$ constitutes the leaf at $k=e$.
As a consequence of \re{MICS} and\re{action}, the leaves of this foliation enjoy the same property as those in the maximally extended non-rotating massive BTZ black hole: they admit an action of the $AN\,$ subgroup of $\SL$.

Until now, we have been working in the whole $AdS_3$ space.
But we know that the identifications (\ref{btzperiod})  induce acausal
 regions containing closed time-like curves passing through every point. The BHTZ subgroup generator $\Xi$
given by eq.(\ref{XiRot}) is time-like in these regions,
whereas it is space-like in the
physical regions. The boundaries between physical and non-physical
regions are the BTZ singularities ${\cal S}$.
In terms of MICS, these singularities are given by the
$\phi$-invariant surfaces :
\begin{equation}\label{singularity1}
q(\tau,u) \equiv M + L_+ L_- \left[ \cos(2 \tau) -u \sin (2 \tau)
\right]=0 \qquad ;
\end{equation}
$q(\tau,u)>0$ corresponds to causally safe regions and $q(\tau,u)<0$ to regions with closed time-like curves.

 
 Finally, let us note that the leaves of the foliation
(\ref{action}), the surfaces of constant $\tau$, are flat\footpourmoi{BIZARRE, bizarre?}. To see
this, notice that the induced metric on the surface $\tau =
\tau_0$ reads as~:
\begin{equation} \label{Mis1}
ds^2 = \frac{d\zeta d\phi}{2 L_+} +  \zeta d\phi^2 \qquad ,
\end{equation}
with $\zeta = 2 q(\tau_0,u)$
 ($\tau_0 \ne m \frac{\pi}{2})$. We
can further perform the change of coordinates $U=\frac{e^{-2 L_+
\phi}}{2 L_+}$ and $V=\frac{\zeta e^{2 L_+  \phi}}{2 L_+}$, in
terms of which the metric becomes~:
\begin{equation}
ds^2 = - dU dV, \qquad \quad  0<U<\infty \qquad, \qquad  -
\infty<V<\infty \qquad ,
\end{equation}
which corresponds to half a Minkowski space. The identifications
according to the bi-action (\ref{BTZRot}) read in these coordinates\footpourmoi{Ou voit-on la-dessus que pour le rotating le quotient est Hausdorff partout?? $U=V=0$ n'est pas un point de l'espace??(car c'est un point fixe!)}
as~:
\begin{equation}
 (U,V) \mapsto (U e^{-4 \pi m L_+}, V e^{4 \pi m L_+}) \qquad .
 \end{equation}
They yield closed time-like curves for $V<0$. As a consequence,
when performing the identifications, the $\tau =$ constant
surfaces exhibit a Misner-space causal structure (see previous section), corresponding to two adjacent regions (if one allows closed time-like curves), or a single one, in Fig.~\ref{MisnerSpaceFig}, showing that in the rotating case the action of the BHTZ subgroup is properly discontinuous everywhere \cite{BHTZ}\footpourmoi{puisqu'ici l'espace complet -meme si on prend les CTC- ne comprend que le demi-Minkowski, ou l'action l'est!}\footpourmoi{Comprendre le passage rotating- non-rotating voir BHTZ, Smoothness of the quotient}.    


\subsection{Horizons and causal structure}

Let us now express the BTZ horizons $\cal H$ in terms of MICS. We use the same tools as in Sect. \ref{Sect-HorNon}. We write a light ray passing through the point ${\bf z_0}$ as~:
\begin{equation}
 z (s) = z_0 \exp (s \lb), \quad s \in \R \qquad ,
 \end{equation} 
with $\lb$ a future pointing null vector obtained by rotation of
$\e$ around the $\t$-axis, expressed as
\begin{equation}
 \lb = Ad(\exp(\kappa \t)) \e \qquad , \qquad 0 \leq \kappa < \pi \qquad .
\end{equation}
This null vector can also be seen as resulting from another null
vector boosted by a Lorentz rotation with axis along $\h$. These
two remarks allow us to parametrize the light rays passing through
a point $z = a^{L_+} n k a^{L_-}$ of $AdS_3$, with
direction specified by $\kappa$, as:
\begin{equation}
\ell_{z}^{\kappa}(s) = a^{L_+} \, n \, k \, \exp\left[s
Ad(\exp(\kappa \t)) \e\right] \, a^{L_-} \qquad .
\end{equation}
The future and past light cones at point $z$ are defined as~:
\begin{equation}
{\cal
  C}_{z}^{\pm} = \{ \ell_{z}^\kappa(s) \, , \, 0 \leq \kappa < \pi, \quad s \in \R^\pm \}.
\end{equation}

The position of $z$ with respect to the BTZ horizons can be
determined according to the number of directions $\kappa$ allowing
the light rays to escape a given singularity, either in the past
or in the future. If this number is zero, $z$ is situated behind
the horizon, whereas if this number is infinite, $z$ is before
the horizon. By continuity, the horizons are defined as the set of
points $z$ for which a finite number of $\kappa$s are able to
escape the singularity for infinite values of $s$, i.e. for which
the equation of the intersections ${\cal C}_{z}^{\pm} \cap \cal
S$:
\begin{equation}\label{hor1}
\beta_{\ell_{z}^\kappa(s)} (\Xi, \Xi) = 0 \qquad ,
\end{equation}
has no solution for a finite $s$. In terms of MICS, this equation
reads as~:
\begin{equation}\label{hor2}
 2 L_+ L_- \,  \beta_{\be}(Ad(l_{z}^\kappa(s)) \h,\h) + M =0 \quad .
\end{equation}
It is a second order equation in $s$, invariant for the
substitution $z \mapsto a.z$, whose coefficients depend on
the coordinates $u$ and $\tau$ of the starting point
$\bz(u,\tau,\phi)$ of the light ray and on its direction $\kappa$.
Requiring that eq. (\ref{hor2}) has no solution for finite $s$,
neither in the past nor in the future, forces the
coefficients of $s$ and $s^2$ to vanish. In this way we obtain the
equations of the inner and outer horizons $\cal H^-$ and $\cal
H^+$~:
\begin{eqnarray}\label{hor3}
{\cal H}_1^+ &:& \tau=m \pi \qquad\qquad , \qquad {\cal H}_2^+ : u=-\tan(\tau)\qquad ,\\
{\cal H}_1^- &:& \tau=\frac \pi 2 + m \pi \qquad , \qquad {\cal H}_2^- : u=\cot(\tau)
\qquad , \qquad \nonumber
\end{eqnarray}
where $m\in\Z$.

Strikingly, the horizons of the rotating BTZ black hole again appear to be closely related to the Iwasawa subgroups $AN$ and $A\bar{N}$ of $\SL$.  
Indeed, one finds that
\beq
 {\cal H} \equiv z. AN \cup z. A\bar{N} \quad ,
\eeq
where $z = \exp (m \frac{\pi}{2} \t)$, for well-chosen $m \in \Z$.

This geometrical discussion is summarized in the
two-dimensional diagram
depicted in Fig. \ref{PenroseRot}\footpourmoi{dans un Penrose ordinaire en rotation, on peut echapper a la singularite si on file vers un autre infini. Est-ce aussi le cas ici?}. The diagram is obtained by considering a
constant $\phi$ section and making the coordinate transformation
$u=\tan(p)$, with $- \frac \pi 2 \le p < \frac \pi 2 $. On this
diagram, each point corresponds to a circle (the orbit of the
point under the action of the one parameter subgroup generated by
$\phi$ translation). The horizons ${\cal H}^\pm$ are depicted by
straight lines inclined at 45 degrees, crossing the $\tau$-axis at
$\tau=m \frac \pi 2$. The singularities, defined in
(\ref{singularity1}), are situated beyond the inner horizons
${\cal H}^-$.

\begin{figure}[ht]
\begin{center}
\includegraphics*[scale=0.75]{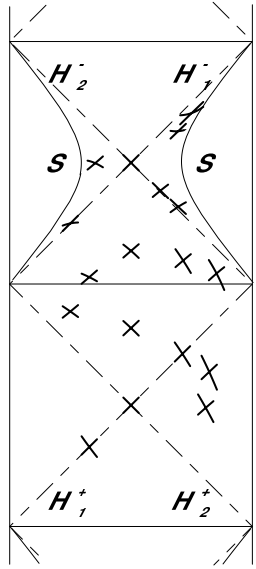}
\caption{The causal structure of the black hole is summarized
in this
diagram. The axis of the time-like coordinate
     $\tau$ is vertical, while that corresponding to the light-like coordinate
     $p$ is inclined at 45 degrees.
     Each point is a circle, the orbit of the isometry subgroup generated by
     $\partial_{\phi} = L_+ \overline{\h} -  L_- \underline{\h}$.
       The projection parallel to $\partial_{\phi}$
     of the light cone generators are depicted at several points.
     Curves always lying inside the projected light cones can be
     lifted to causal curves joining a point ($\tau_0$,$p_0$,$\phi_0$)
     to a point($\tau_1$,$p_1$,$\phi_1$),
     with $\phi_0$ not necessarily equal to $\phi_1$ .\label{PenroseRot}}
\end{center}
\end{figure}

The causal structure of the BTZ spaces is obtained by considering
the two fields of directions
\begin{equation}
\frac{d \tau}{d u} = \frac{-( L_+ L_- u \sin(2 \tau) +  2
q(\tau,u) ) \pm (L_+ + L_- \cos(2 \tau)) \sqrt{2 q(\tau,u)}} {2 (2
q(\tau,u) + L_+^2 u^2)},
\end{equation}
with $q(\tau,u)$, defined in eq.(\ref{singularity1}), positive in
the causally safe region. These direction fields result from a
projection on the tangent subspace, at each point of the $(\tau,\
p)$ coordinate plane, of the light cone generators, parallelly to
the $\partial_{\phi}$ direction. Note that as the vector field
$\partial_\phi$ is not orthogonal to the $(\tau, p)$ coordinate
surface, they are not simply given by the intersections of the
light cones with this plane.

We would like to emphasize that causal curves in the BTZ geometry
are projected on the $(\tau, p)$ coordinate plane onto curves that
never leave the projected cones. Hence, this
diagram
provides the causal structure of rotating BTZ black holes.




 
 

 
\cleardoublepage \chapter{Black holes as quotients of $AdS_l$}\label{ChapBHAdSl}

After the striking observation that black holes could be obtained by performing identifications in $AdS_3$, the natural question arose as to whether a similar construction could be used to construct black holes out of higher-dimensional anti-de Sitter spaces. Some specific cases were first addressed 
in \cite{Aminneborg:1996iz,Holst:1997tm,Banados:1997df,Banados:1998dc}, before the general problem was solved by classifying all possibilities \cite{Madden:2004yc,Figueroa}. The outcome of these analyzes is not only that in higher dimensions no new causally well behaved situations arise, although the variety of isometry subgroups is of course larger, but also that among the generic BTZ black holes, only the non-rotating one has a higher-dimensional counterpart \cite{Holst:1997tm,Figueroa}\footpourmoi{Comprendre la raison precise!! Surtout pourquoi ca marche pour le rotating et pas pour le non-rotating!}\footpourmoi{qu'en est-il de $|J|=M$, extremal ou vacuum? Je pense qu'ils y sont aussi, voir Holst-Peldan, mais il n'y a pas d'horizon stricto sensu pour extremal, et pour la vacuum: $r_-=r_+=0$.}. 

In the three-dimensional case, we have observed that the structure of the non-rotating BTZ black hole was related to the data of solvable subgroups of $\SL=AdS_3$ (see Sect.~\ref{Sect-HorNon}). For higher-dimensional (non-rotating) black holes obtained as discrete quotients of $AdS_l$, $l\geq 3$, we are going to point out that the black hole structure is again closely connected with certain solvable subgroups. Of course, anti-de Sitter spaces are not group manifolds in general. However, their isometry group $SO(2,l-1)$ acts transitively on them, which confers them with a homogeneous (and symmetric) space structure. We will actually show that the closed orbits, in $AdS_l$, of Iwasawa subgroups $AN$ and $A\bar{N}$ of $SO(2,l-1)$, define the black hole singularities, in the same sense as the BTZ ones. This, in particular the existence of event horizons, will be recovered using elementary symmetric space techniques\footpourmoi{Preciser qu'on ne vas pas verifier des trucs comme l'absence de CTC, etc., car cette solution n'est rien d'autre qu'une solution deja connue}.

\section{Anti-de Sitter as symmetric spaces} \label{AdSL}
\footpourmoi{preciser le choix de la decomposition d'Iwasawa + le choix des involutions!! Un choix different ne donne pas les memes orbites fermees, voir texte Laurent!!}
In analogy to the three-dimensional case we already encountered, anti-de Sitter space in $l$-dimensions, can be seen as the surface
\beq \label{CoordAdSl}
  M\equiv u^2+t^2- x^2 - y^2 - x_3^2-\dots-x_{l-1}^2=1
\eeq
embedded in $\R^{2,l-1}$ (we again set the cosmological constant $\Lambda=-1$). It is the maximally symmetric solution of vacuum Einstein's equations in $l$ dimensions with a negative cosmological constant. By construction, its isometry group is $G=SO(2,l-1)$. By letting $H$ being the $SO(1,l-1)$ subgroup of $SO(2,l-1)$ which leaves the vector $(1,0,\dots,0) \in \R^{2,l-1}$ unchanged, $AdS_l$ can be realized as the homogeneous space $G/H$, with elements $[g]$, through the isomorphism
\beq \label{pointAdSl} 
  [g]\to g\cdot
\begin{pmatrix}
1\\0\\\vdots\\0
\end{pmatrix} = \begin{pmatrix}
u\\t\\\vdots\\ x_{l-1}
\end{pmatrix}  ,
\eeq
where the dot means the usual matrix times vector product in $\R^{l+1}$. Let $\mathfrak{g}$ and $\mathfrak{h}$ denote the Lie algebras $\mathfrak{so}(2,l-1)$ and 
$\mathfrak{so}(1,l-1)$ respectively. Elements of $\mathfrak{g}$ are $(n+2) \times (n+2)$ matrices ($n = l-1$)  of the form
\beq
 X = \begin{pmatrix} a & u^T \\
                     u & B \end{pmatrix} \in \mathfrak{g} ,
\eeq
with $a$ and $B$ antisymmetric $2 \times 2$ and $n \times n$ matrices respectively, and $u \in M_{n\times 2}$. With the choice of $H$ here above, the $\mathfrak{so}(1,n)$ subgroup is embedded in $\mathfrak{so}(2,n)$ as
\begin{equation}  \label{eq:mtrH}
\mathfrak{h}=\soun\leadsto
  \begin{pmatrix}
     \begin{matrix}
       0&0\\
       0&0
     \end{matrix}
                       &  \begin{pmatrix}
                     \cdots 0\cdots\\
                \leftarrow v^t\rightarrow
                          \end{pmatrix}\\
    \begin{pmatrix}
       \vdots & \uparrow\\
         0    & v \\
       \vdots & \downarrow
    \end{pmatrix} &  B
  \end{pmatrix},
\end{equation}
where  $v$ is $n\times 1$ and $B$ is a skew symmetric $n\times n$ matrix.
A symmetric space structure (see Sect.~\ref{SectD2}) can be given to $AdS_l$ in the following way.
A complementary space\footnote{Since $G$ and $H$ are semi-simple, the restriction to $\mathfrak{h}$ of the Killing form of $\mathfrak{g}$ is nondegenerate.
From this Killing form, one can define $\mathfrak{q} = \mathfrak{h}^\perp = \{X \in \mathfrak{g} | B(X,H) = 0 \; , \, \forall H \in \mathfrak{h}\}$.
From the non degenerescence, it is clear that $\mathfrak{g} = \mathfrak{q} \oplus \mathfrak{h}$. The relations \re{Incl} can be checked from the ad-invariance of B, $B(ad_X Y, Z) = B(X, ad_Y Z)$.} to $\mathfrak{h}$ in $\mathfrak{g}$, denoted by $\mathfrak{q}$, and such that
\beq  \label{Incl}
[\mathfrak{h},\mathfrak{q}]\subset\mathfrak{q} \quad , \quad [\mathfrak{q},\mathfrak{q}]\subset\mathfrak{h},
\eeq
 is given by
\begin{equation}
\mathfrak{q}\leadsto
 \begin{pmatrix}
     \begin{matrix}
       0&a\\
       -a&0
     \end{matrix}
                       &  \begin{pmatrix}
              \leftarrow w^t\rightarrow \\
                 \cdots 0\cdots\\
                          \end{pmatrix}\\
    \begin{pmatrix}
      \uparrow   & \vdots\\
          w      &  0\\
      \downarrow & \vdots
    \end{pmatrix} & 0
  \end{pmatrix}.
\end{equation}
Therefore the map 
\beq
\sigma=\mbox{id}_{\mathfrak{h}}\oplus(-\mbox{id})_{\mathfrak{q}}
\eeq
 is an involutive automorphism giving rise to a symmetric space decomposition of $\mathfrak{g}$ (see Sect.~\ref{SectD2}):
\beq \label{SymmAdSl}
  \mathfrak{g} = \mathfrak{q} \oplus \mathfrak{h}.
\eeq
As a basis of $\mathfrak{q}$, identified with the tangent space of $M=AdS_{n+1}$ at the identity $[e]$, we will take\footpourmoi{Cela se trouve p379 Laurent}
\beq
 q_0 = E_{1,2} - E_{2,1} \quad , \quad q_i = E_{1,i+2} + E_{i+2,1} \; , \; i=1,\cdots,n \quad,
 \eeq
 where $E_{i,j}$ denotes the $(n+2) \times (n+2)$ matrix with zeros everywhere except for a $1$ at the $i^{\mbox{th}}$ line and $j^{\mbox{th}}$ row.

As a next step, we are going to write an Iwasawa decomposition (see Sect.~\ref{SectD2}) of $\mathfrak{g}$, compatible with the symmetric space decomposition \re{SymmAdSl}\footpourmoi{Unique si on demande la compatibilite?}. If $\mathfrak{h} = \mathfrak{a}_{\mathfrak{h}} \oplus \mathfrak{n}_{\mathfrak{h}} \oplus \mathfrak{k}_{\mathfrak{h}}$ and $\mathfrak{g} = \mathfrak{a} \oplus \mathfrak{n} \oplus \mathfrak{k}$ denote the Iwasawa decompositions of $\mathfrak{h}$ and $\mathfrak{g}$ respectively, these will be such that
\beq\label{ReqDec}
 \mathfrak{a}_{\mathfrak{h}} \subset \mathfrak{a} \quad , \quad \mathfrak{n}_{\mathfrak{h}}\subset \mathfrak{n} \quad , \quad \mathfrak{k}_{\mathfrak{h}} \subset \mathfrak{k},
\eeq 
  where elements of $\mathfrak{h}$ are always understood as embedded in $\mathfrak{g}$. Our choice of Cartan involution $\theta$ will be the following:
\beq
 \theta : \mathfrak{g} \ra \mathfrak{g} : X \ra \theta (X) = - X^T,
\eeq    
giving rise to the Cartan decomposition
\beq
  \mathfrak{g} = \mathfrak{k} \oplus \mathfrak{p}.
\eeq
This choice is such that $[\sigma,\theta]=0$, and ensures that $\theta$ descends to a Cartan involution of $\mathfrak{h}$ (in particular, $\theta(H)  \in \mathfrak{h}$ for all $H\in \mathfrak{k}$. Indeed, if $\theta(H) = H' + Q$, $H'\in \mathfrak{h}$, $Q\in \mathfrak{q}$, then $\theta (\sigma(H)) = H' + Q$ and $\sigma(\theta(H)) = H' - Q$. Since $[\theta,\si]=0$, we have $Q = 0$)\footpourmoi{Voir LaurentJuin p162}.

 The compact part $\mathfrak{k}$ is made
 of  rotations while $\mathfrak{p}$ will contain the boosts. So
\beq\label{PKPK}
  \mathfrak{k}=
\begin{pmatrix}
  \sod\\
&\son
\end{pmatrix} \quad , \quad \mathfrak{p}=
\begin{pmatrix}
 0_{(2)} & u^T\\
u & 0_{(n)}
\end{pmatrix}  ,
\eeq
where elements of $SO(2)$ are represented as
\beq \label{so2}
  \begin{pmatrix}
\cos\mu&\sin\mu\\
-\sin\mu&\cos\mu
\end{pmatrix}
\eeq
and where $u$ is a $n \times 2$ matrix. A common abuse of notation in the text is to identify the angle $\mu$ with the element of $SO(2)$ itself in \re{so2}.
In order to build an Iwasawa decomposition, one has to choose a
maximal abelian
 subalgebra $\mathfrak{a}$ of $\mathfrak{p}$. Since rotations are in $\mathfrak{k}$, they must
  be boosts and the fact that there are only two time-like directions
  restricts
   $\mathfrak{a}$ to a two-dimensional algebra. As a consequence, the real rank of $\mathfrak{so}(2,n)$ is $2$, for all $n \geq 2$.  Up to reparameterization,
   it is thus generated by
 $t\partial_y+ y\partial_t$ and $u\partial_x+x\partial_u$, in terms of the coordinates \re{CoordAdSl}. The corresponding matrices are
\beq\label{A}
   J_1=
\begin{pmatrix}
&0\\
0&0&0&1\\
&0\\
&1
\end{pmatrix}\in\mathfrak{h},
\textrm{ and }
J_2=q_1=
\begin{pmatrix}
0&0&1&0\\
0\\
1\\
0
\end{pmatrix}\in\mathfrak{q}.
\eeq
From here, we have to build root spaces (see \re{RootGen}). In this situation, these are given by
\beq \label{RootPart}
\mathfrak{g}_{(a,b)} = \{ X \in \mathfrak{g} \; | \; [X,J_1] = a J_1 \; \mbox{and} \; [X,J_2] = b J_2\}.
\eeq

From \re{RootPart}, we find (details are worked out in \cite{Laurent})
\begin{equation}
\mathfrak{g}_{(0,0)}\leadsto
\begin{pmatrix}
&&x&0\\
&&0&y\\
x&0\\
0&y\\
&&&& D
\end{pmatrix},
\end{equation}
where $D\in M_{(n-2)\times(n-2)}$ is skew-symmetric,
\begin{subequations}\label{RootNonZero}
\begin{align}
\mathfrak{g}_{(1,0)}&\leadsto W_i=E_{2i}+E_{4i}+E_{i2}-E_{i4},\\
\mathfrak{g}_{(-1,0)}&\leadsto Y_i=-E_{2i}+E_{4i}-E_{i2}-E_{i4},\\
\mathfrak{g}_{(0,1)}&\leadsto V_i=E_{1i}+E_{3i}+E_{i1}-E_{i3},\\
\mathfrak{g}_{(0,-1)}&\leadsto X_i=-E_{1i}+E_{3i}-E_{i1}-E_{i3}
\end{align}
\end{subequations}
with $\dpt{i}{5}{n+2}$ and
\begin{equation}\label{ML}
\mathfrak{g}_{(1,1)}\leadsto M=
\begin{pmatrix}
   0&1&0&-1\\
   -1&0&1&0\\
   0&1&0&-1\\
   -1&0&1&0
\end{pmatrix},
\quad
\mathfrak{g}_{(1,-1)}\leadsto L=
\begin{pmatrix}
   0&1&0&-1\\
   -1&0&-1&0\\
   0&-1&0&1\\
   -1&0&-1&0
\end{pmatrix},
\end{equation}

\begin{equation}\label{NF}
\mathfrak{g}_{(-1,1)}\leadsto N=
\begin{pmatrix}
   0&1&0&1\\
   -1&0&1&0\\
   0&1&0&1\\
   1&0&-1&0
\end{pmatrix},
\quad
\mathfrak{g}_{(-1,-1)}\leadsto F=
\begin{pmatrix}
   0&1&0&1\\
   -1&0&-1&0\\
   0&-1&0&-1\\
   1&0&1&0
\end{pmatrix},
\end{equation}
where the only non-zero part of the last four generators is located in the upper left $4$ by $4$ part. 
Our choice of positivity on the dual space $\mathfrak{a}^*$ is the following: a root $\a = (a,b)$ will be said positive if $a>0$ or $a=0$ and $b>0$. 
This leads to
\begin{equation}
   \mathfrak{n}=\{V_i,W_j,M,L\}.
\end{equation}
This decomposition satisfies the requirements \re{ReqDec}, with $\mathfrak{a}_\mathfrak{h}$ spanned by $J_1$ (see \re{A}), $\mathfrak{k}_\mathfrak{h}=
\begin{pmatrix}
  0_{(2)} \\
&\son
\end{pmatrix}$, and $\mathfrak{n}_\mathfrak{h}$ spanned by $\{W_j, M-L\}$\footpourmoi{v Laurentjuin p163} (note that $\mbox{dim} (\mathfrak{n}) = 2n-2$, while $\mbox{dim} (\mathfrak{n}_{\mathfrak{h}}) = n-1$, and that there are $n-2$ generators $W_i$).
One also gets (see \re{nbarre})
\begin{equation}
   \bar{\mathfrak{n}}=\{Y_i,X_j,N,F\}.
\end{equation}
The analytic connected subgroups of $G$ whose Lie algebras are $\mathfrak{a}$, $\mathfrak{n}$, $\bar{\mathfrak{n}}$ and $\mathfrak{k}$ are denoted by $A$, $N$, $\bar{N}$ and $K$, and a similar notation is used for subgroups of $H$. The subgroups $R\equiv AN$ and $\bar{R} \equiv A \bar{N}$ are solvable subgroups of $G$, sometimes called \ind{Iwasawa subgroups} or \ind{minimal parabolic subgroups}.

\section{Black hole structure and closed orbits of solvable subgroups}

As shown in \cite{Holst:1997tm} for the 4-dimensional case, and in \cite{Banados:1997df, Madden:2004yc,Figueroa} in any dimension, the generalization of the non-rotating BTZ black hole can be obtained by performing identifications in $AdS_l$ along orbits of the Killing vector (up to conjugation)
\beq
  \Xi = r_+(t \p_t + y \p_t).
\eeq
The (chronological) singularities are therefore given by the surface
\beq \label{SingAdSl}
 t^2 - y^2 = 0,
\eeq
while the horizons\footpourmoi{argument Philippe pour ca?} can be shown to be located \cite{Banados:1997df} at 
\beq
 u^2 - x^2 - x^2_3 - \cdots - x_{l-1}^2 = 0.
\eeq 
One may introduce local coordinates $(y_a,\phi)$ in the safe region $||\Xi||^2 > 0$ of anti-de Sitter space in which the black hole structure appears more clearly as (with $x_1 = u$ and $x_2 = x$):
\beq
 x_a &=& \frac{2 \, y_{a-1}}{1 - y^2} \quad , \quad a=1,2,\cdots,l-1 \nn \\
   y &=& \frac{r}{r_+} \sinh(r_+ \phi) \nn \\
   t &=& \frac{r}{r_+} \sinh(r_+ \phi), 
\eeq
with 
\beq
  r = r_+ \frac{1+ y^2}{1-y^2}
\eeq
and $y^2 = \eta_{ab}y^a y^b$ and $\eta_{ab} = \mbox{diag}(-1,1,\cdots,1)$. The ccordinate ranges are $-\infty < \phi < +\infty$ and $-\infty < y^a < +\infty$, with the restriction $-1 < y^2 < 1$. The induced metric has the Kruskal form (see \cite{BHTZ} for the three-dimensional case)
\beq
 ds^2 = \frac{(r + r_+)^2}{r_+^2} dy^a dy^b \eta_{ab} + r^2 d\phi^2.
\eeq
In these coordinates, the identification vector is simply $\Xi = \p \phi$, with $||\Xi||^2 = r^2$, so that the quotient space is obtained by identifying $\phi \sim \phi + 2\pi n$. The Kruskal diagram representing this geometry is nothing other than that of the non-rotating BTZ black hole extended to an arbitrary dimension. As shown in \cite{Banados:1997df}, one may also introduce Schwarzschild coordinates covering the exterior region of the black hole \footpourmoi{et pourquoi pas a l'interieur?? Voir Banados "Constant curvature BH"}.

We are now going to show that the singularities \re{SingAdSl} can be characterized from the symmetric space structure point of view of $AdS_l$. In particular, we will observe that these singularities coincide with the closed orbits of the action of the subgroups $AN$ and $A\bar{N}$ of $G$ on $AdS_l$.
Let us first note that the action of $G$ on $M = AdS_l \simeq G/H$ is realized as
\beq
 \tau : G \times M = G/H : (g,[g']) \ra \tau_g ([g']) = [g g'].
\eeq
The fundamental vector fields for this action are 
\beq
 X^*_{[g]} = \ddto \tau_{\exp(-t X)} ( [g] ), 
\eeq
and can be expressed via the isomorphism \re{pointAdSl} as
\beq \label{FondVect}
   X^*_{[g]} = - X.g.\begin{pmatrix}
1\\0\\\vdots\\0
\end{pmatrix} .
\eeq
This way, the fundamental vector fields for the action of $AN \subset G$ can be found from \re{FondVect}, \re{A}, \re{RootNonZero} and \re{ML}:
\begin{subequations}\label{Gen}
\begin{align}
J_1^*&=-y\partial_t-t\partial_y\\
J_2^*&=-x\partial_u-u\partial_x                                                      \label{eq:Jds}\\
M^*  &=(y-t)\partial_u+(u-x)\partial_t+(y-t)\partial_x+(u-x)\partial_y\\
L^*  &=(y-t)\partial_u+(u+x)\partial_t+(t-y)\partial_x+(u+x)\partial_y\\
W_i^*&=-x_i\partial_t-x_i\partial_y+(y-t)\partial_i\\
V_j^*&=-x_j\partial_u-x_j\partial_x+(x-u)\partial_j \quad ,\quad
i,j=3,\ldots,l-1 \label{eq:Vjs}
\end{align}
\end{subequations}
To determine the closed and open orbits of $AN$ on $AdS_l$, we will use the following result\footpourmoi{LaurentJuin p361}. If $R$ is a Lie subgroup of $G$, the orbit of a point $P \in M=G/H$ under the action of $R$ will be open if and only if
\beq\label{RAction}
 \mbox{Span} \{X^*_P \, | \, X \in R\} = T_P M,
\eeq
i.e. if the fundamental vector fields associated with $R$ span the whole tangent space of $M$ at $P$\footpourmoi{pourquoi peut-on conclure pour toute l'orbite alors qu'on ne regarde que ce qui se passe pres du point $P$ (R?: Laurent p383: espace homogene, donc les espaces tangents en differents points ne sont pas independants?? + je ne pige pas trop les preuves p361}. This is related to the fact that a submanifold is open if and only if it has same dimension as the manifold in which it is embedded.

First consider points in $AdS_l$ with $t-y\neq 0$. At these points, we may see that $J_1^*$, $L^*$ and $M^*$ are three
linearly independent vectors. The vectors $V_i^*$ gives us $l-3$
more. Thus the fundamental vector fields for the action of $AN$ span a $l$-dimensional space at these points, so they belong to an open orbit
of $AN$, according to \re{RAction}.

Next consider points satisfying $t-y=0$. It is clear that, at
these points, the $l$ vectors $J_1^*$, $M^*$, $L^*$ and $W_i^*$
are linearly dependent. Then, there are at most $l-1$ linearly
independent vectors amongst the $2(l-1)$ vectors \eqref{Gen}, thus
the points belong to a closed orbit\footnote{We won't distinguish between closed and non-open here, but a more involved argument can be used to show that the orbit is indeed closed \cite{Laurent}}.

The same can be done with the closed orbits of $A\overline{N}$. The result is that points belong to a closed orbit of $A\overline{N}$ if and only if $t+y=0$. We may re-express this in another way. First note that the $AN$-orbits are trivially $AN$-invariant. So the $K$ part of $[g]=[ank]$ alone fixes the orbit to which $[g]$ belongs. In the explicit parametrization of $K$, we know that the $SO(n)$ part is
  ``killed'' by the quotient with respect to $SO(1,n)$. In definitive, we are left with \emph{at most} one $AN$-orbit for each element in $SO(2)$, the latter being parametrized by an angle $\mu$, see \re{so2}. Finally, the closed orbits of $AN$ and $A\bar{N}$, denoted respectively by ${\cal S}_{AN}$ and ${\cal S}_{A\bar{N}}$ can be written as 
\begin{equation}\label{Sing2}
{\cal S}_{AN} = \pm[AN] \quad , \quad {\cal S}_{A\bar{N}} = \pm[A\bar{N}],
\end{equation}
thus corresponding to $\sin \mu = 0$.

To conclude, we use symmetric space arguments (see Sect.~\ref{SectD2}) to derive an alternative way to show the existence of a black hole, in particular the presence of an event horizon (even if we do not explicitly derive its equation). It is very similar in spirit to the techniques used in the three-dimensional case, adapted to the symmetric space setting. If $E$ is a nilpotent element of $\mathfrak{q} \sim T_{[e]}M$, the set $\{\mbox{Ad}(k)E\}_{k \in K_H}$ is the set of all light-like vectors of $T_{[e]}M$\footpourmoi{Pour n'importe quel $E$, on a tous les rayons lumineux lorsque $k$ varie??}. Then, the light cone at the base point $\vartheta = [e]$ is $\exp_\vartheta( t \mbox{Ad}(k) E) = \exp(t \mbox{Ad}(k)E).\vartheta$, where the exponential in the l.h.s. is the exponential map associated with the symmetric space structure, see Sect.~\ref{AppAffRiem}, while that in the r.h.s. is the usual exponential from a Lie algebra to a Lie group. From these facts and from the isometric action of $G$, the future light cone of an element $[g]$ is obtained as
 \begin{equation}  \label{eq:exprcone}
C^+_{[g]}=\{[g {\rm e}^{-t\mbox{Ad}(k)E}]\}_{%
\begin{subarray}{l}
t\in\R^+\\k\in K_H
\end{subarray}}.
\end{equation}
The denomination ``future'' refers to the fact that it only contains positive $t$. Past light cones correspond to negative $t$.

Let us choose $E=q_0 + q_2$. Then, one may compute $\mbox{Ad}(k)E$, to find
\beq \label{eq:AdkE} 
\mbox{Ad}(k)E=q_0+w_1q_1+\dots+w_{l-1}q_{l-1} \quad , \quad \sum_{i=1}^{l-1} \, w_i^2=1.
\eeq
\footpourmoi{Laurent p392} Let us restrict ourselves to points of $AdS_l$ of the form $P= \kappa.\vartheta$, $\kappa \in K$, which are parametrized by an angle $\mu$, see \re{PKPK} and \re{so2}. 
A light-like geodesic through $P$ is given by
 \begin{equation}
  \kappa \cdot \mbox{e}^{-s\mbox{Ad}(k)E}\cdot\vartheta
\end{equation}
with $k\in SO(l-1)$ and  $s\in\R$.

This geodesic reaches ${\cal S}_{AN}$ and ${\cal S}_{A\bar{N}}$ for
values $s_{AN}$ and $s_{A\overline{N}}$ of the affine parameter,
computed to be
\begin{equation}   \label{eq:tempssingul}
 s_{AN} = \frac{\sin\mu}{\cos\mu - \cos\alpha},\quad \text{and} \quad  s_{A\overline{N}} = \frac{\sin\mu}{\cos\mu + \cos\alpha}
\end{equation}
where we set $\cos\alpha := w_2$, see \re{eq:AdkE} ($-1\leq w_2 \leq 1$).

Because the part $\sin \mu =0$ is ${\cal S}_{AN}$, we may restrict
ourselves to the open connected domain of $AdS_l$ given by $\sin
\mu > 0$. Indeed, we have seen that $\sin\mu=0$ is the equation of
${\cal S}_{AN}$ is the $ANK$ decomposition of an element $g\in G$. In the same way, ${\cal S}_{A\bar{N}}$ is given by $\sin\mu'=0$ in the
$A\overline{N}K$ decomposition.

A point, parametrized by $\mu$, will belong to the interior region of the black hole if, for all $\alpha \in [0,2\pi]$, one of the two solutions \re{eq:tempssingul} is positive. This means that, in any direction, a light ray issued from the point intersects a singularity in the future. The reasoning is similar to the one of Sect.~\ref{Sect-HorNon}. We find that points with $\cos \mu > 0$ belong to the interior region, while points with $\cos \mu < 0$ can escape to infinity. Therefore, this proves the existence of an event horizon, whose one point is given by $\cos \mu=0$. 

\footpourmoi{Demander \`a S. Ross : pourquoi la sgte est-elle donnee par $|| \xi_{AdS_3} ||=0$ pour $AdS_l$? Quels sont les horizons en coordonnees u, t ,x, etc.}
\footpourmoi{Rotating BH as quotients in $AdS_l$ n'existent pas : Figueroa, p.35-36 + Holst-Peldan}
\footpourmoi{Y a-t-il une chance de generaliser cela \`a d'autres espaces?? (lorentziens, solutions des equations d'Einstein)}




 




\cleardoublepage \chapter{Non-formal deformation of BTZ spaces and of the hyperbolic plane}

In this chapter, we will use some of the remarkable geometrical properties of the BTZ black holes put forward in the first chapter to show that they can naturally be embedded in the context of strict deformation quantization. In particular, their foliation into two-dimensional leaves admitting an action of the maximal parabolic subgroup of $\SL$ will make it possible to deform these black holes, in a sense that will be explained in \re{GenDef}. This will be discussed in the second part of this chapter. We will then see that the techniques developed so far could be extended to generalize the construction to a space admitting an action of the whole $\SL$ group : the hyperbolic plane (Sect. \ref{SectHyp}). This chapter assumes some
background material on symplectic and Poisson geometry, as well as on symmetric spaces, as can be found in Appendices \ref{AppSymm} and \ref{PoissSympl}.

\section{Generalities on deformation quantization}\label{GenDef}

\subsection{Weyl quantization and Moyal product}\label{Sect:Moyal-Weyl}

Roughly said, the concept of quantization consists of assigning a quantum system to a classical one. On the classical side, observables are identified with smooth real-valued functions on a Poisson manifold $M$ (the phase space), whose points are the states of the system. These observables form a commutative Poisson algebra $(C^\infty(M), \{.,.\})$. On the quantum side, the observables become linear self-adjoint operators acting on a complex Hilbert space ${\cal H}$ whose rays represent the states. The algebra of operators is now non-commutative, and is endowed with an additional structure provided by the commutator. Furthermore, if $G$ is a symmetry group of the classical system, represented by an action of $G$ by symplectomorphisms on the manifold, then it is required that the observables form a unitary representation of $G$ on ${\cal H}$. In the Heisenberg picture, the states are independent of time, but observables vary. If $A(t)$ is the observable $A$ at time $t$, its evolution is governed by Heisenberg's equation
\begin{equation}\label{Heis}
 \frac{d}{dt} A(t) = \frac{i}{\hbar} [H,A(t)],
\end{equation}
where $H$ is a self-adjoint operator called the Hamiltonian of the system. Comparing \re{Heis} to the classical equation
\begin{equation}\label{ClassEvol}
  \frac{d}{dt} f_t = \{H,f_t\}, 
\end{equation}
where $f_t$ is a function on the evolution space $M \times \R$, we are led to the natural guess that we could obtain quantum mechanics from classical mechanics by means of a map
\begin{equation}
 W_\hbar : C^\infty (M) \lra {\cal B}({\cal H})
\end{equation}
replacing classical observables $f \in C^\infty (M)$ by quantum observables $W_\hbar(f)$ acting on ${\cal H}$. The dynamics will correspond provided 
\begin{equation}
 W_\hbar (\{f,g\}) = \frac{i}{\hbar}[W_\hbar(f) , W_\hbar(g)].
\end{equation}
\footpourtoi{c'est ca Weyl quantiz.?} 
$W_\hbar$ is called the \emph{quantization map}\index{quantization map}, and the general procedure is called \emph{Weyl's quantization}\index{Weyl's quantization}.
\footpourtoi{$W_\hbar$ same as CoursGutt p46?}
Let us consider the quantization of a free particle in $\R^n$. In this case, the Hilbert space consists in the space of square integrable functions on the configuration space $\R^n$,  ${\cal H} = L^2 (\R^n)$. Weyl's quantization map is given by\footnote{At this level, we will simply assume that this integral exists, at least on some suitable subspace of $C^\infty (M)$, see Sect. \ref{FormalDef}}.
\begin{equation}
 (W_\hbar (u)f)(q) = \int_{\R^n \times \R^n} \; e^{\frac{i}{\hbar}(q-q')p'} u\left(\frac{q+q'}{2},p'\right) \, f(q') \, dp' dq' ,  
\end{equation}
\footpourtoi{Weyl correspond \`a un ordering particulier, CoursGuttp46?}
where $\R^{2n}$ is seen as $\R^n \times \R^n = \{q_1,\cdots,q_n,p^1,\cdots,p^n\}= \{q,p\}$, $dp'$ and $dq'$ are some suitable normalization of the Haar measure on $\R^n$, $u \in C^\infty (\R^{2n})$ is a classical observable, and $f\in L^2 (\R^n)$ is a state of ${\cal H}$ on which $W_\hbar (u)$ acts.
One may check for instance that $W_\hbar (q_i)f(q) = q_i f(q)$ and $W_\hbar (p^j)f(q) = \frac{\hbar}{i} \p_{q^j} f(q)$ as expected. The \emph{Moyal-Weyl product}\index{Moyal-Weyl product}, or \emph{Moyal product}\index{Moyal product} in short, is formally defined as 
\begin{equation}
 W_\hbar (u \stm  v) = W_\hbar (u) \circ W_\hbar (v) .
\end{equation}
It can be represented as the following asymptotic expansion:
\begin{equation}\label{DiffMoy}
 u \stm  v = \sum_{n=0}^{\infty} \left( \frac{\hbar}{2i} \right)^n \frac{1}{n!} \, \Omega^{i_1 j_1}\cdots \Omega^{i_n j_n} \p^n_{i_1 \cdots i_n}u \p^n_{j_1 \cdots j_n}v ,
\end{equation}
\footpourtoi{est-ce que ca marche de faire ca?? Resultats? spectre des operateurs?}
where $\Omega$ is the Poisson bi-vector field of $M$, i.e. $\Omega^{ij} \p_i \wedge \p_j=\{.,.\}$. The Moyal product $\stm$ encodes the information about the composition of operators, and so allows the construction of quantities obtained from the operators $W_\hbar (u)$ at the level of the functions $u$. 
\footpourmoi{Gracia-Saz, importance $C^*$ : In QM, the set of possible states is a projective Hilbert space. The observables are self-adjoint (although usually unbounded) operators, forming a non-commutative $C^*$-algebra}

\subsection{Formal Deformation Quantization}\label{FormalDef}

\footpourtoi{Pq g\`en\`eraliser $\R^n$?}

The fundamental idea of the \emph{formal deformation quantization} program \cite{BFFLS}, initiated by Bayen, Flato, Fronsdal, Lichnerowicz and Sternheimer, is to generalize the Moyal-Weyl product to an arbitrary symplectic (or Poisson) manifold. In this context, the framework of quantum mechanics is the same as classical mechanics, observables are the same, and quantization arises as a \emph{deformation} of the algebra of functions on the manifold, from a commutative to a non-commutative one \footpourmoi{voir CoursGutt p47}. The quantum equivalent of the classical evolution relation \re{ClassEvol} 
\footpourtoi{Que peut-on calculer avec ca?}
would then read
\begin{equation}
 \frac{d}{dt} f_t = \{H,f_t\}_*,
\end{equation}
where $\{f,g\}_* = \frac{i}{\hbar} (f \star g - g \star f)$.
\footpourtoi{produit invariant a un sens pour def. formelle?}
Classically, symmetries of the system are represented by a group $G$ acting by symplectomorphisms on $M$,
\footpourtoi{$G = Sp(2n,\R)$?$G=Iso(\R^{2n})$?}
where the action on $C^\infty (M)$ is denoted by $G \times C^\infty (M) \ra C^\infty (M) : (g,u) \ra \a_g(u)$. This leads in this formulation at the quantum level to the requirement that $W_\hbar (\a_g(u)) = U(g)\circ W_\hbar (u) \circ U(g^{-1})$, where $U : G \ra \mbox{End}({\cal H})$ is a unitary representation of $G$. This causes us to impose the following condition:
\begin{equation}\label{GInv}
 \a_g (u \star v) = \a_g(u) \star \a_g(v) \quad , \quad \forall g\in G, \; \forall u,v \in C^\infty (M),
\end{equation}
which is referred to as the \emph{$G$-invariance}\index{G-invariant star product} of the product.

\footpourtoi{pq passer aux series formelles? Gr-Sazp3 Smooth dep in $\hbar$?}
More precisely, in formal deformation quantization, quantum mechanics is formulated in a classical framework as the space of \emph{formal power series} in $\hbar$, with classical observables as coefficients. The set of formal power series in $\hbar$ with coefficients in $C^\infty (M)$ (or in any other algebra) is defined as 
\begin{equation}
    C^\infty (M)[[\hbar]] = \{ \sum_{k=0}^\infty a_k \hbar^k \, \mbox{with} \, a_k \in C^\infty (M)\}.
\end{equation}
For such formal series, the equality $\sum_{k=0}^\infty a_k \hbar^k = \sum_{k=0}^\infty b_k \hbar^k$ means $a_k = b_k$ for all $k$.

A \emph{formal deformation} of a Poisson manifold $(M,\{.,.\}$ is the data of
a
$\C [[\hbar]]$ bilinear product $ \star : C^\infty (M)[[\hbar]] \times C^\infty (M)[[\hbar]] \ra C^\infty (M)[[\hbar]]$ given by 
\begin{equation}\label{DefStar}
 f \star g = \sum_{k=0}^\infty B_k (f,g) \, \hbar^k \quad, \quad \forall f,g \in C^\infty (M),
\end{equation}
where the $B_k$'s are bilinear maps. It has to satisfy the additional properties :
\begin{enumerate}
{\item  Associativity : $ (f \star g) \star h = f \star (g \star h)$,
}
{\item 
 $B_0 (f,g)=f.g$,
}
{\item 
 $B_1 (f,g) - B_1 (g,f)  = \{f,g\}$,
}
{\item  
 $1 \star f = f\star 1 = f$.
}
\end{enumerate}
Moreover, one usually requires the $B_k : C^\infty (M) \times C^\infty (M) \ra C^\infty (M)  $ to be bidifferential operators. The operation \re{DefStar} is called a \emph{star product}\index{star product} on $(M,\{.,.\})$. The first condition is natural since the star product is supposed to keep track of the composition of operators, which is associative, though non-commutative in general. Conditions 2 and 3 express the classical limits : $f \star g_{|\hbar=0} = f.g$ and $\{f,g\}_\star \underset{\hbar \ra 0}{\lra} i \hbar \{f,g\}$. They are usually rephrased through the statement that the star product $\star$ is a deformation of the commutative pointwise product of functions on $C^\infty (M)$ in the direction of the Poisson bracket $\{.,.\}$.
\newline
A question that naturally arose in this context was the following: does any Poisson manifold admit a formal deformation? 
In order to reduce an irrelevant multiplicity of solutions, the problem could be brought down to the study of equivalence classes of such products, where two star products $\star$ and $\star'$ on $C^\infty (M)$ are said to be \emph{equivalent}\index{equivalent star products} iff there exists a linear operator $T : C^\infty (M)[[\hbar] \ra C^\infty (M)[[\hbar]]$ of the form $T f := f + \sum_{i=1}^{\infty} T_i(f) \hbar^i$ such that
\begin{equation}\label{EquivStarProduct}
 f \star' g =  T^{-1} \left(Tf \star Tg \right).
\end{equation}
\footpourmoi{where $T^{-1}$ has to be understood in the sense of formal power series.} With this definition in mind, it can be shown (see e.g. \cite{Cattaneo}) that in any equivalence class of star products, there exists a representative whose first term $B_1$ in the $\hbar$ expansion is skew-symmetric. The subsequent question is then to look for the existence and uniqueness of equivalence classes of star products which are deformations of a given Poisson structure on a smooth manifold $M$. The existence of such products was first proved by De Wilde and Lecomte \cite{DeWildeL} in the symplectic case, where the Poisson structure is defined via a symplectic form. Independently, Fedosov \cite{Fedosov} gave an explicit geometric construction: the star product is obtained by gluing together local expressions obtained via the Moyal formula (see also e.g. \cite{Asakawa:2000bh}). The occurrence of the second de Rham cohomology class of the manifold\footnote{Elements of the $k^{th}$ de Rham cohomology class of a manifold $M$\index{de Rham cohomology}, denoted by $H_{dR}^k(M)$, consist in equivalence classes of closed differential forms of degree $k$, where two forms belong to the same equivalence class if they differ by the addition of an exact form. Recall that a form $\a$ is exact iff $d \a = 0$, and closed if there exists $\b$ such that $\a = d\b$.} in the classification problem can be traced back to \cite{BFFLS} and has been clarified in 
\footpourtoi{signifie quoi $H_{dR}^k(M)[[\hbar]]$?}
subsequent works by different authors \cite{Xu, Deligne, Bonneau, GuttTh, BertelsonCG,NestTs1,NestTs2,NestTs3}, until it came out that equivalence classes of star products on a symplectic manifold are in one-to-one correspondence with elements in $H_{dR}^2(M)[[\hbar]]$. The problem for general Poisson manifolds was longstanding (see e.g. \cite{Melanie} and references therein) until the mid-nineties, when Kontsevich \cite{Kontsevich:1997vb} gave an explicit recipe for the construction of a star product starting from \emph{any} Poisson structure on $\R^d$. This formula can thus be used to define locally a star product on \emph{any} Poisson manifold. The local expressions can be once again glued together to obtain a global star product. We will not enter into the details of this construction, but refer to useful reviews like e.g. \cite{Cattaneo,Felder,Gracia-Saz,Yakimov}
\footpourmoi{Tamassia p17:
Kontsevich showed that order by order in $\hbar$ it is always possible to modify Moyal
product in order to make extra terms vanish when the Jacobi identity for the
coordinates is valid. So Kontsevich's  product is uniquely defined at any order
in the deformation parameter $\hbar$ by the requirement of associativity.}. 






 







\subsection{Strict WKB quantization} \label{SectWKB}
In the context of formal deformation quantization, one does not worry about the convergence of the series \re{DefStar}.
However, situations exist where they appear to be convergent. The most popular example is provided by the Moyal-Weyl product, introduced in Sect. \ref{Sect:Moyal-Weyl}, which has the following integral representation:
\begin{equation}\label{IntMoy}
 (u \stm v)(x) = \hbar^{-2n} \int_{\R^{2n}\times \R^{2n}} \, u(y) v(z) \, e^{-\frac{2i}{\hbar} S^0(x,y,z)} dy dz,  
\end{equation}
where
\begin{equation}
  S^0(x,y,z)  =  \omega (x,y) + \omega (y,z) + \omega (z,x), 
\end{equation}
$\omega$ denoting the symplectic two-form on $\R^{2n}$ (i.e. $\omega (x,y) = x \wedge y$ when $x$ and $y$ are seen as vectors of $\R^{2n}$).
Eq. \re{IntMoy} is sometimes referred to as the \emph{integral Moyal product}\index{integral Moyal product}, while \re{DiffMoy} is called the \emph{differential Moyal product}\index{differential Moyal product}. An asymptotic expansion in $\hbar$ of \re{IntMoy} (for example by setting $\hbar = \lambdabar^2$, $\xi^i = \frac{\sqrt{2}}{\lambdabar} x^i$, etc., and Taylor expanding around $\lambdabar =0$) allows us to reinterpret \re{IntMoy} and extend\footpourmoi{voir Pierrestrict p275:extend by $\C[[\hbar]]$-linearity, voir aussi Gracia-Saz} it into a formal star product\footnote{the appellation "formal" star product will now be used in contrast with "convergent" star product, referring to an integral product of the form \re{IntMoy}} on $C^\infty (\R^{2n})[[\hbar]]$.

 The product represented by \re{IntMoy} enjoys an important property: it is internal on the Schwartz space\footnote{i.e. the space of rapidly decreasing functions} on $\R^{2n}$\cite{Hansen}. Thus, the product of two Schwartz functions is again a Schwartz function. This is the key feature of \emph{strict deformation quantization}\index{strict deformation quantization}, where one deals with \emph{convergent star products}\index{convergent star products}. In this context, and in a suitable functional framework, the product of two functions is again a function, \emph{rather than a formal power series in $\hbar$}.

The precise definition of strict deformation quantization has been introduced by Rieffel \cite{Rieffel1} and gives a setting of deformation quantization which is compatible with \emph{$C^*$-algebras}\index{$C^*$-algebra}\footnote{A $C^*$-algebra is
\begin{itemize}
\item a Banach algebra, i.e. an associative algebra $A$ over the real or complex numbers which at the same time is also a Banach space (i.e. a complete normed vector space). The algebra multiplication and the Banach space norm are required to be related by the following inequality: $ \forall x, y \in A , \|x \, y\| \ \leq \|x \| \, \| y\|$. This ensures that the multiplication operation in $A$ is continuous;
\item a $^*$-algebra, i.e. there exists an antilinear map $^* : A \ra A$ such that 1. $(x + y)^* = x^* + y^*$ for all $x, y \in A$; 2. $(\lambda x)^* = \bar{\lambda} x^*$, $\forall \lambda \in \C$, $\forall x \in A$; 3. $(xy)^* = y^* x^*$, $\forall x, y \in A$; 4. $(x^*)^* = x$,  $\forall x \in A$; 5. $||x^*|| = ||x||$, i.e., the involution is compatible with the norm.
\end{itemize}
It has to satisfy the $C^*$ condition:
\begin{itemize}
\item $||x^*x||=||x||^2$.
\end{itemize}
If all the conditions but the completeness are fulfilled, we have a pre $C^*$-algebra, that can be completed to a $C^*$-algebra by adding the mising elements. Notice that in quantum mechanics, the observables are self-adjoint (although usually unbounded) operators, forming a non-commutative $C^*$-algebra.
}.

Let $(M,\{.,.\})$ be a Poisson manifold, and let ${\cal A}(M)$ be a Poisson $^*$-subalgebra of $C^\infty(M)$. If $M$ is compact, then one may choose ${\cal A}(M) = C^\infty(M)$. If M is not compact, the situation is less clear, and the definition is then formulated in terms of any fixed $^*$-
subalgebra of the algebra of $C^\infty$ functions vanishing at infinity, containing the algebra of compactly supported functions and closed under taking Poisson brackets. A \emph{strict deformation quantization} of $M$ in the direction of $\{.,.\}$ is an open interval $I$ of real numbers containing $0$, together with, for each $\hbar \in I$, a, associative product $\star_\hbar$, an involution $^{*_\hbar}$, and a $C^*$-norm $||.||_\hbar$ (for $\star_\hbar$ and $^{*_\hbar}$) on ${\cal A}$, which for $\hbar =0$ are the original pointwise product, complex conjugation involution, and supremum norm, such that
\begin{enumerate}
\item[(i)] For every $f\in {\cal A}$, the function $\hbar \ra ||f||_\hbar$ is continuous;
\item[(ii)] For every $f,g \in {\cal A}$, $||(f \star_\hbar g - g \star_\hbar f)/i \hbar - \{f,g\}||_\hbar$ converges to $0$ as $\hbar \ra 0$.
\end{enumerate} 
In strict deformation quantization, the parameter $\hbar$ parametrizes a continuous field of $C^*$ algebras (after taking the $C^*$ closure). The second condition formalizes the idea of deforming in the direction of $\{.,.\}$. Since this condition is essentially an infinitesimal condition at $0$, one does not expect strict deformation quantizations for a given Poisson bracket to be unique. 

The notion of invariance can also be implemented in this picture as follows. Let $G$ be a Lie group, and let $\tau$ be an action of $G$ on $(M,\{.,.\})$ by symplectomorphisms. Assume further that the corresponding action $\a$ of $G$ on $C^\infty (M)$ carries ${\cal A}$ into itself. We will say that a strict deformation of ${\cal A}$ is \emph{invariant} under $\a$ if 
\begin{enumerate}
{\item For every $\hbar \in I$, and $\forall g \in G$, the operator $\a_g$ on ${\cal A}$ is an isometric $\star$-automorphism for $\star_\hbar$, $^{*_\hbar}$ and $||.||_\hbar$;}
{\item $\forall f \in {\cal A}$ and $\hbar \in I$, the map $g \ra \a_g (f)$ is a $C^\infty$ function on $G$, for the norm $||.||_\hbar$;}
{\item There is an action $\a$ of the Lie algebra $\mathfrak{g}$ of $G$ on ${\cal A}$, which for each $\hbar \in I$ is by $\star$-derivations of ${\cal A}$ for $\star_\hbar$ and $^{*_\hbar}$, such that $\forall X\in \mathfrak{g}$, $\forall f\in {\cal A}$,
\begin{equation}\label{ActionINF}
 \a_X (f) = \ddto \a_{\exp(tX)}(f).
\end{equation}}
\end{enumerate}
The third condition is nothing other than the infinitesimal version of the expression of the $G$-invariance \re{GInv} of the product.
\footpourtoi{La forme \re{MultGen} est-elle restrictive? Et WKB?}

On a symplectic manifold $(M,\omega)$, it is natural to express a given (continuous) multiplication, $\star$, on functions via a kernel formula of the type
\begin{equation}\label{MultGen}
 (u \star v) (x) = \int_{M\times M} \, K(x,y,z) \, u(y) v(z) \, dy dz \quad,
\end{equation}
for $u,v \in Fun(M)$, $x,y,z \in M$, and where $dy$ and $dz$ represent the Liouville measure\footnote{Every symplectic manifold of dimension $2n$ is canonically oriented and comes with a canonical measure, the \emph{Liouville measure}\index{Liouville measure} (normalized to be $\omega^n / n!$).} on $M$. We will consider kernels of the form
\begin{equation}\label{noyauWKB}
 K = a_\hbar \; e^{\frac{i}{\hbar} S},
\end{equation}
where $S$ is a real-valued smooth function on $M\times M\times M$ called the \emph{phase}\index{phase}, and $a_\hbar$ is a one-parameter family of functions in $C^\infty (M\times M\times M, \R^+)$ called the \emph{amplitude}\index{amplitude} (in formal deformation quantization, the amplitude is generally expressed as a formal power series $a_0 + \hbar a_1 + \cdots$). Convergent star products \re{MultGen} on $M$, with kernels of the form \re{noyauWKB}, are called \emph{strict WKB quantizations}\index{strict WKB quantization}.
\footpourtoi{Inv. de Moyal?$Sp(n,\R) \times /\cap\R^{2n}$?}
 In particular, we observe the the Moyal product \re{IntMoy} constitutes a particular strict WKB quantization, characterized by a unit amplitude. 
 Kernels of the form \re{noyauWKB} were apparently first introduced independently by Karasev, Weinstein and Zakrzewski (see \cite{WeinsteinTr}), where the amplitude was in general represented by a power series in $\hbar$.
\footpourtoi{le plan hyperb tombe dans la categorie des espaces de Weintein?? Que montre(ra)-t-on?} 
WKB-quantization of symplectic symmetric spaces (more precisely, hermitian symmetric space of the non-compact type) has been investigated by Weinstein \cite{WeinsteinTr}, who showed that the phase function $S$ actually carries non-trivial geometric information, and can be interpreted as the symplectic area of some triangle in $M$.
 
In strict WKB quantization, when $(M,\omega)$ admits an action $\tau$ of a Lie group $G$ by symplectomorphisms, the $G$-invariance \re{GInv} is expressed through the invariance of the kernel under the diagonal action of $G$ on $M\times M\times M$:
\begin{equation}
 K(x,y,z) = K(\tau_g(x),\tau_g(y),\tau_g(z)) \quad , \quad \forall g\in G, \forall x,y,z \in M.
\end{equation}
A more precise definition of a $G$-invariant WKB quantization (of a symplectic symmetric space) can be found e.g. in \cite{PierreStrict}.
In what follows, we will essentially focus on explicit integral formulae for WKB quantizations, and assume that they make sense on a well-chosen function space containing the smooth compactly supported functions, on which a deformed norm and involution can be defined.

Let us point out that contrary to
the current status of formal deformation quantization, where there exists a good understanding of the classification and existence of star products,
the situation is much less obvious in strict deformation quantization (see \cite{Rieffel:1997cf}). Actually, the notion of a strict deformation quantization should be regarded as a notion of integrability for a formal solution \cite{connes-2006-}. Then, we have the following type of phenomenon: on the one hand, we have formal solutions, formal deformation quantizations about which a lot is known, but for which, in general, there may not be an integrability result. On the other hand, when we try to pass from formal to actual solutions, there are cases where existence fails, and others where uniqueness fails.
Strict deformation theory in the WKB context was initiated by Rieffel in \cite{Rieffel2} for manifolds admitting an action of an abelian Lie group.
The extension to non-abelian group actions was initiated in \cite{GZh} through the introduction of \emph{universal deformation formulae}\index{universal deformation formula}, where the
case of the group of affine transformations of the real line
 was explicitly described. In the strict (non-formal) setting, universal deformation formulae for Iwasawa subgroups of $SU(1,n)$ have been explicitly given in \cite{BielMassar,BielMassar2}. These were obtained by adapting a method developed in the symmetric space framework in \cite{PierreStrict}. Remarkably, the latter results will reveal deep connections with certain geometrical properties of BTZ black holes we uncovered in Chapter \ref{Chap-BTZ}.

 \section{Deformation of BTZ spaces}\label{SectBTZDef}

 \subsection{Solvable group actions and Poisson structures}\label{SectSGA}
We have seen in Sect. \ref{SubSec-ExtendedBTZ} that the non-rotating massive BTZ black hole space-time admits a maximal extension, that we denoted $\EBTZ$. We also learned that $AdS_3$ could be foliated by two-dimensional leaves, which are stable under the BTZ identifications. It turns out that, in the particular extension we considered, each leaf $\rho=cst.$ admits an action of the solvable (or minimal parabolic) subgroup $AN\,$ of $\SL$. Indeed, points in this extension can be represented as 
\begin{equation} \label{CoordGlobExt}
 z(\rho,\phi,w) =
\tau_{r(\phi,w) \,
 \exp(-\pi/4 \t)}(\exp(\rho\h)) \,\, , \,\,r(\phi,w)=\exp(\phi \h)\exp(w \e) \in AN\,, \quad (w,\rho,\phi)\in \R^3, 
 \end{equation}
where $\tau$ is the twisted action of $\SL$ on itself :
 \begin{equation} \label{TwistedActionBis}
\tau : \SL \times \SL \rightarrow \SL: (g,z) \rightarrow \tau_g(z) = g \, z\,
 \sigma(g^{-1}) \quad,
\end{equation}
$\sigma$ being the external automorphism of $\SL$ corresponding to the Lie algebra automorphism (also denoted by $\sigma$) acting on the generators $\h, \e, \f$ as
\begin{equation}\label{defsigBis}
 \sigma(\h)=\h,\ \sigma(\e)= -\e ,\ \sigma(\f) = -\f
\qquad .
\end{equation}  

There is a natural action, which is expressed using the twisted action $\tau$ as
\begin{equation}\label{LeftActionBla}
 \tau : AN\, \times {\cal O}_\rho \ra {\cal O}_\rho : (r',z) \ra \tau_{r'}(z).
\end{equation}
This is a left action, since $\tau_{r_1 r_2} = \tau_{r_1} \circ \tau_{r_2}$.
Since each leaf is ``almost" identified with the $AN\,$ group itself, there also exists a right action
\begin{equation}\label{RightActionBla}
 \psi : AN\, \times {\cal O}_\rho \ra {\cal O}_\rho : (r',z) \ra \psi_{r'}(z) := \tau_{r r' k}(a),
\end{equation}
with $a \in \ca$, $k = \exp(-\pi/4 \t) \in \ck$, satisfying $\psi_{r_1 r_2} = \psi_{r_2} \circ \psi_{r_1}$.

Each leaf can naturally be endowed with two different Poisson brackets coming from symplectic forms, invariant under the left or right actions of $AN\,$. Since the $AN\,$ group (and the leaves) are two-dimensional, these symplectic forms are nothing other than the canonical volume forms. Parameterizing ${\bf r} \in AN\,$ as ${\bf r} = \exp(a/2 \h) \exp(n \e)$, one may identify the left and right invariant forms $\theta$ and $\sigma$ from 
\begin{equation}\label{rdr}
 {\bf r}^{-1} d{\bf r} = \theta^a T_a \quad , \quad d{\bf r} {\bf r}^{-1} = \sigma^a T_a \quad , \quad a=\h,\e \quad,
\end{equation}
as $\sigma^{\h} = da$, $\sigma^{\e} = e^{a} dn$, for the right-invariant ones, and $\theta^{\h} = da$, $\theta^{\e} = dn$ for the left-invariant.
We thus get the left and right invariant (volume) symplectic forms\footnote{The fact that $\omega_L \ne \omega_R$ expresses that $AN\,$ is not unimodular, i.e. that the left and right Haar measures does not coincide. Indeed, $d\mu_L(g) = da dn = \Delta(g) d\mu_R(g) $, with $\Delta (g) = e^{-a}$, called the modular fonction \cite{Barut, Faraut}.}    
\begin{equation}\label{VolumeAN}
  \omega_L = da \wedge dn \quad , \quad \omega_R = e^a da \wedge dn,
\end{equation}
along with their associated invariant Poisson brackets $\{.,.\}_L$ and $\{.,.\}_R$. Each leaf of the foliation can be interpreted as a symplectic manifold on which there is a left or right action by symplectomorphisms of the $AN\,$ group. This will in particular allow us to study strict WKB quantizations of these spaces.

Note that the same reasoning can be applied for the rotating black hole, and that the subsequent constructions can equivalently be performed in this situation as well.
 
 
\footpourtoi{Lien avec la notion de "regular Poisson structure"!}


 
\subsection{Universal deformation formula for Lie group actions}\label{Sect:UDF}
The aim of this section will be to show how it is possible to exploit the property that the BTZ spaces considered above admit a natural action of a solvable group to construct strict WKB deformations of them.
The idea is simple. Let $G$ be a Lie group acting on a manifold $X$ by a left action $\tau$
\begin{equation}
 \tau : G \times X \ra X : (g,x) \ra \tau_g(x) := g.x \quad , \quad \tau_{gh} = \tau_g \circ \tau_h,
\end{equation}
and let $\a$ be the corresponding left action on space of functions on $X$, denoted by ${\rm Fun}(X)$:
\begin{equation}
 \a : G \times {\rm Fun}(X) \ra {\rm Fun}(X) : (g,f) \ra \a_g[f] \quad , \quad \a_{gh} = \a_g \circ \a_h,
\end{equation}
with $\a_g[f] (x) = f(\tau_{g^-1}(x))$, i.e. $\a_g = \tau^*_{g^{-1}}$  (with this definition, $\a$ is indeed a {\itshape left} action).
Suppose that we have on the manifold $G$ a left-invariant star product $\st LG$, satisfying 
\begin{equation} L^\star_g [u \st LG v]=L^\star_g [u]\st LG
         L^\star_g [v] \qquad , \quad \forall u,v \in {\rm Fun}(G),
         \end{equation}
where $L^\star_g [v]$ denotes the regular left action on ${\rm Fun}(G)$:
\begin{equation}
\forall \ g,\,h\in G,\quad L^\star_h [u]({g})=u(h\,g)\qquad .
\end{equation} 
For fixed $x\in X$, we may define a map $\tilde\alpha^x$ from ${\rm Fun}(X)$ into ${\rm Fun}(G)$:
\begin{eqnarray} {\rm Fun}(X)\rightarrow\kern-1.2em^{\tilde\alpha^{x}}\kern0.5em{\rm Fun}(G):
\ f \mapsto \tilde\alpha^x[f]&&\\ \forall\ g\in G\ :\ \tilde\alpha^x[f]( {g})=f(\tau_{{g}^{-1}}( {x})) &&.
\end{eqnarray}

Then, one may induce an associative product on $X$, denoted $\st {\ \ }X$, defined as :
         \begin{equation}\label{babel}
         \left(u \st {\ \ }X v\right)({x}):=\left(\tilde\alpha^{
         x}[u]\st LG \tilde\alpha^{ x}[v]\right)(e)\qquad, \forall u,v \in {\rm Fun}(X),
         \end{equation}
         $e$ denoting the identity element of $G$.
If $K^G$ denotes the left-invariant kernel of the star product on $G$, \re{babel} can be rewritten as
\begin{equation}\label{babel2}
\left(u \st {\ \ }X v\right)({x}) = \int_{G\times G} \, K^G(e,g,h) \, \a_g[u](x) \a_h[v](x) \, dg dh \quad,
\end{equation}
where $dg dh$ denotes a left-invariant Haar measure on $G \times G$.
Formula \re{babel} (or \re{babel2}) is called a \ind{universal deformation formula}, as introduced in \cite{Rieffel2,Rieffel:1997cf,GZh,BielMassar, BielMassar2,BielBonneau,BielBonneau2,PierreStrict} (see also these references for aspects touching the definition of the right functional framework).  
\footpourtoi{ATT:c'est associatif, mais faut tjs verif. que ca deforme le bon crochet!} It is universal in the sense it is valid for any action of $G$.
It is easy to check that \re{babel} is indeed associative, using the fact that $\st LG$ is associative and left-invariant.

Remark that, by using the inversion map
         \begin{equation} G\times X\rightarrow\kern-1em^i\kern0.5em X\ : \  g\mapsto
         i( {g}):={g^{-1}}\qquad ,
         \end{equation} and its induced action on ${\rm Fun}(G)$:
         \begin{equation}
         \forall g\in G,\quad i^\star [u]( {g}):=u({g^{-1}})\qquad ,
         \end{equation} we may construct a right invariant star product on $G$ starting
         from the left one:
         \begin{equation} u\st RG v:=i^\star \left[i^\star [u]\st LG i^\star
         [v]\right]\qquad .\label{LtoR}
         \end{equation} The right invariance of $\st RG$ results immediately from the
         left invariance of $\st LG$ and the relations on ${\rm
         Fun}(G)$:
         \begin{equation} R^\star _g \left[i^\star [u]\right]=i^\star
         \left[L^\star _{g^{-1}}[u]\right]\quad,\quad L^\star _g
         \left[i^\star [u]\right]=i^\star \left[R^\star _{{g}^{-1}}[u]\right]\qquad .
         \end{equation} Indeed we obtain:
         \begin{equation} R_g ^\star [u]\st RG
         R_g ^\star [v]=i^\star\left[i^\star\left[R_g
         ^\star[u]\right]\st LG i^\star\left[ R_g
         ^\star[v]\right]\right]=R^\star_g \left[u\st RG
         v\right]\qquad ,
         \end{equation}
where $R^\star_g [v]$ denotes the regular right action on ${\rm Fun}(G)$:
\begin{equation}
\forall \ g,\,h\in G,\quad R^\star_h [u]({g})=u(g\,h)\qquad .
\end{equation}

Now suppose that $X$ admits
both a left action $\tau$ and a right action $\sigma$ of $G$ :
\begin{eqnarray}
 \tau &:& G \times X \rightarrow X : ({g},x) \rightarrow \tau_{{g}} (x) := g.x
 \qquad \mbox{with} \qquad\tau_{{ g\,h}}=\tau_{g}\circ \tau_{h}\qquad
 ,\nonumber\\
\sigma &:& G \times X \rightarrow X : ({g},x) \rightarrow
\sigma_{{g}} (x) := x.g
 \qquad \mbox{with} \qquad\sigma_{{\bf g\,h}}=\sigma_{h}\circ \sigma_{g}\qquad
 ,
\end{eqnarray}
and that these actions commute, $\tau_{g}[\sigma_{h}(x)] =
\sigma_{h}[\tau_{g}(x)]$ (if $X=G$, these actions can simply
be the right and left multiplications on $G$). Denote by $\a$ and $\b$ the corresponding left and right actions on $\rm{Fun}(X)$.
 
Then, the associative product on $X$ induced from a {\itshape left} invariant star
product on $G$, will be {\itshape right} invariant, in the sense that
\beq
\b_h[u \st {\ \ }X v] = \b_h[u] \st {\ \ }X \b_h[v] \quad, \quad \forall u, v \in {\rm Fun} (X), \forall h\in G.
\eeq
Indeed,
\begin{eqnarray}
\b_{h} (u \; \st{}{X}\;  v) (x) &=& (u\; \st{}{X}\; v)
(\sigma_{h^{-1}} (x)) \nonumber \\
    \quad     &=& \left(\tilde\alpha^{x.h^{-1}}[u]\st LG  \; \tilde\alpha^{x.h^{-1}}[v]\right)( e) \nn\\
    \quad     &=& \int \, K^L ({ e},{g_1},{g_2}) \, \tilde\alpha^{x.h^{-1}}[u] (g_1)\, \tilde\alpha^{x.h^{-1}}[v] (g_2) \, dg_1 \, dg_2 \nonumber\\
    \quad     &=& \int \, K^L ({ e},{g_1},{g_2})\,
    u\left(\tau_{g_1^{-1}}(\sigma_{h^{-1}} (x))\right)
    v\left(\tau_{g_2^{-1}}(\sigma_{h^{-1}} (x))\right) \, dg_1 \,
    dg_2 \nonumber\\
    \quad      &=& \int \, K^L ({ e},{g_1},{g_2})\,
    u\left(\sigma_{h^{-1}}(\tau_{g_1^{-1}}(x))\right)
    v\left(\sigma_{h^{-1}}(\tau_{g_2^{-1}} (x))\right) \, dg_1 \,
    dg_2 \nonumber\\
    \quad      &=& \int \, K^L ({ e},{g_1},{g_2})\,
     \beta_h[u]\left(\tau_{g_1^{-1}}(x)\right) \beta_h[v]\left(\tau_{g_2^{-1}}(x)\right) \, dg_1 \,
    dg_2 \nonumber\\
    \quad      &=& \int \, K^L ({ e},{g_1},{g_2})\,
     \left(\tilde\alpha^x\left[\beta_h[u]\right]\right)(g_1) \left(\tilde\alpha^x\left[\beta_h[v]\right]\right)(g_2) \, dg_1 \,
    dg_2 \nonumber\\
    \quad      &=& \left( \beta_h [u] \st {\ \ }X \beta_h [v]\right)(x)
    \end{eqnarray}

This will be useful for our purposes since, on the BTZ spaces we consider, we have such commuting left and right actions.





 \subsection{$AN\,-$invariant product on $AN\,$}\label{ANsurAN}
From the previous section, we know that if we were able to construct an $AN\,$-invariant product on $AN\,$, then we would be in a good position to obtain a WKB quantization of every two-dimensional leaf in the BTZ foliation. 

We therefore start with the general form of a WKB-quantization for $AN\,$, with deformation parameter $\lambdabar$:
\begin{equation}\label{AnsatzAN}
 (u \star v)(x) = \int_{AN\, \times AN\,} \, K(x,y,z) \, dy dz \quad , \quad u,v \in {\rm Fun}(AN\,),
\end{equation}
with
\beq\label{AnsatzANW}
 K(x,y,z) =  A_\lambdabar(x,y,x) \, e^{\frac{i}{\lambdabar} S(x,y,z)}.
\eeq
Introducing coordinates on $AN\,$ through $x(a_x,n_x)=\exp(\frac{a_x}{2}\h)\exp(n_x \e)$, the left invariant measure is simply $dx = da_x dn_x$ (see \re{VolumeAN}). 
We then require the following obvious conditions to be fulfilled by \re{AnsatzAN}:
\begin{enumerate}
\item[(i)] Left invariance : $K(x,y,z) = K(r.x,r.y,r.z)$, for all $r \in AN\,$;
\item[(ii)] Associativity : $(u \star v) \star w = u \star (v \star w)$;
\item[(iii)] Existence of a unit : $u \star 1 = u = 1 \star u$ ;
\item[(iv)] Asymptotic behaviour :  $u \star v = u.v - i \lambdabar \{u,v\} + O(\lambdabar^2)$, where $\{u,v\}(x) = \p_{a_x} u \p_{n_x} v - \p_{n_x} u \p_{a_x} v$ denotes the left-invariant Poisson bracket on $AN\,$.
\end{enumerate}
We supplement these by two "technical" assumptions:
\begin{enumerate}
\item[(v)] Hermiticity : $\overline{u \star v} = \overline{v} \star \overline{u}$;
\item[(vi)] Trace condition : $\int \,(u \star v)(x) \, dx = \int \, k[y,z] u(y) v(z) \, dy dz = \int \, k[z,y] u(y) v(z) \, dy dz  $.
\end{enumerate}
Condition (vi) will not be discussed in its whole generality. Rather, we restrict ourselves to invariant two-point kernels of the form
\footpourtoi{Aurait-on pu prendre une condition de trace plus generale?}
\begin{equation}\label{kKern}
 k[x,y] = \delta(a_x-a_y)\, F(n_x-n_y).
\end{equation}
Implementing these conditions on the kernel \re{AnsatzANW} uniquely fixes its form, up to some computational assumptions. The details are worked out in the Appendix \ref{AppendixAN} to lighten the presentation. The final expressions for the phase and amplitude\footpourmoi{Y a-t-il d'autres produits, cf StarP, eq.(4.67)?} are given by:
\begin{equation}\label{SPhase}
         S(x,y,z)= \left\{ \sinh\left[(a_y-a_x)\right]\, n_z + \sinh\left[(a_z-a_y)\right]\, n_x
         + \sinh\left[(a_x-a_z)\right]\, n_y  \right\}\qquad
         ,
         \end{equation}
and
  \begin{equation}\label{AAmpl}
         A_\lambdabar (x,y,z)=\frac{1}{(2\,\pi\,\lambdabar)^2 }
         \frac{{\cal P}(a_y-a_x){\cal P}(a_x-a_z)}{{\cal P}(a_z-a_y)}\cosh(a_z-a_y)\qquad
         ,
         \end{equation}
\footpourtoi{on pourrait virer hermiticite? ${\cal P}$ positive sur axe reel pour que $T^{-1}$ soit definie?}         
where ${\cal P}$ is a non-vanishing complex function, required by (iii) and (iv) to satisfy ${\cal P}(0)=1$, supplemented by the condition ${\cal P}(x)= \overline{{\cal P}(-x)}$ due to (v). The product constructed from \re{SPhase} and \re{AAmpl} will furthermore satisfy the trace condition $(vi)$ and \re{kKern} with 
\begin{equation} F(n_{y}-n_{z})=\frac{1}{2\,\pi\,\lambdabar} \int
         {\cal P}(a) {\cal P}(-a) {\rm e}^{\frac i{\lambdabar}(n_{y}-n_{z})\sinh
         a}da\qquad .
         \end{equation}
For ${\cal P}(a)=1$, we recover the strict quantization of the symplectic symmetric space $M=(SO(1,1) \times \R^2)/\R$ presented in \cite{PierreStrict}. This product is actually invariant under the transvection group (see App.\ref{AppSymm} for definitions) $G=SO(1,1) \times \R^2$, which contains a subgroup isomorphic to the $AN\,$ subgroup of $\SL$ \cite{BielMassar2}\footpourmoi{Ceci reste-t-il vrai pour tout ${\cal P}$?}.
This is actually the case \emph{for all} products obtained here. Indeed, consider the following action on $AN$ (expressed in usual coordinates, see Appendix \ref{AppendixAN}):
\beq
 (a,n) \ra (a + a_\epsilon, n + n_\epsilon {\rm e}^{-a} + m_\epsilon {\rm e}^{a}),
 \eeq
generated by the vector fields $\xi^a = -\p_a$, $\xi^n = {\rm e}^{-a}  \p_n$, $\xi^m = {\rm e}^{a} \p_n$. The left action of $AN$ amounts to set $m_\epsilon=0$. By introducing $\xi^k = \xi^n + \xi^m$ and $\xi^l = \xi^n - \xi^m$, these vector fields are showed to form a representation of the following algebra:
\beq\label{superman}
 [A,K]=L \quad, \quad [A,L]=K \quad , \quad [K,L]=0,
\eeq 
recognized as the Lie algebra  $\mathfrak{so}(1,1) \times \R^2$. The latter can be seen as a contraction of the $\ls$ Lie algebra (for which the last relation in \re{superman} would be replaced by $[K,L]=A$). It is then straightforward to check that both the phase \re{SPhase} and the amplitude \re{AAmpl} are invariant under the additional transformation generated by $\xi^m$.


For ${\cal P}(a)=\cosh a$, the product satisfies
\begin{equation}\label{StronglyCl}
\int \,(u \star v)(x) \, dx = \int \,(u.v)(x) \, dx .
\end{equation}
A (star) product satisfying this condition is said to be \emph{strongly closed}\index{strongly closed star product}.

Finally, let us note that the products that we have discussed here, denoted $\st LG$, are related to the Moyal-Weyl product \re{IntMoy} via the sequence of transformations\footnote{The origin of this relation will be deepened is Sect. \ref{ANBIS}}:
\begin{equation}\label{ANInter}
 (u\st LG v)=T^{-1}\left[T[u] \stm T[v]\right]\qquad ,
  \end{equation}
where:
\begin{equation}\label{Ttransf}
 T[u](a,\, n):=\frac {1}{2\,\pi\,\lambdabar}\int {\rm e}^{-\frac i{\lambdabar}\xi\,n}{\cal P}(\xi){\rm e}^{\frac i{\lambdabar} \sinh(\xi)v}u(a,\, v)\, dv\, d\xi \qquad ,
\end{equation}
slightly generalizing, by an extra multiplication by the (non-vanishing) complex function ${\cal P}$, a similar transformation
first obtained in ref. \cite{PierreStrict}, and expressing that these products are actually all equivalent (in the sense of Sect. \ref{FormalDef}) to the Moyal-Weyl one.

The relation to the Moyal-Weyl product by the explicit intertwiner $T$ \re{Ttransf} allows us to discuss the functional space on which the products
are defined. We recall, from Sect. \ref{SectWKB}, that the Schwartz space $\cs$ is stable with respect to the Moyal-Weyl star product. The
star product given by \re{ANInter} will consequently define an associative algebra structure on the space $T^{-1}\cs \subset
\cs'$, provided $T[f] \in \cs\ \forall \ f\in \cs$. As the Fourier transform is an isomorphism
of the space $\cs$, this will be the case if the function $\cp(a)$ is $C^\infty$, never zero on the real axis and increases at infinity not faster than a power of $\exp(|a|)$. Unfortunately, the functional space so obtained does not yet contain the constant function $1$.
To overcome this difficulty, we forget, for a while, about the dependence of eq. (\ref{Ttransf}) on the variable $a$ and limit ourselves to the $n$ dependence of the function. The mapping $T^{-1}$: 
\begin{equation}\label{InvTtransf}
T^{-1}[a,\,u](\nu):=\frac {1}{2\,\pi\,\lambdabar}\int {\rm e}^{-\frac i{\lambdabar} \sinh(\xi)\,\nu}\cp^{-1}(\xi) {\rm e}^{\frac i{\lambdabar}\xi\,n}\cosh(\xi)\ u(a,\,n)\, dn\, d\xi \qquad ,
 \end{equation}
is well-defined\footnote{Remark that the positivity of $\cp$ on the real axis is also necessary for the existence of $T^{-1}$.} as a linear
injection of $\cs_{(n)}$ (the Schwartz space of functions of the $n$ variable) into the tempered distribution space $\cs'$ \cite{PierreStrict}, and also as an operator $T^{-1}: \cs' \mapsto \cs'$. Consider now the space $\cb_{(n)}$ of smooth bounded functions with all their derivatives bounded in the variable $n \in \mathbb{R}$. This space can be seen as a subspace of $\cs'$. Defining $\ce_{(n)}=T^{-1}[\cb_{(n)}]$, we obtain a deformed algebra
containing constants. Moreover, because $T$ only affects the $n$-variable, $\ce_{(n)}$ also contains the bounded functions in the $a$-variable.
Another way of implementing the constants in our deformed algebra is to consider the unitalization $\CC \oplus T^{-1} \cs $.

 \subsection{Induced product on BTZ spaces}

We now have at hand all the ingredients needed to construct explicit WKB formulae deforming the algebra of functions on each $\rho = \rho_0$ leaf of the foliation in the extension described in Sect. \ref{SectSGA}. These formulae can of course be equivalently interpreted as deforming the pointwise multiplication of functions on the whole extended domain, denoted by $\widetilde{\cal{U}}$, whose intersection with a $\rho = \rho_0$ leaf gives the regions $I \cup II_R \cup III_R$  (see Sect. \ref{SubSec-ExtendedBTZ}), rather than that on a single leaf.

Let $(\rho,\phi,w)$ be the coordinates of a point in $\widetilde{\cal{U}}$. The left action \re{LeftActionBla} of an element $(a,\,n)$ on this point is given by the mapping:
\begin{equation}\label{Lac}
 (\rho,\, \phi,\,w)\mapsto \tau_{(a,n)}\left((\rho,\, \phi,\,w)\right)=(\rho,\,\phi + a,\,w +n\,\exp(-\phi)):= {\cal F}^\mu_L (x^\n, \a^i)\qquad ,
\end{equation}
while the right action \re{RightActionBla} is given by:
\begin{equation}\label{Rac}
(\rho,\, \phi,\,w)\mapsto \psi_{(a,n)}\left((\rho,\, \phi,\,w)\right)=(\rho,\,\phi + a,\,w\,\exp(-a)+n):= {\cal F}^\mu_R (x^\n, \a^i)\qquad ,
\end{equation} 
with $\{x^\m\} = (\rho,\phi,w)$ and $\{x^i\} = (a,n)$.

Let us start from a left invariant product on the group $AN\,$, defined by a kernel $K^L$, see \re{AnsatzANW}, \re{SPhase}, \re{AAmpl}. The induced product (see eq. (\ref{babel})) on the domain $\widetilde{\cal{U}}$ reads, for all $U,V \in {\rm Fun}(\widetilde{\cal{U}})$:
         \begin{eqnarray} (U \overset{R}{\underset{\widetilde{\cal{U}}}{\star}} V)[\rho,\,\phi,\,w]&
         =&\int
         K^L[-a_1,-a_2,\,n_1,\,n_2]\,U[\rho,\,\phi-a_1,\,w-n_1\,\exp(-\phi
         +a_1)]\nonumber\\ && V[\rho,\,\phi-a_2,\,w-n_2\,\exp(-\phi
         +a_2)]\,da_1\,dn_1\,da_2\,dn_2\qquad ,\\
         &=&\int
         K^L[-\phi+\alpha_1,-\phi+\alpha_2,\,(w-\nu_1)\exp(\alpha_1),\,(w-\nu_2)\exp(\alpha_2)]\,U[\rho,\,\alpha_1,\,\nu_1]\nonumber\\
         &&
         V[\rho,\,\alpha_2,\,\nu_2]\,\exp(\alpha_1)\,d\alpha_1\,d\nu_1\,\exp(\alpha_2)\,d\alpha_2\,d\nu_2\qquad
         .
         \end{eqnarray}
  According to the discussion in Sect. \ref{Sect:UDF}, this product is invariant under the right action (\ref{Rac}) of $AN\,$ on $\bbf U$, hence the notation $\st{R}{\bbf U}$. On the other hand, we saw how to obtain a right invariant product on the group $AN\,$ from the kernel $K^L$ (see \re{LtoR}). Explicitly, it reads
\begin{eqnarray} (u\st{R}{G}v)[a_x,\,n_x]&=&\int
K^L[-a_x+a_y,-a_x+a_z,\,(n_x-n_y)\exp(a_y),\,(n_x-n_z)\exp(a_z)]\nonumber\\
 &&u[a_y,\,n_y]
\,v[a_z,\,n_z]\,\exp[a_y]\,da_y\,dn_y\,\exp[a_z]\,da_z\,dn_z
\end{eqnarray}
From the latter, we may induce a product on $\widetilde{\cal{U}}$:
         \begin{eqnarray} (U\overset{L}{\underset{\widetilde{\cal{U}}}{\star}}V)[\rho,\,\phi,\,w]& =&\int
         K^L[a_1,a_2,-\,n_1\,\exp(a_1),-\,n_2\,\exp(a_2)]\,U[\rho,\,\phi-a_1,\,(\,w-n_1)\exp(a_1)]\nonumber\\
         &&
         V[\rho,\,\phi-a_2,\,(w-n_2)\exp(a_2)]\,\exp(a_1)\,da_1\,dn_1\,\exp(a_2)\,da_2\,dn_2\\
         & =&\int
         K^L[\phi-\alpha_1,\phi-\alpha_2,\,\nu_1-w\exp(\phi-\alpha_1),\,\nu_2-w\exp(\phi-\alpha_2)]\,U[\rho,\,\alpha_1,\,\nu_1]\nonumber\\
         &&
         V[\rho,\,\alpha_2,\,\nu_2]\,d\alpha_1\,d\nu_1\,d\alpha_2\,d\nu_2\qquad ,
         \end{eqnarray}
which will be invariant under the left action of $AN\,$ on $\widetilde{\cal{U}}$.

To complete the story, we have further to ensure that the deformations are performed in the direction of the invariant Poisson brackets on the leaves.
From \re{babel} and from the asymptotic behaviour of $\overset{R/L}{\underset{G}{\star}}$, one gets, for $x\in \widetilde{\cal{U}}$,
\beq 
 (U \overset{L/R}{\underset{\widetilde{\cal{U}}}{\star}} V)(x) = U.V (x) - i\, \lambdabar \, \omega_{R/L}^{ij} \, \frac{\p {\cal F}^\n_{R/L}}{\p \a^i}(x^\m,0)\frac{\p {\cal F}^\gamma_{R/L}}{\p \a^i}(x^\m,0)\, \frac{\p U}{\p x^\n} (x) \frac{\p V}{\p x^\gamma} + O(\lambdabar^2), 
\eeq
where $\omega_{R/L}$ denote the invariant symplectic forms on $AN\,$, eq. \re{VolumeAN}. One thus gets
\beq
(U \overset{L}{\underset{\widetilde{\cal{U}}}{\star}} V)(\rho,w,\phi) = U.V (x) -i\, \lambdabar \, (\p_w U \p_\phi V- \p_\phi U \p_w V)+ O(\lambdabar^2), 
\eeq
and
\beq
(U \overset{R}{\underset{\widetilde{\cal{U}}}{\star}} V)(\rho,w,\phi) = U.V (x) - i\, \lambdabar \, e^{-\phi}(\p_w U \p_\phi V- \p_\phi U \p_w V)+ O(\lambdabar^2), 
\eeq
expressing that the deformations are indeed in the direction of the corresponding invariant Poisson bi-vector fields, tangential to the leaves.

The left invariant product $\overset{L}{\underset{\widetilde{\cal{U}}}{\star}}$ is obviously compatible with the quotient leading to the $\EBTZ$ space. Indeed using the global coordinate system (\ref{metricExt2}) the identification yielding the maximally extended space
         from $\widetilde{\cal{U}}$ reads as:
         \begin{equation}
         (\rho,\, \phi,\,w)\equiv (\rho,\,\phi + 2\pi\,\sqrt{M},\,w)\qquad .
         \end{equation}
         Therefore, if the functions $U$ and $V$ are periodic on $\widetilde{\cal{U}}$,
         in the $\phi$ variable, the star product $U\overset{L}{\underset{\widetilde{\cal{U}}}{\star}} V $ will
         also be periodic, contrary to what happens with the right invariant
         star product.

This allows us to define a deformed product {\itshape at the quotient
         level}, i.e. the star product of two functions $f$ and $g$ on
         $\EBTZ$. To this end we adopt a method of images, that is, we periodically extend
          the function on $\widetilde{\cal{U}}$ by defining
         \begin{equation}
         \tilde{f}(\rho,\phi,w) = \sum_{k \in \Z} f (\rho,\phi + k 2\pi\,
         \sqrt{M},w)
         \end{equation}
         and
         \begin{equation}
         f \underset{\widetilde{BTZ}}{\star} g \quad = \quad \tilde{f} \overset{L}{\underset{\widetilde{\cal{U}}}{\star}}
         \tilde{g} \quad .
         \end{equation}

         To give a precise meaning to this formal sum, one proceeds as
         follows. We denote by $\pi : \bbf S_0 \mapsto F_0 $ the quotient
         map onto the $\EBTZ$ space (see Sect.\ref{SubSec-ExtendedBTZ}). If $C^\infty_c$ is the space of smooth
        compact supported functions on $F_0$, we notice that $\pi
         ^\star (C^\infty_c (F_0)) \subset \ce $
         is the space of smooth $\phi$ periodic functions which, for fixed
         value of $\phi$, have compact support in $w$. Now we observe that
         the mappings $T$ and $T^{-1}$ only involve the $w$ variable,
         $\phi$ being a spectator variable. Knowing that the space
         $\cb_{(\phi,\,w)}$ of two-variable ${(\phi,\,w)}$ functions is
         stable under the Moyal-Weyl product (\ref{IntMoy})\cite{Rieffel2}, we
         find that if $U,\, V\in
         \ce_{(\phi,\,w)}=T^{-1}[\cb_{(\phi,\,w)}]$, $U\overset{L}{\underset{\widetilde{\cal{U}}}{\star}} V$ is
         also in $\ce_{(\phi,\,w)}$. The subalgebra of $\ce_{(\phi,\,w)}$
         generated by $\pi^\star (C^\infty_c(F_0))$ is then
         constituted by $\phi$-periodic functions, owing to the
         $\ca$-invariance of the star product. It is therefore identified
         with a function algebra on $F_0$ and on the $\EBTZ$ space.







 
 \section{Towards a non-formal deformation of the hyperbolic plane}\label{SectHyp}
There is a natural extension of the construction we performed in the previous section, allowing us to tackle another problem, that is, that of finding invariant deformations (in a sense we'll make precise below) of the \ind{hyperbolic plane}, or \ind{Poincar\'e disk}. The motivation to tackle this question is at least threefold. The first one relies on the relevance of the hyperbolic plane to the study of \ind{Riemann surfaces}\footnote{A \ind{Riemann surface} is a complex manifold of complex dimension one. A \ind{complex manifold} of complex dimension $n$ is a manifold which is locally isomorphic to $\C^n$. Furthermore, the atlas of charts to $\C^n$ has to be such that the change of coordinates between charts are holomorphic. A function $f : M \ra N$ between two Riemann surfaces $M$ and $N$ is called \emph{holomorphic}\index{holomorphic function} if for every chart $\phi$ in the atlas of $M$ and every chart $\psi$ in the atlas of $N$, the map $\psi \circ f \circ \phi^{-1}$ is holomorphic (as a function from $\C$ to $\C$) wherever it is defined. The composition of two holomorphic maps is holomorphic. Two Riemann surfaces $M$ and $N$ are called \ind{conformally equivalent} if there exists a bijective holomorphic function from M to N whose inverse is also holomorphic (it turns out that the latter condition is automatic and can therefore be omitted).}, through the \ind{uniformization theorem}. It states that, up to conformal equivalence, there exist only three simply-connected (i.e. with $\pi_1(M)= e$) Riemann surfaces: the Riemann sphere $\C \cup \{\infty\}$, the complex plane $\C$, and the Poincar\'e disk ${\mathbb D} =\{ z \in \C\; | \; |z|<1 \}$. For arbitrary Riemann surfaces, it expresses that every Riemann surface $M$ is conformally equivalent to $\Sigma /{\cal G}$, where $\Sigma$ is the Riemann sphere, the complex plane or the Poincar\'e disk, and ${\cal G}$ is a subgroup of the automorphism group\footnote{An automorphism of a manifold $M$ is a \ind{biholomorphism} $f:M\ra M$, i.e. a holomorphic diffeomorphism (which implies that $f^{-1}$ is holomorphic).} of $\Sigma$ admitting a freely discontinuous action on $M$. In addition, $\pi_1(M) = {\cal G}$\cite{FarkasKra,Guffin}. Also, every connected Riemann surface can be turned into a complete 2-dimensional real Riemannian manifold with constant curvature -1, 0 or 1. The Riemann surfaces with curvature -1 are called hyperbolic, whose Poincar\'e disk is the canonical local model. All surfaces with \ind{genus} $g>1$ (see \re{102p}) fall into this class. Therefore, invariant deformations of the Poincar\'e disk could constitute a step towards the deformation of a general class of Riemann surfaces.\footpourmoi{The Riemann surfaces with curvature 0 are called parabolic; $\C$ and the 2-torus are typical parabolic Riemann surfaces. Finally, the surfaces with curvature +1 are known as elliptic; the Riemann sphere is the only example : Deformations exist??}
\footpourtoi{Peut-etre mettre ca en intro?}
The second motivation is to address the question of the existence of a WKB quantization of certain symplectic symmetric spaces, and to understand the underlying geometrical properties of the oscillatory kernel, in particular the reason for associativity (see \cite{PierreStrict,SymplConn}). The third comes from the interpretation of the hyperbolic plane as an Euclidian (Wick-rotated) version of the $AdS_2$ space, and appearing as symmetric D-brane configurations in the euclidean $AdS_3$ space (see Chap. \ref{ChapWZW}).

To obtain an $\SL$-invariant kernel, one could in principle proceed by direct computation, as we did for $AN\,$-invariant kernels (see Sect.\ref{AppendixAN}), but the computations rapidly become difficult to manage. Instead, we will use a less direct, but probably more efficient approach. This technique, from which we will recover the class of $AN\,$-invariant products we already found, will give us a clue about which path to follow to extend the invariance to $\SL$. The latter $\SL$-invariant products will be obtained by twisting the given $AN\,$-invariant products, using a particular intertwining operator $U$, characterized by a two-point kernel (actually, a distribution on the hyperbolic plane). The latter will be shown to be expressed as a superposition of modes, each of them being solution of a second order differential solution. Solving this equation will give a general class of $\SL$-invariant kernels. Actually, as shown in Appendix \ref{AppUnicity}, any $\SL$-invariant kernel on the hyperbolic plane turns out to be characterized by a kernel satisfying this differential equation. Finally, we focus on a particular product and dwell a little upon its geometrical content. In particular, we analyze the geometry associated with the phase of the kernel and make connections with work by Weinstein on WKB quantizations of hermitian symmetric spaces \cite{WeinsteinTr}, and by Bieliavsky on the importance of a particular class of functions, called \emph{admissible}, in this context \cite{PierreStrict}.

\subsection{Basics of the hyperbolic plane}

The hyperbolic plane has a realization as the right complex half-plane, denoted by $\Pi$, with points $X=x+ i \xi$,
$x>0$ (we use the conventions of \cite{Unt2}). It is related to the maybe more usual representation by the upper half plane via $X \ra i X$.

This space is canonically endowed with a Kahlerian   
structure\footnote{Let $J$ be an \ind{almost complex structure} on a manifold $M$, i.e. a tensor field of type $(1,1)$ such that $J(J X) = -X$ for all $X\in \varkappa(M)$. If $M$ is a complex manifold, it has an associated almost complex structure. A Riemannian metric $g$ on a connected manifold $M$ with almost complex structure $J$ is said to be \ind{hermitian} if $g(JX,JY) = g(X,Y)$, for all $X,Y\in \varkappa(M)$, and to give a \emph{Kahlerian structure}\index{Kahler manifold} $(M,g,J)$ if in addition $\nabla J = 0$, where $\nabla$ is the Levi-Civita connection on $(M,g)$. A Kahler manifold is always a complex manifold (because $J$ is said integrable in this case). From a Kahlerian structure $(M,g,J)$, one may define the \ind{Kahlerian symplectic form} as $\omega (X,Y) = g(X,JY)$. Kahler manifolds are thus complex manifolds which can be thought at the same time of as Riemannian and symplectic manifolds in a natural way.} whose line element is $ds^2 = x^{-2}(dx^2 + d\xi^2)$,
with associated measure $d\mu(X) = x^{-2}dx d\xi$. Let $g \mapsto
[g]$ denote the canonical application from the group $G=\SL$
to P$\SL=\SL/\{\pm 1\}$. The group P$\SL$ acts isometrically on
$\Pi$ via :
\begin{equation}
[g].X = \frac{a X
 -b i}{i c X + d} \quad , \, g= \left(
\begin{array}{cc}
a &b\\
c&d
\end{array}
\right) \in G \,.
\end{equation}
The stabilizer of the point $X=1$ being the subgroup $K=SO(2)$ and
$G$ acting transitively on $\Pi$, we can identify $\Pi$ with the
homogeneous space $G/K$ via $G/K \rightarrow \Pi : gK \rightarrow
[g].1$. Also, as explained in Appendix \ref{AppSymm}, the Iwasawa decomposition $G=AN\, K$ induces a global diffeomorphism between the group manifold $R = AN\,$ and the symmetric space of the non-compact type $G/K$, see eq. \re{RGsurK}, allowing us to identify $G/K = AN\,$. Moreover, $\Pi$ can in particular be endowed with a Riemannian symmetric space structure (see Appendix \ref{AppSymm}). \footpourtoi{et symplectic symmetric space structure aussi??} 
The geodesics of the Riemannian space $\Pi$ correspond to straight lines parallel to the $x$-axis and to the euclidean
circles for which one diameter is the $\xi$-axis. For all $Y=y +i \eta \in \Pi$, the geodesic symmetry at $Y$ in the sense of the Riemannian symmetric space structure of $\Pi$, defined by
\beq\label{GeodSymmRiem}
  s_Y X = \frac{y^2}{X - i \eta} + i \eta 
\eeq
is a global isometry of $\Pi$.
Let us note that another commonly used representation of the hyperbolic plane is to realize it as the unit disk  
\beq
{\mathbb D} =\{ z \in \C\; | \; |z|<1 \}
\eeq
in the complex plane.
The map $\Pi \ra {\mathbb D} : X \ra z = \frac{X-1}{X+1}$ is a biholomorphic map \cite{LocSymmSp}, and the transformations
\beq
 z \ra \frac{\a z + \b}{\bar{\b} z +\bar{\a}} \quad , \quad \a,\b\in \C, \; |\a|^2 - |\b|^2 \ne 0
\eeq
are isometries of ${\mathbb D}$, with the corresponding group $G$ given by 
\beq
 \mbox{SU}(1,1) = \{ \left(
\begin{array}{cc}
\a &\b\\
\bar{\b}&\bar{\a}
\end{array}
\right) \; | \; |\a|^2 - |\b|^2 =1 \},
\eeq
which is isomorphic to $\SL$, e.g. through $g_{\tiny{\mbox{SU}(1,1)}} = S^{-1} \, g_{\tiny{\SL}} \, S$, with $S=\frac{1}{\sqrt{2}}\left(
\begin{array}{cc}
1 & i\\
i& 1
\end{array}
\right)$.






\subsection{Hyperbolic plane as adjoint orbit and global Darboux chart}\label{Sect:Darboux}
Another useful representation will be to realize the hyperbolic plane as the adjoint orbit (we used a similar approach in Sect. \re{Sect:TwistedIwasawa})
\begin{equation}
 \mathcal{O} = Ad(G)T \quad , \quad T \in K,
 \end{equation}
through the map
\beq
\psi : G/K \rightarrow \mathcal{O} : [g] \rightarrow \psi([g])=Ad(g)T,
\eeq
which is a $\Z_2$-covering. The advantage of this representation is the following. It is known (see Appendix \ref{PoissSympl}) that the orbits of a
Lie group $G$ in the dual of its Lie algebra, $\mathfrak{g}^*$, the
so-called co-adjoint orbits, constitute a symplectic manifold. The
symplectic two-form is given by
\begin{equation}
 \omega_x (X^*_x,Y^*_x) \overset{\triangle}= <x,[X,Y]> \, ,
 \end{equation}
 where $x$ belongs to the co-adjoint orbit included in $\mathfrak{g}^*$, $X,Y
 \in \mathfrak{g}$, and $X^*_x,Y^*_x$ are fundamental vector fields
 associated to the co-adjoint action of G on the orbit.
 When $G$ is semi-simple, the case at hand here, the Killing form,
 denoted $B(X,Y)$, is nondegenerate and induces an isomorphism between $\mathfrak{g}$ and
 $\mathfrak{g}^*$ given by
 \begin{equation}
 B' : \mathfrak{g} \rightarrow \mathfrak{g}^* : X \rightarrow B(X,\,.\,) \quad.
 \end{equation}
In that case, we can look at the adjoint orbits of $G$ in $\mathfrak{g}$,
which are isomorphic to the co-adjoint orbits. The symplectic form
is then given by
\begin{equation}\label{omega}
\omega_X (Y^*_x,Z^*_x) \overset{\triangle}= B(X,[Y,Z]) \, ,X,Y,Z
\in \mathfrak{g} \, ,
\end{equation}
 where the fundamental vector fields associated to the adjoint
action of $G$ reads
\begin{equation}
 Y^*_X = \ddto Ad(\exp(-t Y)).X \quad.
 \end{equation}

We also point out two useful global diffeomorphisms. First, we have\footnote{Remark there is a difference of a factor of 2 w.r.t. the parameterization above eq. \re{rdr}.} 
\beq \label{anCoord}
 \R^2 \ra \Pi : (a,l) \ra e^{a \h} e^{n \e}.1 = e^{2 a} - i n e^{2 a}\quad, 
\eeq
and as a consequence, we also have the global diffeomorphism $\phi : \R^2={(a,l)}
\simeq AN = G/K \rightarrow \mathcal{O}$ :
\begin{equation}\label{carte}
\phi : \R^2 \ra \mathcal{O} : (a,l) \rightarrow \phi(a,l) = Ad \left( \exp(a \h)\exp(l \e) \right) T
\quad.
\end{equation}
Note that the Kahlerian symplectic form $x^{-2} dx \wedge d\xi$ in the coordinates \re{anCoord}, called the \ind{Iwasawa coordinates} \cite{BielMassar2}, is simply proportional to the left-invariant volume form $da \wedge dn$ on $AN\,$ (see Sect. \ref{SectSGA}). 
From \re{carte}, the 2-form $\omega$ (\ref{omega}) induces a 2-form $\Omega$ on
$\R^2$. By evaluating $\Omega_{(a,l)} (\partial a,\partial l)
\overset{\triangle}= (\phi^* \omega)_{(a,l)}(\partial a,\partial
l)$, one gets
\begin{equation}\label{GlobDarb}
 \Omega = \phi^* \omega = \frac{1}{2!}\Omega_{ij}\,dx^i \wedge dx^j
 = - 2 B(\f,\e) da \wedge dl = dl \wedge da \quad.
 \end{equation}
Thus $\phi$ is a \ind{global Darboux chart} on the orbit.

The evaluation of \re{GlobDarb} goes as follows. First, note that $(\phi^* \omega)_{(a,l)}(\partial a,\partial l)= \omega_{\phi(a,l)} (\phi_* \p_a ,\phi_* \p_l)$, where $\phi_* \p_a$, $\phi_* \p_l \in T_{\phi(a,l)}{\cal O}$. Then, $(\phi_* \p_a)(f) = \p_a (f \circ \phi) = \ddto (f\circ \phi)(a+t,l)$. Thus
\beq
 (\phi_* \p_a)_{\phi(a,l)} &=& \ddto [\phi(a + t,l)] \nn \\
                           &=& \ddto [Ad (e^{(a + t)\h}e^{l \e})T] \nn \\
                           &=& \ddto [Ad(e^{t\h}) \phi(a,l)] \nn \\
                           &=& -H^*_{\phi(a,l)}.
\eeq                          
Similarly, one finds
\beq
 (\phi_* \p_l)_{\phi(a,l)} = -\left(Ad(e^{a \h}\e)\right)^*_{\phi(a,l)}.
\eeq
Thus, by using \re{omega} and the $Ad$-invariance of the Killing form, one successively gets  
\beq
  \Omega_{(a,l)} (\partial a,\partial l) &=& \omega_{\phi(a,l)}(-H^*_{\phi(a,l)}, -(Ad(e^{a \h}\e)^*_{\phi(a,l)}) \nn \\
                                         &=& B( \phi(a,l),[H,Ad(e^{a \h})\e]) \nn \\
                                         &=& B( \phi(a,l),Ad(e^{a \h}) [\h,\e]) \nn \\
                                         &=& B (Ad(e^{a \h}) Ad(e^{l \e}) T, Ad(e^{a \h}) [\h,\e]) \nn \\
                                         &=& B (T, Ad(e^{-l \e}) 2 \e)\nn \\
                                         &=& -2 B(\f ,\e) \quad , \quad \mbox{because} \; T = \e - \f.
\eeq                           

So, as symplectic manifolds, we have
\begin{equation}
 ({\cal O},\omega) \sim (\R^2 ,\Omega) .
\end{equation}
Furthermore, $G$ acts on ${\cal O}$ by symplectomorphisms, and this action is Hamiltonian, that is
$\forall X \in \mathfrak{g}$,  $\exists \lambda_X \in C^\infty(\mathcal{O})$
such that $\omega(X^*,.)=d\lambda_X$ and
$\{\lambda_X,\lambda_Y\}=\lambda_{[X,Y]}$ (see Appendix \ref{PoissSympl}). The functions $\l_X$ are called (dual) \ind{moment maps}. The corresponding action on $\R^2$ is written as
$\tau = \phi^{-1} \circ Ad \circ \phi$ :
 \begin{equation}\label{ActionSurR2}
 \tau : G \times \R^2 \rightarrow \R^2 : (g,(a,l)) \rightarrow
 \tau_g(a,l) = \phi^{-1}(Ad(g)\phi(a,l)) \quad .
 \end{equation}
The fundamental vector fields associated with this action are given
by
\begin{equation}
X^*_x = \ddto \phi^{-1} [Ad(\exp(-tX) \phi(x))] \quad , \quad x=(a,l)\in
\R^2, X\in \mathfrak{g} \quad.
\end{equation}
We also note that $\tau_g^* \Omega = \Omega, \forall g \in G$, since $\tau^* \Omega = \tau^* \phi^* \omega = (\phi \tau)^* \omega = (\phi \circ \phi^{-1}\circ Ad \circ \phi)^* \omega = \phi^* Ad^* \omega = \phi^* \omega = \Omega$. The second and fourth equality stem from $(\phi \tau)^*(\omega) (X) = \omega(d(\phi \circ \tau)X) = \omega((d\phi \circ d\tau)X)=(\phi^*\omega)(d\tau(X))= [(\tau^* \circ \phi^*)\omega](X)$, and the fifth one from the fact that the $G$ acts by symplectomorphisms on the orbit via the adjoint action.
Consequently, $G$ acts by symplectomorphisms on $(\R^2, \Omega)$, and this action is again Hamiltonian, since the functions defined by
\begin{equation}
\lambda_X (a,l) = -B(X,\phi(a,l))
\end{equation}
satisfy $\Omega(X^*,.)=d\lambda_X$ and
$\{\lambda_X,\lambda_Y\}=\lambda_{[X,Y]}$, where $\{.,.\} = \p_a \wedge \p_l$. Explicitly, they
read
\begin{equation}\label{moments}
\lambda_H=\ell\,;\,\lambda_E=\frac{1}{2}e^{-2a}\,;\,\lambda_F=-\frac{1}{2}e^{2a}(1+\ell^2),
\end{equation}
with the corresponding fundamental vector fields
\begin{equation}\label{fundamental}
 H^* = -\partial_a; \, E^* = -e^{-2a}\partial_l;  \,
 F^*= e^{2a}(l \partial_a - (1+l^2) \partial_l) \quad,
 \end{equation}
 which satisfy $[X,Y]^* = [X^*,Y^*]$, see Sect. \ref{Hamilton}. We remark that one can slightly generalize $\l_F$ and $F^*$, without any consequence (i.e. $\{\lambda_X,\lambda_Y\}=\lambda_{[X,Y]}$ and $[X,Y]^* = [X^*,Y^*]$ are still verified) by writing
\beq \label{FStark}
\lambda_F=-\frac{1}{2}e^{2a}(k+\ell^2) \quad \mbox{and} \quad F^*= e^{2a}(l \partial_a - (k+l^2) \partial_l) \quad, k\in \R^+.
\eeq 
The constant $k$ is related to the radius of the orbit ${\cal O}$ under consideration, i.e. by identifying $\Pi$ with ${\cal O} = Ad(G)(E - k F)$. 
 
Finally, note that if we restrict the action of $G$ on $\mathcal{O}$ or
$\R^2$ to an action of $R=AN \subset G$, we get
\begin{eqnarray}
\tau : R \times \R^2 \rightarrow \R^2 : (r',(a,l)) \rightarrow
\tau_{r'}(a,l) &=& \phi^{-1} \circ Ad(r')Ad(e^{aH} e^{lE})T
\nonumber \\
 &=& \phi^{-1} \circ Ad(r' r)T \,\, , \,\, r=e^{aH} e^{lE} \nonumber \\
 &=& \tau_{r'r}(0,0) \quad. 
 \end{eqnarray}
Thus the action of $R$ on $\R^2$ (or $\mathcal{O}$ via the adjoint
action) can be read from the action of $R$ on itself by left
multiplication.


\subsection{Twisting Moyal-Weyl and $AN\,$-invariant products}\label{ANBIS}


Let us focus on the Hamiltonian action \re{ActionSurR2} of $G=\SL$ on $\R^2$, with the corresponding dual moment maps $\l_X$ given in \re{moments} and \re{FStark}. We write the Moyal product on $(\R^2, \Omega)$ as (see \re{IntMoy}, with $\nu = \frac{\hbar}{2i}$)
\begin{equation}
(u \star^0_\nu v)(x) = -\frac{1}{4\nu^2}\int_{\R^2 \times\R^2}
\,\, u(y) v(z) e^{-\frac{1}{\nu}S^0(x,y,z)} \, dy dz \quad,
\end{equation}
with
\begin{equation}
S^0(x,y,z) = \Omega(x,y) + \Omega(y,z) + \Omega(z,x) \,\, , \,\,
x,y,z \in \R^2 \quad .
\end{equation}
We have the corresponding asymptotic expansion :
\begin{equation}\label{asympt}
 u \star^0_\nu v \sim u.v + \nu \{ u,v \} +
 \underset{k=2}{\overset{\infty}\sum}\frac{\nu^k}{k!} \Omega^{i_1
 j_1}\cdots \Omega^{i_k j_k} \partial_{i_1\cdots i_k}u \partial_{j_1\cdots
 j_k}v \quad,
 \end{equation}
 where\footpourmoi{je suppose $\Omega$ normalis\'ee : en fait elle l'est automatiquement.}
 \begin{equation}
  \{ u,v \} = \partial_a u \partial_l v - \partial_l u \partial_a
  v \quad.
  \end{equation}

The Moyal product on $(\R^2,\Omega)$ enjoys the following
property: 
\begin{equation}\label{gcovah}
[\lambda_X,\lambda_Y]_{\star^0_\nu} \overset{\triangle}= \lambda_X
\star^0_\nu \lambda_Y - \lambda_Y \star^0_\nu \lambda_X = 2 \nu
\{\lambda_X,\lambda_Y\} \, , \, \forall X,Y \in \mathfrak{g} = \ls
\quad.
\end{equation}
This can be checked by a direct computation using \re{asympt} and \re{moments}.
The property \re{gcovah} is called the \ind{$G$-covariance} of the Moyal product (see e.g. \cite{BielMassar2}), and here 
relies on the existence of the global Darboux chart $\phi$.

\footpourtoi{Pq reparle-t-on de series formelles maintenant? + general?} 
The $G$-covariance of a star product on a manifold $M$, here $\R^2$, in general allows us to construct a representation of $\mathfrak{g}$ on the
space $C^{\infty}(M)[[\nu]]$ of formal power series in the
parameter $\nu$ with coefficients in $C^{\infty} (M)$ through:
\begin{equation}
\rho_{\nu} : \mathfrak{g} \rightarrow GL(C^{\infty}(\R^2)[[\nu]]) : X
\rightarrow \rho_{\nu} (X)
\end{equation}
with\label{ad0}
\begin{equation}
\rho_{\nu}(X) : C^{\infty}(\R^2)[[\nu]] \rightarrow
C^{\infty}(\R^2)[[\nu]] : u \rightarrow \rho_{\nu}(X)u :=
\frac{1}{2 \nu}[\lambda_X,u]_{\star^0_\nu} \eqdef \frac{1}{2 \nu} ad_{\star^0_\nu} (\l_X).
\end{equation}
Indeed, $\forall u \in C^{\infty}(\R^2)[[\nu]]$,
\beq 
[\rho_\n (X),\rho_\n (Y)]u &=&  \rho_\n (X)\rho_\n (Y) u -\rho_\n (Y)\rho_\n (X)u \nn \\
                       &=& (\frac{1}{2 \n})^2 \left([\l_X,[\l_Y,u]]_{\star^0_\nu} - [\l_Y,[\l_X,u]]_{\star^0_\nu}   \right) \nn \\
                       &=& (\frac{1}{2 \n})^2 \left[ [\l_X,\l_Y]_{\star^0_\nu},u \right]_{\star^0_\nu} \nn \\
                       &=& \frac{1}{2 \n} \left[ \{\l_X,\l_Y\},u \right]_{\star^0_\nu} \nn \\
                       &=& \frac{1}{2 \n} \left[  \l_{[X,Y]},u \right]_{\star^0_\nu} \nn\\
                       &=& \rho_\n ([X,Y]) u ,
\eeq
where we used the Jacobi identity and the fact that the action is Hamiltonian. Furthermore, we also have that $\forall X \in \ls$, $\rho_\n (X) \in Der(C^{\infty}(\R^2)[[\nu]],\star^0_\nu)$, i.e.
\beq \label{MoyDerDef}
 \rho_\n (X) (u \star^0_\nu v) = \rho_\n (X) u \star^0_\nu v + u \star^0_\nu \rho_\n (X) v \quad , \quad  \forall u \in C^{\infty}(\R^2)[[\nu]].
\eeq
For the $\ls$ generators, one finds
\begin{eqnarray}
\rho_{\nu} (\h) &=& -\partial_a \\
\rho_{\nu}(\e) &=& -\frac{e^{-2a}}{2\nu} \sinh(2\nu\partial_l) \\
\rho_{\nu}(\f) &=&
\frac{1}{2\nu}(e^{2a}\left(-\nu^2\sinh(2\nu\partial_\ell)\partial_a^2+
2\nu\ell\cosh(2\nu\partial_\ell)\partial_a-(k+\ell^2)\sinh(2\nu\partial_\ell)
\right)) \quad \label{lambdaF},
\end{eqnarray}
by using the asymptotic expansion of the Moyal product\footpourmoi{voir
27/09/04} re-expressed as
\begin{equation}\label{MoyalAsympt}
u \star^0_\nu v = \sum_{k=0}^{\infty} \, \frac{\nu^k}{k!}
\sum_{p=0}^{k}\, \partial_a^p \partial_l^{k-p} u \,
\partial_a^{k-p} \partial_l^{p} v \, (-1)^{k-p} \,
\frac{k!}{p!(k-p)!} \quad ,
\end{equation}
as well as the relations
\begin{eqnarray}
(\partial_l)^{2k+1}(l \cosh(2\nu \partial_l)) &=&
(2k+1)(\partial_l)^{2k}\cosh(2\nu \partial_l)
+ l (\partial_l)^{2k+1}\cosh(2\nu \partial_l) \\
(\partial_l)^{2k+1}(l^2 \sinh(2\nu \partial_l)) &=& 
(2k+1)(2k)(\partial_l)^{2k-1}(\sinh(2\nu \partial_l)) + \nn \\ && 2 (2k+1) l
(\partial_l)^{2k}\sinh(2\nu \partial_l) + l^2
(\partial_l)^{2k+1}\sinh(2\nu \partial_l).
\end{eqnarray}

Now, suppose, we are looking for a $G$-invariant product on $\R^2$, denoted by $\#$. It would have to satisfy (see \re{GInv} or \re{ActionINF})
\beq
 X^* (u \# v) = X^* u \# v + u \# X^* v \quad, \quad \forall X \in \ls, \; \forall u, v \in C^{\infty}(\R^2)[[\nu]].
\eeq
From \re{MoyDerDef}, we see that we are on the right track, but that the $\rho_\n$ representation is actually a deformation of the $X^*$ action of $\mathfrak{g}$  on $C^{\infty}(\R^2)[[\nu]]$, since
\begin{equation}
 \rho_\nu(X) \overset{\nu \rightarrow 0}\longrightarrow X^* \quad.
 \end{equation}

Suppose we are given an operator $U \in GL(C^{\infty}(\R^2)[[\eta]])$ such that
\begin{equation}\label{Sol}
U \rho_\nu(X) U^{-1} = X^* \quad, \forall X\in \mathfrak{g} \quad.
\end{equation}

This would allow us to get a $G$-invariant $\#$, by setting
\beq
 u \# v = U( U^{-1}u \star^0_\nu U^{-1}v). 
\eeq
Indeed, 
\beq
 X^* (u \# v) &=& X^* U( U^{-1}u \star^0_\nu U^{-1}v) \nn \\
              &=&  U \rho_\nu(X) (U^{-1}u \star^0_\nu U^{-1}v) \nn \\
              &=& U (\rho_\nu(X) U^{-1}u \star^0_\nu U^{-1}v + U^{-1}u \star^0_\nu \rho_\nu(X) U^{-1}v) \nn \\
              &=& U ( U^{-1} X^* u \star^0_\nu U^{-1}v  + U^{-1}u \star^0_\nu U^{-1} X^* v \nn \\
              &=& X^* u \# v + u \# X^* v.
\eeq

As a first step, let us restrict ourselves to invariance under the $AN\,$ group, i.e. we first look for a
transform $U_1$ such that (\ref{Sol}) holds for $X\in \mathfrak{a} \oplus \mathfrak{n}$. In the next section, we will see how to deal with the missing generator $\f$.
Following \cite{PierreStrict,BielMassar2}, with a slight generalization, we introduce the partial Fourier transform
\begin{equation} \label{F}
F(g)(a,\alpha) := \hat{g}(a, \alpha) = \int e^{-i \alpha u} g(a,
u) du \quad, \quad g(a,l)\in C^{\infty}(\R^2)[[\nu]].
\end{equation}
One gets
\begin{eqnarray}
F(\rho_{\nu}(H)f)(a, \alpha) &=& \partial_{a}
\hat{f} (a ,\alpha) \label{A1}\\
F(\rho_{\nu}(E)f)(a ,\alpha) &=& \frac{e ^{-2 a}}{2 \nu} \sinh (2
\nu i \alpha) \hat{f}(a ,\alpha) \, . \label{A2}
\end{eqnarray}
Defining a new representation of $\mathfrak{g}$ on
$\widehat{C^{\infty}(\R^2)}[[\nu]]$ by
\begin{equation}
\hat{\rho}_{\nu}(X) \hat{g} = F(\rho_{\nu}(X)g) \, ,
\end{equation}
we get the following
\begin{eqnarray}
 \hat{\rho}_{\nu}(H)\hat{f} &=& \partial_{a} \hat{f}  \\
 \hat{\rho}_{\nu}(E) \hat{f} &=&  \frac{e^{-2 a}}{2\nu} \sinh (2 \nu i \alpha) \hat{f} \, .
\end{eqnarray}
This is still not exactly what we need, but it suggests to further define the $\Zx_\nu^\PP$-transform\footpourmoi{$\nu = i
\nu$ facteur de 2?}
\begin{equation} \label{Z}
\Zx_{\nu}^{\PP}[g] (a ,b) = \PP(\nu b) \int e^{
-\frac{i}{2\nu} \sinh (2\nu b) u} g(a ,u) du \quad,
\end{equation}
where $\PP$ is an a priori non-vanishing arbitrary function of one variable. Restrictions on it will appear when imposing additional conditions, like the right classical limit or hermiticity (see \ref{ANsurAN}).
For further use, the inverse transformations are
\begin{eqnarray}
[F^{-1}(g(a,b))](a,l) &=& \frac{1}{2\pi} \int e^{i b l}
g(a, b) db \\
(\Zx_{\nu}^{\PP})^{-1}[g] (a ,l) &=& \frac{1}{2\pi} \int \,
\PP^{-1}(\nu b) e^{ \frac{i}{2\nu} \sinh (2\nu b) l}
\cosh(2\nu b) g(a ,b) db \quad.
\end{eqnarray}
One may now verify that
\begin{equation}
\hat{\rho}_{\nu}(X) (\Zx_{\nu} [f]) =
\Zx_{\nu}[X^* f]  \quad , \quad \forall X \in \mathfrak{a} \oplus \mathfrak{n}.
\end{equation}

Thus, by defining $U_1 = (\Zx_\nu^\PP)^{-1} \circ F$, we finally obtain
\begin{equation}\label{SolAN}
\rho_\nu^1(X) \overset{\triangle}= U_1 \rho_\nu(X) U_1^{-1} = X^*
\quad, \forall X\in \mathfrak{a} \oplus \mathfrak{n} \quad.
\end{equation}
 Setting
  \begin{equation}\label{Starr1}
u \star^1_\nu v = U_1[U_1^{-1}u \star^0_\nu U_1^{-1}v] \quad,
\end{equation}
we consequently obtain the $AN\,$-invariant products we were looking for.
 The new representation
 \begin{equation}
\rho_{\nu}^1 : \mathfrak{g} \rightarrow GL(C^{\infty}(\R^2)[[\nu]]) : X
\rightarrow \rho_{\nu}^1 (X)
\end{equation}
can also be written as
\begin{equation} \label{ad1}
\rho_\nu^1(X) = \frac{1}{2 \nu} [\Lambda_X, . ]_{\star_\nu^1} \overset{\triangle}= \frac{1}{2 \nu}
ad_{\star_\nu^1}(\Lambda_X), \quad \mbox{with} \quad \Lambda_X=U_1
\lambda_X\, ,
\end{equation}
as a direct computation shows. We may also verify that
\begin{equation}\label{Homo1bis}
[\Lambda_X,\Lambda_Y]_{\star^1_\nu}  = 2\nu \Lambda_{[X,Y]} \, , \, \forall X,Y
\in \mathfrak{g} = \ls \quad ,
\end{equation}
and
\begin{equation}\label{Homo1}
[\rho_\nu^1(X),\rho_\nu^1(Y)]=\rho_\nu^1([X,Y]) \, \forall X,Y\in
\mathfrak{g} \quad.
\end{equation}

Comparing \re{Starr1} to (\ref{Ttransf}), we obviously have that $U_1
= T^{-1}$.\footpourmoi{Si on fait tout le truc avec les
transform\'ees Z et F, on trouve que les signes dans les
exponentielles des transform\'ees $T$ et $T^{-1}$ sont
intervertis, mais ca ne change rien (voir 5/10/04).}

This technique gives a natural way to recover the formulae for $AN$-invariant products we obtained in Sect. \ref{ANsurAN}. We essentially worked on $\R^2$, but passing the $AN\,$ group manifold is direct because of \re{anCoord} and because the left-invariant measure on $AN\,$ is nothing other than $da \wedge dl$, and thus coincides with the measure on $\R^2$.

Even though apparently less straightforward, this approach yields an explanation to the intertwining \re{Ttransf} with the Moyal-Weyl product and, even more importantly, allows for a natural extension to enhance the invariances to $\SL$, as we will see in the next section.




\subsection{Twisting again to $\SL$-invariance} \label{SectSL}
In the previous section, we have been able to construct a family of $AN\,$-invariant star products, by twisting the Moyal-Weyl product as
\begin{equation} \label{TTransform}
(u \overset{\PP_{\;\, 1}}{\, \star_\n} v)=T^{-1}\left[T[u]\star^0_\n T[v]\right]
\qquad ,
\end{equation}
with
\begin{equation}
[T u](a,n)= \frac{1}{2\pi} \int \, e^{-i \xi' n}\, \PP(\lambda
\xi')\, e^{\frac{i}{2\lambda}\sinh(2\lambda\xi')t} \, u(a,t) \, dt
d\xi'
\end{equation}
and
\begin{equation}
[T^{-1}u](a,\alpha) = \frac{1}{2\pi} \int \,
e^{-\frac{i}{2\lambda} \sinh(2 \lambda \xi)\alpha}\,
\PP^{-1}(\lambda \xi) \,e^{i\xi n} \, \cosh(2 \lambda \xi) \,
u(a,n) \, dn d\xi \quad .
\end{equation}

Here, $\cal{P}$ is an arbitrary positive
function satisfying $\PP(0) = 1$, and the superscript $\PP$ in \re{TTransform} reminds us that for each suitable choice for the function $\PP$, we get another $AN\,$-invariant product.
\footpourtoi{pas trop restrictif??}
 In what follows, we will restrict ourselves to the form $\PP(x) = (\cosh 2x)^n$. The choice $n=1/2$ leads to a
strongly closed product, see \re{StronglyCl}. The notation $T_n$ will be understood as the intertwiner to the $AN\,$-invariant product corresponding to 
$\PP(x) = (\cosh 2x)^n$. The special choice $T_{1/2}$ will also be denoted by $T^T$.

Remember that our present objective is to construct an $\SL$-invariant associative composition law on the hyperbolic plane, which we successively identified (as a symplectic space), first to the group manifold $AN\,$ endowed with its left-invariant volume form (or Haar measure), then to the space $\R^2$ with its canonical measure. In the following, we will extensively use the identification $AN\, \sim \R^2$ \re{anCoord}, sometimes using an obvious abuse of notation identifying an element $r \in AN\,$ with its coordinates $(a_r,l_r) \in \R^2$. 

The $\SL$-invariant associative operation $\#$ will be of the form
\begin{equation}\label{UTransfTw}
f \# g = U\left(U^{-1}f \star^1_\nu U^{-1}g \right) \quad , \forall f,g \in C^{\infty}(AN\,)[[\nu]]
\end{equation}
where the $U$-transform has to enjoy the two following properties:
\footpourtoi{Preciser la premiere condition!!}
\begin{enumerate}
{\item[(i)] it has to be preserved under the left action of $AN\,$, i.e. it must satisfy $L_r^* \circ U \circ L_{r^-1}^* = U$; this property is referred to as the $AN\,$-equivariance ;}
{\item[(ii)]  $U \rho_\nu^1(F) U^{-1} = F^*$.
 This extends the $AN\,$-invariance to the $G$-invariance, see \re{Sol}.}
\end{enumerate}
\footpourtoi{c'est la forme generale pour $U$? L'invariance de $u$ impose $u(z^{-1}x)$?}
Let us express the $U$-transform in terms of a kernel $u(x,y)$, which is a distribution in whole generality:
\beq
 (U f)(x) = \int_{AN\,} \, u(x,y) f(y) d^L y \quad,
\eeq
where $d^L y$ denotes the left-invariant Haar measure on $AN\,$, which reads $da_y dl_y$ in terms of the coordinates $y = (a_y,l_y)$. Condition $(i)$ forces the kernel to be in the following form:
\begin{equation}\label{conv}
(U f)(x) = (u \times f)(x) \overset{\triangle}= \int_{AN\,}
u(z^{-1}x)f(z) d^Lz \quad = \int_{AN\,} u(z^{-1})f(x z) d^Lz  \quad,
\end{equation}
or in coordinates
\begin{equation}\label{convcoord}
(U f)(a,\alpha) = \int \, u(a-b,\alpha - m e^{-2(a-b)}) \, f(b,m)
\, db dm \quad .
\end{equation}
This is a consequence of the following observation, similar to that which led to \re{KL}. The condition $L_r^* \circ U f = U \circ L_r^* f$, $\forall f \in \rm{Fun}(AN\,)$, $\forall r \in AN\,$, imposes that the kernel $u$ satisfies
\beq\label{Invdeu}
 u(r^{-1}x,y) = u(x,ry) \quad , \quad \forall r,x,y \in AN\,.
\eeq 
This is nothing other than the left-invariance of the kernel $u$. This can be rewritten as $u(L_{\rm{e}^{-t X}} x,y) = u(x, L_{\rm{e}^{t X}} y)$, for all $X \in \mathfrak{a} \oplus \mathfrak{n}$. Let us denote by $\bar{X}$ the right-invariant vector field generating left-translations on $AN\,$ : $\bar{X}_x = \ddto L_{\rm{e}^{t X}}\, (x)$, $\forall x \in AN\,$. The infinitesimal form of \re{Invdeu} is then 
\beq \label{InvdeuInf}
 (\bar{X}_x + \bar{X}_y) u(x,y) = 0 \quad , \quad \forall X \in \mathfrak{a} \oplus \mathfrak{n}.
\eeq
Explicitly, in terms of the coordinates $\R^2 \ra AN\, : (a,l) \ra \rm{e}^{a \h} \rm{e}^{l \e}$, the vector fields read
$\bar{\h}_{(a,l)} = \p_a$ and $\bar{\e}_{(a,l)} = \rm{e}^{-2 a} \p_l$. The condition \re{InvdeuInf} tells us that the kernel $u$ depends on two variables instead of four:
\beq
 u(x,y) = u(a_x,l_x, a_y,l_y) = u(a_x-a_y, l_x - \rm{e}^{-2(a_x-a_y)} l_y) = u (y^{-1} x),
\eeq
and hence \re{conv} and \re{convcoord}.

Let us now turn to condition $(ii)$, and see what it imposes on the kernel $u$. By applying
\begin{equation}\label{cond1}
U \rho_\nu^1(F) U^{-1} = F^* \end{equation} to a function $U f$,
one gets
\begin{equation}\label{lhsrhs}
(U \rho_\nu^1(F))({\bf x}) = (F^* U f)({\bf x}) \quad.
\end{equation}

Furthermore, we know that $\rho_\nu^1(F)f  =
[\Lambda_F,f]_{\star^1_\nu}$, for a function $\Lambda_F = T^{-1}
\lambda_F$, see \re{ad1}. Let us focus on the strongly closed $AN\,$-invariant
product, denoted $\overset{T}{\star}$, corresponding to $\PP(x)
 = (\cosh 2x)^{1/2}$. This is not restrictive, since we may easily pass from one $AN\,$-invariant product to another, through the sequence of transformations
\beq
 f \overset{1}{\star} g = T_{12}^{-1} (T_{12} f \overset{2}{\star} T_{12} g) ,
\eeq 
 where $T_{12} = T_{n_2}^{-1} \circ T_{n_1}$, $\overset{1}{\star}$ and $\overset{2}{\star}$ denoting the products constructed from the Moyal-Weyl one with $T_{n_1}$ and $T_{n_2}$ respectively. With this choice, the l.h.s. of
(\ref{lhsrhs}) can be rewritten (with $u({\bf z}^{-1} {\bf x}) :=
u_{{\bf x}}({\bf z})$) as
\begin{eqnarray}
 (U \rho_\nu^1(F) f)({\bf x}) &=& \int \, u_{{\bf x}}({\bf z}) (\rho_\nu^1(F)_{|{\bf z}}f)({\bf z}) \, d^L {\bf z} \\
 &=& \int \, u_{{\bf x}}({\bf z}) \overset{T}{\star} [\Lambda_F,f]_{\overset{T}{\star}}({\bf z}) \, d^L{\bf z} \\
 &=& \int \, u_{{\bf x}} \overset{T}{\star} \Lambda_F \overset{T}{\star} f -  u_{{\bf x}} \overset{T}{\star} f \overset{T}{\star} \Lambda_F  \nn \\
&=& \int \, u_{{\bf x}} \overset{T}{\star} \Lambda_F \overset{T}{\star} f -  \Lambda_F \overset{T}{\star}  u_{{\bf x}} \overset{T}{\star} f  + \Lambda_F \overset{T}{\star}  u_{{\bf x}} \overset{T}{\star} f -  u_{{\bf x}} \overset{T}{\star} f \overset{T}{\star} \Lambda_F   \nn \\
&=& -\int [\Lambda_F, u_{{\bf x}}]_{\overset{T}{\star}} \overset{T}{\star} f - [\Lambda_F, u_{{\bf x}}\overset{T}{\star} f]_{\overset{T}{\star}} \nn \\   &=& -\int [\Lambda_F, u_{{\bf x}}]_{\overset{T}{\star}} \overset{T}{\star} f \nn\\
   &=& -\int \rho_\nu^1(F)_{|{\bf z}} u_{{\bf x}}({\bf z}) . f({\bf z}) \, d^L{\bf z} \quad, \label{lhs}
   \end{eqnarray}
   where we used property \re{StronglyCl} in the second, sixth and seventh equalities, and where $\rho_\nu^1(F) = (T^T)^{-1} \rho_\nu(F) T^T $,
  while the r.h.s. is simply
  \begin{eqnarray}
  F^*_{|{\bf x}} \, \int \, u_{{\bf x}}({\bf z}) f({\bf z}) \, d^L{\bf z} \quad . \label{rhs}
  \end{eqnarray}
Since (\ref{lhs}) and (\ref{rhs}) are valid $\forall f$, one gets
\begin{equation}\label{eqverif}
- \rho_\nu^1(F)_{|{\bf z}} u({\bf z}^{-1}{\bf x}) = F^*_{|{\bf x}}
u({\bf z}^{-1}{\bf x}) \quad .
\end{equation}

This equation can be expressed using the operator $\Box$ defined
as
\begin{equation}\label{BoxSympa}
\left( (\phi_\nu^*)^{-1} F \rho_\nu(F)f \right)(a,b) =
\Box((\phi_\nu^*)^{-1} F f)(a,b) \quad ,
\end{equation} where
\begin{equation}
\phi_\nu (a,l) = (a,\frac{1}{2 i \nu}\sinh 2 i \nu l) \quad
,
\end{equation}
and
\begin{equation}
F(g)(a,b) = \int e^{-i b l} g(a, l) dl \quad
\end{equation}
 as
\begin{equation}\label{BoxBrut}
- \Box_{|\bar{{\bf z}}} W({\bf x},\bar{{\bf z}}) = F^*_{|{\bf x}}
W({\bf x},\bar{{\bf z}}) \quad ,
\end{equation}
with
\begin{equation}\label{DefW}
W({\bf x},\bar{{\bf z}}) = \left( (\phi_\nu^*)^{-1} \circ F
\circ T^T \right)_{|{\bf z}} u({\bf z}^{-1}{\bf x}) \quad .
\end{equation}
$\bar{{\bf z}}$ here denotes the variable conjugated to ${\bf z}$ after
the $T^T$ and $F$ transforms. In \re{BoxBrut}, $\Box$ is a second order differential operator (see Appendix \ref{App-Box}). Solving equation (\ref{BoxBrut})
yields the convolution kernel of an intertwiner between $\#$ and
the strongly closed $AN\,$-invariant product $\st{T}{}$. The
convolution kernel of an intertwiner between $\#$ and a general
$AN\,$-invariant product, with $\PP = (\cosh 2x)^n$ is obtained from
\begin{equation}\label{fromone}
 U_n = U^T \circ (T^T)^{-1} \circ T_n \quad .
 \end{equation}
The kernel of $U_n$ is then given as a superposition of modes,
\beq \label{unkernel}
 u^n ({\bf x}) = \int \, \phi(s) u^{n}_{\lambda,s}({\bf x}) \, ds \quad , \quad  \lambda = i \nu,
\eeq
with 
 \begin{equation}\label{uGen2n}
u^{n}_{\lambda,s}(a,l) =  e^a \, k_s(4\pi e^{2a})\, \int \,
(1+4 \lambda^2 b^2)^{\frac{n-1}{2}} e^{4 \pi i s \sqrt{1+4
\lambda^2 b^2} e^{2a}} \, e^{i b l e^{2a}} \, db \quad,
\end{equation}
where
\begin{equation}\label{kstexte}
k_s(x)=\sqrt{x}\left (A J_{\frac
{\sqrt{k}}{2\,\lambda}}[\sqrt{s^2+\frac m {(8 \pi \lambda)^2}}\,\,
x] + B J_{-\frac{\sqrt{k}}{2\,\lambda}}[\sqrt{s^2+\frac m {(8 \pi
\lambda)^2}}\,\, x]\right).
\end{equation}
In \re{kstexte}, $J_\l$ represents the Bessel functions of first kind, $A$ and $B$ are constants, $k$ and $m$ are constants corresponding to the constant $k$ in \re{FStark}. The first one is associated with the l.h.s. of \re{eqverif}, while the second one corresponds to the r.h.s. (i.e. $F^*= e^{2a}(l \partial_a - (m+l^2) \partial_l)$ in \re{eqverif}).

The derivation of this result is presented in Appendix \ref{App-Box}. An analog computation yields the kernel of the inverse operator $U^{-1}$ 
\begin{equation}
 [U^{-1}f]({\bf x})=\int \, v({\bf z}^{-1}{\bf x}) \, f({\bf z}) d{\bf z} \quad,
 \end{equation}
again as a superposition of modes, given by 
\begin{equation}\label{vGen2n}
v^{n}_{\lambda,s}(a,l) = e^{-a} \, k_s(4\pi e^{-2a})\, \int
\, (1+ 4 \lambda^2 b^2)^{-n/2} e^{4 \pi i s \sqrt{1+4 \lambda^2
b^2} e^{-2a}} \, e^{-i b l} \, db \quad,
\end{equation}



In Appendix \ref{AppUnicity}, we discuss the unicity of our construction, and show that any $\SL$-invariant associative product on the hyperbolic plane can essentially be written as \re{UTransfTw}, with $U$ given by \re{conv} and \re{unkernel}.    
Note that, given an operator $U$, i.e. by fixing $n$ and $\phi(s)$, it is not straightforward to extract the inverse operator $U^{-1}$ from \re{vGen2n}, since one has to guess how to superpose the different modes. There exists however at least one situation where $U$ and its inverse are known, which we describe in the next section.


\subsection{Playing with triangles}

In this section, we would like
to pay some attention to the non-trivial geometric properties
which underlie the associativity of products in WKB form. To begin with, we essentially follow \cite{PierreStrict} and \cite{SymplConn}, Sect.8.

The starting point is
an observation made by Weinstein in \cite{WeinsteinTr} in the context of WKB quantization of \ind{hermitian symmetric spaces}\footnote{These are defined as follows. Let $M$ be a complex manifold, with associated complex structure $J$. $M$ is said to be a \ind{hermitian manifold} if it is endowed with a \ind{hermitian structure}, i.e. a Riemannian metric $g$ such that $g(X,Y) = g(JX,JY)$, $\forall X,Y \in \varkappa (M)$. Hermitian symmetric spaces are hermitian manifolds whose hermitian structure is compatible with the symmetries. If $M=G/K$, $g$ and $J$ are both $G$-invariant. Moreover, the two-form on $M$ defined by $\omega_x (X,Y)= g_x(JX, Y)$, $\forall x \in M$ and $\forall X,Y \in \varkappa (M)$ is a $G$-invariant symplectic two-form. For hermitian symmetric spaces of the of the non-compact type (see Sect. \ref{SectD2}), the diffeomorphism \re{RGsurK} allows us to express $\omega$ as a symplectic form on $AN\,$, invariant under the left action of $AN\,$ on itself. The Poincar\'e disk is the simplest example of a hermitian symmetric space.} (of the non-compact type) $M=G/K$ (see also Appendix \ref{SectD2}). He provided some evidence\footpourmoi{quelles sont-elles??} showing that the phase function $S$ of a WKB kernel \re{noyauWKB} should be closely related to the three point function denoted hereafter $S_A$ and
defined as follows. Given three points $x$, $y$ and $z$ in the symmetric space $(M,s)$ with the property that the equation $s_X s_Y s_Z(x) = x$ admits a (unique)
solution $x$, the value of $S_A(X, Y, Z)$ is given by the symplectic area
\beq \label{PhaseCan}
S_A(X,Y,Z)=-\int_\Sigma\omega,
\eeq
where $\Sigma$ is the geodesic triangle with vertices $x$, $y=s_X (x)$ and $z = s_Y (y) = s_Y s_X (x)$ ($x = s_Z (z)$). The question of geometrically characterizing the amplitude was left open.

A geometrical approach to this question has been initiated in \cite{PierreStrict},
and is based on the definition of a class of three point functions, called hereafter \emph{admissible}\index{admissible function}, on symplectic symmetric spaces (not necessarily hermitian). Let $(M,\omega,s)$ be a symplectic symmetric space. A $k$-cochain on $M$ is a totally skew symmetric real-valued smooth function $S$ on $M^k$ which is invariant under the diagonal action of the symmetries : $S(x_1,\cdots,x_k) = S(s_y(x_1),\cdots,s_y(x_k))$, $\forall x_1, \cdots,x_k,y \in M$. Denoting by $P^k(M)$, the space of $k$-cochains, one has the cohomology operator $\d : P^k(M) \ra P^{k+1}(M)$ defined as
\beq
 (\d S)(x_0,\cdots,x_k) := \sum_j (-1)^j S(x_0,\cdots,\hat{x_j},\cdots,x_k),
\eeq
where $\hat{}$ stands for omission. A 3-cochain $S \in P^3(M)$ is called \emph{admissible}\index{admissible function} if for all $x \in M$, one has
\beq
S(x,y,z) = -S(x,s_x(y),z) \quad ,\quad \forall y,z\in M.
\eeq
A \ind{Weyl triple} is the data of a symmetric space $M$ endowed with an invariant volume form $\m$ together with the data of an admissible 3-cocycle $S$ (i.e. $\d S = 0$, or explicitly $S(x,y,z)=S(x,y,m) + S(y,z,m) + S(z,x,m)$, for all $x,y,z,m \in M$). It is proven\footpourmoi{vrai, meme si je ne passe pas par "`geometrically associative"'?} in \cite{PierreStrict} that if $(M,S,\m)$ is a Weyl triple, then the product defined by 
\beq
 (u \star v)(x) = \int_{M \times M} \, u(y) v(z) \rm{e}^{i S(x,y,z)} \m(y) \m(z)  
\eeq
is associative. This is actually the geometric content of the associativity of the Moyal-Weyl product (see \cite{PierreStrict} for details).

Finally, suppose we are given an admissible three-point function on a symplectic symmetric space $(M,\omega,s)$. The choice of a base point $o$ in $M$ then determines a two-point function $\si$ on $M\times M$:
\beq
 \si (x,y) = S(o,x,y).
\eeq  
Assuming the existence of a smooth \ind{midpoint map}, i.e. a map $M\ra M : x\ra \frac{x}{2}$ such that $s_{\frac{x}{2}}(o) = x$, it can be shown \cite{PierreStrict} that if $\si : M \times M \ra \R$ is a $Stab(o)$-invariant smooth two-point function (where $Stab(o)$ denotes the group leaving the point $o$ fixed) such that 
\beq \label{sisi1}
 \si(x,y) = -\si(y,x) = -\si(x,s_{\frac{x}{2}}(y)) \quad , \quad \forall x,y \in M,
\eeq
 then the three-point function defined by
 \beq \label{sisi2}
 S(x,y,z) = \si(s_{\frac{x}{2}}(y),s_{\frac{x}{2}}(z))
\eeq 
is admissible. This tells us when a given two-point function comes from an admissible three-point function.

\footpourmoi{cela signifie-t-il que toute fct admissible \`a 3 points est necessairement de cette forme?? Sans doute non!! La seule facon d'etendre $\si$ en une fct admissible est-elle celle-l\`a?, cf.11/06/04. Toute fct a trois points qui se reduit \`a une fct \`a 2 points satisfaisant ces conditions est-elle admissible??}
Given a symplectic symmetric space, Weinstein's function $S_A$ turns out to be admissible (wherever it is well-defined)\cite{PierreStrict}. However, the curvature is the obstruction to the cocyclicity of $S_A$\footpourmoi{pas cocyclique n'implique PAS que le produit sans amplitude, avec seulement la phase, n'est pas associatif. C'est seulement une observation}. Therefore, the associativity in general requires the introduction of a non-trivial amplitude. 
As an example, let us consider the $AN\,$-invariant product we constructed in Sect. \ref{ANsurAN}, in particular the phase \re{SPhase}. It can be interpreted as a WKB quantization of the symplectic symmetric space $(\R^2, da \wedge dl,s)$, where the symmetry is given by 
\beq
 s_{(a,l)}(a',l')=(2a-a',2 \cosh(a-a')l - l').
\eeq
see \cite{PierreStrict,SymplConn,ThesePierre}
(see also Sect. \re{Sect:Darboux}).
It turns out, in this case, that $S_A (a_x,l_x,a_y,l_y,a_z,l_z)$ exactly coincides with the phase \re{SPhase} of the invariant products we constructed \cite{PierreStrict,SymplConn}. 

Let us now turn to the inspection of $\SL$-invariant products on the hyperbolic disk. As we mentioned, even if we have the explicit forms of the modes of the kernels at hand, it is not straightforward to write down the explicit form of the composition law, in order to extract the geometrical information contained in the phase. There exists nevertheless one situation where the explicit form is known. It has been derived in a somewhat different context by A. and J. Unterberger, for the composition of symbols in the so-called Bessel calculus \cite{Unt,Unt2}.

The explicit intertwiner $U_U$ between any
$AN\,$-invariant product (but where $\n$ is fixed to  $\frac{1}{8\pi}$) and $\#$
as
\begin{equation}
U_U = G_\mu^{-1} \circ (T_{n=1})^{-1} \circ T_n \quad ,
\end{equation}
with (see (4.16) in \cite{Unt})
\begin{equation}
(G_\mu^{-1}f)(y + i \eta) =4\pi \int \, f(\frac{y}{t} + i \eta)
J_\mu(4 \pi t) dt \quad
\end{equation}
or in our usual $(a,l)$ coordinates \re{anCoord}
\begin{equation}
(G_\mu^{-1}f)(a,l) =4\pi \int \, f(a - \frac{1}{2}\ln t, n t)
J_\mu(4 \pi t) dt \quad .
\end{equation}
The inverse operator is 
\beq
U_U^{-1} = (T_n)^{-1}  \circ T_{n=1} \circ G_\mu  \  \quad ,
\eeq
with (see (4.7) of \cite{Unt})
\begin{equation}
(G_\mu f)(y + i \eta) =4\pi \int \, f(t y + i \eta)
J_\mu(4 \pi t) dt \quad,
\end{equation}
or
\begin{equation}
(G_\mu f)(a,l) =4\pi \int \, f(a + \frac{1}{2}\ln t, \frac{n}{t})
J_\mu(4 \pi t) dt \quad .
\end{equation}

The reason for the appearance of the particular $T_{n=1}$ lies in the fact that, in \cite{Unt}, they exclusively work with a given $AN\,$-invariant product, corresponding
 $\PP = \cosh 2x$, and referred to as Fuchs' product, but we are free to use
a relation similar to \re{fromone} if we wish to start from another $AN\,$-invariant product.

By noting that
\begin{equation}
 ((T_2)^{-1} \circ T_1)f(a,\alpha) = \int \, (1+4 \lambda^2
 p^2)^{\frac{n_1 - n_2}{2}} \, e^{i p (t-\alpha)} \, f(a,t) dt dp
 \quad
 \end{equation}
 we get the convolution kernels for $U_U$ and $U^{-1}_U$ as
\begin{equation}\label{uUnt2}
u_U^{n}(x,y) = -8 \pi  e^{2x}J_\mu (4 \pi e^{2x}) \int \,
(1 + 4 \lambda^2 b^2)^{\frac{n-1}{2}} e^{- i b y e^{2x} } \, db
\quad
\end{equation}
and
\begin{equation}\label{vUnt2}
v_U^{n}(x,y) = 8 \pi e^{-4x}J_\mu (4 \pi e^{-2x}) \int \,
(1 + 4 \lambda^2 b^2)^{\frac{1-n}{2}} e^{-i y b} \, db \quad ,
\end{equation}
We may now compare this with our general solutions \re{uGen2n} and
\re{vGen2n}, with $\lambda = \frac{1}{8\pi}$. We see that \re{uUnt2} is recovered from the single $s=0$ mode, with the additional identifications
$\frac{\sqrt{k}}{2\lambda} = 4 \pi \sqrt{k} = \mu$, $m=1$ and $B=0$ in
\re{uGen2n}. On the other side, \re{vUnt2} is recovered as a
superposition in $s$ of \re{vGen2n}, with amplitude $\phi(s) \sim
\delta'(s)$:
\begin{equation}
 v_U^{n}(x,y) = \int ds \,\,\,
  \phi(s) \,\, v^{n}_{\frac{1}{8\pi},s}(a,l)
\end{equation}
and by letting $\frac{\sqrt{k}}{2\lambda} = 4 \pi \sqrt{k} = \mu$, $m=1$
and $B=0$.

The resulting product $\#$ can then more easily be expressed\footpourmoi{Que represente cette carte geometriquement?? Voir $\#$ comme une superposition de Moyal??} in terms of a map
(see eq (5.1) of \cite{Unt}) 
\beq \label{psi}
\Psi : \R^2
\rightarrow \Pi : (q,p) \ra X=\Psi(q,p) = [(1+q^2 +p^2)^{1/2} + q](1-i p)^{-1},
\eeq 
of which we will need the relations
\beq
 p(X) = \frac{y}{x} \quad, \quad q(X)=\frac{x^2+y^2-1}{2 x} \quad,\quad \forall X=x+i y \in \Pi.
\eeq
For $f,g \in \rm{Fun} (\Pi)$, the explicit expression we get (see \cite{Unt}, eq.(5.26)), evaluated at the point $X=1$ for simplicity (one can go to any other point by using the $\SL$-invariance) is\footpourmoi{Remarque : la mesure invariante sur $\Pi$
dans les coordon\'ees $(p,q)$ est $\frac{dp dq}{\sqrt{1 + q^2
+p^2}}$ et pas $dp dq$... De plus ce n'est pas f qui apparait mais
F}
\begin{equation}\label{dieze}
 (f \# g)(1) = 16 \pi \int \, d^2 x_1 d^2 x_2 F(q_1,p_1) G(q_2,p_2)
 \int_0^\infty \, t^2 J_\mu (4 \pi t) e^{4 \pi i t \varpi} dt
 \quad,
\end{equation}
with $x_i = (q_i,p_i)\in \R^2$, $F(q,p) = (f \circ \Psi)(q,p)$ and
\beq
\varpi (X_1,X_2) = q_1(X_1) p_2(X_2) - q_2(X_2) p_1(X_1).
\eeq
 We would like to analyze some
properties of the kernel\footpourmoi{Est-ce bien ce noyau qu'on doit
consid\'erer, ou faut-il inclure la mesure?}
\begin{equation}\label{noyauK}
K(1,X_1,X_2) = K_\mu (q_1,p_1,q_2,p_2) = K_\mu (\varpi)= \int_0^\infty \, s^2 J_\mu (s)
 e^{i s \varpi} ds \quad ,
\end{equation}
in particular how its phase is related to the $S_A$ in eq.\re{PhaseCan}.

Let us first evaluate $S_A(X,Y,Z)$. Owing to the $\SL$-invariance, we will restrict the computation to $Z=1$ and $X \in \R$, see Fig. \ref{TRI}. We then have to find the corresponding vertices $x$, $y$ and $z$ of $\Sigma$, eq. \re{PhaseCan}. We start by trying to solve $s_X s_1 s_Y x = x$, using the form \re{GeodSymmRiem} for the geodesic symmetries. A straightforward computation shows that the geodesic triangle $\Sigma$ exists iff $\varpi(X,Y)^2 < 1$.\footpourmoi{Weinstein avait-il vu que parfois le triangle n'existait pas? Pour AN, il existe toujours?} This is the first difference with the $AN\,$ case, where $S_A(X,Y,Z)$ is defined everywhere\footpourmoi{se voit comment?}.

\begin{figure}
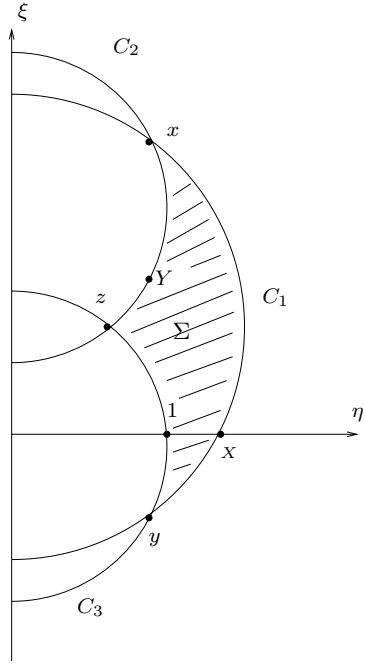

\begin{center}
\input grapp.pstex_t
\label{TRI}
\caption{\small Geodesic triangle $\Sigma$, as defined around eq. \re{PhaseCan},  in the hyperbolic plane with points $\eta + i \xi$, $\eta>0$.}
\end{center}
\end{figure} 

In the case where $\Sigma$ exists, one may then evaluate $S_A(X,Y,1)$ by computing $\int_{\Sigma} \, \frac{dy d\eta}{y^2}$. The result is\footpourmoi{intuitivement, pq a-t-on que $-\pi \leq S_A \leq \pi$?}
\beq \label{SAPi}
 S_A(X,Y,1) = \pi - 2 \arccos[\varpi(X,Y)] := S_A(\varpi(X,Y)).
\eeq
It it worth emphasizing here that it was not totally obvious that $S_A(X,Y,1)$ would depend on $X$ and $Y$ only through the very particular function $\varpi$.

We may now turn to the evaluation of the phase of \re{noyauK}. Using tabulated Laplace transforms, one first computes
\begin{equation}
{\mathcal F}_\m(p)=\int_0^\infty \, e^{-pt} J_\mu (t) dt =
\frac{1}{p^{\mu +1}}
\frac{[1+(1+\frac{1}{p^2})^{1/2}]^{-\mu}}{\sqrt{1+\frac{1}{p^2}}}
\quad ,
\end{equation}
with $p=\varepsilon - i \varpi$, $\varepsilon \ll$ being a small
real part necessary to ensure convergence.
Using the following relations
\begin{equation} 1 + \frac{1}{p^2}  = \left\{ \bearlll
  1- \varpi^{-2} &\;& {\rm if}\ \varpi^2 >1 ,  \\[.3em]
  (\varpi^{-2} -1) \rm{e}^{i s(\varpi)\pi} &\;& {\rm if}\ \varpi^2 < 1 ,  \eear\right .
\end{equation}
and 
\begin{equation} 1 + \sqrt{1 + \frac{1}{p^2}}  = \left\{ \bearlll
 1+ (1- \varpi^{-2})^{1/2} := C(\varpi) \in \R &\;& {\rm if}\ \varpi^2 >1 ,  \\[.3em]
 1+  (\varpi^{-2} -1)^{1/2} \rm{e}^{i s(\varpi)\pi/2} &\;& {\rm if}\ \varpi^2 < 1 ,  \eear\right .
\end{equation}
with $s(\varpi) =\rm{sign}(\varpi)$, one finally gets
\begin{equation}
 {\mathcal F}_\m(-i \varpi)  = \left\{ \bearlll
 \frac{1}{\sqrt{1-\varpi^2}}\,\, e^{i \mu
 \frac{S_A(\varpi)}{2}} &\;& {\rm if}\ \varpi^2 < 1 ,  \\[.3em]
  |\varpi|^{-(\m+2)} (\varpi^2 - 1)^{-1/2} C(\varpi)^{-\m} \rm{e}^{i \frac{\pi}{2} (\m + 1) s(\varpi)} &\;& {\rm if}\ \varpi^2 > 1 .  \eear\right .
\end{equation}

Note that, when $\varpi^2 < 1$, the phase of ${\mathcal F}_\m$ is precisely given by $S_A(\varpi)$, while when $\varpi^2 > 1$, it's absolute value is constant. The kernel $K_\mu$ is now simply obtained by
\beq \label{KMU}
  K_\mu (\varpi) = -\frac{d^2}{d\varpi}{\mathcal F}_\m(-i \varpi).
\eeq

The function $\varpi$ furthermore satisfies two remarkable properties:
\begin{enumerate}
{\item[(i)] $\varpi$ is $K$-invariant, i.e. $\varpi([k].X,[k].Y)=\varpi(X,Y) \quad , \quad \forall k\in K=SO(2), \forall X,Y \in \Pi$;}
 {\item[(ii)] $\varpi (X,Y)= -\varpi (Y,X)  = - \varpi(X,S_{X/2}Y)\quad , \quad  \forall X,Y \in \Pi$.}
\end{enumerate}
These can be checked by direct computation\footnote{For which one may use the explicit relations : if $X=a+ i c$ and $Y=b + i d$, then $\varpi(X,Y) = \varpi (a,c,b,d) = \frac{c}{a}\frac{b^2 + d^2 -1}{2b}-\frac{d}{b}\frac{a^2 + c^2 -1}{2a}$, and $s_{X/2} Y=\frac{a[(1+a^2)+c^2]}{(1+a)^2[b+i d - i \frac{c}{1+a}]} + i \frac{c}{1+a}$.}. From \re{SAPi}, the function $S_A (X,Y,1)= S_A(\varpi(X,Y))$ enjoys the same properties.

The phase of the kernel $K_\m$ can then be determined from \re{KMU}. For $\varpi^2 <1$, one gets
\begin{equation}\label{PhaseK}
 \psi_\mu(\varpi) = \mu \frac{S_A(\varpi)}{2} - \arctan [\frac{ 3
 \mu \sin S_A(\varpi)}{\mu^2 -4 + (\mu^2+2) \cos
 S_A(\varpi)}] \quad .
\end{equation}
The explicit form does not really matter, but the important property is that 
\beq
 \psi_\mu(X,Y,1) = \psi_\mu(\varpi(X,Y)) = -\psi_\mu(-\varpi(X,Y)),
\eeq 
expressing that the phase of the kernel also enjoys properties (i) and (ii). Therefore, according to the discussion around eqs. \re{sisi1} and \re{sisi2}, these two-point functions all come from an \emph{admissible} three-point function, thereby confirming the importance of this class of functions in the geometric understanding of associativity.


\newpage

\cleardoublepage \part{Exact black hole string backgrounds}
\cleardoublepage \chapter{Some features of string theory}\label{ChapStrings}
This chapter is going to present some salient features of string theory, with an emphasis on the role played by conformal field theory. It starts with the very basics and can therefore be overlooked by specialists.  
We will try to emphasize on aspects relevant to the next chapters. 

After discussing the string spectrum, showing in particular how the graviton and the photon emerge as closed and open string modes respectively, we stress the role played by two-dimensional conformal field theory in string theory, and discuss vertex operators and the closed string partition function. Turning to strings in non-trivial background fields, we mention how Einstein's equations come out of the requirement of conformal invariance. This is done by introducing the notion of RG flow and marginal operators. We then turn to the definition of D-branes and present their low-energy effective action, the Dirac-Born-Infeld action. We end by discussing some properties of superstring theories.








\section{Classical bosonic string}\label{Sect-Class}
The dynamics of a classical bosonic string evolving in flat $D$-
dimensional space-time is governed by the \emph{Polyakov action}\index
{Polyakov action}:
\begin{equation}\label{1}
S_P [X^\mu,\gamma_{ab}] = -\frac{1}{4 \pi \alpha'} \int_\Sigma \, d
\tau d\sigma (-\gamma)^{1/2} \gamma^{ab} \p_a X^\mu \p_b X_\mu \quad ,
\, a=(\tau,\sigma) , \mu=0,\cdots,D-1,
\end{equation}
where $(\tau,\sigma)$, $-\infty < \tau < +\infty$, $0 \leq \sigma
\leq \pi$, are coordinates on the surface $\Sigma$ swept out by the
string (the \emph{worldsheet}\index{worldsheet}), the $X^\mu$s
express the embedding of $\Sigma$ in space-time and $\gamma_{ab}$ is
the worldsheet metric tensor, with determinant $\gamma$. $T=\frac{1}
{2 \pi \alpha'}$ is the string tension \cite{JohnsonBook} p32. This
action is the (1+1)-dimensional counterpart of the (0+1)-dimensional
action describing a relativistic point particle of mass $m$ moving in
a flat target space-time with coordinates $X^\mu$, sweeping out a
worldline parametrized by $\tau$:
\begin{equation}\label{1p}
S_1[X^\mu,\eta] = \frac{1}{2} \int \, d\tau (\eta^{-1} \overset{.}{X}^
\mu \overset{.}{X}_\mu - \eta m^2).
\end{equation}
The independent auxiliary field $\eta (\tau)$ is defined on the
particle's world-line and can be thought of as a metric on it : $\eta
(\tau) = [-\gamma_{\tau \tau}(\tau)]^{1/2}$. It is therefore called
\emph{einbein}\index{einbein}. Under reparameterizations $\tau \ra
\tau'(\tau)$ of the world-line, it transforms as $\eta(\tau) \ra
\eta'(\tau') = \eta(\tau) \frac{d \tau}{d \tau'}$ so that $S$ is
reparameterization invariant. The latter fact is associated with the
first class constraint\footnote{For a general account, see \cite
{HenneauxTeitelboim}.} ${\cal H} = p^2 + m^2 = 0$ generating the
gauge transformation $\tau \ra \tau'(\tau)$, where $p^2 = p^\mu p_\mu
$ and $p_\mu = \frac{\p L}{\p \dot{X}^\mu}$ is the canonical momenta
conjugated to $X^\mu$. The einbein is a Lagrange multiplier rather
than a dynamical variable because its equation of motion is
algebraic. Substituting its expression into \re{1p} results in the
familiar action $S_0 [X^\mu] = -m \int \, d\tau \sqrt{-\dot{X}^\mu
\dot{X}_\mu}$.

The world-sheet tensor $\gamma_{ab} (\tau,\sigma)$ in \re{1} simply
generalizes the einbein $\eta(\tau)$ for the propagation of a one-
dimensional object. Variation of $S_P$ w.r.t. $\gamma^{ab}$ yields
\begin{equation}
\frac{\delta S_P}{\delta \gamma^{ab}} = 0 \Rightarrow h_{ab} = \frac
{1}{2} \gamma_{ab} \gamma^{cd} h_{cd} \quad ,
\end{equation}
where $h_{ab} = \eta_{\mu \nu} \p_a X^\mu \p_b X^\nu$ is the induced
metric on $\Sigma$, and hence
\begin{equation}\label{2}
h_{ab} (-h)^{-1/2} = \gamma_{ab} (-\gamma)^{-1/2} .
\end{equation}
Eliminating $\gamma_{ab}$ from \re{1} using \re{2}, we get the \emph
{Nambu-Goto action}\index{Nambu-Goto action}
\begin{equation}\label{3}
S_{NG} [X^\mu ] = -\frac{1}{2 \pi \alpha'}\int_\Sigma \, d\tau d
\sigma \sqrt{-\mbox{det} h_{ab}} ,
\end{equation}
which is proportional to the area of the world-sheet. Thus $S_P$ is
to $S_{NG}$ what $S_1$ is to $S_0$ (the corresponding Lagrangians are
denoted $L_P$, $L_{NG}$, $L_1$ and $L_0$). Both actions $S_{NG}$ and
$S_P$ are invariant under reparameterizations $\tau \ra \tau'(\tau,
\sigma), \sigma \ra \sigma'(\tau,\sigma)$ (2D diff-invariance) :
\begin{eqnarray}
X'^\mu (\tau,\sigma) &=& X^\mu (\tau,\sigma) \\
\frac{\p x'^c}{\p x^a} \frac{\p x'^d}{\p x^b} \gamma'_{cd} (\tau',
\sigma') &=& \gamma_{ab} (\tau,\sigma) \label{4}
\end{eqnarray}
and under global D-dimensional Poincar\'e transformations :
\begin{eqnarray}
X'^\mu (\tau,\sigma) &=& \Lambda^\mu_\nu X^\nu (\tau,\sigma) +
a^\mu \nonumber\\
\gamma'_{ab} (\tau, \sigma) &=& \gamma_{ab} (\tau,\sigma) \label
{5} .
\end{eqnarray}
From the reparameterization invariance, one naturally expects the
counterpart of the constraint ${\cal H} = p^2 + m^2 = 0$ for the
point particle. Indeed, one finds that
\begin{equation}\label{6}
{\cal H}_1 = \Pi_\mu \p_\sigma X^\mu = 0
\end{equation}
and
\begin{equation}\label{7}
{\cal H} = \pi \alpha' (\Pi^2 + \frac{1}{2 \pi \alpha'} (\p_\sigma X)
^2) = 0
\end{equation}
with $$\Pi_\mu = \frac{\p L_P}{\p \dot{X}^\mu} = -\frac{\sqrt{-
\gamma}}{2 \pi \alpha'} \gamma^{\tau \alpha} \p_\alpha X_\mu \, . $$
The constraints \re{6} and \re{7} correspond to reparameterization
invariance in $\sigma$ and $\tau$.\footnote{This can be seen by
computing the Poisson brackets.}
We will now focus on the Polyakov action. In this case, there is an
additional symmetry which has no counterpart in the Nambu-Goto form,
the \emph{Weyl symmetry}\index{Weyl symmetry} :
\begin{eqnarray}
X'^\mu(\tau,\sigma) &=& X^\mu (\tau,\sigma) \nonumber\\
\gamma'_{ab} (\tau,\sigma) &=& \exp(2 \omega(\tau,\sigma)) \gamma_
{ab}(\tau,\sigma) . \label{8}
\end{eqnarray}
It corresponds to a local rescaling of the world-sheet metric, and
can be seen to come from \re{2}, since $\gamma_{ab}$ is fixed only up
to a scale factor. This symmetry plays a special role in the theory
of strings. Remember that, from the point-particle case it was
possible, using reparameterization invariance, to eliminate the
einbein. This is still possible for a string: using diffeomorphism
and Weyl invariance, we may gauge away all the $\gamma_{ab}$
dependence (we have 3 transformations at hand for 3 components). This
would not be possible for higher dimensional objects, for example.

The action \re{1} can be seen as describing a two-dimensional field
theory with $D$ scalar fields $X^\mu(\tau,\sigma)$. The two-
dimensional energy-momentum tensor is defined as\footpourtoi
{discordance entre defs GSW, Polch, confusion $L_P$ et $S_P$}
\begin{equation}\label{9}
T^{ab} = -\frac{4\pi}{\sqrt{-\gamma}} \frac{\delta L_P}{\delta \gamma_
{ab}} = \frac{1}{\alpha'} \left( \p^a X_\mu \p^b X^\mu - \frac{1}{2}
\gamma^{ab} \p_c X^\mu \p^c X_\mu \right) .
\end{equation}
It is conserved as a consequence of invariance under 2-dimensional
diffeomorphisms :
\begin{equation}\label{10}
\nabla_a T^{ab} = 0 \, .
\end{equation}
Furthermore, the invariance under arbitrary Weyl transformations implies
\begin{equation}\label{10b}
T^{ab} \gamma_{ab} \eqdef T^a_a=0 \, ,
\end{equation}
as can be seen directly from \re{9}.
The 3 local symmetries (two reparameterizations and one Weyl scaling)
can be used to bring $\gamma_{ab}$, which has 3 independent
components, to the form
\begin{equation}\label{11}
\gamma_{ab} = {\rm{e}}^{\phi(\tau,\sigma)} \eta_{ab}\, .
\end{equation}
This is called the \emph{conformal gauge}\index{conformal gauge}, in
which $\gamma_{ab}$ is conformally flat. Note that the function $\phi
(\tau,\sigma)$ is fixed by the gauge choice, and the conformal gauge
sometimes refers to the choice $\phi=0$ (this distinction is not
important since the \emph{conformal factor}\index{conformal factor} $
\rm{e}^\phi$ drops out of everything in \re{1} at the classical
level). The Polyakov action, in the conformal gauge, reads
\begin{equation}\label{12}
S[X^\mu; \tau,\sigma] = -\frac{1}{4 \pi \alpha'} \int_\Sigma \, d\tau
d\sigma \eta^{ab} \p_a X^\mu \p_b X_\mu .
\end{equation}
Let us reconsider the Euler-Lagrange equations, derived from \re{12},
in this gauge,
\begin{equation}\label{13}
\frac{\delta S}{\delta X^\mu} = 0 \Rightarrow \Box X^\mu \eqdef \left
( \frac{\p^2}{\p \sigma^2} - \frac{\p^2}{\p \tau^2}\right) X^\mu =0 .
\end{equation}
This is a free two-dimensional wave equation for each $X^\mu$. Eqs.
\re{13} have to be supplemented with the constraints
\begin{equation}
T^{ab} = \frac{1}{\alpha'} \left( \p^a X_\mu \p^b X^\mu - \frac{1}{2}
\eta^{ab} \eta^{cd} \p_c X^\mu \p_d X_\mu \right) = 0 .
\end{equation}
These constraints are equivalent to conditions \re{6} and \re{7}
resulting from reparameterization invariance, just like in the
particle case, where $\frac{\delta S_1}{\delta \eta}=0
\Leftrightarrow P^2 + m^2 = 0$ (because $P_\mu = \frac{\p L_1}{\p \dot
{X}^\mu} = \eta^{-1} \dot{X}_\mu$) and the Weyl invariance. They are
called \emph{Virasoro constraints}\index{Virasoro constraints}.
They read
\begin{equation}\label{15-16}
T_{ab} = 0 \Leftrightarrow
\left\{
\begin{array}{l}
\p_\tau X^\mu \p_\tau X_\mu + \p_\tau X^\mu \p_\tau X_\mu = 0
( \Leftrightarrow {\cal H}=0 )\\
\p_\tau X^\mu \p_\sigma X_\mu = 0 \qquad (\Leftrightarrow {\mathcal
H}_1=0) \\
T^a{}_a = 0
\end{array}
\right.
\end{equation}
In varying \re{12}, we must also impose the following \emph{boundary
conditions}\index{boundary conditions}:
\begin{equation}\label{17}
\left[ \int d\tau \, (\p_\sigma X_\mu \delta X^\mu)\right]^{\sigma=
\pi}_{\sigma=0} = 0 .
\end{equation}
The manner in which one chooses these conditions to be satisfied
depends crucially on whether one deals with closed or open strings.
For \emph{closed strings}\index{closed strings}, the embedding fields
$X^\mu$ are taken to satisfy
\begin{eqnarray}\label{18}
X^\mu(\tau,0)= X^\mu(\tau,\pi) \\
\p_\sigma X^\mu(\tau,0) = \p_\sigma X^\mu(\tau,\pi) .
\end{eqnarray}
The most general solution of \re{13} with boundary conditions \re{18} is
\begin{equation}\label{19}
X^\m (\tau , \si) = x^\m + 2 \a' p^\m \tau + i \sqrt{{\a' \over 2}}
\sum_{n\neq 0} ( {\a^\m_n \over n} e^{-2 i n (\tau - \si)} +
{\tilde \a^\m_n \over n} e^{-2 i n (\tau + \si)}) \, ,
\end{equation}
where $x^\m, \, p^\m, \, \a^\m_n, \, \tilde \a^\m_{n}$ are D-
dimensional constant vectors. To ensure a real solution, one imposes
$\a^\m_{-n} = (\a^\m_n)^*$ and $ \tilde \a^\m_{-n} = (\tilde \a^\m_n)
^*$.
It is convenient to introduce the notation
$\a_0^\m = \tilde \a^\m_0 = p^\m \sqrt{{\a' \over 2}}$. The $\tilde
\a_n^\m$ and $\a^\m_n$ represent the oscillatory modes of the string.

For \emph{open strings}\index{open strings}, there are two canonical
choices of boundary conditions. \emph{Neumann boundary conditions}
\index{Neumann boundary conditions} are defined by
\begin{equation}
\p_\si X^\m_{|\si = 0, \pi} = 0 \, .
\label{20}
\end{equation}
In this case, the string can move freely in space-time. \emph
{Dirichlet boundary conditions} \index{Dirichlet boundary conditions}
are given by
\begin{equation}
\p_\tau X^\m_{| \si = 0, \pi} = 0 \, .
\label{21}
\end{equation}
Integrating condition \re{21} over $\tau$ specifies a space-time
location on which the string ends, and Dirichlet conditions are
equivalent to fixing the endpoints of the string:
\begin{equation}
\d X^\m_{| \si = 0, \pi} = 0 \, .
\label{22}
\end{equation}
We will see later that this latter condition is related to the
existence of a physical object called \emph{D-brane} \index{D-brane}.

For the time being, we will only consider an open string satisfying
Neumann boundary conditions at both its endpoints (N-N boundary
conditions), for which the general solution to \re{13} and \re{20} reads
\begin{equation}
X^\m (\tau, \si) = x^\m + 2 \a' p^\m \tau + i \sqrt{2 \a'} \sum_{n
\neq 0}
({\a_n^\m \over n } e^{-in\tau} cos (n\si))
\, .
\label{23}
\end{equation}
In practice, since the functions $e^{i n \si}$ are not orthogonal on $
[0,\pi]$, one usually introduces a \emph{doubling trick}\index
{doubling trick}, which amounts to extending the solution \re{23} to
the interval $[-\pi,\pi]$, through $X^\mu(\tau, -\si)= X^\mu (\tau,
\si)$, and identifying $\si\sim -\si$.
For further convenience, one also sets $\a_0^\m \triangleq \sqrt{2
\a'} p^\m$.

From Hamiltonian dynamics, one has the following equal time Poisson
brackets,
\beq
\, \{ X^\m(\si), \, \Pi^\n(\si') \} &=& \eta^{\m\n} \d(\si-\si')
\, , \label{24} \\
\, \{ \Pi^\m(\si), \, \Pi^\n(\si') \} &=& 0 \, . \label{25}
\eeq
The Virasoro constraints \re{6} and \re{7}, which read in the
conformal gauge
\beq
\cH &=& { 1 \over 4 \pi \a'} ( (\p_\si X)^2 + ( \p_\tau X)^2) (=0)
\, , \label{27} \\
\cH_1 &=& { 1 \over 4 \pi \a'} \p_\tau X^\m \p_\si X_\m (=0) \, ,
\label{28}
\eeq
are associated with the gauge transformations $\tau \ra \tau'$ and $
\sigma \ra \sigma'$ through \cite{BrinkHenneaux}
\begin{eqnarray}\label{26}
\delta_{(\xi^\tau , \xi^\sigma)} X^\mu &=& \{X^\mu (\sigma) , \int_0^
\pi d\sigma' \, (\xi {\cal H} + \xi_1 {\cal H}_1)(\sigma') \}
\nonumber\\
&\sim& \xi^\tau \dot{X}^\mu + \xi^\sigma X'^\mu \, ,
\end{eqnarray}
which can be seen by using \re{24} and \re{25}.
The Hamiltonian
\begin{equation}
H = \int_0^\pi d\si \cH(\si) \label{29}
\end{equation}
thus generates time translations on the worldsheet, while
\begin{equation}
P = \int_0^\pi d\si \cH_1 (\si)
\label{30}
\end{equation}
generates translations in $\si$.
Using \re{27}, \re{23} and \re{19} in \re{29}, one finds that
\begin{equation}
H = { 1 \over 2}\sum_{n=-\infty}^\infty \a_{-n}. \a_n \label{31}
\end{equation}
for open strings and that
\begin{equation}
H = { 1 \over 2}\sum_{n=-\infty}^\infty (\a_{-n} .\a_n + \tilde \a_{-
n} .\tilde \a_n )\label{32}
\end{equation}
for closed strings, with $\a_{-n} \a_n = \a_{-n}^\m \a_{n}^\n \eta_{\m
\n}$.
It will be useful to introduce \emph{light cone coordinates}\index
{light cone coordinates}\footnote{$\p_\tau = \p_+ + \p_-$,$\p_\si =
\p_+ - \p_-$, $\eta_{++}=\eta_{--}=-1/2$} on the worldsheet:
\begin{equation}
x^{\pm} = \tau \pm \si \, .
\label{33}
\end{equation}
in terms of which the equations of motion read
\begin{equation}
\p_+ \p_- X^\m = 0 \, .
\label{34}
\end{equation}
The solutions \re{19} and \re{23} can be rewritten in the form
\begin{equation}
X^\m (\tau , \si) = X_L^\m (x^+) + X^\m_R(x^-) \, .
\label{35}
\end{equation}
The Virasoro constraints can be recast as
\beq
T_{++}&=& {1 \over 2} (T_{\tau \tau} + T_{\si \si}) = -{1 \over \a'}
\p_+ X_\m \p_+ X^\m = -{1 \over \a'} \dot{X}^\m_L \dot{X}_{L\, \m}
\overset{!}{=}0
\label{36}\, , \\
T_{--}&=& {1 \over 2} (T_{\tau \tau} - T_{\si \si}) = -{1 \over \a'}
\p_- X_\m \p_- X^\m = -{1 \over \a'} \dot{X}^\m_R \dot{X}_{R\, \m}
\overset{!}{=}0
\, \, \, \, \, \, \mathrm{and} \label{37} \\
T_{+-} &=& T_{-+} = 0 \, . \label{38} \eeq
Eq. \re{38} is identically satisfied as a consequence of $T^a{}_a = 0$
\footpourmoi{, and thus does not impose any constraint}.

To end up with classical aspects of the free bosonic string, we draw
attention to a property that is crucial for the analysis of string
theory. In the conformal gauge, the string action \re{12} is still
invariant under some residual local symmetries. It is indeed
invariant under a combination of a Weyl rescaling and a
diffeomorphism $x^\a \rightarrow x^\a + \xi^\a$ satisfying
\begin{equation}
\p^\a \xi^\b + \p^\b \xi^\a = \Lambda^{(x)} \eta^{\a \b}\, ,
\label{39}
\end{equation}
which corresponds to an infinitesimal \emph{conformal transformation}
\index{conformal transformation,} i.e. a transformation that leaves
the metric invariant up to a local scale factor. In light cone
coordinates, this states that any transformation
\begin{equation}
x^+ \rightarrow x'^+ = f_+(x^+) \, , \hspace{1cm} x^- \rightarrow
x'^- = f_-(x^-) \, ,
\label{40}
\end{equation}
which affects $\g_{ab} = \eta_{ab}$ as
\begin{equation}
\g'_{ab} = {\p x^c \over \p x'^a} {\p x^d \over \p x'^b} \eta_{cd} \, ,
\label{41}
\end{equation}
and thus as
\begin{equation}
\g'_{+-} = ({df_+(x^+) \over dx^+} {df_-(x^-) \over dx^-} )^{-1} \eta_
{+-} \, ,
\nn
\end{equation}
can be undone with a Weyl transformation of the form
\begin{equation}
\g'_{+-} = exp ( 2 \o_+(x^+) + 2 \o_-(x^-)) \eta_{+-} \, , \nn
\end{equation}
if $ exp(-2 \o_\pm (x^\pm)) = \p_\pm f_\pm (x^\pm)$.

This symmetry leads to an infinite number of conserved quantities in
the ``+" and ``-" sectors. Indeed, the conservation law $\nabla_a T^
{ab} = 0$ resulting from the diff-invariance of the actions \re{1} or
\re{12} reads, in the conformal gauge ($\nabla_a = \p_a$)
\begin{equation}
\p_- T_{++} = 0 \hspace{2cm} \mathrm{and} \hspace{2cm} \p_+ T_{--}
= 0 \, ,
\label{42}
\end{equation}
where $T_{+-} = T_{-+} = 0$ has been used. It then follows that
\begin{equation}
\p_- (f_+ T_{++}) = 0 \hspace{2cm} \mathrm{and} \hspace{2cm} \p_+
(f_- T_{--}) = 0 \, ,
\label{43}
\end{equation}
resulting in an infinite number of conserved quantities (charges)
\beq
Q_+ &=& \int_0^\pi d\si f_+ T_{++} \nn \, , \\
Q_- &=& \int_0^\pi d\si f_- T_{--} \, , \label{44}
\eeq
since $f_+$ and $f_-$ are arbitrary. These charges are conserved ($\p_
\tau Q_{\pm} = 0$) and generate the transformations \re{40} on the
fields through the Poisson bracket ($\{Q_\pm, X(\si)\} = -f_\pm (x^
\pm) \p_\pm X(\si) $). These properties hold for any conformal field
theory in 2 dimensions. This is the statement that string theory in
the conformal gauge is a conformal field theory, and represents a
very powerful tool in perturbative string theory. We will return to
this in Sect.\ref{Sect-BosCFT}.

According to \re{42}, the energy-momentum tensor may be expanded as
\beq
T_{++} & \triangleq& 4 \sum_m \tilde L_m^{(closed)} e^{-2imx^+} \nn \\
T_{--} & \triangleq& 4 \sum_m L_m^{(closed)} e^{-2imx^-} \, ,
\label{45}
\eeq
for closed strings and
\beq
T_{++} & \triangleq& \sum_m \tilde L_m^{(open)} e^{-imx^+} \nn \\
T_{--} & \triangleq& \sum_m L_m^{(open)} e^{-imx^-} \, ,
\label{46}
\eeq
for open strings, so that the constraints $T_{++} = T_{--} =0$ may be
imposed mode by mode. With these definitions, and with \re{36}, \re
{37}, \re{19} and \re{23}, we obtain
\beq
\tilde L_n^{(closed)} = {1 \over 2} \sum_m \tilde \a_m . \tilde \a_{n-
m} \nn \\
L_n^{(closed)} = {1 \over 2} \sum_m \a_m . \a_{n-m} \label{47}
\eeq
and
\begin{equation}
L_n^{(open)} =\tilde L_n^{(open)} = {1\over 2} \sum_m \a_m . \a_{n-m}
\, ,
\label{48}
\end{equation}
where $\a_m.\a_{n-m} \triangleq \a_m^\m \a_{n-m \, \m}$ (the
prefactors in \re{45} and \re{46} are set so as to get the standard
1/2 normalization factor in
\re{47} and \re{48}).

In particular, one finds from \re{31}, \re{32} and \re{47}, \re{48}
that the total Hamiltonian is given by
\beq
H^{(open)} &=& L_0^{(open)} \, , \label{49}\\
H^{(closed)} &=& L_0^{(closed)} + \tilde L_0^{(closed)} \, . \label{50}
\eeq
In the case of closed strings, the constraint associated with the
reparameterization of
$\si$, eqs \re{28}, \re{30} can be expressed in a relation similar to
\re{32} as
\begin{equation}
P = L_0^{(closed)} - \tilde L_0^{(closed)} \, .
\label{51}
\end{equation}
The equal time Poisson brackets \re{24} and \re{25} translate into
\beq
\{ \a_m^\m, \a_n^\n \} &=& \{ \tilde \a_m^\m, \tilde \a_n^\n \} = i m
\d_{m+n} \eta^{\m\n} \nn \\
\{p^\m, x^\n \} &=& \eta^{\m \n} \, \hspace{1cm} \{ \a^\m_m, \tilde
\a^\n_n\} = 0 \, .
\label{52}
\eeq
Using \re{52}, \re{47} and \re{48}, one finds that the generators $L_n$
(and $\tilde L_n$), called \emph{Virasoro generators} \index{Virasoro
generators}, form, through their Poisson bracket, a \emph{Virasoro
algebra}
\index{Virasoro algebra} (without central extension):
\beq
\{L_n, L_m \} &=& i (m-n) L_{n+m} \nn \\
\{ \tilde L_n, \tilde L_m \} &=& i (m-n) \tilde L_{n+m} \nn \\
\{L_n, \tilde L_m \} &=&0
\label{53}
\eeq
\re{53} is sometimes also called the \emph{Witt algebra} \index{Witt
algebra}.

The appearance of this algebra finds its origin in the residual
invariance \re{40}, see \cite{GSW1}, p.74.

\section{Spectrum of the quantized bosonic string}

There exist many different procedures to describe the quantization of
bosonic string theory. When used correctly, they are all equivalent,
even though the relations between them are non trivial. The \emph
{light cone gauge quantization} \index{light cone gauge quantization}
consists in using the residual gauge symmetry \re{40}, after the
conformal gauge has been chosen, to make a particular noncovariant
coordinate choice in which it becomes possible to solve the Virasoro
constraints. Then, only physical degrees of freedom remain to be
quantized. Lorentz invariance is no longer manifest and has to be
controlled using the generators of the Poincar\'e group.

The \emph{modern covariant quantization} \index{modern covariant
quantization} is rooted in the path integral approach to the
quantization of field theories. It involves the introduction of
Faddeev-Popov ghosts ensuring that we stay on the chosen gauge slice
in the full theory and the identification of BRST symmetries and
currents \cite{GSW1}.

Here we will briefly present a third alternative, the \emph{old
covariant quantization.} \index{old covariant quantization} The $X^\m
$s are interpreted as quantum operators acting on a Hilbert space of
states. The operators $\a^\m_{-n}$ ($\a^\m_n$), $n>0$, create
(destroy) a quantum of left moving oscillator with angular frequency
$n$ along the $X^\m$ direction in the space-time; $\tilde \a_{-n}^\m$
($\tilde \a_{n}^\m$) does the same for the right movers (for closed strings). The
Poisson brackets are promoted to commutators via the substitution $\{\, ,\, \} \rightarrow -i [\, , \, ]$. This gives
\beq
\ [x^\m , p^\n ] = i \eta_{\m\n} \, , \hspace{1cm} [\a_n^\m , \a_m^
\n ] =
[ \tilde \a^\m_n , \tilde \a^\n_m ] = n \d_{m+n} \eta^{\m\n} \, ,
\label{54}
\eeq
coming from
\beq
\ [ X^\m(\tau, \si), \Pi^\n(\tau , \si')] = i \eta^{\m \n} \d(\si-
\si') \, ,
\hspace{1cm} [\Pi^\m(\tau , \si] , \Pi^\n (\tau,\si')] = 0 \, .
\label{55}
\eeq
By setting $a_m^\m = {1 \over \sqrt{m}} \a_n^\m$, $a_m^\m{}^\dagger
= {1 \over \sqrt{m}} \a^\m_{-m}$ (idem for the tilded quantities), we
have
\beq
\ [a_m^\m, a^\m_m{}^\dagger] = 1 \quad , \quad \mu \ne 0.
\label{56}
\eeq
So the Fock space of the quantum theory described by \re{54} is the
tensor product of $2 \times D$ (($1 \times D)$ for open strings)
infinite towers of harmonic oscillators and that of the $D$-
dimensional quantum mechanics coming from the zero modes $x^\m$ and
$p^\m$.

The oscillator ground state $|0\rangle$ is defined as being
annihilated by the $\a^\m_m$, $m>0$. The operators $x^\m$ and $p^\m$
obey the standard Heisenberg commutation relations. Hence, they may
be represented on the Hilbert space spanned by the usual plane wave
basis $|k\rangle = e^{ik.x} $ of eigenstates of $p^\m$. The Fock
space of the string is thus built on the states $|k;0\rangle$ satisfying
\beq
p^\m |k;0\rangle &=& k^\m |k;0\rangle \nn \\
\a^\m_m | k;0\rangle &=& 0 \, , \, \, \, \, m>0
\label{57}
\eeq
These states are themselves obtained from the absolute vacuum $|0;0
\rangle$ by
\beq \label{57p}
|k;0\rangle = e^{ikX} |0;0\rangle ,
\eeq
as can be checked from the commutation relations \re{54}.

For $\m = 0$, the relation $\ [a_m^0, a^0_m{}^\dagger] = - 1 $
signals the fact that a state of the form $a_m^0{}^\dagger | 0\rangle
$ has negative norm. Such a state is called a \ind{ghost}. The
physical space of allowed string states will be a subspace of the
aforementioned Fock space, this subspace being specified by a quantum
mechanical implementation of the Virasoro constraints (these
conditions will be analoguous to the Gupta-Bleuler condition in QED,
in which the classical constraint $\p_\m A^\m = 0$
is replaced by the requirement that the positive-frequency components
of the corresponding quantum operators should annihilate physical
photon states \cite{GSW1}, pp74-82.

At the quantum level, the operators $L_n$ and $\tilde L_n$ are still
defined from \re{47} and \re{48}, where normal ordering is now
understood. Since $\a^\m_{n-m}$ commutes with $\a_m^\m$ unless $n=0$,
the only ordering ambiguity arises in the expression for $L_0$. On
the other hand, the algebra \re{53} gets modified due to quantum
effects \cite{GSW1}, p80. The result is
\beq
\
[L_n, L_m ]&=& i (m-n) L_{n+m} + {c \over 12}(m^3-m) \d_{m+n} \nn \\
\ [ \tilde L_n, \tilde L_m ] &=& i (m-n) \tilde L_{n+m}+ {c \over 12}
(m^3-m) \d_{m+n} \nn \\
\ [L_n, \tilde L_m] &=&0
\label{58}
\eeq
with $c=D$, the dimension of space-time.

The algebra \re{58} is known as a \emph{central extension of the
Virasoro algebra}\index{Virasoro algebra}, or Virasoro algebra in
short (when \re{58} is referred to as a Witt algebra). To impose the
physical state constraints, one would like to require that
\beq
(L_n - a \d_n) | \varphi \rangle = 0 \, , \, \, \, (\tilde L_n - a
\d_n ) |\varphi\rangle = 0 \, , \hspace{2cm} \forall n\, ,
\label{59}
\eeq
for a \emph{physical state}\index{physical state} $|\varphi\rangle$,
where the constant $a$ comes from the normal ordering ambiguity
\footpourtoi{GSW77-95,John41,Szab21}. These conditions are, however,
incompatible with \re{58}, due to the central term. Just as in QED,
the classical constraint $T_{++} = T_{--} = 0 $ is replaced in the
quantum theory by the weaker requirement that the positive frequency
components annihilate a physical state, i.e.
\beq
(L_n - a \d_n ) | \varphi\rangle = 0 \, , \, \, \, (\tilde L_n - a
\d_n ) |\varphi\rangle = 0 \, , \hspace{2cm}\forall n \geq 0 \, ,
\label{60}
\eeq
This ensures that the matrix elements of the operators $L_m$ between
physical states vanish $\forall m$ \cite{GSW1}, p78.

In determining the string spectrum, the Poincar\'e invariance \re{5}
plays an important role, since string excitations should form
irreducible unitary representations of the Poincar\'e group if these
are to be interpreted as particles. The conserved currents
associated with translations and Lorentz transformations can be found
using the Noether procedure \cite{GSW1}, p68. These are respectively
\beq
P^\m_a = {1 \over 2 \pi \a'}\p_a X^\m
\label{61}
\eeq
\beq
J^{\m\n}_a ={1 \over 2 \pi \a'} ( X^\m \p_a X^\n - X^\n \p_a X^\m)
\label{62}
\eeq
and satisfy
\beq
\p_a P^{0\m }= \p_a J^{0\m\n} = 0 \, .
\label{63}
\eeq
The total conserved momentum and angular momentum of a string are
found by integrating these currents over $\si$ at $\tau = 0 $. One has
\beq
P^\m &=& \int_0^\pi d\si \Pi^\m = {1 \over 2 \pi \a'} \int_0^\pi d\si
\dot{X}^\m = p^\m \label{64} \\
J^{\m\n} &=& {1 \over 2 \pi \a'} \int_0^\pi d\si(X^\m \dot{X}^\n -
X^\n \dot{X}^\m) \label{65} \, \, = x^\m p^\n - x^\n p^\m + E^{\m\n} +
\tilde{E}^{\m\n}
\\
& & E^{\m \n} = - i\sum_{n=1}^\infty {1 \over n} (\a^\m_{-n} \a_n^\n -
\a^\n_{-n} \a^\m_n), \label{66}
\eeq
by using the explicit form of the solutions for $X^\mu$.
Since there is no ordering problem in $P^\m$ and $J^{\m\n}$, they can
be interpreted unambiguously as quantum operators \cite{GSW1} p79.
One may then use the canonical commutation relations \re{55} or \re
{54} to verify that they satisfy the Poincar\'e algebra. This comes
from the fact that, in old covariant quantization, i.e. in the
conformal gauge, Lorentz invariance is manifest (the story is
different in light cone quantization, see \cite{GSW1},p.92). It may
also be checked that $\ [L_n, J^{\m\n}] = 0 = [L_n,P^\m] $, so that
physical states are transformed into physical states under Poincar\'e transformations (the physical state conditions are invariant
under Poincar\'e transformations). Thus the physical subspace is
guaranteed to form a representation of the Poincar\'e group.

Before inspecting the first levels of the string spectrum, let us
note that there are still two undetermined constants in \re{58} and
\re{60}: the dimension of space-time, $D$, and $a$, sometimes called
the \ind{intercept}. Their values can be determined by means of
numerous independent arguments. They all converge to $D=26$ and $a=1
$. Some of the 1001 roads to $D=26$ and $a=1$ include: preservation
of Lorentz invariance in light cone quantization \cite{GSW1}, pp.
95-100, ``consistency'' of the string spectrum (e.g. unitarity, no-
ghost theorem) \cite{GSW1}, pp81-86, \cite{JohnsonBook}, p.43,
vanishing of the total conformal anomaly in modern covariant
quantization (giving $D=26$) \cite{GSW1}, p121, funny tricks like $
\sum_{n=1}^\infty n = -\frac{1}{12}$ \cite{GSW1}, p96, etc. We will
not develop these derivations and will take the result as granted.

Let us start with the \emph{open string spectrum}\index{open string
spectrum}. The constraint $L_0 = 1$ states that
\beq
{1\over 2}\a_0^2 + 2 \times {1 \over 2}\sum_{m=1}^\infty \a_{-m}.\a_{+m}
=1
\, .
\label{67}
\eeq
Defining the \emph{number operator}\index{number operator} $N\eqdef
\sum_{n=1}^\infty \a_{-n}.\a_{n}$ counting the number of oscillators
of a given state (because $\ [N, \a^\m_{-m}] = m \a^\m_{-m}$), this
leads to the mass spectrum
\beq
M^2 \triangleq -p^\mu p_\mu = {1 \over \a'}(N-1) \, ,
\label{68}
\eeq
where \re{67} with $\a_0^\m = \sqrt{2 \a'} p^\mu$ has been used.
The ground state, at level $N=0$, has no excited oscillators and is
denoted
\beq |0;k\rangle \, , \label{69} \eeq
$k^\m$ being its center of mass momentum. This state has negative
mass squared, and is called a \emph{tachyon}\index{tachyon}. Its
presence signals that the vacuum is unstable and that we are
perturbing around a local maximum of the potential energy, since the
mass squared of a scalar corresponds to the curvature of the
potential at the critical point. At level 1, the states are
\beq
|k;\zeta \rangle \triangleq \zeta_\m (k) \a_{-1}^\m | 0;k\rangle
\, ,
\label{70}
\eeq
where $\zeta_\m$ is an arbitrary polarization vector for the state.
The mass-shell condition imposes
\beq
M^2 = -k^2 = 0 \, ,
\label{71}
\eeq
while the $L_1$-constraint yields (with \re{54} and \re{57}),
\beq
k . \zeta = 0 \, .
\label{72}
\eeq
Such a state thus has $D-2=24$ independent polarization states, as
the physical constraints remove two of the initial $D$ vector degrees
of freedom. Conditions \re{71} and \re{72} can also be combined into
\beq
k^\m k_\m \zeta_\n - k_\n k^\m \zeta_\m = 0 \iff k^\m (k_\m \zeta_\n
- k_\n \zeta_\m) = 0 \nn
\eeq
which gives the free-field equation of motion
\beq
\p^\m F_{\m\n} = 0\, ,
\nn
\eeq
where $F = dA$ and $A_\m = \int d k \zeta_\m e^{ikx}$.
Let us also consider, at level 1, the state $| \Psi \rangle \equiv L_
{-1} |0;k\rangle$. This state is orthogonal to any physical state: $
\langle \Psi | \varphi\rangle = 0$. It is called a \emph{spurious
state} \index{spurious state}. It also satisfies $L_1 \Psi = 2 L_0 |
0;k\rangle = \a' k^2 |0;k\rangle$. For $k^2=0$, this state is
physical and null ($\langle \Psi | \Psi \rangle =0$). One may
therefore add it to any physical state with no physical consequences.
One should therefore impose an equivalence relationship
\beq
| \varphi \rangle \sim | \varphi \rangle + \lambda | \Psi \rangle \, ,
\hspace{1cm}\lambda \in \R
\, ,
\label{73}
\eeq
which translates into
\beq
\zeta_\m \sim \zeta_\m + 2 \lambda \sqrt{2\a'} k_\m \, .
\label{74}
\eeq
This is a $U(1)$ gauge symmetry (in momentum space); combined with
the fact that $|k;\zeta\rangle$ describes a massless spin 1 particle,
we have thus observed that the level 1 contains a photon field (with 24
degrees of freedom) in $D=26$.
At higher levels, $N\geq 2$, we get an infinite tower of massive states.

We now turn to the \ind{closed string spectrum}. As we have 2 sets of
independent oscillators in this case, the closed string spectrum is
essentially the tensor product of two copies of the open string one.
There is however a subtlety. Let us consider the physical constraints
\re{60} for $n=0$. These give
\beq
(L_0 + \tilde L_0 - 2) | \varphi \rangle &=& 0 \label{75}\\
(L_0 - \tilde L_0) | \varphi\rangle &=& 0 . \label{76}
\eeq
Eq. \re{75} is the counterpart of \re{67} for closed strings and
gives the mass-shell condition
\beq \label{77}
M^2 = \frac{2}{\a'} (N + \tilde{N} -2),
\eeq
where $N$ and $\tilde{N}$ are the number operators for the left and
right sectors. The constraint \re{76} can be understood by noting
that $P=L_0 - \tilde{L_0}$ is the worldsheet momentum, and so
generates translations in the $\si$ coordinate. It reflects the fact
that there is no physical significance as to where we are on the
string, and hence that the physics is invariant under translations in
$\si$. \footpourtoi{$P$ genere transl rigides ou qcques?} This
constraint expresses that
\beq
N = \tilde N \, ,
\label{78}
\eeq
called \ind{level-matching condition}, which amounts to equating the
number of right and left-moving oscillator modes. The mass-shell
condition is thus
\beq
M^2 = {4 \over \a'}(N-1)
\, .
\label{79}
\eeq
At level $N=0$, we again have a \ind{tachyon}:
\beq
M^2 = - {4 \over \a'}
\, ,
\label{80}
\eeq
so that the closed string vacuum is also unstable. The first excited
level, $N= \tilde{N}=1$, consists in states of the form
\beq
| k;\zeta \rangle = \zeta_{\m\n} (\a_{-1}^\m \otimes \tilde \a^\n_
{-1} ) |k;0 \rangle \, .
\label{81}
\eeq
The Virasoro constraints give
\beq
L_1 | k;\zeta \rangle = \tilde L_1 |k;\zeta\rangle = 0 \, ,
\label{82}
\eeq
which are equivalent to
\beq
k^\m \zeta_{\m\n}= 0 = k^\n \zeta_{\m\n} \, .
\label{83}
\eeq
\footpourtoi{vp S16, truc pas clair}
Since $\zeta_{\m\n}$ is the polarization tensor, \re{83} expresses
that only the $D-2 =24$ states of transverse polarization are
physical. To obtain the particle content, we have to extract the
irreducible representations of the $D$-dimensional Poincar\'e group
contained in physical $\zeta_{\m\n}$. These are the symmetric
traceless part identified with the \ind{graviton} $h_{\m\n}$, the
antisymmetric part corresponding to an antisymmetric tensor field $B_
{\m\n}$, sometimes called the \ind{Kalb-Ramond field}, and the trace
associated with a scalar field, the \ind{dilaton} $\phi$.
As in the open string case, physical states are defined only up to
the addition of spurious states, here $L_{-1} |0;k\rangle$ and $\tilde
{L_{-1}} |0;k\rangle$, see eq. \re{73}. The closed string couterpart
of \re{74} is
\beq
\zeta_{(\m\n)} &\sim& \zeta_{(\m\n)} + k_\m \xi_\n + \xi_\n k_\n \nn \\
\zeta_{[\m\n]} &\sim& \zeta_{[\m\n]} + k_\m \g_\n - \g_\n k_\n ,
\label{84}
\eeq
where $\xi_\m$ and $\g_\mu$ are arbitrary vectors orthogonal to the
momentum $k^\m$\footpourtoi{donc pas arbitraires?}. Eq.\re{84} can be
understood as a residual gauge symmetry in momentum space for the
graviton and $B$-field:
\beq
h_{\m\n} &\sim & h_{\m\n} + \p_\m \Lambda_\n + \p_\n \Lambda_\m
\label{85}
\\
B_{\m\n} &\sim & B_{\m\n} + \p_\m A_\n - \p_\n A_\m,
\label{86}
\eeq
with $\Lambda_\n = \int \, \xi_\nu (k) e^{ikx} dk$. The first is
simply diffeomorphisms acting on the space-time metric in the
Minkowski background. The second can be written as $B \ra B + dA$,
suggesting that the physical observable should be the three-form
field strength $H=dB$. Again, the higher levels display an infinite
tower of massive states.

The upshot of the physical string spectrum can be summarized through
the interplay between worldsheet and space-time quantities. At the
lowest level of massless states ($N =1$), open strings correspond to
gauge theory while closed strings correspond to gravity. The higher
levels $N \geq 2$ give an
infinite tower of massive particle excitations. Note that the
massless states are picked
out in the limit $\a' \ra 0$ ($l_s\ra 0$) in which the strings
resemble point-like objects. We shall refer to this \ind{low-energy
limit} as the \ind{field theory limit} of string theory.




\section{CFT approach to the bosonic string}\label{Sect-BosCFT}
Conformal invariance plays an important role in string theory (as
does superconformal invariance in superstring theory, as we will see
later on). This is explicitly revealed by the residual invariances \re
{40} of the Polyakov action, which led us to the observation that
string theory in the conformal gauge is indeed a conformal field
theory. We are now going to make this correspondence manifest, by
reformulating string theory using tools
of two-dimensional CFT.
\subsection{A crash overview of CFT}\label{CrashCFT}
This section assumes a basic knowledge in two-dimensional CFT (of which a short account can
be found say in the 30 first pages of \cite{CoursModave}) and is
essentially aimed at fixing notational conventions. It is not self-contained, as a plethora of very good reviews and courses exist on
the subject \cite{Ginsparg, Bagger, Schellekens,Dots,
Spector} (see also the bibliography of \cite{CoursModave}) as well as
standard references \cite{DiFr,Ketov}. We will only sketch some
principles of 2D CFT.

Consider the two-dimensional Euclidean space, whose metric in complex
coordinates $z$ and $\zb$ is that of the complex plane: $ds^2 = dz d
\zb.$ \emph{Conformal transformations}\index{conformal
transformation} are the set of transformations leaving the metric
tensor invariant up to a scale. These are
\begin{equation}\label{87}
z \ra f(z) \quad , \quad \zb \ra
\bar{f}(\zb) \quad,
\end{equation}
expressing that arbitrary (anti-)holomorphic mappings of the
(compactified) complex plane are conformal transformations.
Infinitesimal conformal transformations correspond to
\begin{equation}\label{88}
z \ra z + \alpha(z) \quad , \quad \zb \ra
\zb + \abar(\zb) \quad,
\end{equation}
with $\a (z) = -\sum_{n=-\infty}^{+\infty} \, \alpha_n z^{n+1}$ and $
\bar{\a} (\zb) = -\sum_{n=-\infty}^{+\infty} \, \bar{\alpha}_n \zb^{n
+1}$. On functions of $z$ and $\zb$, these mappings are generated by
differential operators
\begin{equation}\label{89}
l_n = -z^{n+1} \p_z \quad, \quad \bar{l_n} = -\zb^{n+1} \p_{\zb},
\end{equation}
which satisfy the Witt algebra \re{53}.
Among the set of transformations \re{87}, one may single out \ind
{global conformal transformations}, which are well-defined on the \ind
{Riemann Sphere} (i.e. the complex plane plus a point at infinity).
It can be shown that these trnasformations are given by
\begin{equation}\label{90}
z \ra \frac{a z + b}{c z + d} \quad , \quad \bar{z} \ra
\frac{\bar{a} \zb + \bar{b}}{\bar{c} \zb + \bar{d}} \quad ,
\end{equation}
with 8 complex parameters subject to the constraints $ad-bc=1$ and
$\bar{a}\bar{d}-\bar{b}\bar{c}=1$, thus corresponding to the group
$\mbox{Sl}(2,\C)/\Z_2$ $\times$ $\mbox{Sl}(2,\C)/\Z_2$.
The global conformal transformations of two-dimensional flat space
are obtained by imposing $z^* = \bar{z}$, and can be identified with
2 translations, 1 rotation, 1 dilatation and 2 special conformal
transformations.

Some properties of a two-dimensional CFT can be stated as follows:
\begin{itemize}
\item There exists in the theory a set of fields, called \ind{primary
fields}, which under \re{87} transform as
\begin{equation}\label{91}
\phi (z,\zb) \lra \phi'(w(z),\wb(\zb)) =
\left(\frac{dw}{dz}\right)^{-h}\left(\frac{d\wb}{d\zb}\right)^{-\hb} \,
\phi(z,\zb) \quad,
\end{equation}
where $h$ (resp. $\hb$) is called the holomorphic (resp.
anti-holomorphic) conformal dimension of the field \index{conformal
dimension.} Eq. \re{91}
states that a primary field of conformal dimensions $(h,\hb)$
transforms as the components of a tensor with $h$ covariant
indices $z$ and $\hb$ covariant indices $\zb$. The conformal
dimensions are related to the spin $s$ and scaling dimension
$\Delta$ of the field by $s = h - \hb$ and $\Delta = h + \hb$
\footnote{The variation of a field under a transformation can be
decomposed in $\delta \phi(x) = (\phi'(x) - \phi'(x')) +
(\phi'(x')-\phi(x))$, where the second term encodes the spin and
scaling properties of the field. For a rotation, $z \ra
\mathrm{e}^{i \theta} z$, we find $\phi'(x') = \ex^{i\theta
(h-\hb)} \phi(x)$, while under a scaling $z \ra \mathrm{e}^{
\lambda} z$, we have $\phi'(x') = \ex^{-\lambda (h+\hb)}
\phi(x)$}.

These fields play a crucial role in conformal field theories,
namely because correlation functions involving any field of the
theory can be reduced to correlation functions involving only
primary fields. This is one of the properties rendering conformal
field theories in two dimensions solvable, where a theory is said
to be solved when all correlation functions can be written (at
least in principle). Fields transforming like \re{91}, but only
under global conformal transformations \re{90} are called
\emph{quasi-primary}.
The infinitesimal version of \re{91} is
\begin{equation}\label{92}
\delta_\a \phi(z,\zb) = -(h \a'(z) + \a (z) \p + \hb
\bar{\a}'(\zb) + \bar{\a}(\zb) \pb)\, \phi(z,\zb) \quad,
\end{equation}
with $\p = \p_z$ and $\pb = \p_{\zb}$.

\item The energy-momentum tensor of the theory is conserved and
traceless. Its components are
\begin{equation}\label{92b}
T_{zz} \eqdef T(z) \quad ,\quad T_{\zb \zb} \eqdef \Tb(\zb) \quad ,
\quad T_{z \zb}=T_{\zb z}=0.
\end{equation}
The transformation properties of a field under conformal
transformations are completely encoded in the singular part of its
operator product expansion (OPE) with $T(z)$ and $\Tb(\zb)$. For a
primary field, for instance, we have
\begin{equation}\label{93}
T(z) \phi(w,\wb) \sim \frac{h}{(z-w)^2} \, \phi(w,\wb) +
\frac{1}{z-w} \, \p_w\phi(w,\wb) \quad,
\end{equation}
and a similar relation for the anti-holomorphic part.
The precise meaning of this expression is the following : when
$z\ra w$, the (correlation) function $\langle T(z) \phi(w,\wb)
Y\rangle$ behaves as the function

$\frac{h}{(z-w)^2} \langle
\phi(w,\wb)Y \rangle + \frac{1}{z-w} \p_w \langle \phi(w,\wb)Y
\rangle$, where $Y =\phi_1(w_1,\wb_1) \cdots \phi_n(w_n,\wb_n)$, none
of the
points being coincident in $\langle T(z) \phi(w,\wb) Y\rangle$. The
symbol $\sim$ here means equality up to terms regular as $z \ra w$.

In particular, the energy-momentum tensor has the following OPE with
itself:
\begin{equation}\label{95}
T(z)T(w) \sim \frac{c/2}{(z-w)^4} + \frac{2T(w)}{(z-w)^2} +
\frac{\p T(w)}{z-w} \quad,
\end{equation}
and a similar relationship holds for $\Tb(\zb)$. This expresses the
transformation law of $T$ under $z\ra w(z)$:
\begin{equation}\label{95mp}
T(z) \lra T'(w) = \left(\frac{dw}{dz}\right)^{-2} \, \left[T(z)
-\frac{c}{12}\{w;z\}\right] \quad.
\end{equation}
Here $c$ is a numerical constant characterizing the theory, called
the \ind{central charge} and
\begin{equation}
\{w;z\} \eqdef \frac{d^3 w/dz^3}{dw/dz} -\frac{3}{2}
\left(\frac{d^2w/dz^2}{dw/dz}\right)^2 \quad
\end{equation}
is called the \emph{Schwartzian derivative}.
For $w(z)$ of the form \re{90}, one may check that $\{w(z),z\} = 0$,
showing that $T(z)$ is a quasi-primary field of weight $h=2$, $\bar{h}
=0$.

\item A two-dimensional CFT can be formulated in the \ind{operator
formalism} (as opposed to the path-integral formalism), in which the
OPE of a field with itself determines the commutation relation of the
mode operators of this field. Indeed, through the map $z=e^{\tau - i
\si}$, the usual time-ordering operation of QFT becomes \ind{radial-
ordering} in the complex plane. The equal-time commutators of two
operators $A = \oint_0 a(z) dz$ and $B = \oint_0 b(z) dz$ can then be
represented as
\begin{equation}\label{95p}
[A,B] = \oint_0 dw\oint_w dz a(z) b(w),
\end{equation}
which explicitly relates OPE and commutation relations (see e.g. \cite
{DiFr}).
A quasi-primary field of conformal weight $h$ can, in general, be
mode-expanded as follows:
\begin{equation}\label{95pp}
\phi(z) = \sum_n \phi_n z^{-n-h} \quad , \quad \phi_n = \frac{1}{2\pi
i} \oint dz z^{h+n-1} \phi(z).
\end{equation}
Consider for example the mode expansions of
\begin{equation}\label{96}
T(z) = \sum_{n \in \Z} L_n z^{-n-2} \quad , \quad \Tb(\zb) = \sum_{n
\in \Z} \bar{L_n} \zb^{-n-2}.
\end{equation}
One finds, from \re{95} and \re{95p}, that the mode operators $L_n$
and $\bar{L_n}$ satisfy a Virasoro algebra with central charge $c$.
Furthermore, these are the generators of the local conformal
transformations on the Hilbert space, in the sense that under $z\ra
\a_n z^{n+1} \eqdef z + \a_n(z)$, we have
\begin{equation}\label{97}
\delta_{\a_n(z)} \phi = -\a_n [L_n ,\phi(z)]
\end{equation}

\item The primary fields of the theory are in one-to-one
correspondence with the highest-weight states of a representation of
the Virasoro algebra. This is referred to as the \ind{field-state
correspondence}. To each primary field $\phi$ of conformal dimensions
$(h,\bar{h})$, there corresponds a state $|h,\bar{h}\rangle$ satisfying
\begin{equation}\label{98}
L_{n>0} |h,\bar{h}\rangle = \bar{L_n} |h,\bar{h}\rangle =0 \quad,
\quad L_0|h,\bar{h}\rangle = h |h,\bar{h}\rangle \; , \; \bar{L_0}|h,
\bar{h}\rangle = \bar{h} |h,\bar{h}\rangle .
\end{equation}
Furthermore, a vacuum state exists in the theory which is invariant
under the global conformal group, which translates into
\begin{equation}\label{98p}
L_n |0\rangle = 0 \quad, \quad n=-1,0,1 .
\end{equation}
The field-state correspondence is explicitly represented as
\begin{equation}
|h,\bar{h}\rangle = \lim_{z,\zb\ra 0} \phi_{h,\bar{h}} (z, \zb) |0
\rangle,
\end{equation}
which can be viewed as the creation of an asymptotic state out of the
vacuum at $t\ra -\infty$.

Since the holomorphic and anti-holomorphic components of the Virasoro
algebra \re{58} decouple, one usually restricts the study to a single
sector, and then takes tensor products. As shown in \re{98}, one
chooses the generator $L_0$ to be diagonal in the representation
space, also called \ind{Verma module}. A basis for the other (than
highest-weight) states of the representation, called \ind{descendant
states}, is obtained by applying the raising operators $L_{n<0}$ to
the highest-weight states (h.w.s.) of the representation in the
following way :
\begin{equation}
L_{-k_1} \cdots L_{-k_n} |h\rangle \quad , \quad 1\leq k_1 \leq
\cdots \leq k_n.
\end{equation}
Since $[L_0, L_m] = -m L_m$, such a state has an $L_0$-eigenvalue
equal to $h + k_1 + \cdots + k_n \eqdef h + N$, where $N$ is the
level of the state (not to be confused with the number operator $N$
in string theory). To a Verma module of h.w.s. $|h\rangle$ with
central charge $c$, denoted $V(c,h)$, we associate the \ind
{character} of the module, defined as
\beq\label{97b}
\chi_{(c,h)}(\tau) &=& \mbox{Tr} q^{L_0 - c/24} \quad , \quad q = e^
{2 \pi i \tau}, \tau \in \C \nonumber \\
&=& \sum_n \mbox{dim}(h+n) q^{n+h-c/24} ,
\eeq
where $\mbox{dim}(h+n)$ is the number of linearly independent states
at level $n$ in the Verma module. Knowing the character amounts to
knowing how many states there are at each level.

\item The conformal symmetry imposes constraints on the form of
correlation functions. The variation of a correlation function under
a conformal transformation $z \ra z + \a(z)$ is given by the \ind
{conformal Ward identities}:
\begin{equation}\label{99}
\d_\a \langle \phi_1 \cdots \phi_n \rangle = -\frac{1}{2\pi i}
\oint_C \, d\xi \a(\xi) \langle T(\xi) \phi_1 (z_1) \cdots \phi_n
(z_n)\rangle ,
\end{equation}
where $C$ encircles the position of all fields $\phi_i$. For a global
conformal transformation, the variation \re{99} vanishes, restricting
the functional dependence of the n-point functions.
\end{itemize}

\subsection{String states and vertex operators}\label{StrVert}

In section 4.2, we have had a first look at the different states
contained in the string spectrum. One would now like to find an
appropriate operator-state correspondence which would allow one to
associate quantum fields to quantum states in a one-to-one manner\footpourmoi{Why?}.
The states could then be regarded as being created from the vacuum by
the application of the quantum fields. We will describe how to
formulate this correspondence in the context of string theory. First,
it is convenient to choose the worldsheet metric, in conformal gauge,
to be Euclidian, rather than Lorentzian, by setting $(\sigma,\tau)
\rightarrow (\sigma, \tau_E = i \tau)$, so that the metric is
\beq \label{100} ds^2 = d\tau_E^2+d\sigma^2 \, , \hspace{2cm} -
\infty < \tau_E < \infty, \, \, 0 \leq \sigma \leq \pi . \eeq
From now on, we will forget the subscript ``E'' on $\tau_E$ and $\tau
$ will always be understood to be Euclidian time. In order to be
definite, let us consider the worldsheet diagram of a free closed
string, coming from $\tau =-\infty$ and going out to $\tau= \infty$.
This is the
\emph{cylinder diagram.} \index{cylinder diagram}

One may perform the conformal transformation
\beq z = e^{2(\tau-i\sigma)}, \, \, \bar{z} = e^{2(\tau + i \sigma)}
\, , \label{101}\eeq
which maps the cylinder onto the complex plane $\CC \cup \{ \infty \}
$, or Riemann sphere. Surfaces of equal $\tau$ on the cylinder become
circles of equal radii on the complex plane. The string from the
infinite past $\tau = -\infty$ is mapped to the origin, while the
string in the infinite future $\tau = + \infty$ is mapped to the
point at $\infty$. If we wish to project both the outgoing and
incoming strings to a finite distance, we may again use conformal
invariance. Inserting (\ref{101}) in (\ref{100}) has the effect
\beq ds^2 = d\tau^2 + d\sigma^2 \, \rightarrow \, \arrowvert z
\arrowvert^{-2} dz d\bar{z} \,
\label{102} \eeq
so to go to the complex plane, one simply chooses a conformal factor $
\arrowvert z \arrowvert^2$. If instead one chooses $\arrowvert z
\arrowvert^2 {R^2 \over (R^2 + \arrowvert z \arrowvert^2)^2}$, then
the metric becomes the standard round one on the sphere and the
incoming strings become finite points.

For a free open string, an analog of this process will relate the
strip of the open string to the upper half plane, or to the disc. The
open strings are mapped to points on the real axis, which is
equivalent to the boundary of the disc.

In this picture, with $z = e^{\tau - i\sigma}$ and $\sigma \in [0,
\pi]$, the \emph{doubling trick} \index{doubling trick} amounts to
extend the upper half plane to the entire complex plane, and identify
$z \sim z^* = \bar{z}$.

One can go further by considering interactions between strings, let
us say closed for simplicity (for open strings, see e.g. \cite
{Mohaupt} p.37), which are described by worldsheets which connect a
given initial configuration of strings to a final configuration
(closed strings are simpler in this context, see \cite{GSW1} p31,
\cite{Mohaupt} p20).

Again, an appropriate choice of conformal factor can be made to map
the tubes, corresponding to the external strings, to points on the
surface of the sphere. This would then correspond to a \emph{tree-
level string diagram} \index{tree-level diagram}. Adding loops would
require the modification of the topology of the compact closed
surface on which the string states are inserted. For a given number
of holes (or external states), the topologies of such surfaces are
classified by their \ind{genus} $g \geq 0$, or equivalently, by their
Euler number,
\beq \chi = {1 \over 4 \pi} \int_{\Sigma_g} d^2 \sigma \sqrt{+\gamma}
R^{(2)} (\gamma) = 2-2g \, \label{102p} .\eeq
Here $g=0$ is the two-sphere, $g=1$ is the torus. The general genus $g
$ surface $\Sigma_g$ is obtained from the sphere by attaching $g$
handles. The handles play the role of the loops in Feynman diagrams.
A scattering amplitude involving a given number of external states
will then roughly be given by a sum over $g$. This, in a few words,
is the general philosophy, and we will expand this further in Sect.~\ref{Sect-ClosedStringPartition}.

We have thus seen that, thanks to the invariance of the Polyakov
action under conformal rescalings of the worldsheet, we are able to
map worldsheets having tubes (in the closed string case), extending
into the far past and the far future, to compact worldsheets, where
the original holes corresponding to the external states are now
closed up. The external string states now appear as points. To keep
track of the properties of the corresponding string state, an object,
called the \emph{vertex operator} \index{vertex operator,} has to be
inserted at that point. Note that the appellation vertex operator has
this particular meaning in the context of string theory and is less
general than in CFT or Mathematics for example, even though these
notions are related.

To make this statement more precise, let us perform the change of
coordinates (\ref{101}) in the mode expansions (\ref{19}) and (\ref
{23}), after Wick rotation. This will allow us to reinterpret the
mode expansions as Laurent series in the complex plane. In the closed-
string case, we find
\beq
\partial X^\m (z) &=& -i \sqrt{{\alpha' \over 2}} \sum_{n=-\infty}^
\infty \alpha_n^\m z^{-n-1} \nn \\
\bar{\partial} X^\m (\bar{z}) &=& -i \sqrt{{\alpha' \over 2}}
\sum_{n=-\infty}^\infty \tilde \alpha_n^\m \bar{z}^{-n-1} \, . \label
{103}
\eeq
These relations can be inverted by using the Cauchy integral formula
to give
\beq \alpha_{-n}^\m &=& \sqrt{{1 \over \alpha'}} \oint {dz \over 2
\pi } z^{-n} \partial X^\m(z) \nn \\
\tilde \alpha_{-n}^\m &=& \sqrt{{1 \over \alpha'}} \oint {d\bar{z}
\over 2 \pi } \bar{z}^{-n} \bar{\partial} X^\m(\bar{z} ) \, , \label
{104} \eeq
where the contour integrations encircle the origin $z=0$ of the
complex plane with counterclockwise orientations.
But, as we have seen, string states are built by the application of oscillators to states $| k,0\rangle $, which correspond to states
with momentum $k^\m$ and no oscillator excited, see \re{57p}.



%
%
%
%
%
%

States built from the $\a^\m_{-n}$'s  are thus related, from (\ref{104}) and the residue formula\footnote{$\oint_C {\phi(z) \over (z-z_0)^\m }dz = 2 \pi i {\phi^{(n-1)}(z_0) \over (n-1)!} $, $\phi(z)$ analytic in $C$.}, to $\partial^{(n)}X^\m(0)$, corresponding to an insertion of a point-like operator at $z=0$. 

From this point of view, the state with no oscillators excited will correspond to the insertion of 
\beq  
\int d^2z :e^{ikX} : \, . \label{105} 
\eeq
The integration over the whole sphere expresses that the result should not depend upon the parametrization or, stated in another way, that the state should be independent of the particular insertion point \cite{JohnsonBook}, p.50. 

The normal ordering signs $: \, \, \, : $ can be understood in the usual fashion as an operator normal ordering, where oscillators with positive frequencies (destruction operators) have to be moved to the right \footnote{We will not insist on the distinction between operator and conformal normal ordering, even though they are in general different and that in principle the second one has to be used. This is because they coincide for systems called free fields, which is the case here in particular. For the WZW models which we will be dealing with in the next chapter, it turns out that both definitions "almost" coincide. For a discussion on normal ordering, see e.g. \cite{CoursModave}}. As  another example, the closed string level  one  vertex operator, corresponding to the emission or absorption of  the state $\zeta_{\mu\nu} \a^\m_{-1} \tilde \a_{-1}^\n | k, 0\rangle $ representing $h_{\m\n}, \, B_{\m\n}$ and $\phi$ is 
\beq 
\int d^2z : \zeta_{\mu\nu} \partial X^\m \bar{X}^\n e^{ikX} : \, . 
\label{106}
\eeq

\subsection{Bosonic string as a CFT, and the Weyl invariance}\label{Weyl}

The Polyakov action in conformal gauge, eq(\ref{12}), is nothing other than 26 free scalar boson fields $X^\m$ in 2 dimensions in flat space. This system is a well-known conformal field theory, and propagators as well as OPEs of bosonic fields can be obtained using standard techniques (see \cite{DiFr} p128-161, \cite{Spector} p35). The energy momentum tensor, using the coordinates (\ref{101}) is 
\beq 
T(z) &=& - {1\over \a'} : \partial X^\m \partial X_\m : (z) \nn \\
\bar{T}(\bar{z}) &=& - {1\over \a'} : \bar{\partial} X^\m \bar{\partial} X_\m : (\bar{z})
\, . 
\label{107}
\eeq
Note that in field theory in general, and in CFT in particular, the background metric is generally supposed to be fixed (and usually flat), and not considered as a dynamical field (namely, there is no kinetic term in the action associated with it). The energy-momentum tensor takes its canonical form in terms of the Lagrangian density
\beq
T^{\a\b}_c = -\eta^{\a\b} \cal{L} + 
\frac{\p \cal{L}}{\p (\p_\a \phi)}
\p^\b \phi 
\, , 
\label{108}
\eeq
where $\phi$ denotes the collection of fields of the theory (here $\phi = \{ X^\m\}$). The \emph{canonical energy-momentum tensor} \index{canonical energy-momentum tensor} is the conserved current associated with translation invariance \cite{DiFr}, p45. However, a commonly used trick in 
CFT for actions of the form (\ref{1}) (viewed as a 2D field theory) relies on the fact that the Lagrangian is a scalar density, which allows us to forget about the (``fixed'') status of the background metric and allow it to vary temporarily\footnote{When this is the case, the action is then invariant under arbitrary diffeomorphisms. The situation is then sensibly different from ordinary QFT, where the background metric is fixed to $\eta^{\m\n}$, and where the action is only invariant under Poincar\'e transformations.}\footpourmoi{Discussion Philippe: on peut ecrire une theorie des champs ou $g$ n'est pas dynamique, mais ou l'action est invariante sous diffeos, c-a-d contient un terme $\sqrt{-g}$. On a alors une famille de theories, une theorie pour chaque choix de systeme de coordonnees. Travailler avec $\eta$ correspond a un choix particulier. Pour Maxwell, il y a plus, car l'action est aussi invariante pour $g_{ab} \ra \phi g_{ab}$, ce qui peut se formuler par l'invariance de Weyl de Maxwell, meme si $g$ n'est pas dynamique.}. This leads to definition \re{9} for the energy-momentum tensor, usually evaluated in flat space. 

The advantage of this definition is that the energy-momentum tensor is then identically symmetric. Notice that both definitions in general do not coincide. Since we always have the freedom to modify the energy-momentum tensor by adding the divergence of a tensor $B^{\rho\m \n}$, with $B^{(\rho \m) \n} = 0$ (this does not affect the classical conservation law nor the Ward identity, \cite{DiFr}), one can show that for systems with rotation invariance, $T^{ab}_c$ can be brought into the so-called Belinfante form $T^{ab}_B$, which is (classically) symmetric \cite{DiFr}, p49. 

It also turns out that, as far as scalar fields are concerned, both definitions (\ref{9}) and (\ref{108}) coincide, as can be checked from (\ref{1}) or (\ref{12}), which makes the distinction irrelevant in our case. 

Let us point out a relationship between Weyl and conformal symmetry in 2D flat space, when only scalar fields are involved. We particularize to 2 dimensions. Since $T^{\m\n}$ is then symmetric and generates translations, under $x^a \rightarrow x^a + \epsilon^a(x^b)$ the action varies as 
\beq 
\delta S \div \int d^2 \sigma T^{ab} \partial_a  \epsilon_b \, .
\label{109}
\eeq 
This does not vanish identically, since an ordinary QFT is not invariant under general coordinate transformations. Let us now consider conformal transformations. The $\epsilon^a$ is a conformal Killing vector which satisfies
\beq 
\partial_a \epsilon_b + \partial_b \epsilon_a = (\partial_c \epsilon^c) g_{ab} \, , 
\label{110m1}
\eeq
where $g_{ab}$ is the (Lorentzian or Euclidian) flat metric. Eq (\ref{109}) thus becomes, using the symmetry of $T^{ab}$: 
\beq 
\delta S \div \int d^2\sigma T^a{}_a (\partial_c \epsilon^c) \, . 
\label{110}
\eeq
Thus, the (off-shell) tracelessness of the energy-momentum implies the invariance of the action under conformal transformations. 

We already observed in eq. \re{10} that the trace of the 2D energy-momentum tensor corresponding to the bosonic string vanishes, using its explicit form \re{10}. This is a general property for theories exhibiting Weyl invariance. A Weyl transformation in 2 dimensions reads  
\beq 
\delta_w \gamma_{ab} &=&  2 w \gamma_{ab} \nn \\
\delta_w X^\m &=& 0  \hspace{2cm} \m = 0, \dots, D-1 \, . 
\label{111}
\eeq 
The index $\m$ labels the $D$ scalar fields. This is nothing other than the infinitesimal form of eq. \re{8}. To be rigorous, Weyl transformations only make sense in a theory where the metric is dynamical\footpourmoi{Ph n'est pas d'accord, voir une footpourmoi precedente}. But as we might expect, things are special in 2 dimensions. Weyl invariance of a general action
$S = \int d^2x \cL (\g, X^\m)$ implies (with ${\d\cL \over \d \g_{ab} }Ê= \frac{1}{2} \sqrt{-\g} T^{ab}$)
\beq 
0 = \d_W S = \int d^2x (\sqrt{-\g} T^{ab} w \g_{ab}) \, , 
\label{112}
\eeq
and thus
\beq T^a{}_a = 0 \, , 
\label{113}
\eeq
and vice-versa.

Now, since the Polyakov action \re{1} is constructed using $\g_{ab}$ in a covariant way, one is naturally led to add a kinetic term for $\g_{ab}$ to \re{1}, which is simply the Einstein-Hilbert action in two dimensions:
\beq 
\chi = {1 \over 4 \pi} \int_\Sigma d^2 \sigma \sqrt{-\g} R^{(2)} \, , 
\label{114}
\eeq
where $R^{(2)}$ is the two-dimensional Ricci scalar\footnote{We assume here for simplicity that $\Sigma$ has no boundary, otherwise see \cite{Pol1}, pp15-30.} on $\Sigma$. The additional term \re{114} is the only possible one respecting the original symmetries \re{4}, \re{5} and \re{8} \cite{Pol1}.
The action $S_P' = S_P - \lambda \chi$ looks like general relativity coupled to 26 scalar fields. The equation of motion for $\g_{ab}$ then reads 
\beq
R_{ab} - {1\over 2} \g_{ab} R^{(2)} = T_{ab} \, . 
\label{115}
\eeq
However, in 2 dimensions, the l.h.s. of \re{115} vanishes identically (see e.g. \cite{GSW1}p.179), and there is indeed no dynamics associated with the metric. It is nothing more than an auxiliary field, and there is no clash between the facts that we let it vary on one side, and it is non-dynamical on the other side. As discussed around equation \re{102p}, the quantity $\chi$ depends only on the topology of the worldsheet, and so will only matter when comparing worldsheets of different topologies, i.e. when discussing perturbation theory and interactions.

In summary, starting from the covariant Polyakov action, its Weyl invariance implies the tracelessness of the 2D energy-momentum tensor. Because we are in two dimensions, there is no dynamics associated with $\g_{ab}$, which corresponds to an auxiliary field, whose degrees of freedom can furthermore be removed using the invariances at hand. Using the latter, one can go to the conformal gauge, where the resulting action can be identified as describing a two-dimensional CFT (26 free bosons). The string spectrum is obtained from this CFT, after further imposing the Virasoro constraints, which will restrict the physical spectrum to a subspace of that of the free-bosons CFT.

The basic OPEs for closed strings are given by (see e.g. \cite{Spector}p59, \cite{DiFr,JohnsonBook})

\beq 
T(z) \p X^\m (w) &\sim& {\p X^\m (w) \over (z-w)^2} + {\p (\p X^\m)(w) \over (z-w)}
\label{116} \, , \\
T(z) : e^{ikX} : (w, \bar{w}) &\sim& {\a' k^2 \over 4} {:e^{ikX}:(w,\bar{w}) \over (z-w)^2} + {\p_w:e^{ikX}: (w,\bar{w}) \over (z-w)} \, , 
\label{117}\\
\bar{T}(\bar{z}) : e^{ikX} :(w, \bar{w}) &\sim& {\a' k^2 \over 4} {:e^{ikX}:(w,\bar{w}) \over (\bz-\bo)^2} + {\p_\bo :e^{ikX}: (w,\bar{w}) \over (\bz-\bo)} \, , 
\label{118}\\
\p X^\m(z) \p X^\n(w) &\sim& - {\a' \over 2}{ \eta^{\m0} \over (z-w)^2} \, , 
\label{119}\\
\p X^\m(z):e^{ikX} : (w,\bo) &\sim& - {i \a' \over 2} {k^\m \over (z-w)} :e^{ikX}:(w,\bo) 
\, , 
\label{120} 
\\
:e^{ik_1X(z)}: :e^{ik_2X(w)}: &\sim& |z-\o|^{\a'k_1 k_2} :e^{ik_1X(z) +ik_2X(w)}:
\, . 
\label{121}
\eeq
The relationships \re{116}-\re{118} state that $\p X^\m (z)$, $\pb X^\mu (\zb)$ and $:e^{ikX}:(z,\zb)$ are primary fields of conformal dimensions $(h,\bar{h})$ equal to $(1,0)$, $(0,1)$ and $(\frac{\a' k^2}{4},\frac{\a' k^2}{4})$ respectively. Eq. \re{119} is equivalent to the commutation relations \re{54} satisfied by the modes $\a^\m_n$, see Sect. \ref{CrashCFT}. Indeed, in radial quantization, equal time commutators become contour integrals, from which we may find, using \re{103} and \re{119}

\beq 
\, -\frac{\a'}{2} [\a^\m_n, \a^\n_m] &=& {1\over (2\pi i)^2} \oint   d w\oint  dz
z^n w^m  \p X^\m(z) \p X^\n (w) \nn\\
&=&   {1\over (2\pi i)^2} \oint   d w\oint  dz
z^n w^m  \p (-{\a' \over 2}) {\eta^{\m\n} \over (z-w)^2} \nn\\
&=& -\frac{\a'}{2}{1 \over 2 \pi i} \oint dw^m \eta^{\m\n}  n w^{n-1} \nn\\
 &= & -\frac{\a'}{2} {1 \over 2 \pi i} \oint dw^{m+n-1} \eta^{\m\n}  n  = -\frac{\a'}{2} n \eta^{\m\n} \delta_{n+m,0} \nn \eeq


These OPEs allow one to make the field-state correspondence mentioned in Sects. \ref{CrashCFT} and \ref{StrVert} more precise. The absolute vacuum state $|0\rangle$, with no oscillator excited and zero momentum corresponds to the unit operator (field) $I(z,\zb)$. Since the OPE $\p X^\m (z) I(z,\zb)$ has no singular term, it follows from \re{104} that 
\beq 
\a^\m_{-n} | 0\rangle = \sqrt{{2 \over \a'}}\, \,  \lim_{w,\bo \rightarrow 0} \,  \oint {dz \over 2 \pi} z^{-n} \p X^\m (z) I(w,\bo) = 0 \, , \, \, \, n \leq 0 \, . 
\label{123}
\eeq
Tachyon ground states with momentum $k$, $|k;0\rangle$, are related to highest weight states created from the vacuum by the operator $:e^{ikX}:(w,\wb)$
\beq 
| k,0\rangle = \, \, \lim_{z,\bz \rightarrow 0}Ê :e^{ikX} : (z,\bz) | 0\rangle \, , 
\label{124}
\eeq 
in the usual field-state correspondence. In the language of Sect. \ref{CrashCFT}, we find from \re{47} and \re{48} that $L_{n>0} |k;0\rangle = 0$, and, as a consequence, these states form a continuum of highest weight states of the Virasoro algebra, with weights  $h=\frac{\a' k^2}{4}$. But we also have to impose the physical constraints \re{75} and \re{76}, i.e. $\frac{\a' k^2}{4} \overset{!}{=}1$, see also \re{80} ($M^2 = -k^2$). Thus, the tachyon vertex operator is a primary field of dimensions $(1,1)$. It follows from \re{120} and \re{104} that
\beq 
\a^\m_{-n} | k,0\rangle &=& \sqrt{{2 \over \a'}} \oint {dz \over 2 \pi} z^{-n} \p X^\m (z) 
: e^{ikX}: (0,0) | 0\rangle \nn \\
 &= &  \sqrt{{2 \over \a'}} \oint {dz \over 2 \pi} z^{-n-1}(- {i \a' \over 2} ) k^\m
 : e^{ikX} : (0,0) | 0\rangle \, . 
\nn
\eeq
 If $n<0$, this vanishes as it should. If $n=0$, we find
 \beq 
\a^\m_0 | k,0\rangle \triangleq p^\m \sqrt{{\a'\over 2}} |k,0\rangle &=& 
\sqrt{{\a'\over 2}}  {1 \over 2 \pi i} \oint {k^\m \over z} | k,0\rangle \nn \\
&=& \sqrt{{\a'\over 2}}  k^\m | k,0\rangle \, , \nn
\eeq 
as expected.

For higher level states, the constraints \re{75} and \re{76} impose that all vertex operators, corresponding to physical string states, be Virasoro primary fields with conformal dimensions $(1,1)$. We will return to this property when discussing strings in curved backgrounds.

\section{Closed string partition function and modular invariance}\label{Sect-ClosedStringPartition}
We already started a heuristic discussion in Sect. \ref{StrVert} on how interactions are implemented in string theory. A scattering amplitude involving $M$ external states is supposed to have the following form:
\beq 
A(1,\dots,M) = \sum_{g=0}^\infty K^{M-\chi} A(1,\dots,M)_g\, , 
\label{125}
\eeq
where $A(1,\dots,M)_g$ is the $g$-loop contribution. This is a perturbative expansion in the \ind{string coupling} $K$. Since a factor of $K$ is required at each interaction vertex, an interaction process on $\Sigma_g$ involving $M$ external states will involve $M-2+2g$ fundamental string interactions and will be assigned a factor $K^{M-\chi}$, see \re{125}. This is because a tree diagram with $M$ external states has $M-2$ interaction vertices, while each loop (measured by $g$) adds 2 more. Because there is a single coupling, the expansion in the coupling coincides with the expansion in loops. The contribution $A(1,\dots,M)_g$ is the correlation function of the integrated vertex operators on $\Sigma_g$, corresponding to the states involved in the process (examples of such operators are \re{105} and \re{106}). This correlation function is computed in the worldsheet quantum field theory defined by the Polyakov action \re{1}. The result is then interpreted as a scattering amplitude of strings in space-time, with the in- and out-states represented by the vertex operators. Explicitly, one gets for \re{125}
\beq 
A(1,\dots, M) = \sum_{g=0}^\infty K^{M-\chi}N_g \int {\cal{D}} X^\m 
{\cal{D}}\g_{ab} e^{-S_p [X^\m,\g_{ab}]}V_1\dots V_M \, , 
\label{126}
\eeq 
where we are still working in Euclidean signature on the worldsheet (see Sect. \ref{CrashCFT}). $N_g$ are normalization factors needed to define the path integral. The $V_i$s are the integrated vertex operators of the physical states 
\beq 
V_i = \int_{\Si_g} d^2 z \sqrt{g} V_i(z,\bz) \, .
\label{127}
\eeq
For example, for $i=$tachyon, $V_i(z,\zb) = :e^{ikX}:(z,\zb)$. We are not going to enter into a full analysis of the features of \re{126} (for a detailed account of the Polyakov path integral, see \cite{Pol1}, Chap.3 for example). Rather, we will focus on a particular example: the closed-string one-loop amplitude. Before this, let us pause for a moment on the ${\cal{D}}\g_{ab}$ integration. Here, we have to deal with a theory exhibiting local gauge symmetries, those corresponding to 2D diffeomorphisms ($Diff$) and Weyl invariance ($Weyl$). The correct way to deal with \re{126} is the Faddeev-Popov procedure (for a short account, see \cite{Zee}), which might be better known from electromagnetism, consisting of fixing a gauge choice and factoring out the integration over the gauge group. This way avoids the overcounting arising from the fact that configurations related by $Diff \times Weyl$, in the case of string theory, represent the same physical configuration. This method implies the introduction of additional fields which supplement the fields $X^\mu$, called (anti-)ghosts. It turns out that there is an obstruction to the decoupling of the infinite Weyl factor: it can only happen when the conformal anomaly is cancelled. This is precisely the case in 26 dimensions, where the c-number anomaly in the Virasoro algebra cancels out if the ghost contributions are included. This can be shown to be equivalent to the decoupling of the Weyl factor \cite{GSW1}, p124. Another point of view consists in accepting the Weyl factor as a new, purely quantum degree of freedom. This leads to \ind{non-critical string theory} or \ind{Liouville string theory}. We will not consider these theories here.

Finally, we must note that the ${\cal{D}}\g_{ab}$ integration has not completely disappeared, as could have been expected from the gauge-fixing procedure. This is because the gauge-fixing
\beq\label{128}
\g_{ab} = \d_{ab} \quad \mbox{or} \quad \g_{ab} = e^{\phi} \d_{ab}
\eeq
can only be imposed locally. Let us see what happens in the case of the \ind{closed-string one-loop amplitude}. This has the physical interpretation of computing the vacuum energy. There are no external strings and the worldsheet is topologically a torus (for a discussion on general Riemann surfaces, see \cite{Pol1}p150, \cite{Mohaupt}p25). Start with the $(\si,t)$ plane with the identifications of point
\beq 
(\si, t) \sim (\si, t) + 2 \pi (m,n) \, , \, \, m,n \in \Z \, , 
\label{129}
\eeq
with $\g_{ab} (\si,t) $ and $X^\m (\si,t)$ periodic. A theorem (\cite{Pol1}, p147) states that it is not possible to bring a general metric to unit form by a $Diff \times Weyl$ transformation leaving invariant the periodicity conditions \re{129}, but it is possible to bring it to the form 
\beq 
g_{ab} (z) dx^a dx^b \triangleq |d\si + \tau dt|^2\, , \, \, \tau \in \C 
\, .
\label{130}
\eeq
Schematically, because of global periodicity constraints, one sees that $\int {\cal{D}}\g_{ab} F[\g_{ab}] \ne \mbox{Vol}(Diff \times Weyl) F[\g_{ab}]$, $F$ being an arbitrary functional\footpourmoi{Philippe: preciser}. The periodicity conditions being fixed, one can only bring the metric in a form $g_{ab}(\tau) \ne \d_{ab}$. We will thus rather have
\beq\label{131}
 \int {\cal{D}}\g_{ab} F[\g_{ab}]_{|\underset{\large{t \sim t +2 \pi m}}{\si \sim \si +2 \pi n}} =  \mbox{Vol}(Diff \times Weyl) \int d\tau F[g_{ab}(\tau)]_{|\underset{\large{t \sim t +2 \pi m}}{\si \sim \si +2 \pi n}}.
\eeq
After having fixed the gauge invariance, we are thus left with an ordinary integral over $\tau$. In other words, not all metrics on the two-torus are 
$Diff \times Weyl$ equivalent: a one-parameter family of inequivalent metrics  parametrized by $\tau$ exits. The parameter $\tau$ is called \ind{complex structure} or \ind{modulus} of the torus. Reparameterizations and Weyl transformations cannot change the complex structure, but can map $g_{ab}(\tau)$ to the unique representative of the complex structure class which has constant curvature, since the space of constant curvature metrics on a two-dimensional closed compact surface is isomorphic to the space of complex structures \footpourtoi{et pour cordes ouvertes?}. 

Alternatively, we may always bring the metric in unit form, but then there is  no guarantee that this respects the original periodicity. The complex structure $\tau \eqdef \tau^1 + i \tau^2$ specifies the shape of the torus in this case, which cannot be changed by $Diff \times Weyl$, and
\beq\label{132}
 \int {\cal{D}}\g_{ab} F[\g_{ab}]_{|\underset{\large{t \sim t +2 \pi m}}{\si \sim \si +2 \pi n}} = \mbox{Vol}(Diff \times Weyl)\int d \tau F[\d_{ab}]_{|\underset{\large{\si \sim \si+ 2 \pi n \tau^1 +2 \pi m}}{t \sim t +2 \pi n \tau^2}} .
 \eeq 
The same reasoning can be extended to more general string diagrams. For genus $g$ surfaces, the integral over $\tau$ is replaced by an integral over the space ${\cal M}_g$ of complex structures ($dim_\C {\cal M}_g = 3g - 3$, $g>1$).

Let us note that we also have a reciprocal phenomenon for $g<2$: there are reparameterizations (global conformal isometries) which do not change the metric, implying an overcounting of equivalent contributions to the amplitude. This yields a multiplicative factor, which is the volume of the group of conformal isometries, and which has to be canceled by the normalization factors (see \cite{Pol1}p149, \cite{GSW1} p39, \cite{Mohaupt} p27).

We may finally turn to the explicit form of the partition function (or one-loop string amplitude). The metric \re{130} is invariant under $\tau \ra \tau^*$ and degenerate for $\tau \in \R$, so we can restrict attention to $Im(\tau)>0$. Thus the identifications are
\beq 
\si &\sim& \si + 2 \pi m + 2 \pi n \tau^1 \nn \\
t &\sim& t + 2 \pi n \tau_2\, , \, \, \tau_2 >0 \, .
\label{133}
\eeq
\footpourtoi{faire dessin S32}
Introducing $w = \si + i t $, the torus is represented in the upper half plane as a parallelogram with base length equal to $1$, height equal to $\tau^2$, and with $\cos \a = \tau^2/\tau^1$, $\a$ characterizing its ``inclination". There is some additional redundancy, since the values $\tau + 1$ and $-1/\tau$ generate the same set of identifications as $\tau$ (see \cite{Pol1}, p148). Repeated applications of these 2 transformations 
\beq 
T : \, \tau' = \tau + 1 \, , \, \, \, \d : \, \tau' = - {1 \over \tau} 
\label{134}
\eeq
generates
\beq
\tau' = { a\tau + b \over c \tau + d}\hspace{2cm} a,\, b, \, c, \, d \, \in \Z \quad , \quad  ad-bc = 1\, . 
\label{135}
\eeq
We may also consider the transformation
\beq
\left( \begin{array}{c}
\si \\
\tau 
\end{array}
\right)
= 
\left( \begin{array}{cc}
d & b \\
c & a
\end{array}
\right)
\left( \begin{array}{c}
\si' \\
\tau' 
\end{array}
\right)
\, , 
\label{136}
\eeq
which takes the metric \re{130} for $(\si,\tau)$ into a metric of the same form in $(\si',\tau')$, but with modulus $\tau'$. This is a diffeomorphism of the torus. It is one-to-one as a consequence of $ad-bc=1$, and it preserves the periodicity \re{129}. However, it cannot be obtained from the identity by successive infinitesimal transformations; it is a so-called \ind{large coordinate transformation}. They form the group $\mbox{SL}(2,\Z)$. The group on the $\tau$-plane is $\mbox{PSL}(2,\Z) \eqdef \mbox{SL}(2,\Z)/\Z_2$, because $\tau'$ is left unchanged if all the signs of $a,b,c,d$ are reversed. The group of transformations \re{136}is called the \ind{modular group}. Using \re{135}, it can be shown that every $\tau$ is equivalent to exactly one point of the \ind{fundamental modular domain} ${\cal F}$ for the upper half-plane mod $\mbox{SL}(2,\Z)$:  
\beq 
-{1\over 2} \leq Re(\tau) \leq {1\over 2} \, , \, \, |\tau| \geq 1 \, . 
\label{137}
\eeq
The fundamental region ${\cal F}$ is a representation of the moduli space of $(Diff \times Weyl)$-inequivalent metrics or inequivalent flat tori.

We are now finally ready to turn to the evaluation of the one-loop closed string diagram. It could be evaluated from \re{126}, for $M=\chi=0$, and no vertex operator inserted, and by going through the path integral on $X^\mu$ (see \cite{Pol1}p210). By fixing $\g_{ab}$ to $\d_{ab}$, the $\tau$-dependence will appear in the periodicity conditions on $X^\m$, and the $\int {\cal D} \g_{ab}$ integration will be trade to a $\int d\tau$ integration. This derivation is rather technical and we will skip it (see \cite{DiFr} p.340, and subtleties due to the residual conformal group are discussed in \cite{Pol1} p.149, \cite{GSW1} p.39). Instead, we will use the more intuitive operator formalism to compute the partition function. Recall for instance that in the path integral formalism, the partition function for a quantum physical system with one degree of freedom can be written as (see e.e. \cite{Pol1}, Appendix A):
\beq 
\mbox{Tr} \; \exp(\hat{H}T) &=& \sum_i exp(-E_i T) \nn \\
 &\eqdef& \int dq \langle q | e^{-\hat{H}T} | q\rangle  \nn \\
 &=& \int dq \, \langle q,T|q,0 \rangle_E \nn \\
 &=& \int dq \; \int [{\cal D}q]_{q,0}^{q,T} \, e^{-S_E} \nn \\ 
 &=& \int {\cal D} q_p e^{-S_E} \, , 
 \label{138}
 \eeq
where $\hat{H}$ is the Hamiltonian of the system $\ [{\hat{p}^2 \over 2m} + V(\hat{q}) ]$, the sum running over all eigenvectors of $\hat{H}$ with eigenvalues $E_i$. $S_E$ is the Euclidean action, and ${\cal D} q_p$ denotes the integral over all periodic paths on $[0,T]$. 

By analogy, let us consider the propagation of a closed string, evolving in a loop. The space and time directions are seen to run along the real and imaginary axes. A fixed point in the string, which is assumed to lie horizontally in the $w$-plane, propagates upwards for a time $t=2 \pi \tau^2$, and translates to the right by an amount $\si = 2 \pi \tau^1$. Since time translation is generated by the worldsheet Hamiltonian $H$, while the shift along the string is generated by the worldsheet momentum $P$, we find for the contribution of a torus with parameter $\tau = \tau^1 + i \tau^2$:
\beq 
Z = \mbox{Tr} \{ e^{-2 \pi \tau_2 H }e^{2 \pi i \tau_1 P }\} \, . 
\label{139}
\eeq
We found in \re{50} and \re{51} the expressions $H=L_0 + \tilde{L_0}$ and $P=L_0 - \tilde{L_0}$. However, due to ordering ambiguities in quantizing the theory, we had to perform the shift $L_0 \ra L_0 -1$, $\tilde{L_0} \ra \tilde{L_0} -1$. The latter point may alternatively be understood as follows \cite{DiFr}, p.139, \cite{Pol1}, p.52. We used, in Sect. \ref{StrVert}, two representations of the string worldsheet: the cylinder, with coordinates $\sigma$,$\tau$ (Euclidian time), and the complex place, with coordinates
\beq\label{139p} 
z=e^{-i w},\quad \zb=e^{-i \wb}, \quad w=\si + i \tau.
\eeq
Notice we use $0 \si \leq 2 \pi$ for convenience here. We know from \re{95p} the transformation law of the energy-momentum tensor under conformal transformations. So, we have
\beq
T_{ww}&=& (\p_w z)^2 T_{zz} + {c \over 24} \nn \\
T_{\bo \bo} &=& (\p_\bo \bz)^2 T_{\bz \bz}+ { \tilde c \over 24} \nn \\
T_{w \bo} &=& T_{z \zb} = 0 \, . 
\label{140}
\eeq
Using \re{15-16},\re{27}-\re{30}, and the conformal transformation \re{139p}, we get that the Hamiltonian in the $w$ frame is
\beq 
H =  L_0 + \tilde L_0 - {(c + \tilde c) \over 24} \, , 
\label{141}
\eeq
and for the momentum
\beq 
P =  L_0 - \tilde L_0 - {(c - \tilde c) \over 24}=  L_0 - \tilde L_0  \, , 
\label{142}
\eeq
because here $c = \tilde c = D$. 
This actually computes the zero-point energy of the cylinder, assuming that the plane has a vanishing vacuum energy density \cite{DiFr}, p.138. Note that for $D=26$, \re{141} reproduces the ordering factor $a=1$ which we already encountered. 

Thus \re{139} can be recast into  
\beq 
Z = (q \bar{q})^{-D/24} \mbox{Tr}(q^{L_0} \bar{q}^{\tilde L_0})\, , \hspace{1cm} q = e^{2 \pi i \tau}\, ,
\label{143}
\eeq
where the trace is to be taken over momenta and over all excited oscillator states. With the expressions \re{47} for $L_0$ and $\tilde L_0$, it is possible to evaluate the trace explicitly (see \cite{JohnsonBook}p.92, \cite{VanP}p.24, \cite{Pol1}p.210, \cite{SzaboStrings}p.43, \cite{SchellStr}p.86), which breaks up into a sum over occupation numbers and an integral over momentum.
The result, up to irrelavant factors, is
\beq 
Z(\tau)  \sim (\tau_2)^{-D/2} |\eta(\tau)|^{-2D} \, ,
\label{144}
\eeq
where
\beq
\eta(\tau) = q(\tau)^{1/24} \prod_{n=1}^\infty (1-q^n)
\label{145}
\eeq 
is the \ind{Dedekind eta function}.
By adding the ghost contribution to the partition function, one gets (\cite{Pol1}, p.212)
\beq 
Z(\tau)  \sim (\tau_2)^{-D/2} |\eta(\tau)|^{-2(D-2)} \, .
\label{146}
\eeq
The factor $D-2$ can also be understood from light cone gauge quantization, where only the $D-2$ transverse degrees of freedom are quantized.
Finally, we have to integrate over $\tau$ (with $d\tau d\bar{\tau} = 2 d\tau^2 d\tau^2$). Also, we have to take into account the global conformal transformations, which here consist of the translations of the torus. Since this group has finite volume, we need not fix it, and we may divide it out by hand. Therefore the integration reads  
\beq 
\int {d\tau d\bar{\tau} \over (2 \pi)^2 \tau_2} 
\label{147}
\eeq
for the torus defined by \re{133}, the denominator being the area of the torus, which is the volume of the global conformal group of the torus, $U(1)^2$.
We thus finally get
\beq 
Z  \overset{.}{=} \int_{{\cal F}} {dz d\bz \over \tau_2} (\tau_2)^{-D/2} |\eta(z)|^{-2(D-2)}
\, . 
\label{148}
\eeq 
We have here chosen to integrate over the fundamental domain ${\cal F}$. However we could have equally chosen the image of ${\cal F}$ under arbitrary combinations of the $S$ and $T$ transformations \re{134}. When integrating over each such region, we should get the same answer, since otherwise the result would depend on the choice we make among the equivalent regions. The requirement that we always get the same answer is called \ind{modular invariance}. It is important since it means that we are correctly integrating over all inequivalent tori, and that we are counting each such torus only once. Incidentally, let us note that this argument yields an additional support for the value of the critical dimension: it is indeed only for $D=26$ that \re{148} is modular invariant. This follows from the fact that the measure 
\beq 
{dz d\bz \over \tau_2^2} \, ,
\label{149}
\eeq 
which is nothing other than the invariant measure on the upper half plane (or \ind{Poincar\'e disc}), as well as
\beq
 \tau_2 |\eta(\tau)|^4
\eeq
are actually separately modular invariant, which stems from the modular transformations of $\eta(\tau)$:
\beq 
\eta(\tau + 1 ) &= &e^{i\pi /12 }\eta(z) \label{150} \\
\eta (- 1/ \tau) &=& (-i\tau)^{1/2} \eta(z) \label{151}.
\eeq
Let us rewrite \re{148}, for $D=26$, as
\beq 
& &\int_{{\cal F}} {d\tau d\bar{\tau} \over (\tau_2^2)} (\tau_2)^{-12} (\eta(z) \bar{\eta}(\bar{\tau}))^{-24} \nn \\
&=&  \int_{{\cal F}} {d\tau d\bar{\tau} \over (\tau_2^2)} (\tau_2)^{-12} (q \bar{q})^{-1}
\prod_{n=1}^\infty (1- q^n)^{-24} \prod_{m=1}^\infty (1- \bar{q}^m)^{-24} \nn \\
& \triangleq &  \int_{{\cal F}} {d\tau d\bar{\tau} \over (\tau_2^2)} Z_p (\tau , \bar{\tau})
\, . 
\label{152}
\eeq
Now note that 
\beq 
\prod_{n=1}^\infty (1- q^n)^{-1} &=& \sum_{N \geq 0} p(N) q^N \nn \\
&=& 1 + q + 2 q^2 + 3 q^3 + 5 q^4 + 7 q^5 + 11q^6 + \dots \nn 
\eeq 
is the generating function for the number of partitions \cite{DiFr}, p.193. Similarly, we find
\beq 
\prod_{n=1}^\infty (1- q^n)^{-24} &=& (1 + 24 q + {24.25 \over 2} q^2 + \dots)(1+ 24 q^2 + \cdots)\cdots 
 \nn \\
 &=& 1 + 24 q + \frac{24.27}{2} +\cdots \nn 
 \eeq
$Z_p(q)$ deserves the name \ind{partition function} since, when expanded in powers of $q$ and $\bar{q}$, the powers in the expansion times $4/\a'$ refer to the mass squared level of excitations in the left and right sectors, while the coefficient in the expansion gives the degeneracy at that level.





\section{Strings in non-trivial background fields}\label{SectNTBG}
Let us return for a moment to Sect. \ref{Sect-Class}, where we first studied the classical propagation of a particle and a string in flat space-time. Turning to curved space-time is almost a trivial operation for the particle case, since it amounts to replacing $\eta_{\m\n}$ by any curved metric in \re{1p}. This will lead to the geodesic equations of the given background. For strings, we are thus tempted to consider the minimal modification to \re{1}:
\beq
S = - \frac{1}{4 \pi \a'} \int d^2\si (-\g)^{1/2} \g^{ab} g_{\m\n}(X^\a)\p_a X^\m \p_b X^\n. 
\eeq
In conformal gauge, this describes a 2D field theory, called \ind{non-linear sigma model}, describing $D$ bosonic fields with field dependent couplings $g_{\m\n}(X^\a)$. Looking back at what we already uncovered about strings, some embarrassment could spring to mind: since the quantized string possesses a graviton in its spectrum, we could reasonably expect that strings generate dynamically curved backgrounds. Also we would have to be sure that \re{152} is compatible with the way strings generate gravitons. Actually, we will see that consistency of the theory restricts the allowed $g_{\m\n}$. This is the striking difference between points and strings: it is not possible to make sense of string propagation in an arbitrary gravitational background, as it was for a particle. An analogous situation can be found in electromagnetism: one may for instance use the photon picture to describe a process in the background of strong magnetic fields. The latter obviously have to satisfy Maxwell's equations, and are referred to as a coherent state of photons, to remind us that these cannot be reconstructed using only the perturbative photon description \cite{JohnsonBook}. Loosely stated, the restriction on the allowed curved backgrounds would be that these could be interpreted as a coherent state of gravitons. This statement will be made more precise in  Sect. \ref{SVOMO}.

The crucial ingredient for the consistency of string theory turns out to be the local Weyl invariance, since the construction of states, vertex operators and amplitudes relies heavily on having a CFT on the worldsheet, see Sect. \ref{StrVert}. Weyl invariance is also mandatory to allow for all degrees of freedom of the auxiliary metric $\g_{ab}$ to be removed. If it is broken, we are left with an additional degree of freedom, the Liouville field.

Therefore we will first discuss the relation between Weyl and conformal invariance, and then discuss how the preservation of conformal invariance {\itshape on the worldsheet} imposes restrictions on the non-trivial {\itshape space-time} backgrounds.

\subsection{Tracelessness of $T_{\mu\nu}$}
This section complements the discussion of Sect.\ref{Weyl} on the relation between Weyl, scale and conformal invariance. We will always assume that translation and rotation invariance hold, so that the energy-momentum tensor is conserved and symmetric \cite{DiFr}, Chap.2. It can then be shown that conformal invariance is equivalent to the tracelessness of the energy-momentum tensor: we saw one side of the implication in \re{110}, while the other one is shown by introducing an improved tensor, see \cite{RivaCardy}. It is also of widespread belief that scale invariance implies conformal invariance for systems invariant under translations and rotations in 2 dimensions \cite{DiFr}, p.107. This is under some general assumptions, that we will always hold to be true in the following. It is however interesting to analyze the reasoning. Scale, rotation and translation invariance imply the weaker condition
\beq 
\langle T^a{}_a (x)  T^b{}_b (0) \rangle = 0 \, . 
\label{153}
\eeq 
This expresses that the operator $T^a_a$ has zero expectation value and zero standard deviation in the ground state \cite{DiFr}, p109. There exists nevertheless at least one example where \re{153} does not imply $T^a_a = 0$, i.e. $\langle T^a_a \phi \rangle = 0$, for all fields, see \cite{RivaCardy}.

Weyl invariance, on the other hand, is rigorously defined only in theories where the metric is dynamical; we mentioned in Sect.\ref{Weyl} the trick consisting of allowing $g_{\a\b}$ to vary before setting it to its fixed value, from which we may define Weyl invariance in ordinary QFT. In 2 dimensions, there is no such distinction because gravity is non-dynamical, anyway. It implies that the energy-momentum tensor obtained by varying the action w.r.t. the metric is traceless, see \re{112} and \re{113}. Thus Weyl invariance is, in our situation, equivalent to conformal invariance.

\subsection{RG flow and marginal operators}\label{I.4.2}
Let us consider a field theory in $D$ dimensions defined in terms of a scalar field $\phi$, with Lagrangian $\cL (\phi)$, involving coupling constants $\l_i$. The field may be Fourier decomposed as
\beq 
\phi(x) = \int {d^d k \over (2 \pi)^d}\tilde \varphi(k) e^{ikx} \, . \label{153p}
\eeq
Now look at the Euclidian path integral
\beq 
Z(\Lambda) = \int_\Lambda {\cal D}\phi e^{- \int d^d x  \cL(\phi)} \, , 
\label{154}
\eeq
depending on some cut-off $\Lambda$. The notation $\int_\Lambda$ instructs us to integrate over only those field configurations $\phi(x)$ such that $\phi(k) =0$ for $|k|=(\sum_{i=0}^{D-1}k_i^2)^{(1/2)} > \Lambda$. This amounts to putting on blurry glasses with resolution $L=1/\Lambda$, such that we do not see or admit fluctuations with length scales less than $L$ \cite{Zee}. The reason for this is that quantum field theory should be regarded as an effective low energy theory, valid up to some energy scale $\Lambda$, beyond which we are ignorant of the physics. Roughly said, the \ind{renormalization group flow equations}(RG flow equations) of a given QFT measures how the physical coupling constants depend on the energy scale we are working on. This comes from a renormalization procedure known as Wilson-Kadanoff renormalization scheme \cite{Peskin}. The first step consists in integrating out the Fourier components $\tilde \phi (k)$ such that $\Lambda/s <|k| < \Lambda$, the so-called fast (or wriggly) modes, where $s$ is some dilation factor ($s>1$). To this end, we write
\beq \label{155}
\varphi(k) = \varphi_s(k) + \varphi_f(k)\, , \nn 
\eeq
with 
\beq 
\varphi_s(k) = \left\{ 
\begin{array}{ll}
\varphi(k) &  \, \, |k| \leq 1/s \\
  0 & \, \, |k| > 1/s 
\end{array}
\right.
\label{156}
\eeq
and 
\beq 
\varphi_f(k) = \left\{ 
\begin{array}{ll}
\varphi(k) &  \, \, \Lambda/s \leq |k| \leq \Lambda \\
  0 & \, \, |k| < \Lambda/s
\end{array}
\right.
\label{157}
\eeq
and we plug this into \re{154}. We obtain 
\beq 
Z = \int_{\Lambda/s}  {\cal D} \phi_s e^{-\int d^dx \cL (\phi_s)} \int \cd \phi_f e^{-\int d^dx \cL_1(\phi_s,\phi_f)} ,
\label{158}
\eeq
where all the terms in $\cL_1(\phi_s,\phi_f)$ depend on $\phi_f$. For example, in the $\lambda \phi^4$ theory, $\cL(\phi) = {1\over 2} (\p_\m \phi)^2 + {1\over 2} m^2 \phi^2 + {\lambda_4\over 4!} \phi^4$, we would find \cite{Peskin}
\beq 
\cL_1 (\phi_s , \phi_f ) = {1 \over 2} (\p_\m \phi_f)^2 +  {1 \over 2} m^2 \phi_f^2 + \lambda_4 ( {1 \over 6}
\phi_s^3 \phi_f +  {1 \over 4 } \phi_s^2 \phi_f^2 +  {1 \over 6} \phi_s \phi_f^3 +  {1 \over 4!} \phi_f^4)
\label{159} 
\eeq
by noticing that terms of the form $\phi_s \phi_f$ vanish, since Fourier components of different wavelengths are orthogonal.
Now, imagine doing the integral over $\phi_f$, and call the result
\beq 
e^{-\int d^d x \d_0 \cL(\phi_s)} \triangleq \int \cd \phi_f e^{-\int d^dx \cL_1 (\phi_s,\phi_f)} \, .
\label{160}
\eeq
This equality is of course formal, since in practice the integral over $\phi_f$ can only be done perturbatively, order by order, assuming that the relevant couplings are small. However, with \re{160}, we may write
\beq
S = \int_{\Lambda /s} \cd \phi_s e^{-\int d^dx [ \cL(\phi_s + \d \cL( \phi_s)]} 
\, . 
\label{161}
\eeq
We have thus re-expressed the theory in terms of the smooth field $\phi_s$ only. The corrections $\d \cL(\phi_s)$ compensate for the removal of the large-$k$ Fourier components, by supplying the interactions among the remaining $\phi_s$ that were previously mediated by fluctuations of the $\phi_f$.

Suppose we want to compare \re{154} and \re{161}. We would then like to change $\int_{\Lambda/s}$ in \re{161} to $\int_\Lambda$. Let us make a scale transformation on the slow-mode action \re{161}: 
\beq 
k \, \rightarrow \, k' = s k \hspace{1cm} \mathrm{or} \hspace{1cm} \, x \, \rightarrow \, x' = x / s \, , 
\label{162}
\eeq
so that $|k'| < \Lambda$. In general such a transformation also affects the fields 
\beq 
\phi(x) \, \rightarrow \, \phi'(x/s) = s^\Delta \phi(x) \, \, \, \, \mathrm{or} \, \phi'(k) = s^{\Delta-d} \phi(k)
\, , 
\label{163}
\eeq
where $\Delta$ is the \ind{scaling dimension} of the field.

For the time being, let us ignore $\d \cL (\phi_s)$ to keep the discussion simple. This amounts to assuming that the couplings (e.g. $\l_4$ in \re{159}) are so small that they have a negligible effect on the fast mode integration \re{160} (and just contribute to a number)\footpourtoi{Check S41!}. Then,
\beq 
\int d^d x \cL (\phi_s) &=& \int d^d x' s^d [ {1 \over 2} s^{-2} (\p' \phi_s)^2 + {1\over 2} m^2 \phi_s^2 + {\lambda_n
\over n!} \phi_s^n ] \nn \\
&=& \int d^dx' [ {1\over 2} s^{d-2 -2\Delta} (\p' \phi'_s )^2 + {1\over 2} m^2 s^{d-2\Delta} \phi'{}_s^2+  
{\lambda_n\over n!}Ês^{d-n\Delta} \phi'_s ] 
\, ,
\label{164}
\eeq
where we considered an interaction  term ${\lambda_n\over n!} \phi_s^n$ to be slightly more general. 
For $\Delta = {1\over 2}d-1$ (called Gaussian scaling), we are in position to compare (\ref{161}) to (\ref{154}), by redefining the new couplings
\beq 
m'^2 &=& m^2 s^2 \nn \, , \\
\lambda'_n &=& \lambda_n s^{d + {n\over 2} (2-d)} = \lambda_n s^{d- \Delta_n} \, ,
\label{165}
\eeq 
where $\Delta_n = n \Delta$ is the scaling dimension of the operator $\phi^n$.
This shows that, as one becomes interested in physics over larger distance scales (in passing from $\Lambda$ to $\Lambda/s$), we can take \re{161} and write it as in \re{154} (with the {\itshape same} cut-off), except that the couplings have to be replaced by their primed versions \re{165}. We will anticipate a little here, by pointing out that \re{165} is a simple example of a general procedure, which leads to a set of coupled flow equations for the evolution of the couplings with the energy scale. For a general theory with coupling constants $g_i$, and when also including higher order contributions (\re{165} is only the zero order contribution), they take the form
\beq \label{166}
{dg_i \over d \ln(s)} = \b_i (g_j) \, .
\nn
\eeq
These are the \ind{renormalization group flow equations}, and $\beta_i$ is commonly referred to as the \ind{beta function} associated with the parameter $g_i$. A \ind{fixed point} $g_j^*$ of the renormalization group is a point in parameter space that is unaffected by the renormalization procedure. It is thus characterized by a vanishing beta function: $\b_i(g_j^*) = 0$. A fixed point of the RG transformations defines a theory that has scale invariance at the quantum level.

The simplest example where the whole procedure can be carried out exactly is the \ind{free boson} or \ind{Gaussian model}, for which $\l_{n\geq 3} =0$. In this case, we see from \re{159} and \re{158} that the fast and slow modes are decoupled. Therefore, integrating the fast modes in \re{160} can be performed exactly and produces only an irrelevant multiplicative constant in front of the partition function. From \re{165}, it is clear that $m=0$ gives a fixed point of the renormalization group, provided the scaling dimension is $\Delta =  \frac{1}{2} d - 1$.

Let us return to \re{165}, that we obtained in the zero order approximation, i.e. by neglecting $\d \cL(\phi_s)$. Since $s>1$, we see that the operators with $d-\Delta_n <0$ will have a coupling constant which gets smaller and smaller as $s$ grows, i.e. as we look over larger distance scales. These operators are said to be  \emph{irrelevant}\index{irrelevant operator}, as they can eventually be neglected. Conversely, the operators with $d-\Delta_n >0$ are called \emph{relevant}\index{relevant operator}. Operators for which $d-\Delta_n =0$ are called \emph{marginal}\index{marginal operator}.

From another point of view, let us start from a fixed-point action $S_0 [\phi]$ (e.g. that of a massless free boson), and see what happens when we go away from this point. In its vicinity, a generic action $S[\phi]$ may be expressed as
\beq 
S[\phi] = S_0[\phi] + \sum_i g_i \int d^d x O_i (x) 
\, , 
\label{167}
\eeq
where the $O_i$s are some local operators, expressible in terms of the fields $\phi$ (e.g. $\phi^n$ as above). The couplings must be small if we are close to the fixed point. Again, we assume that the couplings are so small that they are negligible when treating the fast mode integration, i.e. we do not consider a term $\d \cL (\phi_s)$. In this approximation, we may follow the procedure described above to find
\beq
g_i' = g_i s^{d-\Delta_i}\, , 
\label{168}
\eeq
where $\Delta_i$  is the scaling dimension of the operator $O_i $ ($O'_i(x) = s^{\Delta_i} O_i (sx)$). In differential form, 
(\ref{168}) may be expressed as
\beq {dg_i \over d \ln(s)} = g_i (d- \Delta_i) 
\label{169}
\eeq
or 
\beq 
L {d g_i \over dL} = (d-  \Delta_i) g_i \, , 
\label{170} 
\eeq 
where $L = \Lambda^{-1} $ and $\Lambda- \d \Lambda = \Lambda (1 - \d s)$. Going being the zero order approximation, we have to consider the contribution of $\d \cL(\phi_s)$, which will yield an additional term, called dynamical, in contrast to the geometrical term in the r.h.s. of \re{169} or \re{170}. It can be evaluated perturbatively by expanding the exponential in the r.h.s. of \re{160} in powers of the perturbing couplings, up to the desired order (see e.g. \cite{Peskin}, p.397, \cite{DiFr}, p.86). Eqs. \re{169} and \re{170} will then be modified to the more general form \re{166}, where $\b_i$ also encodes information on the dimension $d$ and the scaling dimensions of the fields.

\subsection{String vertex operators as marginal operators}\label{SVOMO}
We discussed at the beginning of Sect. \ref{SectNTBG} the importance of Weyl invariance for string theory, or equivalently the necessity for the corresponding two-dimensional field theory to be conformally invariant. Since the latter requirement is equivalent (up to some details) to the scale invariance of the theory, we will be able to take over some terminology introduced in the preceding section, and adapt it to the string theory situation.

In conformal gauge, the flat space action defines the free field theory of 26 bosons on the worldsheet, which is conformally invariant at the quantum level. We know from Sect.\ref{StrVert} that the insertion of states is equivalent to the insertion of operators into the theory, so that
\beq 
S \, \rightarrow \, S' = S + \d \lambda \int d^2 z \, O(z,\bar{z}) \, . 
\label{168p}
\eeq
This is, of course, nothing other than \re{167} in 2 dimensions. Recalling that the scaling dimension of a field is related to its conformal weights $(h,\bar{h})$ by $\Delta = h + \bar{h}$, \emph{marginal operators}\index{marginal operator} will correspond to dimensions $(1,1)$, which is required to preserve conformal (scale) invariance of the action at least at the classical level \cite{Ginsparg}. As before, operators with $h + \bar{h}<(>)2$ will be said relevant (irrelevant).

We thus see that, for the action to remain conformally invariant, a necessary condition is to consider only insertions of marginal operators. But marginality is exactly equivalent to the mass-shell conditions \re{59}, $(L_0 -1) |\varphi\rangle = (\tilde L_0 -1) |\varphi\rangle =0$. So, things seem to fit together well: we may consistently add string vertex operators into the game. However, it is not guaranteed that a marginal operator is still marginal in the infinitesimally deformed theory. In other words, it is not clear that we may run finitely away from $S$ in \re{168p}, for a finite parameter $\l$. Only those marginal operators which stays marginal under deformation generate finite deformations of a conformal field theory. These are called \emph{exactly marginal operators}\index{exactly marginal operator} or \emph{truly marginal operators}\index{truly marginal operator}.

Let us further describe how gravity is described in string theory, starting with our ansatz \re{152} for a string propagating in a curved background.
First we write
\beq 
g_{\m\n} (X) = \eta_{\m\n} + h_{\m\n} (X) \, , 
\label{169p}
\eeq
where $h_{\m\n}(X)$ is supposed to be a small perturbation. Then we may rewrite 
\beq 
S_p[ g_{\m\n}] = S_p [ \eta_{\m\n}] + V[ \zeta_{\m\n}] \, , 
\label{170p}
\eeq 
with $h_{\m\n}(X) = \zeta_{\m\n}(k) e^{ikX}$, and $V[\zeta_{\m\n}]$ corresponding to the graviton vertex operator with polarization tensor $\zeta_{\m\n} (k)$. Writing \re{162} is thus consistent with the way string theory incorporates gravity: $S_P[g_{\m\n}]$ is obtained by deforming the flat space action by the graviton vertex operator. From \re{170p}, we may also gain a better vision of what we mean by ``a coherent state of gravitons". Indeed, it allows us to relate amplitudes computed with both actions as
\beq \label{171}
 \langle V_1 \cdots V_M \rangle_g = \langle V_1 \cdots V_M \, e^{V[\zeta_{\m\n}]} \rangle_{\mbox{flat}}.
\eeq
The operator $e^{V[\zeta_{\m\n}]}$ generates a coherent state of gravitons in flat space. To see this, we may think of the simple harmonic oscillator in Quantum Mechanics, with creation operators $a^\dagger$ and destruction operators $a$ satisfying $[a,a^\dagger] =1$. Coherent states are defined as states with minimal Heisenberg uncertainty \footpourtoi{Pq?}. They are eigenstates of annihilation operators and can be constructed as $e^{z a^\dagger} |0\rangle$ \cite{JohnsonBook}. They are not eigenstates of the number operators, but are superpositions of states with all possible occupation numbers. In \re{171}, $V[\zeta_{\m\n}]$ plays the role of $a^\dagger$, and shows that amplitudes in a curved background can be computed equivalently by inserting the vertex operator for a coherent state of gravitons into the flat space amplitude.

Now, thanks to the physical mass-shell condition, we know that the graviton vertex operator is marginal. But actually, we have to require it to be exactly marginal. This demands that \re{152} is a conformally (scale) invariant field theory, and amounts to the vanishing of the \ind{beta functional} associated with the field-dependent coupling $g_{\m\n}(X)$\footpourtoi{Comment S45}. The concept of beta functional, instead of beta function, comes out because the theory has a coupling functional $g_{\m\n}(X)$, which can be seen as a continuously infinite number of couplings. Since a theory with $n$ couplings has $n$ beta functions, here we will get a beta functional $\b^g_{\m\n} (X^\rho)$, depending on the same degrees of freedom that enter the coupling \cite{GSW1}, p.172. It can be computed perturbatively, order by order in $\a'$\footpourtoi{v. Johnp59, GSW 168, Mohauptp32}.
The one-loop beta functional can be computed (see e.g. \cite{GSW1,ForsteStrings}) as
\beq 
\b^g_{\m\n} (X^\rho) = -{1 \over 2 \pi} R_{\m\n} (X^\rho) \, .
\label{172} 
\eeq
This is a striking result, since demanding Weyl (or conformal) invariance of the two-dimensional field theory implies that the target-space metric satisfies vacuum Einstein's equations! Of course, \re{172} receives higher loop corrections (higher order terms in $\a'$), which can be interpreted as stringy corrections to general relativity.
  
Including the remaining massless modes of the closed string spectrum (antisymmetric tensor field $B$ and dilaton $\phi$), we are led to the action
\begin{equation}\label{173}
 S = \frac{1}{4 \pi \a'} \int_\Sigma \, d^2\sigma \left( \sqrt{-\gamma} \gamma^{ab} \p_a X^\mu \p_b X^\nu g_{\m\n} + \epsilon^{ab} \p_a X^\mu \p_b X^\nu B_{\m\n} + \a' \phi  R^{(2)}\right).
\end{equation}
The coupling for $B_{\m \n}$ is rather straightforward, while for the dilaton, a coupling to the two dimensional Ricci scalar is the simplest way of writing a reparameterization invariant coupling when there is no index structure. Correspondingly, there is no power of $\a'$ in this coupling, as it is already dimensionless. This action brings to light the prominent role played by the dilaton. Let us indeed return to the string partition function,eq. \re{126}, with $V_i = 1$ and with the full action \re{173} inserted. Shifting the dilaton by a constant $a$ has the effect of shifting the total action \re{173} by a constant proportional to the Euler number. For the corresponding partition function, this is equivalent to rescaling the coupling by $e^a$:
\begin{equation}
 Z = \sum_g \kappa^{-\chi(g)} \int \, {\cal D}X {\cal D} \gamma \, e^{-S} \lra \sum_g (\kappa e^a)^{-\chi(g)} \int \, {\cal D}X {\cal D} \gamma \, e^{-S} .
\end{equation}
This indicates that the string interaction coupling constant is not an arbitrary external parameter, but is fixed by the vacuum expectation value of a space-time dynamical scalar field in the theory (for details, see \cite{Mohaupt}, p35, or \cite{GSW1}, p183).


The new couplings $B_{\m \n}$ and $\phi$ give rise to additional beta functionals, that are required to vanish:
\begin{eqnarray}\label{174}
 \beta_{\m\n}^G  &=& \a' (R_{\m\n} + 2 \nabla_\m \nabla_\n \phi - \frac{1}{4} H_{\mu \k \sigma}H_\n^{\k \sigma}) + O(\a'^2) = 0 \nonumber\\
  \beta_{\m\n}^B  &=& \a' (-\frac{1}{2} \nabla^\k H_{\k \m\n} + (\nabla^\k \phi) H_{\k \m\n}) + O(\a'^2) = 0 \nonumber\\
  \beta^\phi  &=& \a' (\frac{D-26}{6 \a'} - \frac{1}{2} \nabla^\m \nabla_\m \phi + \p_\m \phi \p^\m \phi - \frac{1}{24} H_{\k \m\n} H^{\k \m\n}) + O(\a'^2) = 0 
 \end{eqnarray}
with $H=dB$. A remarkable fact is that these equations can be interpreted as space-time field equations for the background fields. They can indeed be derived from the following \emph{low energy effective action}\index{low energy effective action}
\begin{equation}\label{175}
 \frac{1}{2 \k^2} \int \, d^D X \sqrt{-G} e^{-2 \phi} \left[ R + 4 \p_\mu \phi \p^\m \phi - \frac{1}{12} H^2 - \frac{2(D-26)}{3 \a'} + O(\a')\right].
\end{equation}
This action describes the interaction of the massless modes of the bosonic closed string, in the decoupling limit $\a' \ra 0$. This action can be put in the more usual form $\int \, d^D X \sqrt{-G} \left[ R + \cdots \right]$ by absorbing a suitable power of $e^{-\phi}$ in the definition of the space-time metric. One then obtains the action expressed in the Einstein frame, in opposition to the string frame (see \cite{Mohaupt} p35).

A background will be called an \emph{exact string background}\index{exact string background} if the non-linear sigma model \re{173} is conformally invariant to all orders in $\a'$. We will be dealing with such backgrounds in the next chapter. In these situations, algebraic and conformal field theoretic techniques are to be used to establish the conformal invariance.

For further purposes, we mention that a symmetry exists in the string background fields equations, pointed out in \cite{Buscher,Buscher:1987qj}. This duality maps any solution of the low energy string equations \re{174} with a translational symmetry to another solution. Given a solution $(g_{\m\n},B_{\m\n},\phi)$ that is independent of one coordinate, say $x$, then $(\tilde{g}_{\m\n},\tilde{B}_{\m\n},\tilde{\phi})$ is also a solution where
\begin{eqnarray}\label{Buscher}
\tilde{g}_{xx} &=& 1/g_{xx} \quad \tilde{g}_{x\a}=B_{x\a}/g_{xx} \nonumber \\
\tilde{g}_{\a\b}&=& g_{\a\b} - (g_{x\a} g_{x\b} - B_{x\a} B_{x\b})/g_{xx}   \nonumber\\
\tilde{B}_{x\a}&=& g_{x\a}/g_{xx} \quad \tilde{B}_{\a\b} = B_{\a\b} - 2 g_{x[\a}B_{\b]x}/g_{xx} \nonumber\\
\tilde{\phi} &=& \phi - \frac{1}{2} \ln g_{xx}.
\end{eqnarray}

This is called \ind{Buscher's duality} or sometimes \ind{T-duality}.




\section{D-branes}
We are going to draw some attention to \emph{D-branes}\index{D-brane}, which we caught a glimpse of when analyzing the possible boundary conditions for open strings. We saw that choosing Dirichlet boundary conditions for $D-p-1$ space-like coordinates at one end (say $\si =0$) of the string:
\beq\label{176}
 X^i = x^i \quad \mbox{at} \quad, i =p+1, \cdots ,D 
\eeq
implies that this open string end is allowed to move in a hypersurface embedded in space-time, called \ind{Dp-brane}, where $p$ refers to the number of space-like directions.

Useful references for this wide subject, of which we discuss only certain aspects, include \cite{DaiLP, Leigh, ForsteStrings, JohnsonBook, Vancea, Abouel,Callan:1988wz,Pol1,Pol2}. We will unfortunately have to leave aside topics like the boundary state formalism \cite{JohnsonBook, Vancea, DiVecchia1,DiVecchia2} and the identification of D-branes as non-perturbative string states \cite{Pol1,Pol2, SzaboStrings, JohnsonBook, ForsteStrings}.  

\subsection{Bosonic T-duality}
One route to get a feeling for the nature of D-branes is the notion of \ind{T-duality}, which appears when studying the spectrum of strings in a space-time where one or more directions have been compactified. It manifests in very specific ways according to the type of strings we are dealing with. 
\subsubsection{Closed strings}
We first consider closed strings, where we compactify one of the space dimensions (say the 25th) along a circle of radius $R$:
\beq 
X^{25} \sim X^{25}+ 2 \pi R \, .
\label{177}
\eeq 
The conjugate momentum corresponding to $X^{25}$ must be quantized as 
\beq
p^{25}= {n\over R} , \, n \in \Z 
\, 
\label{178}
\eeq 
as a consequence of the fact that the generator of translations along the compact direction  $e^{ipa}$ must reduce to the 
identity for $a = 2 \pi R$ \cite{JohnsonBook}.

If we run along the string, $\si \rightarrow \si + \pi$, then from (\ref{19}), re-expressed as
\beq \label{178p}
X^\m = x^\m + \sqrt{2 \a'} (\a_0^\m + \tilde \a_0^\m) \tau - \sqrt{2 \a'} (\a_0^\m - \tilde \a_0^\m) \si + \mbox{oscillators},
\eeq
we may rewrite
\beq 
X^\m \, \rightarrow \, X^\m + \pi \sqrt{2 \a'} (\a_0^\m - \tilde \a_0^\m) \, .
\label{179}
\eeq
For the uncompactified directions, the single-valuedness of the embedding was encoded in the solution \re{19}, to give
\beq\label{180}
\a_0^\m = \tilde \a_0^\m= \sqrt{\frac{\a'}{2}} p^\m.
\eeq
For  the compact direction, (\ref{177}) and (\ref{179}) imply
\beq 
\sqrt{2\a'} (\a_0^{25}-\tilde \a_0^{25} ) = 2  R \o \, , \, \, \, \o \in \Z \, , 
\label{181}
\eeq
where $\o$ is the \emph{winding number} \index{winding number} of the string along $X^{25}$. The space-time momentum (\ref{64}) of the string, with (\ref{178p}), is 
\beq 
p^\m = {1 \over \sqrt{2 \a'}} (\a_0^\m + \tilde \a_0^\m) \, , 
\label{182}
\eeq
and thus (\ref{182}) and (\ref{170}) yield 
\beq 
\a_0^{25} + \tilde \a_0^{25} = \sqrt{2 \a'} {n\over R} \, , \, \, n \in \Z \, .
\label{183}
\eeq 
From (\ref{181}) and (\ref{183}), we have
\beq 
\a_0^{25} = \sqrt{{\a'\over 2}} \underbrace{( {n\over R} + { \o R\over \a'})}_{:= p_L =p_-^{25}} \hspace{1cm} \mathrm{and}
\hspace{1cm} 
\tilde \a_0^{25} = \sqrt{{\a'\over 2}} \underbrace{( {n\over R} - { \o R\over \a'})}_{:= p_R = p_+^{25}}
\, , 
\label{184}
\eeq 
from which we may compute the mass spectrum in the (24 + 1) uncompactified dimensions ($M^2 = - p_\m p^\m, \, \m = 0 , \dots , 24)$: 
\beq 
M^2 = {2 \over \a'}Ê(N + \tilde N - 2 )  + ({ n \over R})^2 + ( { \o R \over \a'})^2 \, . 
\label{185}
\eeq
With respect to \re{77}, we see that the spectrum has been enriched by the appearance of two kinds of particles: the usual Kaluza-Klein (KK) modes, familiar from dimensional reduction, together with some new excitations, called 
\ind{winding modes},
because they can be thought of as generated by the winding of the closed strings around the compact direction. 

Now notice that \re{185} is invariant under the exchange
\beq 
n \leftrightarrow \o \, \, \, \mathrm{and} \, \, \, R \leftrightarrow R' = { \a \over R} 
\, . 
\label{186}
\eeq
This is called a
\ind{T-duality} 
transformation and $R'$ is the compactification radius of the T-dual theory. Dual theories are identical in the fully interacting case as well, as both the partition function and correlators can be shown to be invariant under T-duality \cite{JohnsonBook}, pp.100-103.
Substituting \re{186} into \re{184}, we get the action of T-duality on the zero modes
\beq 
\a_0^{25}\rightarrow \a_0^{25}\, , \hspace{2cm} \tilde \a_0^{25}\rightarrow - \tilde \a_0^{25}\, . 
\label{187}
\eeq
This transformation has to be extended to all the oscillators
\beq 
\a_n^{25}Ê\rightarrow \a_n^{25}\, , \hspace{2cm} \tilde \a_n^{25}Ê\rightarrow - \tilde \a_n^{25}\, . 
\label{188}
\eeq
in order for the operators $\tilde L_n$ in \re{47} and consequently the physical subspace defined by \re{60} to be left invariant. 

Eqs. \re{187} and \re{188} allow us to define the action of T-duality directly on the string coordinate $X^{25}$. From \re{187} and \re{188}, we see by writing
\beq 
X^{25} &=& {1 \over 2} (X^{25}_- + X^{25}_+ ) \, , \label{189} \\
X_-^{25} &=& x_-^{25} + \sqrt{2 \a'} (\tau - \si) \a_0^{25 }Ê+ i \sqrt{2 \a'} \sum_n 
{\a_n^{25} \over n }Êe^{-2 i n (\tau-\si)} \nn \\
X_{+}^{25} &=& x_+^{25} + \sqrt{2 \a'} (\tau + \si) \a_0^{25 }Ê+ i \sqrt{2 \a'} \sum_n 
{\tilde \a_n^{25}\over n }Êe^{-2in(\tau+\si)} \nn 
\eeq
that the T-dual coordinate $\hat{X}^{25}$ can be taken to be
\beq
\hat{X}^{25} = {1 \over 2} (X_-^{25} - X^{25}_+)
\label{190},
\eeq
since we must only impose that $x^{25}_+ \rightarrow - x_+^{25}$ under T-duality, as a consequence of $p_+^{25} \rightarrow - p_+^{25}$ (\re{187} and \re{188}), in order to keep the commutation relation $[x^{25},p^{25}]=i$ unchanged \cite{OoguriStrings}. 

Therefore, the T-duality transformation acts on the right sector as a parity operator changing the sign of the right moving coordinate $X_+$ and leaving unchanged the left moving one $X_-$. 

Let us glance at the behavior of the spectrum \re{185} at particular values of $R$. As $R \rightarrow \infty$, all of the winding states become infinitely massive and therefore cannot be excited, while the KK (momentum) states go over to a continuum. This fits with intuition since we recover the fully uncompactified result. 
In the case $R \rightarrow 0$, all the momentum states become infinitely massive, a phenomenon well-known from Kaluza-Klein reduction in field theory. But at the same time, the pure winding states ($n=0$, $\o \neq 0$ in \re{185}) form a continuum, since it costs less energy to wind a string around a small circle.  
The presence of this continuum can be interpreted as the reappearance of an effective uncompactified dimension in the $R\ra 0$ limit.
\footpourtoi{T-dual BG always correspond to the same CFT, see \cite{Rocek:1991ps,TheseRib}?}

The T-duality symmetry becomes even richer if one considers the compactification of more than one dimension (see e.g. \cite{JohnsonBook}, p104, \cite{Pol1}, p249, \cite{ThIsrael}), called toroidal compactification. Also, the notion of T-duality has a generalization in non-trivial background configurations. It can be seen as a symmetry of the two-dimensional non-linear sigma model \re{173} when the latter exhibits an invariance under constant shifts in a given direction $x$ (i.e. the Lie derivatives of $g$, $H=dB$ and $\phi$ w.r.t. $\p_x$ vanish), whose effect on the background field corresponds to Buscher's duality transformations \re{Buscher}, see \cite{Rocek:1991ps,Giveon:1991jj,Giveon:1994fu} (see also \cite{ForsteStrings}).
These can be interpreted as generalizing the duality $R \leftrightarrow \a/R$, since, if one compactifies the $x$ direction, the square root of the metric component $g_{xx}$ measures the radius of the $x$ direction.

\subsubsection{Open strings}
For open strings, the story is different, since the string coordinate does not satisfy any periodicity requirement on $\sigma$. This implies that, in its compactified version, there are only momentum modes, while the winding sector is absent. So, when $R \rightarrow 0$, the states with non-zero internal momentum go to infinite mass, but there is no new continuum of states arising from winding: we are left with a theory in one dimension less. A puzzle seems to appear when considering a theory with both closed and open strings, since in the $R \rightarrow 0 $ limit the closed strings live in $D$ space-time dimensions, but the open strings only in $D-1$. This mismatch can be solved by requiring that, in the T-dual picture, open strings can still oscillate in $D$ dimensions, while their endpoints are fixed on a 25-dimensional hyperplane called $D24$-brane (24 referring to the number of spatial dimensions). Let us see more precisely how this comes about. Write the open string mode expansion \re{23} as
\beq 
X^\m &=& {1 \over 2}(X^\m_+ + X^\m_-) \, , \hspace{1cm} \mathrm{with} \nn \\
X^\m_+ &=& x^\m  - c^\m + \a_0^\m \sqrt{2 \a'} (\tau + \si) + i \sqrt{2 \a'} \sum_n {\a_n^\m \over n} e^{-in(\tau + 
\si ) }\, , \nn \\
X^\m_- &=& x^\m  + c^\m + \a_0^\m \sqrt{2 \a'} (\tau - \si) + i \sqrt{2 \a'} \sum_n {\a_n^\m \over n} e^{-in(\tau - 
\si ) }Ê\, , 
\label{191}
\eeq
where $c^\m$ is an arbitrary number. We now extend the definition of the T-dual coordinate \re{190} to the open string case:
\beq 
\hat{X}^{25}  = c^{25}- \a_0^{25}\sqrt{2 \a'} \si - \sqrt{2 \a'} \sum_n {\a_n^{25} \over n}e^{-in\tau} \sin(n\si)
\, . 
\label{192}
\eeq
Notice that there is no $\tau$-dependence in the zero mode sector. Since the oscillator terms vanish at $\si=0, \pi$, we see that the endpoints do not move in the $\hat X^{25}$-direction. This is of course expected, since from \re{191} and \re{190} it is clear that $\p_\tau X^{25}Ê= - \p_\si \hat X^{25} $ and $\p_\si X^{25} = - \p_\tau \hat X^{25}$: under T-duality, Neumann and Dirichlet boundary conditions are exchanged (see \re{20} and \re{21}) \cite{DiVecchia1}. More precisely, we have that 
\beq 
\hat{X}^{25} (\pi) - \hat{X}^{25} (0) &=& - \sqrt{2 \a'} p^{25} \sqrt{2 \a'}\pi \nn \\
&=& -{2 \pi \a' n \over R}Ê= - 2 \pi n \hat{R} \, . 
\label{193}
\eeq
This means that the values of the coordinates $\hat X^{25}$ at the two ends are equal up to an integral multiple of the periodicity of the dual dimension. Consequently, in the T-dual theory, the two endpoints of the open string are attached to the same $D$-brane. 

The previous discussion generalizes trivially if we compactify $D-p-1$ directions on circles. We would then find that in the T-dual picture the strings' endpoints are fixed on a $(p+1)$-dimensional hyperplane, called \emph{$Dp$-brane}\index{$Dp$-brane}. From the previous construction, it also follows that each $Dp$-brane can be transformed into a $Dp'$-brane through an appropriate sequence of compactifications and T-dualities \cite{DiVecchia1}. 




\subsection{D-branes as dynamical objects}
Since the closed string sector contains gravity, one would expect that a $Dp$-brane cannot be rigid but must have dynamics \cite{DaiLP}. Indeed, some open string states propagating along the hypersurface can be interpreted as collective motions of the hyperplane, which can fluctuate in shape and position. This can be seen by looking at the massless spectrum of the theory, interpreted in the dual coordinates. Consider for simplicity the situation where a single coordinate is dualized. As for \re{185}, we may  obtain the mass spectrum (in $D-1$ dimensions)
\beq 
M^2 = { 1 \over \a'}(N- 1) + ({n \hat{R} \over \a'})^2 
\, . 
\label{194}
\eeq
Generically, the massless states in the T-dual theory (compactified on $\hat R$) will be \cite{JohnsonBook, DiVecchia1}
\beq 
&\a_{-1}^\m& | k\rangle \, , \hspace{1cm} \m = 0 , \dots , 24 \, (\m = 0, \dots , p) \nn \\
&\a_{-1}^m& | k\rangle \, , \hspace{1cm} m = 25 \,  (m = p+1, \dots , 25) \nn 
\eeq
Starting from the massless excitations of an open string with Neumann boundary conditions in all directions (described by a $D$-dimensional abelian gauge potential), $A_\m$, $\m= 0,\cdots,25$, see eq. \re{74}, compactifying one ($D-p-1$) dimension(s) and T-dualizing, the vector potential $\hat{A}_\m$ splits in a 25(p+1)-dimensional vector $\hat{A}^b ( \xi^a) = \a_{-1}^b |k\rangle$, $a,b=0,\cdots,24(p)$ and $1 (D-p-1)$ scalar field(s) $\hat{\phi}^m(\xi^a)$, where $\xi^a$ denote the coordinates on the worldvolume of the $D24$($Dp$)-brane. In general, a Dp-brane will break the space-time symmetry as 
\beq
SO(1,25) \lra SO(1,p) \times SO(25-p),
\eeq
and its massless degrees of freedom will be given by the set $\{\hat{A}^\a (\xi), \hat{\phi}^m(\xi) \}$. The most natural interpretation of the T-dual version of the abelian gauge field is that the longitudinal coordinates $\hat{A}^\a$ still describe a gauge field living on the Dp-brane, while the $D-p-1$ scalars coming from the transverse components appear as the transverse coordinates of the Dp-brane. They describe the shape of the brane as it is embedded in the dual space-time. The $\hat{\phi}^m(\xi)$ are analogous to the $X^\m (\tau,\si)$ for the string. Meanwhile their quanta describe fluctuations in that background. This is the same phenomenon we found for the description of space-time in string theory. We started with strings in a flat background, and discovered that a massless closed string state corresponds to fluctuations of the geometry. Likewise we first found a flat hyperplane, and then discovered that a certain open string mode corresponds to fluctuations of its shape \cite{JohnsonBook}.

\subsection{Effective action for $Dp$-branes}\label{SectDBIBROL}

We saw that when compactifying closed and open string theory on a $(k=D-p-1)$-torus, and sending the radii to 0, the spectrum of massless modes of the T-dualized theory consists of the usual closed string modes propagating in 26 dimensions with $(26-k)$ Neumann open string modes $A^\a(\xi)$ ($U(1)$ gauge bosons) supplemented by $(D-p-1)$ Dirichlet modes $\phi^m(\xi)$ corresponding to scalar fields propagating in $(p+1)$ dimensions. The latter represent the vibration of the Dp-brane.

The degrees of freedom of the branes are thus expressed in terms of open strings ending on them. As we already stressed, consistency of string theory is tightly knit to the conformal invariance of the corresponding 2$D$ field theory, and this imposes constraints on the allowed non-trivial background fields that we insert into the action, as in \re{173}. The story is the same for open strings in the presence of Dp-branes: the open strings couple to the Dp-brane's degrees of freedom $\{A^b(\xi^a),\phi^m(\xi^a)\}$, as well as to the possibly non-vanishing massless fields of the closed string spectrum\footnote{Notice that we could in principle also include massive, higher spin fields present in the string spectrum}, defining a $2D$ non-linear sigma model \cite{Leigh}: 
\beq 
S&=& S_g + S_B + S_\phi + S_A \, , 
\label{195}Ê\\
S_g &=& {1 \over 4 \pi \a'}\int_\Si d^2z g_{\m\n} \p_\a X^\m \p^\a X_\m
\nn \\
S_B &=&={1 \over 4 \pi \a'}\int_\Si d^2z \epsilon^{\a\b} B_{\m\n} (X) \p_\a X^\m \p_\b X^\n \nn \\
S_\phi &=&  {1 \over 4 \pi \a'}\int_\Si d^2z(- {1\over 2} \a') \sqrt{h} R^{(2)} \phi(X) + {1 \over 2 \pi \a'} \int_{\p \Si} d\tau (- {1 \over 2} \a') K\phi(X) \nn \\
S_A &=& {i \over 2 \pi \a'}\int_{\p\Si} d\tau A_\a(\xi^a) \p_\tau \xi^\a \, , \nn 
\eeq 
where $K$ is the extrinsic curvature of the boundary $\p \Si$ and where the last part corresponds to the coupling of the $U(1)$ gauge field to the boundary of the worldsheet, included in the Dp-brane. The Dirichlet boundary conditions satisfied by $X^\m$ are encoded in
\beq 
X^m_{| \p \Si} = \phi^m(\xi^a) \, , \hspace{1cm}m = 1, \dots , D-p-1 
\label{197}
\eeq
The equations of motion for the space-time degrees of freedom are obtained by requiring conformal invariance of the action \re{196}. The $\b$-functional for the closed string modes are left unchanged (because the bulk theories are the same), while those for the open string modes are equivalent to the equations of motion derived from the so-called 
\ind{Dirac-Born-Infeld action} (detailed computations may be found, for example, in the original papers \cite{Callan:1988wz,Abouel}), which in the static gauge $X^a = \xi^a$ reads \cite{JohnsonBook}, p.136: 
\beq 
S^{DBI}_{Dp} = - T_p \int d^{p+1} \xi e^{-\phi} \sqrt{\hat g_{ab} + \hat B_{ab} + 2 \pi \a' F_{ab}} -= T_p \int d^{p+1}x \: L_{BI}
\label{198}
\eeq 
with
\beq 
\hat g_{ab}&=& g_{\m\n} \p_aX^\m \p_bX^\n \, , \hspace{1cm} a=0,\dots, p \nn \\
\hat B_{ab}&=& B_{\m\n}\p_aX^\m \p_bX^\n \, , \nn \\
F_{ab} &=& \p_aA_b - \p_b A_a \nn 
\eeq
$T_p$ is the brane tension. In contrast to the closed string action, \re{175}, this action is true to all orders in $\a'$, although only for slowly varying field strengths (i.e. disregarding higher derivatives of $F$). This action is invariant under the space-time gauge transformation 
\beq
 \d B = d \zeta \quad , \quad \d A = -\zeta/2\pi \a',
\eeq
suggesting that the physical, gauge-invariant, quantity be $B+ 2 \pi \a' F$ (in the following, we will often set $2\pi \a' = 1$). 

The DBI action
is also invariant under the following transformations of the dynamical
variables :
\begin{itemize}
\item{$x^i \rightarrow x^i + \xi^i, \quad \delta_{\xi}X^{\mu} =
\xi^i \partial_i X^{\mu} , \quad \delta_{\xi}A_i = \xi^j
\partial_j A_i + A_j \partial_i \xi^j$}
\item{$A_i \rightarrow A_i + \partial_i \lambda$}.
\end{itemize}

The associated Bianchi identities, satisfied off-shell, are :
\begin{equation}
 X^{\mu}_{,i} \frac{\delta L_{BI}}{\delta X^{\mu}} + F_{ij} \frac{\delta
 L_{BI}}{\delta A_j} = 0 \quad \mbox{and} \quad \partial_i \frac{\delta
 L_{BI}}{\delta A_i} = 0
\end{equation}

With $K_{ij} = g_{ij} + B_{ij} +
         F_{ij}$ and $K = \mbox{det}(K_{ij})$, the equations of motion (for
         Abelian $F$-field) derived from the DBI action are :
         \begin{eqnarray}
         \partial_k(\sqrt{- K} K^{(kj)} X^{\mu}_{,j})g_{\mu \lambda} +
         \sqrt{-K} K^{(kj)} \Gamma_{\mu \nu,\lambda}
         X^{\mu}_{,j}X^{\nu}_{,k} + \sqrt{-K} K^{[kj]} H_{\mu \nu \lambda}
         X^{\mu}_{,j}X^{\nu}_{,k} = 0
         \nonumber\\
         \partial_i(\sqrt{- K} K^{[ij]}) = 0 \label{eqA} \quad,
         \end{eqnarray} with $K^{(kj)}= \frac{1}{2} (K^{kj} + K^{jk})$ and
         $K^{[kj]}=\frac{1}{2}(K^{kj} - K^{jk})$, $H_{\mu \nu \lambda}
         =\frac{1}{2}(B_{\mu \nu, \lambda}+B_{\nu \lambda,\mu}-B_{\mu
         \lambda,\nu})$ and $\Gamma_{\mu \nu,\lambda} = \frac{1}{2}(g_{\mu
         \lambda,\nu} + g_{\nu \lambda,\mu}-g_{\mu \nu \lambda})$.
         Furthermore, in the presence of space-time Killing vectors $\Xi$, such
         that ${\cal L}_{\Xi} B = d \widetilde{\alpha}$
, the following current is conserved
         on-shell:
         \begin{equation}\label{intprem} J^i = \Xi^{\mu} \frac{\partial L_{BI}}{\partial
         X^{\mu}_{,i}} - \alpha_j \frac{\partial L_{BI}}{\partial A_{j,i}}
         \:,\quad \partial_i J^i \approx 0 \: ,
         \end{equation} with $\alpha_j = \widetilde{\alpha}_{\mu} X^{\mu}_{,j}$ .





\section{Superstrings}

Despite all its nice features, the bosonic string cannot tell the whole story: its spectrum of quantum states contains a tachyon, which signals an unstable vacuum, and furthermore has no fermions. These two shortcomings can be cured by \emph{superstring theory}\index{superstring theory}. The latter admits a consistent truncation of its spectrum of physical states (GSO projection) which namely removes the tachyonic degrees of freedom of the bosonic spectrum, while at the same time renders the particle spectrum in the target space supersymmetric. We will only skim over  this wide subject and just present its basic ingredients; useful references include \cite{GSW1,Pol1,JohnsonBook, OoguriStrings, SzaboStrings, ForsteStrings}. We may take as starting point the generalization of the Polyakov action in conformal gauge \re{12}: 
\beq
S = {1 \over 4\pi\a'} \int_\Si d^2 \si (\p_\a X^\m \p^\a X_\m - i \bar{\psi}^\m
 \rho^\a \p_\a \psi_\m) 
 \, . 
 \label{199}
 \eeq 
 This defines a one-dimensional object with fermionic coordinates on the worldsheet: $\psi_\a^\m (\si , \tau)$ are Majorana worldsheet fermions, where $\a  = 1,2$ is a spinor index, and represent vectors (though the index $\m$) of the space-time Lorentz group $SO(D-1,1)$. The symbol $\rho^\a$ represents two-dimensional Dirac matrices satisfying $\{\rho^\a,\rho^\b\}=-2\eta^{\a\b}$ (see \cite{GSW1} for details)
 
 This action can actually be obtained following a philosophy similar to that which led us to \re{12} \cite{}: start from the superparticle, introduce auxiliary fields ($\g_{ab}$ and its fermionic partner), generalize it to the string, and turn the algebraic equations of motion for the auxiliary fields into constraints to impose on the physical phase space. The equations of motion of the bosonic and fermionic coordinates are the Klein-Gordon and Dirac equations in two dimensions: 
\beq 
\p_a \p^a X^\m = 0 \, , \hspace{1cm}\rho^a \p_a \psi^\mu \, , 
\label{200}
\eeq
which in light cone coordinates and with a decomposition of $\psi^\m_a$ into its chiral components $\psi^\m_\pm$ in a suitable basis of gamma-matrices can be rewritten as 
\beq 
\p_- \p_+ X^\m = 0\, , \hspace{1cm}\p_- \psi^\m_+ = 0 = \p_+ \psi^\m_- 
\, . 
\label{201}
\eeq
The superconformal gauge action \re{199} has global worldsheet supersymmetry
\beq
\d X^\m = \bar{\epsilon} \psi^\m\,, \hspace{1cm}Ê \d \psi^\m = -i \rho^a \p_a X^\m 
\, , 
\label{202}
\eeq
with $\epsilon$ a constant anti-commuting spinor. Correspondingly, there is a conserved supercurrent
\beq
J_{(\a)a} = {1 \over 2}Ê\rho^b \rho_a \psi^\m_{(\a)} \p_b X_\m \, , \hspace{1cm}Ê\p^a J_{\a a}Ê= 0 
\label{203}
\eeq 
The equations of motion \re{201} must be supplemented by the constraints coming from the e.o.m. of the auxiliary fields: 
\beq 
T_{\pm \pm} &=& \p_\pm X^\m \p_\pm X_\m + {i \over 2} \psi^\m_\pm
\p_\pm \psi_{\pm \, \m}Ê = 0 \nn \\
J_{\pm \pm }Ê&=& \psi^\m_\pm \p_\pm X_\m = 0 \, , 
\label{204}
\eeq
If we now focus on closed superstrings for definiteness, one would like to perform a mode expansion of the fermionic coordinates. There are actually two types of boundary conditions a fermion can have: it can be periodic, and hence have integer moding, in which case it is said to be in the \emph{Ramond sector}\index{Ramond sector} (R) , or it can instead be anti-periodic, have half-integer moding and said to be in the \emph{Neveu-Swartz sector} \index{Neveu-Schwartz sector} (NS) \cite{JohnsonBook},p.115. We thus have
\beq
\psi^\m (\si + \pi,\tau ) &=& \psi^\m(\si,\tau) \hspace{1cm} (R) \nn \\
\psi^\m (\si + \pi,\tau ) &=& -\psi^\m(\si,\tau) \hspace{1cm} (NS)
\label{205}
\eeq 
and the corresponding mode expansions: 
\beq 
\psi^\m_-(\si, \tau) &=& \sqrt{2\a'} \sum_{n\in \Z} d^\m_n e^{-2inx^-} \hspace{1cm} (R)\nn \\
\psi^\m_-(\si, \tau) &=& \sqrt{2\a'} \sum_{n\in \Z+1/2} b^\m_n e^{-2inx^-} \hspace{1cm} (NS) 
\label{206}
\eeq
along with their left-moving counterparts. The modes of the generators in \re{204}, defined by\footnote{with the change of coordinates $z = e^{2ix^-}, \, \bar{z} = e^{2ix^+}$}
\beq
T(z) &=& \sum_n L_n z^{-n-2} \nn \\
G(z) &=& \sum_{n/r} G_{n/r} z^{-n/r-3/2} \, ,\quad  n\in \Z \; (R) \, , \quad r\in \Z + 1/2 \; (NS) , \nn
\eeq
along with their anti-holomorphic counterparts, may be expressed in terms of those of $X^\m$ \cite{} and $\psi^\m$. In this way, the constraints may be rewritten mode by mode \cite{JohnsonBook,OoguriStrings}. Upon canonical quantization, the modes satisfy
\beq
 \{d^\m_n,d^\n_m\} = \d_{m+n} \d^{\m\n} \; (R) \quad , \quad \{b^\m_r,b^\n_s\} = \d_{r+s} \d^{\m\n} \; (NS),
\eeq
which implies that the modes $L$ and $G$ satisfy an \ind{$N=1$ superconformal algebra}:
\begin{eqnarray}
[L_n, L_m ] &=& i (m-n) L_{n+m} + \frac{c}{12} (m^3-m) \d_{m+n} \nonumber \\
\{ G_r,G_s \} &=& 2 L_{r+s} +  (4r^2-1)\d_{r+s} \quad , \quad [L_m , G_r ] = \frac{1}{2}(m-2r) G_{m+r}, \nonumber \\
\end{eqnarray}
with $c=3D/2$, the total contribution to the conformal anomaly, and similarly for anti-holomorphic generators, which together form the $N=(1,1)$ superconformal algebra. The appearance of this algebra reflects, in the bosonic case, the presence of a residual symmetry of the gauge-fixed action \re{199}, the superconformal symmetry.  

The quantization of the superstring can be carried out along the same lines as for the bosonic string, up to some important subtelties (mainly having to do with boundary conditions on the fermions, and also because of additional operations like the GSO projection) that we will not discuss here \cite{GSW1}. 

The critical dimension turns out to be $D=10$ in this case (see \cite{GSW1},pp.206,213,235). The closed-superstring spectrum can again be deduced from the open superstring one, by combining the left and right movers, each of them containing an NS and a R sector.

For the complete massless spectrum, one finds two possibilities, according to the relative chiralities of the left and right-movers. These correspond precisely to the 2 types of $N=2$ massless supermultiplets in 10 dimensions, which are the building blocks of \emph{type IIA}\index{type IIA } and \emph{type IIB}\index{type IIB} supergravity theories. The corresponding superstrings are accordingly called \emph{type IIA} and \emph{type IIB}. The massless spectra of these two closed string theories are in perfect correspondence with those of the two supergravities. Actually, by focusing on the massless sector of type II string theories (i.e. taking the $\a' \rightarrow 0 $ limit), one may find that the dynamics can be encoded in terms of a low energy effective action, corresponding to the supergravity action \cite{JohnsonBook,Mukhi}.

Another two superstring theories in 10 dimensions are called \emph{heterotic strings}
\index{heterotic strings}. They arise by fusing the left movers of the bosonic string and the right movers of the fermionic string. The worldsheet theory is then an $N=(1,0)$ superconformal field theory. For this to be possible, the left-moving bosonic string has to be compactified on a torus from 26 to 10 dimensions. Only two 16-tori are consistent (with modular invariance), and they give rise to gauge fields forming Yang-Mills multiplets of $SO(32)$ and $E_8\times E_8$ respectively. The resulting theories are $N=1$ superstrings in 10 dimensions: the \emph{$SO(32)$} \index{$SO(32)$ heterotic superstrings} and \emph{$E_8\times E_8$ heterotic superstrings}
\index{$E_8\times E_8$ heterotic superstrings}. 

The bosonic massless states consist of the usual graviton, $B_{\m\n}$ and dilaton, with an additional $E_8\times E_8$ or $SO(32)$ gauge boson $A^{\m a}$ \cite{JohnsonBook}. The heterotic low-energy effective action can be found to be that of $N=1$, $D=10$ supergravity coupled to $E_8\times E_8$ or $SO(32)$ vector multiplets (see \cite{JohnsonBook}, p.176, \cite{Callan:1985ia}).
\footpourtoi{Comprendre lien entre \cite{JohnsonBook}176, \cite{Callan:1985ia} et \cite{Kiritsis:1994ta}(3.14)!} 

 
 Finally, the last of the five consistent superstring theories is the type I superstring which is a theory of unoriented string worldsheets and contains open strings. 
 
 There are, of course, many very exciting subjects worth discussing (see e.g. \cite{Mohaupt,Uranga}), but it is now time to stop this first contact with strings. 

\cleardoublepage \chapter{Wess-Zumino-Witten models} \label{ChapWZW}
\footpourtoi{13-05-06}
We have seen that exact string backgrounds, as defined in Sect.~\ref{SVOMO}, are associated with two-dimensional conformal field theories on the string worldsheet. This imposes strong constraints on the allowed backgrounds for consistent string propagation: they must satisfy equations of the form \re{174}\footpourtoi{forme, car pour heterotique c'est un peu different}, \emph{and} the whole set of higher order $\alpha$' corrections. Of course, the exactness of a string background has to be inferred deviously, to ensure exact conformal invariance at all orders. One method to achieve this consists in using algebraic tools of conformal field theory. These will allow us to establish the conformal invariance of a class of models, called \emph{Wess-Zumino-Witten (WZW) models}\index{WZW models}, describing, in their ungauged versions, strings propagating on group manifolds. This model is therefore of direct interest in the study of strings on $AdS_3$ and BTZ black holes, which will hereby be automatically promoted to exact string theory backgrounds. 
WZW models are a class of conformal field theories which have the property that they are actually invariant under a larger symmetry algebra, namely an affine Lie algebra. The conformal invariance of the model is actually rooted in this larger symmetry, as we will show. We then discuss a particular class of D-branes in these models, called symmetric, because they correspond to conformally-invariant boundary conditions and preserve the maximal amount of the original affine symmetry. We show that these boundary conditions completely encode the geometry of these branes.
We turn to the description of the gauged WZW models, which provide a Lagrangian description for the so-called coset models, allowing to reach a new class of conformal field theories.
WZW models enjoy the interesting property that they allow deformations by exact marginal operators, allowing to reach an large class of conformal backgrounds. We end up by discussing such deformations.
The aspects of WZW models presented throughout this chapter reflect the principal motivation of Chapter 6, where we will essentially focus on the geometrical, semi-classical interpretation of the exact string backgrounds obtained, and on some D-brane configurations they may support.


\section{Ungauged WZW Models}
\subsection{Affine extensions and Virasoro algebras}\label{AffVir}
\footpourtoi{14-05-06}
Let $G$ be a finite dimensional compact connected Lie group \footpourtoi{compact, connected:pas notre cas?}. The \emph{loop group}\index{loop group} of $G$ is the group of mappings of the circle $S^1 = \{z \in  \mathbb{C} : |z| = 1\}$ into $G$. Its consists of maps
\begin{equation}
\gamma : S^1 \ra  G : z \ra \gamma(z) \in G \quad,
\end{equation} 
with the group operation defined by pointwise multiplication, i.e. given two maps $\gamma_1, \gamma_2$, the product of $\gamma_1$ and $\gamma_2$ is $\gamma_1 . \gamma_2 (z) \eqdef \gamma_1(z) \gamma_2(z)$.    
The infinite dimensional Lie algebra $\hat{\mathfrak{g}}_0$ of the loop group of $G$ can be obtained from the finite dimensional algebra $\mathfrak{g}$ of $G$, 
\begin{equation}
[T^a, T^b] = i f^{ab}_c T^c
\end{equation}
by writing
\begin{equation}
\gamma(z) = \exp (-i \sum_{a=1}^{dim (\mathfrak{g})} T^a \theta_a(z)),
\end{equation}
where $\theta_a(z)$ are $dim (\mathfrak{g})$ functions defined on $S^1$. Expanding these functions into modes 
\begin{equation}
\theta_a(z) = \sum_{n=-\infty}^{\infty} \theta_a^{-n} z^n ,
\end{equation}
we can introduce generators 
\begin{equation}\label{genT}
t^a_n = T^a z^n
\end{equation}
 such that 
\begin{equation}
\gamma(z) = \exp (-i \sum_{a,n} t^a_{-n} \theta_a^n).
\end{equation}
The $\theta_a^n$s appear as an infinite set of parameters for the loop group, and the $t^a_n$s as an infinite number of generators satisfying the following algebra
\begin{equation}\label{loop}
 [t^a_m , t^b_n] = i f^{ab}_c t^c_{m+n} .
\end{equation}
This is the \emph{loop algebra}\index{loop algebra} $\hat{\mathfrak{g}}_0$  of $G$. It is a particular case of an (untwisted) \emph{affine Lie algebra}\index{affine Lie algebra}, also referred to as \emph{affine Kac-Moody algebra}\index{affine Kac-Moody algebra} in the physics literature (where they are sometimes simply called Kac-Moody algebras; however, the name Kac-Moody is in fact attached to a more general construction). 


Let us now consider the algebra of Diff$(S^1)$, the infinitesimal diffeomorphisms of the circle. If the circle is parameterized by an angular variable $\phi \in [0,2\pi]$, these are expressed as
\begin{equation}\label{fphi}
 \phi \ra \phi + \epsilon(\phi).
\end{equation}
A complete basis of generators is then given by
\begin{equation}\label{genV}
 l_n = i e^{i n \phi} \p_\phi = -z^{n+1} \p_z ,
 \end{equation}
  with $z=e^{i \phi}$. They satisfy a Virasoro algebra without central charge:
\begin{equation}\label{Witt}
 [l_n, l_m] = (n-m) l_{n+m} .
 \end{equation} 
We already encountered this algebra: it showed up in the context of string theory for the first time when we noticed that string theory in the conformal gauge was still invariant under the residual symmetries given by eq.\re{40}. The latter being just two copies of \re{fphi}, this is the reason for the appearance of the algebra \re{53} (with $l_n \ra i l_n$). Although the $x^\pm$s in \re{40} are not angular variables a priori, they become angular variables upon imposing the equations of motion, since the mode expansions \re{19} and \re{23} contain only integer modes (i.e. $\exp(i n x^\pm), n \in \mathbb{Z}$)(\cite{GSW1},p74).

Now, the loop algebra $\hat{\mathfrak{g}}_0$ and the algebra of Diff$(S^1)$, denoted ${\cal V}_0$, are clearly related : using the explicit forms \re{genT},\re{genV} for the generators, one finds
\begin{equation}\label{TL}
[l_m, t^a_n] = -n t^a_{n+m}.
\end{equation} 
Thus $\hat{\mathfrak{g}}_0$ and ${\cal V}_0$ form a semi-direct sum.

As we saw in the particular case of string theory (see eqs. \re{53},\re{58}), going through the quantization procedure may modify the original classical symmetry algebra by introducing a \emph{central extension}\index{central extension} (\cite{GO},p311). This possibility also arises in the case of algebras of the form \re{loop}. For $G$ compact and simple, it is possible to show that, up to redefinitions of the generators, the Jacobi identity fixes the allowed form of the central extension (see \cite{GO}, p311).
\footpourtoi{tte algebre affine est-elle une extension d'une algebre simple?}
 An (untwisted) \emph{affine Lie algebra}\index{affine Lie algebra} $\hat{\mathfrak{g}}$ is then defined by the following commutation relations\footnote{Note that some definitions include the operator $L_0$ in the definition of an affine Lie algebra, see \cite{DiFr}}:
\begin{equation}\label{affine}
[T^a_m, T^b_n] = i f^{ab}_c T^c_{m+n} + k m \delta^{ab} \delta_{m+n}.
\end{equation} 
The algebra \re{affine} is an \emph{affine extension}\index{affine extension} of the simple Lie algebra $\mathfrak{g}$. 
$k$ is a central term which commutes with all $T^a_m$, and is a real constant in each representation. It is called the \emph{level}\index{level} of the affine Lie algebra. Note from \re{affine} that the zero modes $T^a_0$ generate a subalgebra isomorphic to ${\mathfrak{g}}$, and that the generators $T^a_n$ transform in the adjoint representation of ${\mathfrak{g}}$.
If $G$ is non-compact, $\delta^{ab}$ in \re{affine} is then replaced by $\eta^{ab} = B(T^a,T^b)$, where $B$ denotes the Killing form of ${\mathfrak{g}}$.

\footpourtoi{certain? forme quadratique peut tjs etre diagonalisee, DiFrp493}{\footpourtoi{\bf Check!}} 

An already familiar affine Lie algebra is the one generated by the modes of a free boson:
\begin{equation}
[a_n, a_m] = n \, \delta_{n+m,0}
\end{equation}
This is the affine extension $\hat{u}(1)$ of the $u(1)$ algebra generated by $a_0$. We encountered $D$ copies of this algebra in the study of string theory in flat target space-time, see eqs. \re{52},\re{54}. We will shortly see that more general affine Lie algebras arise naturally in the context of WZW models, and that these could be seen as generalizations of string propagation in flat space-time.
Let us note that central extensions can already appear \emph{at the classical level}. This was shown in the case of Virasoro algebras in \cite{BrownHenneaux} through the study of asymptotic symmetries of three-dimensional anti-de Sitter space. Central extensions may also appear classically for affine Lie algebras. This happens in particular for WZW models. We will comment on this in a subsequent section.


\footpourmoi{
{\itshape £ Unitarity : Lust (11.10) + p229, GO p308, DiFr p576: Representations integrables, highest-weight, contraintes sur k}}

\subsection{Classical action and equations of motion}


The action of the WZW model is expressed in terms of a field $g$, which takes its values in some Lie group $G$. It describes, as we will see shortly, the propagation of a string on the group manifold $G$. Therefore $g$ can be seen as a map 
\begin{equation}\label{g}
g : \Sigma \ra G : (z,\zb) \ra g(z,\zb) ,
\end{equation} 
expressing how the string worldsheet $\Sigma$ is embedded in the group manifold (we use complex coordinates on a Euclidean worldsheet, $z=\tau_E + i\sigma$, $\zb=\tau_E - i \sigma$, related to eq. \re{101} by an exponential map, see also eq. (1.7) of \cite{KZ}). It plays the role of the $X^\mu$s of the flat case analysis. 
The WZW action reads \cite{Novikov:1982ei,Witten:1983ar,Polyakov:1983tt,KZ,GepWitt, DiFr}

\begin{equation}\label{WZAction}
S[g] = \frac{1}{4 \lambda^2} \int_\Sigma \, h^{ij} \sqrt{h} \mbox{Tr}\left( \p_i g^{-1} \p_j g \right) \, d^2x   \, + k \, \Gamma[g] \quad,\quad \lambda^2 = \frac{4 \pi}{k} , 
\end{equation} 
with 
\begin{equation}\label{WZTerm}
\Gamma[g] = \frac{i}{24 \pi} \int_B \, d^3y \, \varepsilon_{ijk} \mbox{Tr}\left( g^{-1}\p^i g g^{-1}\p^j g g^{-1}\p^k g \right).
\end{equation} 
The trace appearing in the action is to be taken in the (unitary) representation to which $g$ belongs, and is normalized in such a way that it is independent of the particular representation (\cite{DiFr}, p618); $h_{ij}$ is a metric on $\Sigma$.
The term \re{WZTerm} is called the \emph{Wess-Zumino term}\index{Wess-Zumino term}. In two dimensions, it is defined by an integral over a three-dimensional manifold $B$, whose boundary is identified with the two-dimensional space $\Sigma$ : $\p B = \Sigma$ (in \re{WZTerm}, the map \re{g} is actually extended to a map $\tilde{g}: B \ra G$, such that $\tilde{g} = g$ at $\p B$).
\footpourtoi{Comment on extension,$k$:DiFr620,GO328,MO1p4 SL(2)}
\footpourtoi{NonAbel470 + GOp329! + Zee:Dirac String}
\footpourtoi{15-05-06}

Although the Wess-Zumino term is expressed as a three-dimensional integral, it can nevertheless be interpreted as an integral over the two-dimensional string worldsheet $\Sigma$ (\cite{Witten:1983ar}, p469). As will be developed in more details in Sect. \ref{WZWBackground}, \re{WZTerm} can actually be written as
\begin{equation}\label{pullbackHcompact}
\Gamma = \int_B \, g^* H \quad, 
\end{equation} 
where the integrand is the pull-back of the closed bi-invariant three-form $H = \frac{1}{3!} f_{abc} \theta^a \wedge \theta^b \wedge \theta^c$. $f_{abc}$ denote the structure constants of $\mathfrak{g}=\mbox{Lie}(G)$ (the indices are raised and lowered using the Killing metric) and $\{\theta^a\}$, $a=1 \cdots dim(G)$,  is the set of left-invariant one-forms (see \cite{GeoSpindel}).
 Because $dH = 0$ (as can be seen using the Jacobi identity and the Maurer-Carten structure equations), it can be locally written as $H = dB$, and, using Stokes' theorem, $\Gamma$ can be brought to the form \footpourtoi{Existe forme explicite generale pour B?}
\begin{equation}
\Gamma = \int_\Sigma \, g^* B .
\end{equation} 
Note that $\omega = -f_{abc} \theta^a \wedge \theta^b \wedge \theta^c$ is equally good, this choice changing the sign in front of the Wess-Zumino term in \re{WZAction} (again, we will see in Sect. \ref{WZWBackground} that these two choices are related to the left and right parallelisms on the group manifold $G$).
The outcome of this discussion is that the variation of the Wess-Zumino term under $g\ra g+ \delta g$ is a two-dimensional functional, because the variation of its density can be written as a total derivative and
\begin{equation}
\int_B \, d^3y \, \varepsilon_{ijk} \, \p^k (\cdots) = \int_{\p B=\Sigma} \, \varepsilon_{ij} (\cdots) \quad.
\end{equation} 
Then, using the cyclic property of the trace and the antisymmetry of $\varepsilon_{\alpha \beta \gamma}$, the variation of $\Gamma$ can be rewritten as
\begin{equation}
\delta \Gamma = \frac{i}{8\pi} \int_\Sigma \, d^2x \, \varepsilon_{ij} \, \mbox{Tr} \left(g^{-1} \delta g \p^i (g^{-1} \p^j g) \right).
\end{equation} 
Variation of the complete action \re{WZAction} leads to the equations of motion\footnote{In the conventions of \cite{DiFr}, we have $h_{z \zb}=1/2$, $\p^z=2 \p_\zb$ and $\varepsilon_{z \zb} = i/2$. The last definitions differs from usual ones, where $\varepsilon_{\mu \nu}$ is a pseudo-tensor, and thus a numerical invariant, but multiplied by the Jacobian of the transformation. The latter is here included in the definition of $\varepsilon_{\mu \nu}$, see \cite{DiFr}, p113} (with $\p = \p_z$ and $\pb = \p_\zb$).
\begin{equation}\label{EOM1}
 \p (g^{-1} \pb g) = 0 .
\end{equation} 
Since 
\begin{equation}\label{RelCour}
\p (g^{-1} \pb g) = g^{-1} \pb (\p g g^{-1}) g ,
\end{equation} 
 the equations of motion \re{EOM1} imply that 
\begin{equation}\label{Cons}
 \p \bar{J} = 0 \quad , \quad \pb J = 0 ,
\end{equation} 
with the currents\footpourtoi{conserved? DiFr619(Dual?),W2}
\begin{equation}\label{currents}
 J = -k \p g g^{-1} \quad , \quad \bar{J} = k g^{-1} \pb g
\end{equation} 
thus being holomorphic and anti-holomorphic respectively. By virtue of this, they may be expanded in Laurent series as:
\footpourtoi{cette decomposition suppose a priori que les J ont un poids 1?}
\begin{equation}\label{ModeCourants}
 J(z) = \sum_{n=-\infty}^{n=+\infty} J_n \, z^{-n-1} \quad , \quad \bar{J}(\zb) = \sum_{n=-\infty}^{n=+\infty} \bar{J}_n \, \zb^{-n-1} .
\end{equation} 
The solution of the classical field equation is simply 
\begin{equation}
g(z, \zb) = f(z) \bar{f}(\zb) \quad,
\end{equation} 
for arbitrary functions $f(z)$ and $\bar{f}(\zb)$. 
The most important property of the action \re{WZAction} is its invariance with respect to an infinite-dimensional affine Lie algebra of the type we encountered in the preceding section. This action remains unchanged under the transformations
\begin{equation}\label{AffSymm}
g(z,\zb) \ra \Omega(z) g(z,\zb) \bar{\Omega}(\zb)
\end{equation} 
where $\Omega(z)$ and $\bar{\Omega}(\zb)$ are two arbitrary matrices valued in $G$. This may be checked by considering the infinitesimal transformations
\begin{equation}\label{TransfInf}
\Omega(z) = 1+ \omega(z) \quad , \quad \bar{\Omega}(\zb) = 1 + \bar{\omega}(\zb)
\end{equation} 
under which $g$ varies as
\begin{equation}\label{deltaom}
 \delta_\omega g = \omega g \quad , \quad \delta_{\bar{\omega}} g = - g\bar{\omega}.
\end{equation} 
The variation of the action under $g\ra g + \delta g$ is
\begin{equation}
\delta S \propto \int \, d^2x \mbox{Tr} \left( g^{-1} \delta g [\p(g^{-1} \pb g)] \right),
\end{equation} 
which vanishes using \re{deltaom}, \re{RelCour} and an integration by parts. The invariance may also be ensured using the following remarkable relationship satisfied by \re{WZAction} \cite{Polyakov:1983tt}:
\begin{equation}
S[ g h^{-1} ] = S[g] + S[h] + \frac{1}{16 \pi}\int \, d^2x \mbox{Tr} \left(g^{-1} \pb g h^{-1} \p h \right) ,
\end{equation} 
see for example \cite{Nair}, p15. This relation is called the \emph{Polyakov-Wiegmann identity}\index{Polyakov-Wiegmann identity}.

Finally, one may derive from \re{WZAction} (where an auxiliary worldsheet metric $h_{\a \b}$ has been restored) the components of the two-dimensional energy-momentum tensor (the numerical prefactor is for further convenience):
\begin{equation}\label{TWZW}
 T_{\alpha \beta} = -\frac{4 \pi}{\sqrt{h}} \frac{\delta S}{\delta h^{\alpha \beta}} = -\frac{k}{2} \mbox{Tr} (\p_\alpha g \p_\beta g^{-1} ).
\end{equation} 
Using coordinates $\{X^\mu\}$ on the group manifold, we may express the left and right invariant one-forms (see \cite{GeoSpindel}) as
\begin{equation}\label{InvForms}
 \theta = g^{-1} dg = \theta^a T_a = g^{-1} \p_\mu g\, dX^\mu \quad, \quad \sigma = dg g^{-1} = \sigma^a T_a = \p_\mu g g^{-1}\, dX^\mu ,
\end{equation} 
where the set $\{T_a\}$ is a basis for $\mathfrak{g} = \mbox{Lie}(G)$, with structure constants $f^a_{bc}$. The currents \re{currents} can then be rewritten as  (including $k$ in the invariant vielbeins)
\begin{equation}\label{currents2}
 J = -\sigma_\mu^b \,  \p X^\mu \, T_b \eqdef J^b \, \,T_b \quad, \quad \bar{J} = \theta_\mu^b \,  \pb X^\mu \, T_b \eqdef \bar{J}^b \, \, T_b .
\end{equation} 
We then find from \re{TWZW}, \re{currents} and \re{currents2} that
\footpourtoi{autres comp. de T nulles car inv. de Weyl!}
\begin{equation}\label{EnImpWZW}
T_{zz} = \frac{1}{2k} J^a J^b \beta_{ab} \eqdef T(z)     \quad    , \quad T_{\zb \zb} = \frac{1}{2k} \bar{J}^a \bar{J}^b \beta_{ab} \eqdef \bar{T}(\zb)
\end{equation} 
where we used $\mbox{Tr} (T_a T_b) = 2 \beta_{ab}$. The last equalities in the preceding equations hold because of \re{Cons}.
Comparing \re{Cons} to \re{34}, \re{ModeCourants} to \re{103}, and \re{EnImpWZW} to \re{107}, we see that the group manifold is a simple generalization of the flat case, which is recovered by taking $G=U(1)^D$. The $J_n$s generalize the creation and annihilation operators of the flat string, with $J_0$ and $\bar{J}_0$ playing the role of angular momenta of the string's center of mass.

\subsection{Currents as generators of affine symmetry}
\footpourtoi{16/17-05-06}
\footpourtoi{voir Notes MO1 pour les qttes conservees ds WZW}

Classically, the invariance of the action under a continuous symmetry implies the existence of a conserved current and of a conserved charge. At the quantum level, the symmetry is reflected into constraints relating different correlation functions. The consequence of a symmetry of the action and the measure on correlation functions is generally expressed through the so-called \emph{Ward identities}\index{Ward identities} (see e.g. \cite{DiFr},p.43).
We are not going to derive the Ward identities following from the symmetry \re{AffSymm} (see e.g. \cite{DiFr}). Rather, we will state them and analyze their implications. Before starting, let us recall how things work for the perhaps more familiar conformal symmetry in two dimensions (see Sect. \ref{CrashCFT}). Under an infinitesimal conformal transformation $z \ra z + \a(z)$, $\zb \ra \zb + \bar{\a}(\zb)$, the variation of a correlation function  involving a string $X = \phi_1 (z_1,\zb_1) \cdots \phi_n (z_1,\zb_n)$ of fields is encoded in the \emph{conformal Ward identities}\index{conformal Ward identities} (see \cite{DiFr},p.118, \cite{CoursModave}):
\begin{equation} \label{WardConf}
\delta_{\a,\abar} \langle X \rangle = -\frac{1}{2\pi i}\oint_C \,
 dw \a (w) \langle T(w)X\ \rangle
  + \frac{1}{2\pi i}\oint_C \,
d\wb \abar (\wb) \langle \bar{T}(\wb)X\ \rangle  \quad ,
\end{equation}
where $C$ is a contour encircling all the points appearing in the correlation function, $T$ and $\bar{T}$ being the holomorphic and anti-holomorphic parts of the energy-momentum tensor. Because of the ``factorization" in eq.\re{WardConf}, one may restrict to the holomorphic part of the transformation.
The variation of a single field can be written
\begin{equation}\label{VarField}
 \delta_{\a(z)}\phi(z) = -\frac{1}{2\pi i}\oint_{C_z} dw \,
 \a(w) T(w)\phi_(z) \quad .
 \end{equation}
It is thus completely determined by the singular terms of its OPE with $T(z)$ (for details, see \cite{CoursModave}). In the operator formalism of conformal field theory\footnote{In this formalism, the variable to be used is $z=e^{2(\tau + i \sigma)}$, so that time ordering become radial ordering. It is related to the choice of the preceding section through the exponential map, which does not affect the form of the relations we derived.}, this is usually rewritten as
\begin{equation}
 \delta_{\a(z)} \phi(z) 
  = -[Q_\a , \phi(z)] \label{91BIS} \quad ,
  \end{equation}
  where
  \begin{equation}\label{ConfCharge}
   Q_\a \eqdef \frac{1}{2\pi i} \oint dw \, \a(w)
 T(w) \quad
 \end{equation}
is called the \emph{conformal charge}\index{conformal charge}. This justifies the usual statement that the energy-momentum generates conformal transformations. 

The story is the same for theories invariant under transformations \re{AffSymm}. The variation of correlation functions under an infinitesimal transformation \re{TransfInf} is encapsulated in the so-called \emph{affine Ward identities}\index{affine Ward identities} (see \cite{DiFr}, p.622):
\begin{equation} \label{WardAffine}
\delta_{\omega,\bar{\omega}} \langle X \rangle = -\frac{1}{2\pi i}\oint_C \,
 dw \,\, \beta_{ab} \,\omega^a \langle J^b(w)X\ \rangle
  + \frac{1}{2\pi i}\oint_C \,
d\wb \,\,\beta_{ab} \,\bar{\omega}^a \langle \bar{J}^b(\wb)X\ \rangle  \quad ,
\end{equation}
where
\begin{equation}
\omega = \omega^a T_a \quad , \quad \bar{\omega} = \bar{\omega}^a T_a .
\end{equation} 
This expresses the fact that $J(z)$ and $\Jb(\zb)$ generate the affine transformations of the fields.
The transformation law of the currents follows from \re{currents} and \re{deltaom}, and gives, for example,
\begin{equation}
 \delta_\omega J = [\omega, J] - k \p \omega
\end{equation} 
and
\begin{equation}
 \delta_\omega J^c = i f_{ab}^{\;\;c} \omega^a J^b - k \p \omega^c .
\end{equation} 
This transformation completely fixes the OPE of the components of the currents as
\begin{equation}\label{OPECourants}
 J^a (z) J^b (w) \sim \frac{k \beta^{ab}}{(z-w)^2} + i f^{ab}_{\;\;c} \frac{J^c(w)}{z-w} .
\end{equation} 
This is sometimes referred to as a \emph{current algebra}\index{current algebra} in the literature. From \re{OPECourants}, the mode expansion \re{ModeCourants} and \re{95p}, one finds that the modes $J^a_n$ satisfy 
\begin{equation}\label{affine2}
 [J^a_n, J^b_m] = i f^{ab}_c \, J_{n+m}^c + k n \beta^{ab} \delta_{n+m,0} \quad,
\end{equation} 
i.e. the commutation relations of an affine Lie algebra $\mathfrak{g}$ at level k (see \re{affine}). Similar relations hold for the modes $\Jb^a_n$ yielding another copy of the affine algebra \re{affine2}. Furthermore, we deduce from $\delta_{\bar{\omega}} J = 0$, that
\begin{equation}
 [J^a_n, \Jb^b_m] = 0 .
\end{equation} 
The occurrence of two independent conserved currents generating independent affine Lie algebras is the fundamental property of the WZW model.  We outlined in Sect. \ref{AffVir} the fact that affine Lie algebras are closely related to the Virasoro algebra, which signals the presence of the conformal symmetry in a theory (i.e. through the fact that the fields of the theory form a representation of the Virasoro algebra, the commutator being defined within the operator formalism). In the next section, we will make the explicit link between two algebras, showing at the same time the conformal invariance of the WZW model.

\subsection{Sugawara construction and conformal invariance}\label{Suga}

We are going to show that the affine symmetry of the WZW model actually implies its conformal invariance. In the classical theory, the energy-momentum tensor is expressed quadratically in terms of the currents \re{EnImpWZW}. Let us assume that a normal-ordered version of \re{EnImpWZW} holds in the quantum theory:
\begin{equation}\label{TSuga}
T(z) = \gamma \beta_{ab} \normord{J^a J^b} (z) \quad ,
\end{equation}
where $\gamma$ is a constant to fix, with a similar relation for $\Tb (\zb)$. A comment is in order here.

In the context of conformal field theories, the ordering prescription which is commonly used is the so-called
\emph{conformal normal ordering}\index{conformal normal ordering}, which in general differs from 
the more familiar operator normal ordering, consisting of moving
annihilator operators to the right. The OPE of two
fields can be written in general as

\beq A(z) B(w) = \sum_{n=-\infty}^{N} { \{AB\}_n (w) \over
(z-w)^n} \label{N189} \eeq The conformal normal ordering of the two
fields, denoted by $\normord{AB} $, is by definition

\beq \normord{AB} (w) \eqdef \{AB \}_0 (w) \label{N190} , \eeq that
is, the term of order $(z-w)^0$ in the OPE. This in general does not coincide with the definition

\beq\label{ONO} \normord{A_m B_n}   =\left\{
\begin{array}{ll}
                      A_m B_n  & \mbox{for $n>0$} \,,  \\[1mm]
                      B_n A_m & \mbox{for $n \leq 0$ }\,.
                      \end{array}
                      \right.
\label{N191} \eeq where
\beq A_m &=& {1 \over 2\pi i } \oint_0 dz z^{m + h_A - 1}
A(z) \nonumber \\ B_n &=& {1 \over 2\pi i } \oint_0 dz z^{m + h_B -
1} B(z) \label{Modes} \eeq are the modes of the fields $A(z)$ and
$B(z)$ (see \re{95pp}). 
The singular terms in an operator product are called
\emph{contraction}\index{contraction} and are denoted by
\beq \Cont{A(z)B(w)} \eqdef \sum_{n=1}^N { \{AB \}_n (w) \over
(z-w)^n} \label{N193} . \eeq Again this definition is, in general,
different from the QFT one, where contraction generally means
propagator (two-point function). \newline
 This is well illustrated in the case of the field $T(z)$.
Indeed, with \re{N193} and \re{96}, one has
\beq \Cont{T(z)T(w)} = { c/2 \over (z-w)^4} + { 2 T(w) \over
(z-w)^2} + {\partial T(w) \over (z-w)} , \label{194bis} \eeq 
while (see \cite{CoursModave} for example)
\beq \langle T(z) T(w) \rangle = {c/2 \over (z-w)^4 } \label{N195}.
\eeq From \re{N190}, the conformal normal ordering can be
represented using contour integration by

\beq \normord{AB} (w) = {1 \over 2\pi i} \int {dz \over (z-w)}
A(z) B(w) \label{196} \eeq where, as usual, radial ordering is
understood in the r.h.s. \newline

The OPE \re{N189} is then \beq A(z) B(w) = \Cont{A(z)B(w)} +
\normord{AB}(w) + O(z-w) \label{199bis}\eeq

Note that, when the contraction coincides with the propagator,
this formula is exactly the one we get in usual QFT, where the
ordering is the familiar operator normal ordering \re{ONO}. A
field for which the contraction contains only one singular term
(necessarily coinciding with the 2-point function) is called a
\emph{free field}\index{free field}. In this situation, both orderings are
equivalent, and Wick's theorem may be applied as such. When the
field are not free, Wick's theorem must be adapted (see
\cite{DiFr} p188). One then finds that the normal-ordered modes defined by
\begin{equation}
\normord{AB}(z) = \sum_n z^{-n - h_A - h_B} \normord{AB}_n
\end{equation} 
satisfy (see \cite{DiFr} p. 175) \beq
\normord{AB}_m = \sum_{n \leq -h_A} A_n B_{m-n} + \sum_{n > -h_A}
B_{m-n}A_n \label{N198} . \eeq

For WZW models, it turns out that both definitions \emph{also} coincide, as for free fields. 
By expanding \re{TSuga} in modes following \re{N198}, one obtains
\begin{equation}\label{LnSuga}
 L_n = \gamma \beta_{ab} \left( \sum_{m\leq -1} J^a_m J^b_{n-m} + \sum_{m > -1} J^b_{n-m} J^a_m \right),
\end{equation}    
\begin{equation}
\end{equation} 
where we expect from \re{TL} that the currents be primary fields of conformal weight 1.
If we take $\beta_{ab}$ in diagonal form, then for $n\ne 0$, $J^a_m$ and $J^a_{n-m}$ commute, and so the order of the terms is irrelevant. For $n=0$, the above expression shows that the term with the larger subindex must be placed at the furthest right position. But this is just the definition \re{N191} of the usual normal ordering of modes. It could also have been deduced from \re{OPECourants}. Indeed, although \re{OPECourants} has more than one singular term, the normal ordering of the expression $\beta_{ab} J^a J^b$ only requires the subtraction of a single one, since $\beta_{ab} f^{ab}_{\;\;c}=0$. This consequently agrees with the QFT normal ordering, and amounts us to place positive modes to the right.

Actually, demanding that the currents be Virasoro primary fields determines the constant $\gamma$ uniquely. To find its value, one may compute the OPE of \re{TSuga} with $J^a$, using an adapted version of Wick's theorem (see \cite{DiFr},p.624). But from the previous discussion, one may equally work with the more familiar operator ordering, by acting with $\gamma [L_m, J^a_n]$ on a state $| \varphi \rangle$ such that $J^a_P | \varphi \rangle \ne 0$ and $J^a_{P+1} | \varphi \rangle = 0$\footnote{The $J^a_n$ can be viewed as lowering and raising operators, and states will be constructed by acting with $J^a_{n < 0}$ on a highest-weight state \cite{DiFr}, p.628}. Whatever the method, some manipulations yield
\begin{equation}
 \gamma = \frac{1}{2(k + c_\mathfrak{g})} \quad,
\end{equation} 
where 
\begin{equation}
  f_{abc} f_d^{\;bc}= 2 c_\mathfrak{g} \delta_{ad}  
\end{equation} 
defines the \emph{dual Coxeter number}\index{dual Coxeter number} of $\mathfrak{g}$.
As a next step, we may compute the commutation relations satisfied by the modes \re{LnSuga} of the energy-momentum tensor. One finds
\begin{equation}
 [L_n,L_m] = (n-m) L_{n+m} + \frac{k \;dim(\mathfrak{g})}{12(k+c_\mathfrak{g})} (n^3-n) \delta_{n+m,0} \quad
\end{equation} 
i.e. a Virasoro algebra with a central charge
\begin{equation}
 c = \frac{k\; dim(\mathfrak{g})}{(k+c_\mathfrak{g})}.
\end{equation} 
Of course, the barred modes satisfy the same algebra, with the same value of the central charge. This result indirectly establishes the conformal invariance of the WZW model: it has a conserved and traceless energy-momentum tensor, whose modes, satisfying a Virasoro algebra, generate the conformal transformation of the fields. What we have actually shown is that the Virasoro algebra belongs to the enveloping algebra of the affine Lie (or current) algebra $\hat{\mathfrak{g}}$, a result known in the physics literature as the \emph{Sugawara construction}\index{Sugawara construction}.


\subsection{WZW background}\label{WZWBackground}

We have just put forward the fact that WZW models, as two-dimensional conformally invariant CFTs, represent exact string backgrounds.
To identify the background fields in which the string propagates, we are going to compare \re{WZAction} to \re{173}.

The first part 
\begin{equation}\label{Cin}
\int_\Sigma \, \mbox{Tr}\left( \p_i g^{-1} \p^i g \right) \, d^2x
\end{equation} 
of the WZW action can be developed by plugging $g^{-1} \p_i g = \theta^a_\alpha T_a \p_i X^\alpha$ (see \re{InvForms}) into \re{Cin}, using  
Tr$(T_a T_b) = 2 \beta_{ab} = 2 B(T_a,T_b)$, where $B$ is the Killing form of $\mathfrak{g}$. We obtain
\begin{equation}
 2 \int_\Sigma \, g_{\mu \nu} \p_i X^\mu \p^i X^\nu\, d^2x \quad,
\end{equation} 
with 
\begin{equation}\label{gMet}
 g_{\mu \nu} = \beta_{ab} \,\, \theta^a_\mu \, \theta^b_\nu  
\end{equation} 
being the components in a coordinate basis of the bi-invariant Killing metric of $G$.
On the other hand, the Wess-Zumino term
\begin{equation}
\int_B \, d^3y \, \varepsilon_{ijk} \mbox{Tr}\left( g^{-1}\p^i g g^{-1}\p^j g g^{-1}\p^k g \right)
\end{equation}
can be rewritten, using the antisymmetry of $\varepsilon_{ijk}$, as 
\begin{equation}
\frac{1}{2} \int_B \, d^3y \, \varepsilon_{ijk} \mbox{Tr}\left( [g^{-1}\p^i g, g^{-1}\p^j g] g^{-1}\p^k g \right) .
\end{equation} 
With $g^{-1} \p_i g = \theta^a_\alpha T_a \p_i X^\alpha$ and $\mbox{Tr} ([T_b,T_c],T_a) = 2 f_{abc}$ (indices are lowered and raised using the Killing metric, implying that the structure constants are completely antisymmetric)\footpourmoi{,see Appendix \ref{GeomLie},pa30)}, this leads to 
\begin{equation}\label{pullbackH}
\int_B \, d^3y \, \varepsilon^{ijk} \, f_{abc} \theta^a_\lambda \, \theta^b_\mu \, \theta^c_\nu \, \p_k X^\lambda \, \p_i X^\mu \, \p_j X^\nu.
\end{equation} 
The three-form
\begin{equation}\label{H1}
 H = \frac{1}{3!}  f_{abc} \theta^a \wedge \theta^b \wedge \theta^c \eqdef \frac{1}{3!} H_{\mu \nu \rho} dx^\mu \wedge dx^\nu \wedge dx^\rho 
\end{equation} 
(sometimes also denoted by $H \propto B([\theta,\theta],\theta)$ or $B(\theta, d\theta)$, the explicit form being recovered using the Maurer-Cartan equations) is closed, as can be seen from the Maurer-Cartan structure equations and the Jacobi identity, and can thus locally be written as 
\begin{equation}\label{H2}
 H = dB \quad , \quad H_{\mu \nu \rho} = 3 \p_\lambda B_{\alpha \beta}. 
\end{equation} 
Eq. \re{pullbackH} is nothing other than the explicit expression of \re{pullbackHcompact}.
Using again the antisymmetry of $\varepsilon$, we get 
\begin{equation}
3 \int_B \, d^3y \, \varepsilon^{ijk} \, \p_k (B_{\mu\nu} \, \p_i X^\mu \, \p_j X^\nu) = 3 \int_\Sigma \, d^2x \, \varepsilon^{ij} \,  B_{\mu\nu} \, \p_i X^\mu \, \p_j X^\nu \quad, 
\end{equation} 
where, in the last equality, we made use of the Stokes' theorem.

The action can finally be expressed as
\begin{equation}\label{gB}
 \frac{k}{2\pi} \int_\Sigma \, d^2x \, (g_{\mu\nu} + B_{\mu\nu}) \, \p X^\mu \pb X^\nu , 
\end{equation}
with backgrounds \re{gMet},\re{H1} and \re{H2}.

The WZW model thus describes string propagation on a curved space, whose metric is the Killing metric of some semi-simple Lie group $G$, in presence of a magnetic field $B$ whose field strength $H=dB$ is proportional to the Cartan three-form of $G$.

Of course, one can also show that the equations of motion \re{EOM1}, can be brought to the form \cite{MemMichel}
\begin{equation}\label{Eq}
  \p\pb X^\nu + \Gamma_{\nu\rho}^\mu \p X^\nu \pb X^\rho + \frac{1}{2} H_{\rho \nu}^\mu \p X^\nu \pb X^\rho = 0 \quad ,
\end{equation} 
which can be obtained from \re{173} for a vanishing dilaton. Using \re{currents2}, \re{EOM1} is equivalent to 
\begin{equation}
\p\pb X^\nu + l^\mu_b \theta^b_{\beta,\alpha} \p X^\alpha \pb X^\beta = 0 ,
\end{equation} 
where $\{l_b\}$ is the dual basis of $\{\theta^a\}$ (left-invariant vector fields). Explicit computations reveal that $l^\mu_b \theta^b_{\beta,\alpha} = \Gamma^\mu_{\beta\alpha} + \frac{1}{2} H^\mu_{\beta\alpha}$, where the $\Gamma$s are the coefficients of the Levi-Civita connection associated with the Killing metric, and $H$ is defined as before.

Eq. \re{Eq} allows us to understand the role played by the Wess-Zumino term, i.e. by the field $H$. Recall that for a group manifold, we may define three ``natural" metric connections. The first two are the connections corresponding to the right and left parallelism, generated by left and right invariant vector fields respectively, denoted by $\{ l_a \}$ and $\{r_a\}$, $a=1\cdots dim(G)$. They are defined by (see \cite{GeoSpindel})
\begin{equation}\label{leftright}
 \overset{(r)}{\nabla}_{l_a} l_b \eqdef \overset{(r)}{\omega^c}_{ab}\, l_c \eqdef 0 \quad , \quad \overset{(l)}{\nabla}_{r_a} r_b \eqdef \overset{(l)}{\omega^c}_{ab}\, r_c \eqdef 0.
\end{equation} 
These two connections have a vanishing curvature, but their torsions do not vanish (and have opposite signs). The last connection is defined as
\begin{equation}
 \overset{(0)}{\nabla}= \frac{1}{2}(\overset{(r)}{\nabla} + \overset{(l)}{\nabla})
\end{equation} 
and satisfies
\begin{equation}\label{zero}
 \overset{(0)}{\nabla}_{l_a} l_b = \frac{1}{2} f^c_{ab} l_c \eqdef \overset{(0)}{\omega^c}_{ab}\, l_c \quad , \quad \overset{(0)}{\nabla}_{r_a} r_b = -\frac{1}{2} f^c_{ab} r_c \eqdef \overset{(0)}{\omega^c}_{ab} \, r_c .
\end{equation} 
This is the Levi-Civita connection on the group manifold: it is metric compatible and torsion free. Its Riemann curvature tensor is non-vanishing and is given by
\begin{equation}
 R^a_{bcd} = \frac{1}{4} f^a_{\,\,b i} \, f^i_{\,\,cd} .
\end{equation} 
Let us now work in a left-invariant basis. Then, from \re{leftright} and \re{zero}, we have
\begin{equation}
 \overset{(0)}{\omega^c}_{ab} - \overset{(r)}{\omega^c}_{ab} = \frac{1}{2} f^c_{\,\,ab}
\end{equation} 
Similarly, in a right-invariant basis we may write
\begin{equation}
\overset{(0)}{\omega^c}_{ab} - \overset{(l)}{\omega^c}_{ab} = -\frac{1}{2} f^c_{\,\,ab}
\end{equation} 
Now, recalling that the difference of two connections is a tensorial object, and denoting by $\overset{(0)}{\omega^\mu}_{\nu \rho} := \Gamma^\mu_{\nu \rho}$ the coefficients of the Levi-Civita connection in a coordinate basis, we see that
\begin{equation}\label{LRConn}
\overset{(l/r)}{\omega^\mu}_{\nu \rho} = \Gamma^\mu_{\nu \rho} \pm \frac{1}{2} f^\mu_{\nu \rho},
\end{equation} 
where the upper (lower) sign corresponds to the left (right) parallelism. By inspecting \re{LRConn} and \re{Eq}, and because $H_{\mu\nu\rho} = f_{\mu\nu\rho}$, we notice that the H-field acts as a \emph{parallelizing torsion}\index{parallelizing torsion}. According to the sign in front of the Wess-Zumino term, it thus selects the curvature-free connection associated with left or right parallelism.


\footpourtoi{18/19-05-06}
\subsection{Further remarks}\label{further}
We end this first contact with WZW models with some comments.

\begin{itemize}
{\item We have observed that the WZW model possesses an affine symmetry generated by the holomorphic and anti-holomorphic currents $J$ and $\Jb$, whose modes satisfy the affine Lie algebra \re{affine2}. It turns out that this algebra already emerges at the classical level, within the Hamiltonian formalism. Indeed, starting from \re{gB}, and imposing the canonical equal-time Poisson brackets                 
\begin{equation}
 \{X^\mu(\sigma), \Pi_\nu(\sigma')\}_{PB} = \delta^\mu_\nu\, \delta(\sigma-\sigma') \quad,
\end{equation} 
with $\Pi^\nu$ denoting the conjugate momenta to $X^\mu$, a somewhat lengthy computation starting from \re{currents2} (see \cite{MemMichel} for details) reveals that 
\begin{equation}\label{Schw}
\{ J^a (\sigma), J^b (\sigma')\}_{PB} = f_c^{\,\, a b} \, J^c(\sigma) \delta (\sigma- \sigma') + k \beta^{ab} \p_\sigma \delta(\sigma - \sigma') 
\end{equation} 
Expanding in modes as
\begin{equation}
J^a (x^-) = \sum_n J^a_n (\tau) e^{i n \sigma} \quad ,
\end{equation} 
one gets
\begin{equation}
\{ J^a_n , J^b_m \}_{PB} = f_c^{\,\, a b} \, J^c_{n+m} + k n \beta^{ab} \delta_{n+m}.
\end{equation} 
The c-number term appearing in \re{Schw} with the derivative of a delta function is called a \emph{Schwinger term}\index{Schwinger term}, and is responsible for the occurrence of a classical central extension of the current algebra.  
}

{\item The shift $k \ra k + c_{\mathfrak{g}}$ in the pre-factor of the energy-momentum tensor that we noticed in Sect. \ref{Suga}, can be given another interpretation. Its origin can be traced back by computing the effective action of the WZW model \cite{Tseytlin:1992ri}. The notions of \emph{effective action}\index{effective action} and \emph{effective potential}\index{effective potential} show up when, starting from a classical action (e.g. the WZW action \re{WZAction}), one would like to encode quantum fluctuations, i.e. perturbative loop corrections. In this situation, counterterms which shift the values of the coupling constants (whose roles are played by the background fields $g$ and $B$ in WZW models) have to be included in the Lagrangian. The net effect is that the vacuum expectation value $\langle \phi \rangle$ of the quantum field $\phi$ of the theory departs from its classical value, obtained by minimizing the potential energy. There exists however, in the full quantum theory, a function whose minimum gives the exact value of $\langle \phi \rangle \eqdef \phi_{cl}$, as well as a functional $\Gamma [\phi_{cl}]$ (not to be confused with the WZ term \re{WZTerm}!) satisfying       \begin{equation}
 \frac{\delta }{\delta \phi_{cl}} \Gamma[\phi_{cl}] = 0 .
\end{equation} 
These are, respectively, called \emph{effective potential}\index{effective potential} and \emph{effective action}\index{effective action} (see  \cite{Peskin,Zee} for example, or any other textbook on QFT). In general, these two quantities will satisfy
\begin{equation}
 \Gamma[\phi_{cl}] = S[\phi_{cl}] + O(\hbar) \quad , \quad V_{eff}(\phi_{cl}) = V(\phi_{cl}) + O(\hbar) ,
\end{equation} 
where $S$ and $V$ denote the classical action and potential. For WZW models, the role of $\hbar$ is roughly played by $1/k$, and the classical limit would correspond to a large level $k$. It turns out that, in the case of the WZW model, the effective action can be computed exactly.
\footpourtoi{calcul Domenico?} The result is \cite{Tseytlin:1992ri}:
\begin{equation}\label{shift}
\Gamma[g] = \frac{k + c_{\mathfrak{g}}}{k} S[g].
\end{equation}
The expression \re{shift} for the effective action implies that the exact target space metric and B-field
of the corresponding conformal sigma model are given by their classical expressions (see \re{gB}), modulo the replacement $k \ra k + c_{\mathfrak{g}}$.
Stated in yet another way, the pre-factor $1/k$ in front of the energy-momentum tensor receives a finite multiplicative renormalization from quantum effects, which could be calculated in perturbation theory, yielding the expansion $(1/k)(1 - c_{\mathfrak{g}}/k + \cdots)$ of the exact value $1/(k + c_{\mathfrak{g}})$ \cite{Fuchs}.}

{\item So far, we have left aside important questions related to the peculiar nature of the Wess-Zumino term \re{WZTerm}. Can such a term always be defined? Is the extension from $\Sigma$ to $B$ unique? Remember that the string worldsheet $\Sigma$ can be mapped onto the complex plane (plus the point at infinity), or equivalently onto the Riemann sphere $S^2$. Then the field $g$ can be seen as mapping $S^2$ into the group manifold $G$. Therefore the \emph{homotopy groups}\index{homotopy groups} $\pi_n (G)$ enter consideration (see e.g. \cite{Nakahara}).
\footpourtoi{relation avec homology groups??(Fuchs177, Braaten-Appendice)} The
elements of $\pi_n (G)$ are the equivalence classes of continuous maps of the n-sphere $S^n$
into $G$. Two such maps are equivalent if their images can be continuously deformed into each other. If all images of $S^n$
 in G are contractible to a point, then the n-th homotopy group of G is trivial, $\pi_n (G)$ = 0. A non-trivial $\pi_n (G)$ indicates the
presence of non-contractible n-cycles\footnote{A \emph{cycle}\index{cycle} is a n-dimensional submanifold without
boundary; a \emph{non-contractible}\index{non-contractible} one is also not a boundary itself.} in G.  So, homotopy is quite a
fine measure of the topology of a group manifold G. For example, $\pi_n (G)= \mathbb{Z}$ implies there
is a non-contractible n-cycle in G that generates $\pi_n (G)$, and a map $S^n \ra G$ can ``wind
around" this cycle any number $\in \mathbb{Z}$  of times. 

Therefore, a condition for the Wess-Zumino term to be defined, i.e. for $g$ to be extended to a map (again denoted $g$) of $B$ into $G$, where $\p B = S^2$, is that 
\begin{equation}
 \pi_2 (G) = 0.
\end{equation}
This holds true for any compact connected Lie group, and for non-compact groups with trivial ($\R^n$) topology. \footpourtoi{CHECK!}
On the other hand, if the WZW action is to describe a local theory on the worldsheet, then the physics should be independent of which particular choice of the three-dimensional extension $B$ of $S^2$ is used.
\footpourtoi{A preciser $\pi_3(G)=0$?!! Voir Nair p13-18,Waltonp5!}
 This requires that the path integral $e^{-S}$ is single-valued. The difference between any two values of $\Gamma_B$ can be represented as an integral of $\Gamma$ over $S^3$, since $B' -B$ is homotopically equivalent to $S^3$. Now, it can be observed \cite{Nair,Walton} that $\Gamma_{S^3} [g]$ is precisely $2\pi i$ times the winding number $Q[g]$ of the map $g: S^3 \ra G$, which, for any $g$, is an integer\footnote{For $G=SU(2)\simeq S^3$, the map $Q[g]$ counts how many times the target sphere $S^3$ is covered by the map $g$ as we cover the original $S^3$ once \cite{Nair},p16. This is also valid for any Lie group $G$ having an $SU(2)$ subgroup, since a theorem of Bott states that any continuous mapping of $S^3$ into a general simple Lie group can be continuously deformed into a mapping of $S^3$ into an $SU(2)$ subgroup of $G$.}. There are different situations. If $\pi_3 (G) = 0$,
then all the extensions $g$ can be mapped continuously on one another, and $\Gamma[g]$ is independent of the extension chosen \footpourtoi{winding number = 0?} \footpourmoi{{\large{\bf CHECK : $\Gamma_{S^3} [g] = 0??$ as the winding number vanishes?}}}.
 If $\pi_3 (G) = \mathbb Z$, as it is the case for any compact simply-connected simple Lie group, then $\exp(- k \Gamma[g])$ will be independent of how the extension into the three-space is made, provided the level $k$ is an integer.

}

\end{itemize}

Historically, the WZW model has been introduced in \cite{Witten:1983ar} in the context of non-abelian bosonization in two dimensions.
It had been known \cite{Coleman:1974bu,Mandelstam:1975hb} that, in 1+1 dimensions, the theory of a massless Dirac fermion is equivalent to that of a free massless scalar field (see also \cite{GO}, Sect.2). \emph{Quantum equivalence}\index{quantum equivalence} of the two theories here means equivalence of their correlation functions \cite{DiFr}\footpourtoi{DiFrp647}. This implies directly that the two theories have the same symmetry algebra (i.e. the same Ward identities associated with the symmetry) and the same primary-field spectrum (which is encoded in the two-point correlation functions).  
The procedure of representing fermionic quantities in terms of bosonic ones is called \emph{bosonization}\index{bosonization}. In \cite{Witten:1983ar}, this procedure was generalized to a system of $N$ free massless Dirac or Majorana fermions, where the quantum equivalence with bosonic $SU(N)$ and $O(N)$ WZW models at level $k=1$ was shown \footpourmoi{{\large{\bf CHECK!!}OK:KZp86!},GepWittp503, DiFrp646,NonAbel}. Algebraic aspects of the WZW model have then been developed in \cite{KZ,GepWitt}, stressing the importance of the Sugawara construction in proving the conformal invariance of the model. The relevance of the WZW model to string theory seems to have first been pointed out in \cite{GepWitt}. The role played by the torsion (see Sect. \ref{WZWBackground}) in the conformal invariance of the model has been emphasized in \cite{Braaten:1985is}.


\footpourmoi{
{\itshape £ Comment on Dirac string singularities: GepWittp522, WittenNonAbelianp470

Limites Bordalo p34

 Homologie, WZ term, cycles : appendix Braaten-CurtGeometroTorsion !}
}

\section{Symmetric D-branes in WZW models}\label{BWZW}
We have seen in the previous section that WZW models share many features of string theory in flat space-time. The analogies we already encountered are recalled in Table \ref{FlatWZW}. We would now like to address the question of the existence of D-branes in group manifolds. We know from the previous chapter that D-brane configurations in an arbitrary background have to satisfy the Dirac-Born-Infeld equations (see Sect. \ref{SectDBIBROL}), at least in some limit (that of slowly varying electric fields on the brane). Nevertheless, we will tackle the problem another way around, exploiting again the similarities with the flat case analysis. This will lead us to a particular class of D-branes in group manifolds, called \emph{symmetric D-branes}\index{symmetric D-branes} \cite{AleksSchom,Stanciu,StD0,Sttalk,Stanciu3,Klimcik:1996hp,Felder:1999ka}.

\begin{table}[h]
\begin{center}
\begin{tabular}{|c|cc|}
\hline 
   & Flat case & WZW models \\
 \hline & & \\
currents & $\partial X^\mu ,  \, \bar \partial X^\mu$ & $J, \, \bar J$ \\
& & \\
e.o.m. & $\partial \bar \partial X^\mu$ =0 & $\partial \bar J = 0, \, \bar \partial J =0$ \\
& & \\
closed string modes & $\alpha^\mu_n , \, \tilde \alpha^\mu_n$ & $J^a_n, \, \bar J^a_n $ \\
& & \\
energy-momentum tensor & $T(z) \sim \normord{\partial X . \partial X}$ &
$T(z) \sim \normord{J.J}$ \\
 & & \\
Virasoro generators & $L_n \sim \sum_m \normord{\alpha_m \alpha_{n-m}}$ & 
$L_n \sim \sum_m \normord{J_m. J_{n-m}}$ \\
& & \\
 \hline
 \end{tabular}
 \end{center}
\caption{ \label{FlatWZW} \small Analogies between strings in flat space and WZW models. The anti-holomorphic counterparts of the energy-momentum tensor and of the Virasoro generators are omitted. }
\end{table}

\subsection{Gluing conditions}

We have already seen that D-branes are defined by open string boundary conditions. These can be of two types: Neumann or Dirichlet. In terms of complex coordinates $z=e^{\tau + i \sigma}$ and $\zb = e^{\tau - i \sigma}$, a Dp-brane corresponds to the following boundary conditions at $z=\zb$ (i.e. $\sigma=0,\pi$)\footpourtoi{M=Lie alg autom?}:  
\begin{eqnarray}\label{GCFlat}
 \mbox{Neumann\;\; :\;\;} \p X^\mu &=& \;\;\pb X^\mu \quad , \quad \mu=0,\cdots,p \nonumber\\
 \mbox{Dirichlet \;\;:\;\;} \p X^i &=& -\pb X^i \quad , \quad i=p+1,\cdots,D  .
\end{eqnarray}
This can be written in more compact form as
\begin{equation}\label{GluingFlat}
 \p X = M \pb X \quad \mbox{at} \; z=\zb,
\end{equation}
\footpourtoi{et si M pas diag:les vp -1 donnent D-?(Stanp12) + Dessin}
where $M$ is a diagonal $D\times D$ matrix with $(p+1)$ $``+1"$ entries and $(D-p-1)$ $``-1"$ entries.   
The conformal field theory associated with open strings is defined on the upper half-plane, as we saw in the previous chapter. The boundary conditions imposed on the currents $\p X$ and $\pb X$ at the boundary $z=\zb$ translate into the following condition relating the holomorphic and anti-holomorphic energy-momentum tensors: 
\begin{equation}\label{TTbar}
 T(z) = \Tb (\zb) \quad \mbox{at} \quad z=\zb. 
\end{equation}
This actually defines what is called a \emph{boundary conformal field theory}\index{boundary conformal field theory} (BCFT). In a half-plane, the admissible diffeomorphisms must respect the boundary, taken as the real axis. Thus only real analytic changes of coordinates satisfying $\epsilon(z)= \bar{\epsilon}(\zb)$ for $z=\zb$ real are allowed. \footpourtoi{BCFTZub19: flow across bound?}The condition \re{TTbar} expresses the absence of momentum flow across the boundary and enables one to extend the definition of $T$ to the lower half-plane by $T(z) := \Tb(z)$ for $\mbox{Im}(z) <0$. 
As a consequence of \re{GCFlat} and \re{TTbar}, only one set of oscillators survives in the mode expansions, and there is only one copy of the Virasoro algebra. 
\footpourtoi{PQ pas de facteur $4 \pi \alpha'$ devant bord??}
Conditions \re{GluingFlat} can be slightly extended by starting from the following action, describing an open string in non-trivial background fields $g$ and $B$, whose ends couple to the massless open string mode $A$:
\begin{equation}\label{Open}
S = -\frac{1}{4 \pi \alpha'} \int_\Sigma \, d^2x  \left( g_{m n} \p_a X^m \p^a X^n + B_{m n} \p_a X^m \p_b X^n \epsilon^{ab} \right) 
     + \int_{\p \Sigma} \, d\tau A_m (X^n) \p_\tau X^m .
\end{equation}      
The boundary conditions at $\p\Sigma$ derived from this action are
\begin{eqnarray}\label{boundopen}
 \p_\tau X^i  &=&0 \qquad\qquad\; \mbox{Dirichlet} \nonumber\\ 
	g_{\mu \nu} \p_\sigma X^\nu &=& {\cal F}_{\mu \nu} \p_\tau X^\nu \quad \mbox{Neumann}	
\end{eqnarray}
with 
\begin{equation}
 {\cal F}_{\mu \nu} = B_{\mu \nu} + 2 \pi \alpha' F_{\mu \nu} \quad , \quad F=dA.
 \end{equation}
These can be rewritten in a more compact form as
\begin{equation}\label{GluingCurved}
 \p X^m = C^m_n \pb X^n \quad ,
 \end{equation}
 where the $-1$ eigenvalues of the matrix $C$ correspond to Dirichlet boundary conditions, while for the Neumann boundary conditions, $\mu,\nu=0,\cdots,p$,  we have
\begin{equation}
 C^\mu_\nu = (D^{-1}.A)^\mu_\nu \quad, \quad A = g- {\cal F}, D=g + {\cal F}.
 \end{equation}
It is clear that the field ${\cal F}$ is only defined along Neumann boundary conditions, i.e. in the worldvolume of the Dp-brane. This is because $A$ and $F$ are associated with the open strings' endpoints, and consequently only exist in the directions where these are allowed to move.

A natural generalization of \re{GluingFlat} and \re{GluingCurved} in the case of group manifolds is provided by
\begin{equation}\label{GluingWZW}
 J = r(\Jb) \quad \mbox{at} \; z=\zb,
\end{equation}
where $r$ is a Lie algebra automorphism, $r \in \mbox{Aut}(\mathfrak{g})$. These are called \emph{gluing conditions}\index{gluing conditions}.  They ensure that the resulting D-brane configuration will be conformally invariant, since the condition \re{TTbar} will be respected (as follows from the Sugawara construction and from the fact that $B(\psi(X),\psi(Y)) = B(X,Y)$, with $B$ the Killing form of $\mathfrak{g}$, $\forall \psi \in \mbox{Aut}(\mathfrak{g})$, $\forall X,Y \in \mathfrak{g}$). Actually, the conditions \re{GluingWZW} preserve one half of the original infinite-dimensional symmetry of the current algebra \cite{Stanciu}. The D-brane corresponding to these gluing conditions are therefore called \emph{symmetric D-branes}\index{symmetric D-branes}.

\subsection{Geometry of symmetric D-branes}
Let us determine the geometry of the D-brane configurations described by the gluing conditions \re{GluingWZW} (for a detailed account, see \cite{Stanciu}). Returning to the coordinates $(\sigma,\tau)$, \re{GluingWZW} can be rewritten as
\begin{equation}
 i (1- r \circ Ad_g) g^{-1} \p_\sigma g = (1 + r \circ Ad_g) g^{-1} \p_\tau g \quad,
\end{equation}
implying that $g^{-1} \p_\tau g \in \mbox{Im}(1- r \circ Ad_g)$. Then, there exists $Y \in \mathfrak{g}$ such that
\begin{equation}\label{dtaug}
 \p_\tau g = Y g - g r(Y).
\end{equation}
Eq. \re{dtaug} shows that the vector $\p_\tau g$, which must be tangent to the D-brane worldvolume at $\sigma=(0,\pi)$, belongs to $T_g C_g^R$, where
\begin{equation}\label{TCC}
C_g^R = \{ h\, g \, R(h^{-1}) \; , \; h \in G \},
\end{equation}
with $R$ the Lie group automorphism corresponding to $r$ ($R(e^{tY}) = e^{t \,r(Y)}$).
When $R=\mbox{Id}$, \re{TCC} represents regular conjugacy classes of $G$. When $R$ is an inner automorphism, \re{TCC} describes a translated conjugacy class. When $R$ is an outer automorphism, \re{TCC} describes a twisted conjugacy class, which we already encountered in the first part of this work.
Defining the \emph{twisted centralizer}\index{twisted centralizer} of an element $h\in G$ as
\begin{equation}
 {\cal Z}^R (g) = \{h \in G \, | \, h g R(h^{-1}) = g\} ,
\end{equation}
the twisted conjugacy class $C_g^R$ can be described as the homogeneous space
\begin{equation}
 C_g^R \simeq G/{\cal Z}^R (g) ,
\end{equation}
since it can be seen as an the orbit of $g$ under the \emph{twisted action}\index{twisted action} of $G$ on itself: $ G \times G \ra G : (g,h) \ra h g R(h^{-1})$ (see also Sect \ref{Sect:TwistedIwasawa}).

Provided that the Killing metric restricts nondegenerately to $C_g^R$, the tangent space at $g$ splits into the tangent space to the conjugacy class and its orthogonal complement, which can be identified with the tangent space to ${\cal Z}^R (g)$ \cite{Stanciu3}:
\begin{equation}
 T_g G = T_g C_g^R \overset{\perp}{\oplus} T_g {\cal Z}^R (g).
\end{equation}
At the identity, this can also be expressed as
\begin{eqnarray}
\mathfrak{g} &=& Im (I - r \circ Ad_g) \overset{\perp}{\oplus} Ker (I - r \circ Ad_g) \nonumber \\
         &=& T_e^\parallel \, G \overset{\perp}{\oplus} T_e^\perp \, G
\end{eqnarray}

\footpourmoi{
{\itshape £ More on geometry of twisted conjugacy classes : TheseRibp41! (espace tangent, etc.) + These Bordalo } }

\subsection{Boundary WZW model}
\footpourtoi{20-23/05/06:marBen,Carg}
\footpourtoi{24-05-06}
We have defined symmetric D-branes in WZW models through gluing conditions relating the chiral currents $J$ and $\Jb$, at $z=\zb$. An obvious question one could ask is: do \re{GluingWZW} arise as boundary conditions from the original action \re{WZAction}? The answer is clearly no. The reason for this is that boundary conditions of course only come out of the variation of the action of an open string. But then comes the point: an open string worldsheet $\Sigma$ has a boundary, in contrast to a closed string one. Thus one cannot find a three-dimensional extension $B$ such that $\p B = \Sigma$. To cure this problem, one considers, in addition to \re{WZTerm}, rewritten as
\begin{equation}\label{WZ1}
 \int_{B'} \, \tilde{g}'^* H
\end{equation}
the following term:
\begin{equation}\label{WZ2}
 - \int_D \, g'^* \omega
\end{equation}
which has to be added to \re{WZAction} to give
\begin{equation}\label{BWZW3}
 S = \int_\Sigma \, \mbox{Tr} (g^{-1} dg \wedge *g^{-1}dg) + \int_{B'} \, \tilde{g}'^* H - \int_D \, g'^* \omega,
\end{equation}
where the first term of \re{WZAction} has been reformulated in the language of differential forms, the exterior derivative being understood as acting on the worldsheet coordinates\footnote{Using complex coordinates on the worldsheet (such that the components of the metric tensor are $g_{z \zb}=1/2$, $g_{zz}=g_{\zb\zb}=0$), the integration measure is $(1/2) dz d\zb$. The integration of the exterior product of two one-forms $a$ and $b$ on $\Sigma$ is $\int_\Sigma \, a \wedge b = \int_\Sigma \, d^2x  \epsilon^{ij} a_i b_j$, with $a = a_z dz + a_\zb d \zb$ and $\epsilon^{z\zb} = -\epsilon^{\zb z} = 2 i$. In two dimensions, the Hodge dual $* dx^i$ of a one-form $dx^i$ is $* dx^i = \epsilon^{ij} dx_j$.  The first term is therefore equal to $2\int_\Sigma \, d^2x  \mbox{Tr}\left(g^{-1} \p g g^{-1} \pb g \right)$. These conventions amount to those in \cite{WHF}, where $g_{z \zb}=1$ and $\epsilon^{z\zb} = i$  }.   
This action is called the \emph{boundary WZW model}\index{boundary WZW model}. Let us pause for a moment to explain the different notations. The boundary action we just introduced is to describe open strings ending on a D-brane, which in the case of interest will be represented geometrically by twisted conjugacy classes $C$. $D$ is a two-dimensional submanifold chosen in such a way that 
\begin{equation}
 \p (\Sigma') \eqdef \p (\Sigma \cup D) = \emptyset.
\end{equation}
It then does make sense to define a three-dimensional extension $B'$ such that $\p B' = \Sigma'$. The maps $g'$ and $\tilde{g}'$ simply extend the map $g$ \re{g} as $g':\Sigma' \ra G$ and $\tilde{g}': B'\ra G$. The submanifold $D$ has to satisfy $g'(D) \subset C$.
The D-brane configuration will be characterized in this setting by a two-form $\omega$ defined on its worldvolume, satisfying 
\begin{equation}
 d \omega = H_{|C} := dB_{|C}, 
\end{equation}
where the subscript indicates the restriction to the D-brane worldvolume\footnote{Of course, the condition is trivially satisfied if one deals with D1-branes, since the worldvolume is then two-dimensional}. Since $d(\omega - B)_{|C} = 0$, one can locally define the one-form potential $A$ such that
\begin{equation}
 F \eqdef dA = \omega - B_{|C}. 
\end{equation}
The two topological terms \re{WZ1} and \re{WZ2} can then successively be rewritten as
\begin{eqnarray}
\int_{B'} \, \tilde{g}'^* H &=& \int_{\; \Sigma \cup D} g'^* B \nonumber \\
                             &=& \int_\Sigma g^* B + \int_D g'^* B
\end{eqnarray}
and
\begin{eqnarray}
- \int_D \, g'^* \omega &=& -\int_D g'^* B - \int_D g'^* dA \nonumber \\
												&=&  -\int_D g'^* B - \int_{\p D} g'^* A \nonumber \\
												&=&  -\int_D g'^* B + \int_{\p \Sigma} g'^* A ,
\end{eqnarray}
so that the boundary action can be brought into the form (omitting multiplicative factors)
\begin{equation}\label{BoundWZW}
 S = \int_\Sigma \, \mbox{Tr} (g^{-1} dg \wedge *g^{-1}dg) + \int_\Sigma g^* B + \int_{\p \Sigma} g^* A. 
\end{equation}
This action precisely coincides with the action \re{Open}, and describes an open string propagating in the WZW backgrounds, ending on a D-brane where it couples to a one-form living on the brane. Comparing \re{Open} and \re{BoundWZW} allows us to identify the two-form $\omega$ with ${\cal F}$, where we have set $2 \pi \alpha' =1$.

We have seen that the geometry of symmetric D-branes is completely encoded in the gluing conditions \re{GluingWZW}. We will now show that this is also the case for the two-form field $\omega$ living on its worldvolume. Let us first point out that the gluing conditions are not, strictly speaking, boundary conditions derived from an action. The reason is of course that the currents are Lie-algebra valued objects, taking values in the tangent space to the group manifold at the identity. However, it is possible to turn them into boundary conditions by an appropriate translation on the group manifold \cite{Stanciu}. Plugging \re{currents2} into \re{GluingWZW}, one gets
\begin{equation}\label{boundgluing}
 \p X^n = \widetilde{R}^n_m \pb X^m ,
\end{equation}
with
\begin{equation}
 \widetilde{R}^n_m = -r^n_b \, R^b_a \, \theta^a_m
\end{equation}
being the point-dependent matrix of boundary conditions. At a given point, a Dirichlet boundary condition corresponds to a $-1$ eigenvalue of $\widetilde{R}$, meaning that the directions normal to the worldvolume of the D-brane are spanned by the corresponding eigenvectors of $\widetilde{R}$.
By an appropriate redefinition of the coordinates $X^m$, \re{boundgluing} reduces to 
\begin{eqnarray}
\p_\tau X^i  &=& 0 \qquad\qquad\; \mbox{Dirichlet} \nonumber\\ 
	{\cal A^+}^\mu_\nu \p_\sigma X^\nu &=& {\cal A^-}^\mu_\nu \p_\tau X^\nu \quad \mbox{Neumann}	,
\end{eqnarray}
with ${\cal A^+} = 1 + \widetilde{R}$ and ${\cal A^-} = 1 - \widetilde{R}$. Since ${\cal A^+}$ is invertible in Neumann directions, the second equation may be rewritten  
\begin{equation}
  \p_\sigma X^\nu = \left({\cal A^+}.{\cal A^-}\right)^\mu_\nu \p_\tau X^\nu.
\end{equation}
Comparing this to the second equation of \re{boundopen}, we get (remembering that $\omega = {\cal F}$)
\begin{equation}\label{omega1bis}
 \omega_{\mu \nu} = g_{\mu \alpha} \left({\cal A^+}.{\cal A^-}\right)^\a_\nu ,
\end{equation}
where the indices run only \emph{on the Neumann directions}.
The two-form $\omega$ can be expressed in a more geometrical way as \cite{AleksSchom, StD0, Bordalo:2001ec}
\begin{equation}\label{omegaomega}
 \omega = B(g^{-1}dg, \frac{1+ r \circ Ad_g}{1 - r \circ Ad_g} g^{-1} dg), 
\end{equation}
its action on two vectors $u,v \in T_g C^R_g \sim g \, \mbox{Im}(1 - r \circ Ad_g)$ being given by
\begin{equation}
\omega_g (u,v) = B(g^{-1}u, \frac{1+ r \circ Ad_g}{1 - r \circ Ad_g} g^{-1} v).
\end{equation}
\footpourtoi{pas fait le dernier: $d \omega_{|_{C^R_g}} = H_{|_{C^R_g}}?$}
It can be checked that $\omega$ satisfies all the required properties. First, it is well defined, since $(1 - r \circ Ad_g)$ is invertible on $T_e C^R_e$. Secondly, $\omega$ is single-valued, even if $(1 - r \circ Ad_g)^{-1} g^{-1} v$ is possibly not, since two choices differ by an element of $\mbox{Ker}(1 - r \circ Ad_g)$, which is orthogonal to $g^{-1}u \in \mbox{Im}(1 - r \circ Ad_g)$. Then, $\omega$ is indeed a two-form, as $\omega_g (u,v) = -\omega_g (v,u)$. Finally, we may verify that $d \omega_{|_{C^R_g}} = H_{|_{C^R_g}}$ \cite{StD0}.

We thus conclude that a symmetric D-brane configuration is described by a (twisted) conjugacy class endowed with a two-form $\omega$ which is uniquely determined in terms of the gluing conditions \re{GluingWZW}. This implies, in particular, that if one makes a certain gauge choice for the B-field in the bulk, then the field $F$ on a given D-brane is uniquely determined in terms of $\omega$, and the pull-back on the worldvolume of the D-brane of that B-field.

\footpourtoi{$\pi_2 (C_g^R)=?$ for non-compact? for BTZ?}
Let us end this section by making two remarks.
First, we discussed in Sect.\ref{further} the possible ambiguities in the definition of the three-dimensional extension $B$ of $\p \Sigma$. The situation will be sensitively more subtle for the boundary actions \re{BWZW3} or \re{BoundWZW}. Once $D$ has been fixed, the choice of $B$ is again parameterized by $\pi_3 (G)$, as in the closed string case. On the other hand, the choice of $D$ will be parameterized by $\pi_2 (C_g^R)$. Looking at the last term of \re{BoundWZW}, this implies that $\int_{S^2} F$ will be quantized \cite{BDS,TheseRib,StD0} when $\pi_2 (C_g^R)$ does not vanish.
Subtleties appear relating to the fact that the only gauge invariant quantities in the boundary action are $H$ and $\omega = B_{|_C} + F$ (i.e. the gauge-invariant combination in the DBI action), so that it may seem ambiguous to express a quantization condition in terms of a quantity that is not gauge invariant. In particular, there are situations where $F$ could be gauged away by an appropriate gauge choice for $B$ \cite{BDS,Forste:2002uv,Forste:2001gn}. 
Without entering the details of this tricky question, let us remark that this apparent paradox can be accounted for by noticing that $H$ is closed, but not necessarily exact, so $B$ may not be defined globally. In this case, the boundary action has to be supplemented by an additional term corresponding to the integral of $B$ over a two-dimensional surface surrounding the singularities of $B$.\footpourtoi{pas clair!! + voir BDS} This modification ensures that the quantization conditions are the same for any gauge choice for $B$ (see \cite{TheseRib},p45, for an explicit example).

Also, it has been checked that symmetric D-branes configuration corresponding to regular (twisted) conjugacy classes in compact groups do satisfy the DBI equations of motion. \footpourtoi{ThRibaultpp52-53 + 3.1.2} The spectrum of quadratic fluctuations reveals that these solutions are stable in the space $G\times \R$ (for a flat time direction). It also turns out that the energy spectrum derived from the DBI action coincides, up to some details, with the exact CFT description of these branes, suggesting that the DBI action may be valid outside its believed range of applicability \cite{Bordalo:2001ec}.

\footpourmoi{

Commentaires W6 verso. Literature is somehow messy. 

Comment on symmetry preserved 

DBI: invariances de jauge (Johnson)(dB et combinaison)


 {\itshape £ TheseRib, TheseBord 67-83, papiers Stanciu:transformer gluing en boundary conditions puis identifier!, mes notes (Interlude on D-branes), p4}


Arbitraire quand on choisit surfaces, again implies quantization conditions supplementing those occurring in the ordinary WZW action. These are of the same kind as the one we discussed in Sect.\ref{}, but are however a little bit trickier. For a discussion, see Ribault, Bordalo, Stanciu.

Solve DBI, quadratic fluctuations. Parlent de F et B. This is for compact groups (qu'est-ce qui change pour non-compacts: voir Bord-Rib-Schweig)

}

 \section{Gauged WZW models}
The WZW action functional \re{WZAction} is invariant under transformations \re{AffSymm}, thus in particular under the global $G\times G$ action. Given a Lagrangian with a (global) symmetry, it is usually possible to gauge this symmetry, introducing gauge fields and constructing a gauge invariant extension of the original Lagrangian. In the context of WZW models, one would like to consider an extension of the original action which would be invariant under 
\begin{equation}\label{leftrightgauged}
 g(z, \zb) \ra h_L (z,\zb)  g(z,\zb)  \left(h_R (z,\zb)\right)^{-1},
\end{equation}
\footpourtoi{$H_R \ne H_L$ possible?} 
where $h_{L/R}$ represent embeddings of left/right subgroups $H_{L/R} \subset G$ into $G$. It turns out that, generically, such a gauge invariant extension of \re{WZAction} does not exist, unless one is restricted to so-called \emph{anomaly-free subgroups}\index{anomaly-free subgroup}\cite{WHF}, which are such that 
\begin{equation}\label{anomalyfree}
\mbox{Tr} (T_{a,L} T_{b,L} - T_{a,R} T_{b,R}) = 0.  
\end{equation}
\footpourtoi{papier Wilch-Henneaux aussi?}
Here, $T_{a,L/R}$ represent embeddings of the generators of the Lie algebras of $H_{L/R}$ into $\mathfrak{g}$. This restriction is rooted in the peculiar nature of the Wess-Zumino term \re{WZTerm}, which does not allow for an invariant extension under a general gauging, whereas the kinetic term in \re{WZAction} can always be gauged by replacing ordinary derivatives by covariant ones \cite{WHF}. 
The action for the \emph{gauged WZW model}\index{gauged WZW model} is\footnote{there is a global factor of 2 w.r.t. \cite{WHF}, eq. (A.18)}
\begin{equation}\label{gaugedWZAction}
S(g,A) = \frac{k}{16 \pi} \int_\Sigma \, \mbox{Tr}(g^{-1} d_A g \wedge *g^{-1} d_A g) + i \Gamma(g,A),
\end{equation}
where $d_A$ is the gauge-invariant extension of the exterior derivative, and\footnote{Note that, in the scanned version of \cite{WHF}, there is a missing term in eq. (A.16).} 
\begin{eqnarray}\label{gaugedWZTerm}
        &&\Gamma(g,A) = \Gamma(g) - \frac{1}{16\pi} \int_\Sigma \, \mbox{Tr} (A_L \wedge dg g^{-1} + A_R \wedge g^{-1} dg + A_R\, g^{-1} \wedge A_L\, g) \\
        &=& \frac{1}{48 \pi} \int_B \, Tr(g^{-1}dg \wedge g^{-1}dg \wedge g^{-1}dg) \nn  \\
        &-& \frac{3}{48 \pi} \int_\Sigma \, \mbox{Tr} (A_L \wedge dg g^{-1} + A_R \wedge g^{-1} dg + A_R\, g^{-1} \wedge A_L\, g) \nonumber.
\end{eqnarray}
In the last equation, $A_L$ and $A_R$ are the gauge fields associated with $H_L$ and $H_R$. These are Lie algebra-valued one forms:
\begin{equation}\label{ALR}
 A_{L/R} = A^a T_{a,L/R} dz + \bar{A}^a T_{a,L/R} d\zb := A_{L/R} dz + \bar{A}_{L/R} d\zb , 
 \end{equation}
with an obvious abuse of notation. The gauge-invariant exterior derivative is given by
\begin{equation}
 d_A g = dg + A_L g - g A_R .  
\end{equation}
It may then be checked that \re{gaugedWZTerm} is invariant under the following variation of the fields
\begin{eqnarray}\label{GaugeTransf}
 \delta g &=& \epsilon^a(z,\zb) (T_{a,L} g - g T_{a,R}) \nonumber \\
 \delta A_L &=& -d u_L + [u_L, A_L] \nonumber \\
  \delta A_R &=& -d u_R + [u_R, A_R] \qquad,\qquad u_{L/R}=\epsilon^a (z,\zb) T_{a,L/R},  
\end{eqnarray}
if condition \re{anomalyfree} is satisfied\footnote{The first term of \re{gaugedWZAction} is identically invariant under \re{GaugeTransf}, while \re{gaugedWZTerm} is only when \re{anomalyfree} holds, which can be checked from \re{gaugedWZTerm} and \re{GaugeTransf}, using the fact that $ \delta \mbox{Tr} (g^{-1}dg)^3 = 3 d \mbox{Tr}(u_L (dg g^{-1})^2 - u_R (g^{-1} dg)^2)$.}. The gauged action can be rewritten in a more easy to handle way as
\begin{equation}
 S(g,A) = k \left (S(g) + \frac{1}{4\pi} \int_\Sigma \, d^2x \mbox{Tr}\left( A_L \pb g g^{-1} - \bar{A}_R g^{-1} \p g + \frac{1}{2} (A_L \bar{A}_L + A_R \bar{A}_R) - g^{-1} A_L g \bar{A}_R \right)\right) ,
\end{equation}
where $S(g)$ is the ungauged WZW action \re{WZAction}. The most commonly used gaugings correspond to the choices 
\begin{equation}
 h_L = h_R \in H\quad \mbox{axial} \qquad , \qquad h_L = h_R^{\;-1} \in H \quad \mbox{vector},
\end{equation}
respectively called \emph{axial} and \emph{vector} gaugings, \index{axial gauging}\index{vector gauging} for which the gauge fields \re{ALR} are $A_L = A_R = A$ (axial) or $A_L = -A_R = A$ (vector). 
\footpourtoi{gauged = CFT? Preuve? Calcul tenseur en-imp. avec $\Gamma$?}

The gauged WZW models represent, besides their ungauged cousins, another class of conformal field theories. For every anomaly-free subgroup of $G\times G$ (with generators satisfying \re{anomalyfree}), one has a corresponding gauge-invariant generalization of the WZW action which, upon quantization, leads to a conformal field theory \cite{WHF}. Gauged WZW models were extensively studied at the end of the nineties \cite{Gawedzki:1988hq,Gawedzki:1988nj,Altschuler:1987zb,Bardakci:1987ee,Karabali:1989dk,Schnitzer:1988qj,Karabali:1988au,Guadagnini:1987ty}, mainly with the aim of giving a Lagrangian description of the \emph{Goddard-Kent-Olive (GKO) construction} \index{GKO construction} of \emph{coset models} \index{coset model} \cite{GKO} (whose first traces date back to \cite{Bardakci:1970nb,Halpern:1971ay}). The GKO construction can be expressed very roughly as follows (for more details, see \cite{GO,DiFr}).
\footpourtoi{Int de chemin : V Tseytlin}
Let $\mathfrak{\hat{g}}$ be an affine algebra with a subalgebra $\mathfrak{\hat{h}}$. Assume for simplicity that $\mathfrak{g}$ and $\mathfrak{h}$ are both simple. We can choose an orthonormal basis for $\mathfrak{g}$ which includes as a subset an orthonormal basis for $\mathfrak{h}$, say $J^a_0$, $a=0,\cdots, \mbox{dim} \mathfrak{h}$. The Sugawara construction \re{TSuga} can be applied to both $\mathfrak{\hat{g}}$ and $\mathfrak{\hat{h}}$, to obtain Virasoro generators $L^{\mathfrak{g}}_n$ and $L^{\mathfrak{h}}_n$. They have different prefactors and different $c$-numbers in general. Defining
\begin{equation}
 L^{\mathfrak{g}/\mathfrak{h}}_n = L^{\mathfrak{g}}_n - L^{\mathfrak{h}}_n \quad,
\end{equation}
it may be seen that the $L^{\mathfrak{g}/\mathfrak{h}}_n$s satisfy a Virasoro algebra, with a central charge given by the difference of the central charges of the constituent models: 
\begin{equation}
c^{\mathfrak{g}/\mathfrak{h}} = \frac{k\; dim(\mathfrak{g})}{(k+c_\mathfrak{g})} - \frac{k\; dim(\mathfrak{h})}{(k+c_\mathfrak{h})} \quad,
\end{equation}
where $k$ is the level of $\mathfrak{\hat{g}}$ and $c_\mathfrak{h}$ the dual Coxeter number of $\mathfrak{h}$. An important remark is that the coset construction allows for a description of models with a central charge less than unity, in contrast to WZW models\footnote{For a WZW model, we always have $c > \mbox{rank} \mathfrak{g}$, see e.g. \cite{GO} p363}. It is furthermore expected that any solvable CFT can be described by some coset model \cite{DiFr}.
\footpourtoi{Pq solution degeneree; oblige de fixer la jauge?El-Magn?}
From the point of view of string theory, a number of new exact backgrounds can in principle be constructed starting with gauged WZW models, a famous example being the two-dimensional black hole (see \cite{Witten:1991yr} and references therein) obtained as a gauged $\SL/U(1)$ model. A procedure to extract a background field interpretation from \re{gaugedWZAction} is the following. As the Lagrangian is quadratic in the gauge fields, these can be integrated out using their equations of motion, which are algebraic (stated another way, the path integral over the gauge fields is Gaussian and can be performed).
\footpourtoi{Influence du choix de jauge sur le BG obtenu?? Idem \`a dualité pr\`es?}
As such, the resulting solution is degenerate due to the gauge invariance \re{leftrightgauged}. We may then use the latter to fix the gauge, reducing the number of fields from $\mbox{dim}\, \mathfrak{g}$ to $\mbox{dim}\, \mathfrak{g}/\mathfrak{h}$. The metric and $B$-field can be identified by comparing the resulting action to the non-linear sigma model \re{173}. These fields have to be supplemented by a dilaton field coming from the measure in the path integral when integrating over the gauge fields \cite{Witten:1991yr,Kiritsis:1991zt,Buscher:1987qj}. A more convenient strategy consists in determining it from the one-loop equations. It should be noticed that the gauged action \re{gaugedWZAction} does not enjoy the same property \re{shift} as the ungauged model, and that the background fields will be affected by quantum corrections. Therefore, the backgrounds are only valid to first order in $\a'$ or in $1/k$. Techniques exist however which allow us to determine the backgrounds to all orders, at least in principle \cite{Tseytlin:1994my,Tseytlin:1992ri,DVV,Bars:1992sr,Bars:1992dx,Sfetsos:1992yi,Sfetsos:1993ka,Bars:1993zf}. As an example, the exact metrics for the $SU(2)/U(1)$ and $\SL/U(1)$ gauged models appear to be different from the standard $G$-invariant metrics $G/U(1)$, being deformed due to the presence of a non-constant dilaton \cite{Tseytlin:1994my}.


\footpourmoi{
 Exemple, pour $SU(2)$, donner le truc au premier ordre, puis a tous les ordres en k (Tseytlin): deformation de la metrique sur la sphere $S^2$. 
Faire l'exemple explicite de $SU(2)$?




Parafermions: voir Bagger, DiFr; Spector pour causalite des OPE (aussi Fateev-Zamolodchikov?)!
Pq les parafermions, qui n'obeissent ni commutation ni anti-commutation ne sont pas locaux?


 
 
 
 
 
 
 Cosets, parafermions : Bagger (statistique : Spector) }
 
 
 \section{Marginal deformations of WZW models}
As we discussed in sections \ref{I.4.2} and \ref{SVOMO}, it may be possible, given a conformally invariant model, to deform it using truly marginal operators of the theory, so as to obtain a new class of conformal field theories. This is of direct interest in particular in the context of string theory, since marginal deformations of WZW models could allow us to reach a wide variety of new exact string theory backgrounds. We will therefore consider the following
deformation of the action \re{WZAction}:
\begin{equation}
   S_\lambda = S_0 + \lambda \int \, d^2 z \; {\cal O}(z,\zb),
\end{equation}
where $\lambda$ is a parameter being switched on continuously. A necessary condition for the operator ${\cal O}(z,\zb)$ to be exactly marginal is obviously that it is marginal, i.e. of conformal weights $(1,1)$. In WZW models, such operators are naturally present, and appear to be truly marginal under additional conditions.

\subsection{Symmetric deformations}\index{symmetric deformation}\label{SectSymm}
The basic OPEs for a WZW model include
\begin{eqnarray}\label{OPEWZW}
 T(z) J^a (w) &\sim& \frac{J^a (w)}{(z-w)^2} + \frac{\p J^a(w)}{z-w} \nonumber\\
 \Tb(z) \Jb^a (w) &\sim& \frac{\Jb^a (\wb)}{(\zb-\wb)^2} + \frac{\pb \Jb^a(\wb)}{\zb-\wb} \quad ,
\end{eqnarray} 
which express that the currents  $J^a$ and $\Jb^a$ have conformal weights $(1,0)$ and $(0,1)$ respectively. Therefore, any operator of the form
\begin{equation}\label{OWZW}
 {\cal O}(z,\zb) = \sum_{a,b} c_{ab} J^a(z) \Jb^b(\zb)
\end{equation}
\footpourtoi{Q : Ch-Schw p293}
will represent a marginal operator in the WZW model. A necessary and sufficient condition for \re{OWZW} to be exactly marginal was found in \cite{Chaudhuri:1989qb}, and reads  
\begin{equation}\label{ChSchw}
 \sum_{a,b} c_{ad} c_{be} f^{abg} = 0 \quad \forall \; d,e,g.
\end{equation}
To show that it is necessary, one has to compute the change in the conformal dimension of ${\cal O}(z,\zb)$ due to adding the term $\lambda {\cal O}(z,\zb)$ to the Lagrangian, and observe that there is a non-zero contribution at order $\lambda^2$ unless \re{ChSchw} is satisfied\footnote{Actually, the original derivation of \cite{Chaudhuri:1989qb} assumes that the group on which the WZW model is based is compact, i.e. such that the components $\beta_{ab}$ of the (diagonalized) Killing metric are all positive. For more general groups, it seems that this condition is no longer necessary, see eq. (11) of \cite{Chaudhuri:1989qb}. }. This can be done by computing the two-point function of ${\cal O}(z,\zb)$ perturbatively in $\lambda$. To ensure that \re{ChSchw} are sufficient, it may be checked that, whenever \re{ChSchw} holds, ${\cal O}(z,\zb)$
retains dimension $(1,1)$ to all orders in $\lambda$.

One would now like to identify the background corresponding to a symmetrically deformed WZW model. Besides the simple case of a free boson compactified on a circle, for which switching on a deformation ${\cal O}(z,\zb) = \p X \pb \bar{X}$ simply changes the compactification radius (see \cite{Forste:2003yh} for example), it is not completely straightforward to extract the geometrical content of the deformed background.
We are going to focus on a trivial solution of \re{ChSchw}, consisting of constructing ${\cal O}(z,\zb)$ from currents belonging to the Cartan subalgebra of $\mathfrak{g}$, which all commute. Furthermore, we will restrict ourselves to rank 1 algebras.

Various approaches exist to determine the background fields of a symmetric deformation. A first technique was developed by Hassan and Sen in reference \cite{HassanSen}, where they conjectured that \emph{$O(d,d)$ transformations}\index{$O(d,d)$ transformations} of the background fields of any WZW model correspond to marginal deformation of the WZW theory by an appropriate combination of left and right moving currents belonging to the Cartan subalgebra. More precisely, one has to
identify a coordinate system in which the background fields are
independent of $d$ space dimensions, and where metric and $B$ field are
written in a block diagonal form. In this way the following $2d \times 2d$ matrix
is defined:
\begin{equation}
  M = \left( \begin{tabular}{c|c}
      $\hat{g}^{-1}$ &  $-\hat{g}^{-1} \hat{B}$ \\ \hline
      $\hat{B} \hat{g}^{-1}$ & $\hat{g} - \hat{B}\hat{g}^{-1}\hat{B}$
    \end{tabular}
  \right),
\end{equation}
where $\hat g $ and $\hat B$ are the pull-backs of the metric and
B-field on the selected directions. Then the action
of the $O(d,d)$ group on these fields and dilaton is given by:
\begin{align}
  M & \to M^\prime = \Omega M \Omega^{t}, \label{eq:OddMetric}\\
  \Phi & \to \Phi^\prime = \Phi - \frac{1}{4} \log \left( \frac{\det \hat
g}{\det \hat g^\prime} \right),\label{eq:OddDilaton}
\end{align}
where $\hat g^\prime$ is the metric after the
transformation~\eqref{eq:OddMetric} and $\Omega \in O (d,d)$. The components of the fields $g$ and $B$ which do not appear in $M$ remain unchanged under the transformation. It must be emphasized that these transformation rules are valid at the
lowest order in $\alpha^\prime$ (but at all orders in the
deformation parameters). So, although the model is exact, as we
learn from the CFT side, the field expressions that we
find are only true at leading order in $\alpha^\prime$.
We may also note that not all $O(d,d)$ transformations will lead to a non-trivial change in the backgrounds. First, matrices of the form
\begin{equation}
 \Omega_2 = \left( \begin{tabular}{cc}
      $(A^T)^{-1}$ &  0 \\ 
      0 & $A$
    \end{tabular}
  \right) \quad , \quad
  \Omega_3 = \left( \begin{tabular}{cc}
      $1_d$ &  0 \\ 
      C & $1_d$
    \end{tabular}
  \right),
\end{equation}
with $A$ and $C$ constant matrices in $d$ dimensions, and $C$ antisymmetric generate transformations of the form $\hat{g} \ra A \hat{g} A^T$, $\hat{B} \ra A \hat{B} A^T$ and $\hat{B} \ra \hat{B} + C$. These are implemented by general coordinate transformations of the form $X'^m = A^m_n X^n$ and gauge transformations of $B_{mn}$ with gauge parameter $\Lambda_m = C_{mn} X^n$. Then, an element $\Omega_1$ in the $O(d) \times O(d)$ subgroup of $O(d,d)$ can be parameterized as
\begin{equation}\label{omega1}
\Omega_1 = \frac{1}{2} \left( \begin{tabular}{cc}
      $R+S$ &  $R-S$ \\ 
      $R-S$ & $R+S$
    \end{tabular}
  \right),
\end{equation}
\footpourtoi{Pq coset?}
with $R,S \in O(d)$. The diagonal part $R=S$ corresponds to rotations in $d$-dimensions. Matrices of the form $\Omega_1$ modulo the diagonal subgroup form a coset $O(d) \times O(d)/O(d)$. A general $O(d,d)$ transformation may be expressed as the product of an element of this coset and elements of the group generated by $\Omega_2$ and $\Omega_3$ \cite{Sen:1991zi}. Thus, the non-trivial transformations correspond to elements of the coset $O(d) \times O(d)/O(d)$, which may be labelled by an element of $O(d)$ by making the choice of representative $S=R^T$. 

When the $d$ spectator coordinates are compact, a new background obtained by a transformation of coordinates is not equivalent to the old background \emph{if we assume the same periodicity for the new and old coordinates}. For angular variables, one can indeed end up with backgrounds exhibiting conical singularities, which can be removed using an appropriate change of coordinates. As a simple example, suppose that the resulting background is $ds^2 = (1/C^2) dr^2 + r^2 d\theta^2$, $C\ne1$, with $r \in \R^+$ and $\theta \in [0,2,\pi[$. This metric has a conical singularity at $r=0$, which can easily be removed through the change of coordinates $r'= r C$, $\theta' = \theta/C$, where $\theta' \in [0,2\pi[$. 

Let us get familiar with this procedure by considering a simple example, which will not be chosen at random since it will be closely related to our analysis in the next chapter. Let us therefore consider a WZW model based on the group $SU(2)$, with elements parametrized as
\begin{equation}\label{sphan}
 {\bf g}(\beta,\psi,\phi) = {\rm e}^{i{\frac{\psi - \varphi}{2}} \sigma_3} {\rm e}^{i \beta \sigma_2}
  {\rm e}^{i \frac{\psi + \varphi }{2} \sigma_3}.
\end{equation}
This model describes strings propagating on a three-sphere $S^3$, with background metric and B-field written as
\begin{eqnarray}\label{BackSU2}
ds^2 = d\beta^2 + \sin^2 \beta d\phi^2 + \cos^2 \beta d\psi^2 \\
 H = dB = \pm \sin 2\beta d\beta \wedge d\psi \wedge d\phi , 
\end{eqnarray}
where the sign in front of $H$ depends on the choice of orientation of the volume form in the $SU(2)$ group manifold. 
The coordinates $\psi$ and $\phi$ have a periodicity of $2 \pi$, so that the metric has no coordinate singularity at $\beta =0$ and $\beta = \pi$.
Since the background fields \re{BackSU2} are independent of $\psi$ and $\phi$, we may form the $4 \times 4$ matrix $M$, and act with an $O(2,2)$ rotation, whose non-trivial part can be obtained from \re{omega1} with
\begin{equation}
 R = S^{-1} = e^{\a \t} \in SO(2).
\end{equation}
In order to avoid conical singularities, this transformation has to be followed by a change of coordinates rescaling the angular variables to restore their original periodicity, given by
\begin{equation}
\left(
\begin{array}{cccc}
a^{-1} &  0 & 0 & 0 \\
0 &  b^{-1} & 0 & 0 \\
0 &  -a c & a & 0 \\
b c  &  0 & 0 & b \\
\end{array}
\right), \end{equation}
with $a=\cos \alpha$, $b=\cos \alpha -\sin \alpha$ and $c=-\tan
\alpha$. \footpourtoi{Geom de ce BG?} The final form of the background is then obtained as
\begin{eqnarray}\label{BGSU2Deforme}
ds^2 &=& d\beta^2 + \frac{\kappa^2 \cos^2\beta}{\cos^2\beta + \kappa^2\sin^2\beta} d\psi^2
+
\frac{\sin^2\beta}{\cos^2\beta + \kappa^2\sin^2\beta} d\phi^2 \label{MetricDefSU2}\\
B &=& \frac{\cos^2\beta}{\cos^2\beta + \kappa^2\sin^2\beta} d\phi \wedge d\psi
\label{BDefSU2}\\
\mbox{e}^{2 \Phi} &=& \frac{\kappa}{\cos^2\beta + \kappa^2\sin^2\beta} \quad,
\label{DilDefSU2}
\end{eqnarray}
where $\kappa=\frac{1}{1-\tan \alpha}$. The undeformed background is obtained for $\kappa =1$.
\footpourtoi{Pq pas de dilaton dans cette action?}
The deformed model can be written as
\begin{equation}\label{DefKappa}
 S_\kappa = \frac{k}{2\pi}\int \, d^2z \left( \p \beta \pb \beta + \frac{\sin^2 \beta}{\Delta_\kappa} \p \phi \pb \phi + \frac{\kappa^2 \cos^2\beta}{\Delta_\kappa} \p \psi \pb \psi + \frac{\cos^2 \beta}{\Delta_\kappa} (\p \phi \pb \psi - \p \psi \pb \phi) \right),
\end{equation}
with 
\begin{equation}
\Delta_\kappa = \cos^2 \beta + \kappa^2 \sin^2 \beta
\end{equation}
To identify \re{DefKappa} with the symmetric deformation, we note that an infinitesimal variation away from the original model $\kappa =1$ can be recast as
\begin{equation}
 S_{\kappa = 1 + \delta \kappa} = S_{\kappa =1} - \frac{k}{\pi} \delta \kappa \int \, d^2z  J^3 \Jb^3 ,
\end{equation}
where $J^3$ and $\Jb^3$ are the components of the chiral currents \re{currents}, defined by
\begin{equation}
 J^i = \frac{1}{2} \mbox{Tr}(i \sigma_i \p g g^{-1}) \quad , \quad \Jb^i = \frac{1}{2} \mbox{Tr}(i \sigma_i g^{-1} \pb g) ,
\end{equation}
which read explicitly  
\begin{equation}
 J^3 = \sin^2 \beta \p \phi + \cos^2 \beta \p \psi \quad ,  \quad \Jb^3 = \sin^2 \beta \pb \phi - \cos^2 \beta \pb \psi. 
\end{equation}
\footpourtoi{Pq ce ne sont pas les $J$ qui restent marginaux??}
We have thus obtained the symmetric deformation of the $SU(2)$ model by the truly marginal operator ${\cal O}(z,\zb) = J^3(z) \Jb^3(\zb)$.
For a generic value of the deformation parameter $R$, one gets
\begin{equation}
 S_{\kappa + \delta \kappa} = S_{\kappa} - \frac{k}{2\pi} \delta \kappa^2 \int \, d^2z  J^3_\kappa \bar{J^3_\kappa} ,
\end{equation}
where the following chiral and anti-chiral currents have been defined
\begin{equation}\label{cukappa}
  J^3_\kappa = J^3/\Delta_\kappa \quad , \quad  \bar{J^3_\kappa} = \Jb^3/\Delta_\kappa.
\end{equation}
The equations of motion for $\psi$ and $\phi$ derived from $S_\kappa$ simply state that these currents are conserved:
\begin{equation}
 \pb  J^3_\kappa = 0 \quad , \quad \p \bar{J^3_\kappa} = 0.
\end{equation}
\footpourtoi{Lien entre courants conserves holo et symetrie affine? :voirQuestions. Pq sont-ce des ch. primaires $(1,0)$?} 
The presence of these currents can be interpreted in the following way \cite{HassanSen}. The $O(2,2)$ transformation involves two of the target space coordinates on which the background fields do not depend. Global shifts in these coordinates are therefore commuting isometries of the backgrounds. The conservation of the currents \re{cukappa} expresses that this global symmetry is extended into a $\widehat{U(1)}_L \times \widehat{U(1)}_R$ symmetry generated by these currents, corresponding to local shifts with only holomorphic or anti-holomorphic dependences on the worldsheet coordinates. Therefore the theory contains a pair of chiral currents, and hence, a marginal operator. The integrability of the marginal perturbation, which was proved in \cite{Chaudhuri:1989qb} to all orders in the deformation parameter, starting from the WZW model, ensures that the new theory $S_{\kappa + \delta \kappa}$ also contains chiral currents and can be further perturbed. As a result, the symmetric deformation breaks the original $\widehat{SU(2)}_L \times \widehat{SU(2)}_R$ symmetry of the $SU(2)$ WZW model into $\widehat{U(1)}_L \times \widehat{U(1)}_R$.

Another way of implementing the symmetric deformation has been put forward in \cite{Giveon:1993ph}, for deformations of the $SU(2)$ model. There it is shown that the whole deformation line can be realized as an axially and vectorially gauged $\left(SU(2) \times U(1)\right)/U(1)$ WZW model, in which the embedding of the dividing group has a component in both factors. Let us work this out explicitly. We write a general element of $SU(2) \times U(1)$ as
\begin{equation}
g = \left(
 \begin{array}{ccc}
\cos\beta e^{i \psi} &  \sin\beta e^{i \phi} & 0 \\
-\sin\beta e^{-i \phi} & \cos\beta e^{-i \psi} & 0 \\
0& 0 & e^{i X}
\end{array}
\right) 
\end{equation}
The corresponding gauged WZW action \re{gaugedWZAction} is
\begin{equation}
S \propto \int \, d^2z \left( \p \beta \pb \beta + \sin^2 \beta \p \phi \pb \phi + \cos^2\beta \p \psi \pb \psi + \cos^2 \beta (\p \phi \pb \psi - \p \psi \pb \phi ) +\frac{1}{2} \p X \pb X \right) + \int \, d^2 z L_A ,
\end{equation}
with
\begin{equation} 
L_A =  \mbox{Tr} \left(\bar{A_R} g^{-1} \p g - A_L \pb g g^{-1} + \bar{A_R} g^{-1} A_L g - \frac{1}{2}(A_L \bar{A_L} + A_R \bar{A_R})  \right).
\end{equation}
The gauge fields associated with the subgroup $U(1) \subset SU(2) \times U(1)$ are expressed as
\begin{equation}
 A_{L/R} = T_{L/R} A(z,\zb) \quad , \quad \bar{A}_{L/R} = T_{L/R} \bar{A}(z,\zb)  \quad,
 \end{equation}
 where $T_{L/R}$ are matrix representations of the
 Lie algebra of $U(1)$ embedded in a representation of $SU(2) \times U(1)$.
For an axial gauging, one has to choose the following embeddings:
\begin{equation}
 T_L = \mbox{diag}(i,-i,i \lambda) = T_R \quad , \quad \lambda \in \R. 
\end{equation}
From the expression $e^{\a(z,\zb) T_L} g e^{-\a(z,\zb) T_R}$, we see that a possible gauge choice is $\psi = 0$. After integrating out the gauge fields $A$ and $\bar{A}$, and making this gauge choice, we can read the background corresponding to the gauged model, which coincides precisely with \re{BGSU2Deforme}, by setting $\kappa^2 =\frac{\lambda^2}{\lambda^2 + 2}$ and reintroducing the variable $\psi$ as $\psi = X/\lambda$. However, with this approach, we cover only half of the deformation line, that is $\kappa^2 < 1$. To obtain the other half of the line, one has to gauge the vector current, and perform a T-duality transformation \cite{Giveon:1993ph}. By considering the following embeddings:
\begin{equation}
 T_L = \mbox{diag}(i,-i,i \lambda) = -T_R \quad , \quad \lambda \in \R, 
\end{equation}
we find that a possible gauge choice is $\phi = 0$. The background we then get is (by relabelling $X \ra \phi$):
\begin{eqnarray}
ds^2 &=& d\beta^2 +  (1/2 + \frac{\lambda^2}{4 \sin^2\beta}) d\phi^2 +  \cot^2\beta d\psi^2 + \lambda \cot^2\beta  d\phi d\psi \\
B &=&  0 \\
e^{-\Phi} &=& \sin\beta 
\end{eqnarray}
The metric has a scalar curvature independent of the parameter $\lambda$ : $R = \frac{-4}{\sin^2\beta}$. If we dualize this solution along the $\phi$-direction according to \re{Buscher}, we again recover \re{BGSU2Deforme}, with $\kappa^2 = \frac{\lambda^2 + 2}{\lambda^2} > 1$ and $\phi \ra \frac{\lambda}{2} \phi$.

The are two particular points in the moduli space of the deformed model \re{BGSU2Deforme} : at the end of the deformation line, $\kappa=0$ (resp. $\kappa=\infty$) where we are left with an axially (resp. vectorially) gauged $SU(2)/U(1)$ model times a free $U(1)$ boson. The resulting backgrounds are dual in the sense of \re{Buscher}, and are referred to as the \emph{bell}\index{bell} geometry. This shows in particular that symmetric deformations allow us to link  the WZW models and their gaugings continuously.
\footpourtoi{preciser, voir ThIsr,ForsteRog,etc.}

To end this section, we must stress that we have essentially focused on the sigma-model interpretation of the deformed models, illustrated by a simple example, the $SU(2)$ model. These constructions can be extended to more general WZW models, and can also be described in terms of general conformal field theory considerations, where e.g. the partition function of the deformed model can in principle be written down (see \cite{ForsteRoggenkamp} and references therein, also \cite{Forste:2003yh,3DBH,Israel:2003ry}).









\subsection{Asymmetric deformations}\index{asymmetric deformation}
The spectrum of possible deformations of WZW models can be further enlarged, if one considers their $N=1$ supersymmetric extension. In particular, a given WZW model based on $\mathfrak{g}$ can be embedded in heterotic string theory, by adding $dim\; \mathfrak{g}$ left-moving free fermions transforming in the adjoint representation of $\mathfrak{g}$, while leaving the right-moving sector unchanged. As a result, we end up with a left-moving $N=1$ current algebra and a right-moving $N=0$ one (for details, see \cite{Israel:2004cd,Israel:2004vv,Giveon:2003wn}). We are therefore allowed to consider the following deformation operator
\begin{equation}
  {\cal O}(z,\zb) =\sum_{a=1}^r \textsc{h}_a J^a (z) \bar{I}^a (\zb) \quad ,
\end{equation}
where $\textsc{h}_a$ are the deformation parameters, $J^a (z)$ belong to the $r$-dimensional Cartan subalgebra of $\mathfrak{g}$ and $\bar{I}^a (\zb)$ are right-moving currents belonging to the Cartan subalgebra of the heterotic gauge group\footnote{We suppose that the WZW model under consideration is part of an exact string background, with critical total central charge. Because the right-moving sector has only bosons, for which the contribution of the ghosts is $c_{gh}=-26$, and because the theory is to have only 10 dimensions to accommodate with the left-moving sector, a right-moving current algebra with total central charge $c=16$ is present, representing the internal (compactified) bosons, see \cite{JohnsonBook},p170-172. The index $"a"$ on  $\bar{I}^a (\zb)$ is thus completely outside the Lie algebra $\mathfrak{g}$.}. These are normalized as
\begin{equation}
 \bar{I}^i (\zb) \bar{I}^j (\wb) \sim \frac{k_G h^{ij}}{2 (\zb - \wb)^2} \quad , \quad i,j=1,\cdots,\mbox{rank (gauge group)},  
\end{equation}
with $h^{ij} = f^{ik}_{\;\;l} f^{lj}_{\;\;k}/g^*$, $f^{ik}_{\;\;l}$ and $g^*$ being the structure constants and dual Coxeter numbers of the heterotic gauge group. 
According to the discussion at the beginning of the previous section, this operator is truly marginal by construction. As for the symmetric deformation, we will be interested in the background field interpretation of the deformation. We will again illustrate one way of proceeding for asymmetric deformations of the $SU(2)$ model \cite{Israel:2004vv,Israel:2004cd,Orlando:2005im,Horowitz:1995rf,Kiritsis:1994ta}. 
\footpourtoi{Pq le BG est-il exact? V.ThIsrp289, ElMagnp4,\cite{Orlando:2005im}} \footpourtoi{{\bf pq terme cin pour $\varphi$??}}
The current $\bar{I} (\zb)$ belongs in this case to a $U(1)$ subalgebra of the gauge group, and has a representation in terms of a free boson as $\bar{I} = i \pb \varphi$, where $\varphi (z,\zb)$ is now interpreted as an internal degree of freedom. The deformed action can then be recast as a 4-dimensional non-linear sigma model, whose backgrounds are independent of the internal coordinate $\varphi$. Using Kaluza-Klein reduction, the background fields are found to be
\begin{subequations}
  \label{eq:asym-deform}
  \begin{align}
    G_{\mu \nu } &= \mathring G_{\mu \nu } - 2 \sum_{i=a}^r \textsc{h}^2 J^3_{\mu}
    J^3_{\nu} \label{eq:asym-metric},\\
    B_{\mu \nu } &= \mathring B_{\mu \nu}, \\
    A_{\mu } &= \textsc{h} \sqrt{\frac{2k}{k_G}} J_\mu^3,
  \end{align}
\end{subequations}
where $\mathring G_{\mu \nu}$ and $\mathring B_{\mu \nu }$ are the initial, unperturbed background fields. 
It is interesting to point out that there is, in this case, also an endpoint on the deformation line, corresponding to a limiting value of the deformation parameter. One observes \cite{Israel:2004vv} that, in this limit, a direction decompactifies and factorizes from the three-dimensional geometry:
\begin{equation}
 S^3 \lra \R \times S^2 \quad \mbox{as}\quad \textsc{h}^2 \ra \textsc{h}^2_{max}.  
\end{equation}
Therefore, an asymmetric deformation allows us to reach the \emph{geometric} coset $S^2 \sim SU(2)/U(1)$ (for more general homogeneous spaces interpreted as exact string backgrounds, see \cite{Israel:2004cd,Israel:2004vv}; the realization of $S^2$ and $AdS_2$ as exact CFTs had been pointed out in \cite{Lowe:1994gt,Berglund:1996dv,Johnson:1995kv}). Let us finally mention that asymmetric deformations can also be obtained as a gauged model, and that this again in principle allows to extract the partition function and spectrum of the deformed model \cite{Israel:2004vv,Johnson:1995kv,Israel:2003cx,Quella:2002fk}.







\footpourmoi{}


\cleardoublepage \chapter{$\SL$ black holes : deformations and branes} 

We are now in a position to apply the machinery developed in the previous chapter to the case of $AdS_3$ space, whose relevance to string theory now appears clear: it is part of exact string theory backgrounds, through the $\SL$ WZW model.

Three-dimensional anti-de Sitter space provides a good laboratory for
studying many aspects of gravity and strings, including black-hole physics.
Locally anti-de Sitter three-dimensional black holes are obtained by
performing identifications in the original $AdS_3$ under discrete isometry
subgroups~\cite{BTZ,BHTZ,BRS,Bieliavsky:2003de}.
Those black holes (BTZ) have mass and angular momentum.
Generically, two horizons (inner and outer) mask the singularity, which
turns out to be a chronological singularity rather than a genuine curvature
singularity.

The first aspect we would like to discuss is related to the fact that the two-dimensional sigma-model description of the $AdS_3$ plus Kalb--Ramond (B-)
field background, allows for exact conformal deformations, driven by
integrable marginal operators~\cite{Chaudhuri:1989qb,HassanSen,Giveon:1994ph,Forste:2003km,Israel:2003ry,Israel:2003cx,Israel:2004vv,Israel:2004cd}.
In general, a subgroup of the original isometry group survives along those
lines. Identification under discrete isometries is thus legitimate and
provides a tool for investigating new and potentially interesting ``deformed
BTZ'' geometries. The latter may or may not be viable black holes, whereas
black holes may also appear by just deforming $AdS_3$ without further
surgery~\cite{Horne:1991gn}.
The aim of the work \cite{3DBH} was to clarify those issues, and reach a global
point of view on the geometries that emerge from the $\SL$, by
using the above techniques.  This will allow us to introduce new three
dimensional black hole backgrounds that, in general, involve the presence of
an electric field. For these theories, we give a complete CFT description.
 In particular, the usual black string
background~\cite{Horne:1991gn} will appear in this terms as a special
vanishing-field limit.  Carrying on identifications \emph{\`a la} BTZ
on these geometries will let us obtain more black string and/or black hole
backgrounds, generalizing the one in~\cite{Banados:1992wn} and
in~\cite{Horne:1991gn}, for which we again provide a CFT
description. Not all the backgrounds could be adapted to support the
discrete identifications. This will be stated in terms of a consistency
condition that has to be satisfied in order to avoid the presence of naked
causal singularities.

Another point we would like to address is the search for D-brane configurations in some of the backgrounds mentioned above. In $AdS_3$ space, symmetric D-branes have been studied extensively from different points of view (see e.g. \cite{BachPetr,Lee:2001gh,Stanciu3,Petropoulos:2001qu,Hikida:2001yi,Rajaraman:2001cr,Rajaraman:2001ew,Giveon:2001uq,Deliduman:2002bf}). We will show in particular that, due to the non-trivial topology of the BTZ black holes, there exist in some backgrounds symmetric D1-brane configurations winding around the compact direction. Using a marginal deformation, we will relate these configurations to similar-looking D0-branes in the two-dimensional black hole \cite{Yogendran,Sugawara}.


%


%
\section{Black holes from deformed anti-de Sitter}\label{TaMereEnShort}
We will start with a quick overview of the symmetric and asymmetric deformations of
 the $SL(2,\mathbb{R})$ WZW model, which are found throughout the literature. These
results enable us to recast the
three-dimensional black string solution of~\cite{Horne:1991gn} in sec. \ref{sec:blackstring}, as a
patchwork of marginal deformations of the $SL(2,\mathbb{R})$ WZW model. In this way, we
clarify the role of the mass and charge parameters of the black
string. Section~\ref{sec:two-parameter} is devoted to a two-parameter
deformation of $SL(2,\mathbb{R})$. This leads to a new family of black
strings, with NS--NS and electric field. We study the causal structure of
these black holes as well as their mass and charges.
 They exhibit genuine curvature singularity hidden behind horizons. In
Sec.~\ref{sec:btz} we proceed with discrete identifications as a
solution-generating procedure (\emph{\`a la} BTZ), applied to the
deformed AdS$_3$ -- wherever it is allowed by residual symmetries. After
having stated the consistency conditions that need to be fulfilled in order to avoid
naked singularities we find that, depending on the Killing vector used for
the identifications, two types of chronological singularities are possible:
a time-like one, protected by two horizons, or a space-like one with only
one horizon.

\subsection{Symmetric and asymmetric deformations}

\subsubsection{Symmetric deformations}\label{sec:backgr-fields-symm}
 Symmetric deformations of the $\SL$ WZW model have been extensively studied in the literature (see e.g. references in \cite{Israel:2003cx,Israel:2004vv}). Metric and antisymmetric
tensor for the undeformed model read (in Euler coordinates, see App. \ref{antidss}):
\begin{subequations}
  \begin{align}
    d s^2&= L^2\left[d\rho^2 +
    \sinh^2 \rho \, d\phi^2 - \cosh ^2 \rho  \, d\tau ^2 \right],     \\
    H_{[3]} &= L^2\sinh 2\rho
      d\rho \land d\phi  \land d\tau,
  \end{align}
\end{subequations}
with $L$ related to the level of $SL(2,\mathbb{R})_k$ as usual:
$L=\sqrt{k+2}$. In the case at hand, 
\footpourmoi{Mixing left and right currents of different
kind in the bilinear is forbidden because it generates anomalies: PAS NECESSAIREMENT!!}
lines of symmetric deformations arise due to the presence of
time-like ($J^3$, $\bar J^3$), space-like ($J^1$, $\bar J^1$,
$J^2$, $\bar J^2$), or null generators
\cite{Forste:2003km,Forste:1994wp,Israel:2003ry}. The residual
isometry $U(1) \times U(1)$ can be time-like $(L_3, R_3 )$,
space-like $(L_2, R_2 )$ or null $(L_1 + L_3, R_1 + R_3 )$,
depending on the deformation under consideration.


The \emph{elliptic deformation} is driven by the $J^3\bar J^3$
bilinear. At first order in $\alpha^{\prime}$, the background
fields are given by\footnote{The extra index ``3" in the
  deformation parameter $\kappa$ reminds that the deformation refers here to $J^3
  \bar J^3$.}:
\begin{subequations}
  \begin{align}
    ds^2&= k \left[d\rho^2 + \frac{\sinh^2 \rho \, d\phi^2 -\kappa_3^2 \cosh
        ^2 \rho \, d\tau ^2
      }{\Theta_{\kappa_3} (\rho )} \right],\label{eq:J3J3-metric}\\
    H_{[3]} & = k \frac{\kappa_3^2\sinh 2\rho}{\Theta_{\kappa_3} (\rho )^2 }
    d\rho \land d\phi  \land d\tau,\\
    { \mathrm e}^{\Phi} &= \frac{\Theta_{\kappa_3} (\rho )}{\kappa_3}.
  \end{align}
\end{subequations}
where $\Theta_{\kappa_3} (\rho ) = \cosh ^2 \rho -\kappa_3 \sinh^2 \rho$.  At extreme
deformation ($\kappa_3^2 \to 0$), a time-like direction decouples and we are left
with the axial\footnote{The deformation parameter has two T-dual branches.
  The extreme values of deformation correspond to the axial or vector
  gaugings. The vector gauging leads to the \emph{trumpet}\index{trumpet}.  For the
  $SU(2)_k / U(1)$, both gaugings correspond to the \emph{bell}\index{bell}.}
${SL(2,\mathbb{R})_k / U(1)_{\text{time}}}$. The target space of the latter
is the \emph{cigar}\index{cigar} geometry (also called Euclidean two-dimensional black
hole):
\begin{eqnarray}
{\mathrm e}^{\Phi}&\sim & \cosh^2 \rho,\\
ds^2&=&k \left[ d\rho^2+\tanh^2\rho \, d\phi ^2 \right],
\end{eqnarray}
($0\leq \rho < \infty$ and $0\leq \phi \leq 2\pi$).


Similarly, with $J^2\bar J^2$ one generates the \emph{hyperbolic
deformation}. This allows us to reach the Lorentzian two-dimensional
black hole times a free space-like line. Using the coordinates
defined in Eq.~(\ref{sphan}), we find:
\begin{subequations}
\label{eq:J2J2-deform}
  \begin{align}
    ds^2&= k \left[- dt^2 + \frac{\sin^2 t \, d\varphi^2
    + \kappa_2^2 \cos ^2 t  \, d\psi^2}{\Delta_{\kappa_2} (t)}\right], \label{eq:J2J2-metric}\\
    H_{[3]} & = k \frac{\kappa_2^2\sin 2t}{\Delta_{\kappa_2} (t)^2 }
      dt \land d\psi  \land d\phi,\\
  { \mathrm e}^{\Phi} &= \frac{\Delta_{\kappa_2} (t)}{\kappa_2},
  \end{align}
\end{subequations}
where $\Delta_{\kappa_2} (t) = \cos ^2 t + \kappa_2^2 \sin^2 t$.  This coordinate patch
does not cover the full $\mathrm{AdS}_3$. We will expand on this line in
Sec.~\ref{sec:blackstring}.


Finally, the bilinear $\left(J^1 + J^3\right)\left(\bar J^1 + \bar
  J^3\right)$ generates the \emph{parabolic deformation}. Using Poincar\'e
coordinates
(Eqs.~(\ref{eq:ads-poinc-tr})--(\ref{eq:ads-poinc-volume}))\footnote{Note
  that $x^± = X ± T$.} we obtain:
\begin{subequations}
\label{eq:null-deform}
  \begin{align}
    ds^2&= k \left[\frac{du^2}{u^2}+
    \frac{dX ^2- dT^2}{u^2 + 1/\nu}
    \right], \label{eq:null-metric}\\
    H_{[3]} & = k \frac{2u}{\left(u^2 + 1/ \nu \right)^2}
      du \land dT  \land dX,\\
  { \mathrm e}^{\Phi} &= \frac{u^2 + 1/\nu }{u^2}.
  \end{align}
\end{subequations}
The deformation parameter is $1/\nu$. At infinite value of the
parameter $\nu$, we recover pure AdS$_3$; for $\nu \to 0$, a whole
light-cone decouples and we are left with a single direction and a
dilaton field, linear in this direction.

The physical interpretation of the parabolic deformation is far reaching,
when AdS$_3$ is considered in the framework of the \textsc{ns5/f1}
near-horizon background, $AdS_3 \times S^3 \times T^4$. In this physical
set-up, the parameter $\nu$ is the density of \textsc{f1}'s (number of
fundamental strings over the volume of the four-torus
$T^4$)~\cite{Israel:2003ry,Kiritsis:2003cx}\footnote{Our present convention
  for the normalization of the dilaton results from
  Eq.~(\ref{eq:OddDilaton}). It differs by a factor $-2$ with respect to the
  one used in those papers.}. At infinite density, the background is indeed
$AdS_3 \times S^3 \times T^4$. At null density, the geometry becomes
$\mathbb{R}^{1,2} \times S^3 \times T^4$ plus a linear dilaton and a
three-form on the $S^3$.

\subsubsection{Asymmetric deformations}\label{sec:backgr-fields-asymm}

 As for the symmetric deformations, three
asymmetric deformations are available: the elliptic, the
hyperbolic and the parabolic.

The \emph{elliptic deformation} is generated by a bilinear where
the left current is an $SL(2,\mathbb{R})_k$ time-like current. The
background field is magnetic and the residual symmetry is
$U(1)_{\text{time}} \times SL(2,\mathbb{R})$ generated by $\{L_3,
R_1, R_2, R_3 \}$ (see App. \ref{antidss}). The metric reads (in
elliptic coordinates):
\begin{equation}
  d s^2= \frac{k}{4} \left[ d\rho^2 + \cosh^2 \rho  d\phi^2 -
    \left( 1 + 2 \textsc{h}^2\right) \left( d t + \sinh \rho d
      \phi \right)^2 \right],
  \label{dsnecogo}
\end{equation}
where $\partial_t$ is the Killing vector associated with the
$U(1)_{\text{time}}$. This AdS$_3$ deformation was studied in
\cite{Rooman:1998xf} as a \emph{squashed anti-de Sitter} and in
\cite{Israel:2003cx,Israel:2004vv} from the string theory point of view. It has
curvature
\begin{equation}
  \mathcal{R}=-\frac{2}{k}(3 - 2\textsc{h}^2). \label{curnecogo}
\end{equation}
Here, it comes out as an \emph{exact string solution} (provided $k\to
k +2$) together with an NS three-form and a magnetic field:
\begin{subequations}
  \begin{align}
  H_{[3]} &= d B - \frac{k_g}{4} A \land d A =
  - \frac{k}{4}\left( 1+ 2\textsc{h}^2\right) \cosh \rho d\rho \land d\phi \land d
  t,  \label{adsmagH}
  \\
    A &= \textsc{h} \sqrt{\frac{2k}{k_g}} \left(d t + \sinh \rho d\phi
  \right).
    \label{adsmag}
  \end{align}
\end{subequations}
For $\textsc{h}^2>0$ (unitary region), the above metric is pathological because it has
topologically trivial closed time-like curves passing through any
point of the manifold. Actually, for $\textsc{h}^2=1/2$ we recover exactly
the Goedel space, which is a well-known example of pathological
solution of Einstein--Maxwell equations.

The \emph{hyperbolic deformation} can be studied in a similar
fashion, where the left current in the bilinear is an
$SL(2,\mathbb{R})_k$ space-like current. In hyperbolic
coordinates:
\begin{equation}
  d s^2= \frac{k}{4}\left[ d r^2 - \cosh^2 r d\tau^2 +
    \left( 1-2\textsc{h}^2\right) \left( d x + \sinh r d\tau
    \right)^2\right],
  \label{dsnecoma}
\end{equation}
where $\partial_x$ generates a $U(1)_{\text{space}}$. The total
residual symmetry is $U(1)_{\text{space}} \times
SL(2,\mathbb{R})$, generated by $\{L_2, R_1, R_2, R_3 \}$, and
\begin{equation}
  \mathcal{R}=-\frac{2}{k}\left(3+2\textsc{h}^2\right).\label{Rel}
\end{equation}
The complete string background now has an NS three-form and an
electric field:
\begin{subequations}
  \begin{align}
    H_{[3]} &= \frac{k}{4} \left(1-2\textsc{h}^2\right) \cosh r d r \land
      d\tau \land d x,\label{adselH}\\
    A &= \textsc{h} \sqrt{\frac{2k}{k_g}} \left( d x + \sinh r d\tau
  \right).\label{adsel}
  \end{align}
\end{subequations}

The background at hand is free of closed time-like curves. The
squashed AdS$_3$ is now obtained by going to the AdS$_3$ picture
as an $S^1$ fibration over an AdS$_2$ base, and modifying the
$S^1$ fiber. The magnitude of the electric field is limited at
$\textsc{h}_{\text{max}}^2 = 1/2$, where it causes the degeneration of the
fiber, and we are left with an AdS$_2$ background with an electric
monopole; in other words, a geometric coset $SL(2,\mathbb{R})/
U(1)_{\text{space}}$.

The string spectrum of the above deformation is accessible by
conformal-field-theory methods. It is free of tachyons and a whole
tower of states decouples at the critical values of the electric
fields. Details are available in~\cite{Israel:2004vv}.


Finally, the \emph{parabolic deformation} is generated
by a null $SL(2,\mathbb{R})_k$ current times some internal
right-moving current. The deformed metric reads, in Poincar\'e
coordinates:
\begin{equation}
  d s^2 =k\left[\frac{ d u^2 }{u^2}+ \frac{d x^+ d x^-
    }{u^2}-2\textsc{h}^2 \left(\frac{d x^+}{u^2}\right)^2 \right],
  \label{dsemdef}
\end{equation}
and the curvature remains unaltered $\mathcal{R}=-6/k$.
This is not
surprising since the resulting geometry is the superposition of
AdS$_3$ with a gravitational plane wave. The residual symmetry is
$U(1)_{\text{null}} \times SL(2,\mathbb{R})$, where the $U(1)_{\text{
null}}$ is generated by $\partial_- = -L_1-L_3$.

The parabolic deformation is somehow peculiar. Although it is continuous,
the deformation parameter can always be re-absorbed by a redefinition of the
coordinates\footnote{This statement holds as long as these coordinates are
  not compact.  After discrete identifications have been imposed (see Sec.
  \ref{sec:btzas})), $\textsc{h}$ becomes a genuine continuous parameter.}: $x^+ \to
x^+ / |\textsc{h}|$ and $x^-\to x^- \abs{\textsc{h}}$. Put another way, there
are only three truly different options: $\textsc{h}^2 = 0,  1$. No limiting geometry emerges in the case at hand.

As expected, the gravitational background is accompanied by an NS
three-form (unaltered) and an electromagnetic wave:
\begin{equation}
  A = 2 \sqrt{\frac{2k}{k_g}} \textsc{h} \frac{d x^+}{u^2}.\label{adsem}
\end{equation}

A final remark is in order here, which holds for all three asymmetric
deformations of $SL(2,\mathbb{R})$. The background electric or magnetic
fields that appear in these solutions (Eqs.~(\ref{adsmag}), (\ref{adsel})
and (\ref{adsem})) diverge at the boundary of the corresponding spaces.
\footpourtoi{localized charges??}
Hence, these fields cannot be considered as originating from localized
charges.

\subsection{The three-dimensional black string revisited}
\label{sec:blackstring}

The AdS$_3$ moduli space contains black hole geometries. This has been known
since the most celebrated of them -- the two-dimensional $SL (2, \R )/U(1)$
black hole -- was found by Witten~\cite{Witten:1991yr,Dijkgraaf:1992ba}.
Generalizations of these constructions to higher dimensions have been considered
in~\cite{Horne:1991gn,Gershon:1991qp,Horava:1991am,Klimcik:1994wp}. The
three-dimensional black
string~\cite{Horne:1991gn,Horne:1991cn,Horowitz:1993jc} has attracted a lot of
attention, as it provides an alternative to the Schwarzschild black hole in
three-dimensional asymptotically flat geometries\footnote{Remember that the
  \emph{no hair} theorem doesn't hold in three
  dimensions~\cite{Israel:1967wq,Heusler:1998ua,Gibbons:2002av}.}. In this
section we want to show how this black string can be interpreted in terms of
marginal deformations of $SL ( 2, \R)$

In \cite{Horne:1991gn} the black string was obtained as an $\left( SL (2,
  \R) \times \R \right) /\R$ gauged model. More precisely, expressing $g \in
SL(2,\R) \times \R$ as:
\begin{equation}
  g = \begin{pmatrix}
    a & u & 0 \\
    -v & b & 0 \\
    0 & 0 & {\rm e}^x
  \end{pmatrix},
\end{equation}
the left and right embeddings of the $\R$ subgroup are identical and given by:
\begin{align}
    \epsilon_{L/R} :& \,  \R \to \SL \times \R \\
    \lambda &\mapsto \begin{pmatrix}
      {\rm e}^{\frac{1}{\sqrt{\lambda^2 + 2}}} & 0 & 0 \\
      0 & {\rm e}^{-\frac{1}{\sqrt{\lambda^2 + 2}}}   & 0 \\
      0 & 0 & {\rm e}^{\frac{\lambda}{\sqrt{\lambda^2 + 2}}}
    \end{pmatrix}
\end{align}
From the discussion in Sec.~\ref{sec:backgr-fields-symm}, we see that
performing this gauging is just one of the possible ways to recover the $J^2
\bar J^2$ symmetrically deformed $\SL$ geometry. More specifically, since
the gauged symmetry is axial ($g \to h g h$), it corresponds (in our
notation) to the $\kappa_2 < 1$ branch of the deformed geometry\footnote{The $R \gtrless 1$
  convention is not univocal in literature.} in Eq. \eqref{eq:J2J2-metric}.
One can find a coordinate transformation allowing us to pass from the usual
black string solution
\begin{subequations}
\label{eq:black string}
  \begin{align}
    \di s^2 &= \frac{k}{4}\left[-\left(1-\frac{1}{r}\right) \di t^2 + \left(1-\frac{\mu^2}{
      r}\right) \di x^2 + \left(1-\frac{1}{r}\right)^{-1}
      \left(1-\frac{\mu^2}{r}\right)^{-1} \frac{\di r^2}{r^2}\right],\label{eq:black string-met} \\
    H &= \frac{k}{4} \frac{\mu}{r} \di r \land \di x \land \di t, \\
    \mathrm{e}^{2 \Phi} &= \frac{\mu}{r}
  \end{align}
\end{subequations}
to our (local) coordinate system, Eq.~\eqref{eq:J2J2-deform}. The attentive
reader might now be puzzled by this equivalence between a one-parameter
model such as the symmetrically deformed model and a two-parameter one such
as the black string in its usual coordinates (in
Eqs.~\eqref{eq:black string} we redefined the $r$ coordinate as $r\to r/M$
and then set $\mu = Q/M$ with respect to the conventions
in~\cite{Horne:1991gn}). One point that it is interesting to make here is that
although, out of physical considerations, the black string is usually
described in terms of two parameters (mass and charge), the only physically
distinguishable parameter is their ratio $\mu = Q/M$ that coincides with our
$\kappa_2$ parameter. In Sec.~\ref{sec:two-parameter} we will introduce a
different (double) deformation, this time giving rise to a black hole
geometry depending on two actual parameters (one of which being related to
an additional electric field).


As we remarked above, the axial gauging construction only applies
for $\mu <1 $, while, in order to obtain the other $\kappa_2 > 1$
branch of the $J^2 \bar J^2$ deformation, one should perform a
vector gauging. On the other hand, this operation, that would be
justified from a CFT point of view, is not natural when one
takes a more geometrical point of view and writes the black string
metric as in Eq.~\eqref{eq:black string-met}. In the latter, one
can study the signature of the metric as a function of $r$ in the
two regions $\mu^2 \gtrless 1 $, and find the physically sensible
regions (see Tab.~\ref{tab:bs-signature}).

%

\begin{table}[h]
\begin{tabular}{|c|c|c|c|c|c|c|} \hline \multirow{2}{*}{$\mu$}&
    \multirow{2}{*}{name} & $dt^2 $ & $dx^2 $ & $dr^2$ &
    \multirow{2}{*}{range} & \multirow{2}{2.5cm}{CFT
      interpretation} \\
    \cline{3-5} && $ - \left( 1- \frac{1}{r} \right) $ & $1- \frac{\mu^2}{
      r}$ & $ \left( 1- \frac{1}{r} \right)^{-1} \left( 1- \frac{\mu^2}{r}
    \right)^{-1} $ & &\\ \hline \hline \multirow{3}{*}{$\mu^2 >1$}& $\left(
      c^+ \right)$ & $-$ & $+$ & $+$ &$ r
    > \mu^2$ & $J^3 \bar J^3$, $\kappa_3>1$  \\
    \cline{2-2}\cdashline{3-3} \cline{4-7} &$\left( b^+ \right)$ & $-$ & $-$
    & $-$ &$1< r< \mu^2$ &\\ \cline{2-3} \cdashline{4-4} \cline{5-7} &$\left(
      a^+ \right)$ & $+$ & $-$ & $+$ & $0< r<1$ & $J^3 \bar J^3$, $\kappa_3<1$
    \\ \hline \hline \multirow{3}{*}{$\mu^2 < 1$}& $\left( a^- \right)$ & $+$
    & $-$ & $+$ & $0 < r < \mu^2$ & \multirow{3}{*}{$J^2 \bar J^2$, $\kappa_2 <
      1$} \\ \cline{2-2}\cdashline{3-3} \cline{4-6} &$\left( b^- \right)$ &
    $+$ & $+$ & $-$ &$\mu^2 < r < 1 $ & \\ \cline{2-3} \cdashline{4-4}
    \cline{5-6} &$\left( c^- \right)$ & $-$ & $+$ & $+$ &$ r > 1$ & \\
    \hline
  \end{tabular}
   \caption{Signature for the black string metric as a function of $r$, for
    $\mu^2 \gtrless 1 $.}
  \label{tab:bs-signature}
  \end{table}

Our observations are the following:
\begin{itemize}
\item The $\mu^2 < 1$ branch always has the correct $\left( -, +, + \right)
  $ signature for any value of $r$, with the two special values $r = 1 $ and
  $r = \mu^2 $ marking the presence of the horizons that hide the
  singularity in $r=0$.
\item The $\mu^2> 1$ branch is different. In particular we see that there
  are two regions: $\left(a^+\right)$ for $0<r<1$ and $\left( c^+ \right)$
  for $r > \mu^2$ where the signature is that of a physical space.
\end{itemize}
One fact deserves to be emphasized here: it should noted that while for
$\mu^2 < 1$ we obtain three different regions of the same space, for $\mu^2 >
1$ what we show in Tab.~\ref{tab:bs-signature} are really three different
spaces and the proposed ranges for $r$ are just an effect of the chosen
parameterization. The $\left( a^+ \right), \kappa_3 < 1 $ and $\left(
  c^+\right), \kappa_3 > 1 $ branches are different spaces and not different
regions of the same one and one can choose in which one to go when
continuing to $\mu > 1$.

But there is more. The $\mu^2 > 1 $ region is obtained via an
analytic continuation with respect to the other branch, and this
analytic continuation is precisely the one that interchanges the
roles of the $J^2$ and the $J^3$ currents. As a result, we pass
from the $J^2 \bar J^2$ line to the $J^3 \bar J^3$ line. More
precisely the $\left(c^+\right)$ region describes the ``singular"
$\kappa_3 > 1 $ branch of the $J^3 \bar J^3 $ deformation
(\textit{i.e.} the branch that includes the $r=0$ singularity) and
the $\left(a^+\right)$ region describes the regular $\kappa_3<1$
branch that has the \emph{cigar} geometry as $\kappa_3 \to 0$. Also notice that the regions $r<0$ have to be excluded in
order to avoid naked singularities (of the type encountered in the
Schwarzschild black hole with negative mass). The black string
described in~\cite{Horne:1991gn} covers the regions
$\left(a^-\right), \left(b^-\right), \left(c^-\right),
\left(a^+\right)$.

Our last point concerns the expectation of the genuine $\mathrm{AdS}_3$
geometry as a zero-deformation limit of the black string metric, since the
latter turns out to be a marginal deformation of AdS$_3$ with parameter
$\mu$. The straightforward approach consists in taking the line element in
Eq.~\eqref{eq:black string-met} for $\mu = 1$. It is then puzzling that the
resulting extremal black string geometry \emph{is not} $\mathrm{AdS}_3 $.
This apparent paradox is solved by carefully looking at the coordinate
transformations that relate the black string coordinates $(r,x,t)$ to either
the Euler coordinates $(\rho, \phi, \tau)$ (\ref{euler}) for the $J^3 \bar
J^3$ line, or the hyperbolic coordinates $(y,x,t)$ (\ref{sphan}) for the
$J^2 \bar J^2$ line. These transformations are singular at $\mu = 1$, which
therefore corresponds neither to $\kappa_3 =1$ nor to $\kappa_2 =1$.  Put another way,
$\mu = 1$ is not part of a continuous line of deformed models but marks a
jump from the $J^2 \bar J^2 $ to the $J^3 \bar J^3 $ lines.

The extremal black string solution is even more peculiar.
Comparing Eqs. (\ref{eq:black string}) at $\mu = 1$ to Eqs.
(\ref{eq:null-deform}), which describe the symmetrically
null-deformed $SL(2,\mathbb{R})$, we observe that the two
backgrounds at hand are related by a coordinate transformation,
provided $\nu = -1$.

The black string background is therefore entirely described in
terms of $SL(2,\mathbb{R})$ marginal symmetric deformations, and
involves all three of them. The null deformation appears, however,
for the extremal black string only and at a negative value of the
parameter $\nu$. The latter is the density of fundamental strings,
when the deformed AdS$_3$ is considered within the \textsc{ns5/f1}
system. This might be one more sign pointing towards a possible
instability in the black string \cite{Gregory:1993vy,Gregory:1994bj}.

Notice finally that expressions~\eqref{eq:black string} receive
$1/k$ corrections. Those have been computed
in~\cite{Sfetsos:1992yi}. Once taken into account, they contribute
to making the geometry smoother, as usual in string theory\footpourmoi{Precisely?}.

\subsection{The two-parameter deformations}
\label{sec:two-parameter}

\subsubsection{An interesting mix}
\label{sec:an-interesting-mix}

A particular kind of asymmetric deformation is what we will call  \emph{double deformation}\index{double deformation} in the
following \cite{Israel:2003cx,Kiritsis:1995iu}. At
the Lagrangian level this is obtained by adding the following marginal
perturbation to the WZW action:
\begin{equation}
  \delta S = \delta \kappa^2 \int \di^2 z \: J \bar J + \textsc{h} \int \di^2 z \: J \bar
  I;
\end{equation}
$J$ is a holomorphic current in the group, $\bar J$ is the corresponding
anti-holomorphic current and $\bar I$ an external (to the group)
anti-holomorphic current (\emph{i.e.} in the right-moving heterotic sector
for example). A possible way to interpret this operator consists in thinking
of the double deformation as the superposition of a symmetric -- or
gravitational -- deformation 
 and of an antisymmetric one
-- the electromagnetic deformation. This mix is consistent because if we
perform the $\kappa $ deformation first, the theory keeps the $U(1) \times U(1) $
symmetry generated by $J$ and $\bar J$ that is needed in order to allow for
the $\textsc{h}$ deformation. Following this trail, we can read off the background
fields corresponding to the double deformation by using at first one of the
methods outlined in Sec.~\ref{sec:backgr-fields-symm} and then applying the
reduction in Eq.~\eqref{eq:asym-deform} to the resulting background fields.
The final result consists of a metric, a three-form, a dilaton and a gauge
field. It is in general valid at any order in the deformation parameters $\kappa
$ and $\textsc{h}$ but only at leading order in $\alpha^\prime$ due to the presence of the
symmetric part.
Double deformations of $\mathrm{AdS}_3$ where $J$ is the time-like
$J^3$ operator have been studied in~\cite{Israel:2003cx}. It was
shown there that the extra gravitational deformation allows us to get
rid of the closed time-like curves, which are otherwise present in
the pure $J^3$ asymmetric deformation (Eq.~(\ref{dsnecogo})) --
the latter includes Goedel space. Here, we will  focus instead on
the case of double deformation generated by space-like operators,
$J^2$ and $\bar J^2$.

\subsubsection{The hyperbolic double deformation}

In order to follow the above  prescription for reading the
background fields in the double-deformed metric, let us start with
the fields in Eqs.~(\ref{eq:J2J2-deform}). We can introduce those
fields in the sigma-model action. Infinitesimal variation of the
latter with respect to the parameter $\kappa^2$ enables us to
reach the following expressions for the chiral currents
$J^2_\kappa \left( z \right)$ and $\bar J^2_\kappa \left( \bar z
\right)$ at finite values of $\kappa^2$:
\begin{align}
  J^2_\kappa \left( z \right) &= \frac{1}{\cos^2 t + \kappa^2 \sin^2
    t}\left(\cos^2 t \: \partial \psi -\sin^2t
    \: \partial \varphi \right), \\
  \bar{J}^2_{\kappa} (\bar z) &= \frac{1}{\cos^2 t + \kappa^2 \sin^2 t}
  \left(\cos^2 t \: \partial \psi + \sin^2 t \: \partial \varphi
  \right).
\end{align}
Note in particular that the corresponding Killing vectors (that
are clearly $\partial_\varphi $ and $\partial_\psi $) are to be
rescaled as $L_2 = \frac {1}{\kappa^2}
\partial_\psi - \partial_\varphi $ and $R_2 = \frac {1}{\kappa^2} \partial_\psi + \partial_\varphi $. Once the
currents are known, one just has to apply the construction sketched in
Sec.~\ref{sec:backgr-fields-asymm} and write the background fields as
follows:
\begin{subequations}\label{SolDoub}
  \begin{align}
    \frac{1}{k} \di s^2 &= - \di t^2 + \cos^2 t \frac{ \left( \kappa^2 - 2
        \textsc{h}^2 \right) \cos^2 t + \kappa^4 \sin^2 t }{\Delta_\kappa (t)^2} \di \psi^2 -
    4 \textsc{h}^2 \frac{\cos^2 t \sin^2 t}{\Delta_\kappa (t)^2} \di \psi \di \varphi +
         \nonumber \\ & \hspace{6cm}+ \sin^2 t \frac{ \cos^2 t + \left( \kappa^2 - 2 \textsc{h}^2
          \right) \sin^2 t}{\Delta_\kappa (t)^2} \di \varphi^2 \label{tmetdef} \\
    \frac{1}{k} B &= \frac{\kappa^2 - 2 \textsc{h}^2 }{\kappa^2} \frac{\cos^2 t}{ \Delta_\kappa (t) } \di \varphi \land
    \di \psi
    \\
    F &= 2 \textsc{h} \sqrt{\frac{2 k}{k_g}} \frac{\sin \left( 2 t\right)}{\Delta_\kappa
      (t)^2}
    \left( \kappa^2 \di \psi \land \di t + \di t \land \di \varphi \right) \\
    \mathrm{e}^{- \Phi } &= \frac{\sqrt{\kappa^2 - 2 \textsc{h}^2}}{\Delta_\kappa (t)}
  \end{align}
\end{subequations}
where $\Delta_\kappa ( t ) = \cos^2 t + \kappa^2 \sin^2 t$ as in
Sec.~\ref{sec:backgr-fields-symm}. In particular the dilaton that can be
obtained by imposing the one-loop beta equation is proportional to the ratio
of the double deformed volume form and the $\mathrm{AdS}_3 $ one.

An initial observation about the above background is needed here.  The
electric field is bounded from above since $\textsc{h}^2 \leq \frac{\kappa^2}{2}$. As
usual in string theory, tachyonic instabilities occur at large values of
electric or magnetic fields, and we already observed that phenomenon in Sec.
\ref{sec:backgr-fields-asymm}, for purely asymmetric ($\kappa^2 = 1$)
deformations. At the critical value of the parameter $\textsc{h}$, one dimension
degenerates and the $B$-field vanishes. We are left with a two-dimensional
space (with non-constant curvature) plus an electric field.

The above expression~(\ref{tmetdef}) here of the metric provides only a
local description of the space-time geometry.  To discuss the global
structure of the whole space it is useful to perform several coordinate
transformations. Firstly let us parametrize by
$\kappa^2=\lambda/(1+\lambda)$ the deformation parameter (with $\kappa <1$
for $\lambda>0$ and $\kappa >1$ for $\lambda<-1$) and introduce a radial
coordinate {\it \`a la } Horne and Horowitz:
\begin{equation}
  r=\lambda +\cos^2t,
\end{equation}
which obviously varies between $\lambda$ and $\lambda +1$. The
expression of the metric (\ref{tmetdef}) becomes in terms of this
new coordinate:
\begin{multline}
  \di s^2 = -\left[\left( 2 \textsc{h}^2 \left( 1 + \lambda \right)^2 - \lambda \right) +
      \frac{\lambda \left( \lambda - 4 \textsc{h}^2 \left( 1 + \lambda \right)^2
        \right) }{r} + \frac{ 2\lambda^2 \textsc{h}^2 \left( 1 + \lambda
        \right)^2))}{r^2}\right]\di \psi^2 + \\
    - \left( 1 + \lambda \right) \left[ 2 \textsc{h}^2 \left( 1 + \lambda \right) +
      1 - \frac{ \left( 1 + \lambda \right) \left( 1 + 4 \textsc{h}^2 \left( 1 +
            \lambda \right)^2 \right) }{r} + \frac{ 2 \left( 1 + \lambda
        \right)^3 \textsc{h}^2 }{r^2} \right] \di \varphi^2 + \\
    + 4 \textsc{h}^2 \left( 1 + \lambda \right)^2 \left[ 1 - \frac{ 1 + 2 \lambda
      }{r} + \frac{\lambda \left( 1 + \lambda \right) }{r^2} \right] \di \psi \di\varphi + \frac
    1{4 \left( r - \lambda \right) \left( r - \lambda - 1 \right) }\di r^2 .
\end{multline}
This expression looks close to the one discussed by Horne and
Horowitz. It also represents a black string. However, it depends
on more physical parameters as the expression of the scalar
curvature shows:
\begin{equation}
  \mathcal{R} = 2\frac{2 r \left( 1 + 2 \lambda \right) - 7  \lambda \left( 1 + \lambda
    \right) - 2  \textsc{h}^2 \left( 1 + \lambda \right)^2}{r^2} .
\end{equation}
Obviously this metric can be extended beyond the initial domain of
definition of the $r$ variable. But before we discuss it, it is interesting
to note that the Killing vector ${\rm \bf k}=(1+\lambda)\,\partial_\psi + \lambda\,\partial_\phi \propto
R_2 $ is of constant square length
\begin{equation}
  \mathbf{k}.\mathbf{k} = \lambda \left( 1 + \lambda \right) - 2 \textsc{h}^2
  \left( 1 + \lambda \right)^2 := \omega .
\end{equation}
Note that as $\textsc{h}^2$ is positive, we have the inequality $\omega <\lambda \left( 1 +
  \lambda \right)$. Moreover, in order to have a Lorentzian signature, we must impose
$\omega > 0$. The fact that the Killing vector $\mathbf{k}$ is
space-like and of constant length makes it a candidate to perform
identifications. We shall discuss this point at the end of this
section.

The constancy of the length of the Killing vector $\mathbf{k}$
suggests making a new coordinate transformation (such that
$\mathbf{k}=\partial_x$) :
\begin{subequations}
  \begin{align}
    \psi &= \left( 1 + \lambda \right)  x + t ,\\
    \varphi &= t + \lambda x ,
  \end{align}
\end{subequations}
which leads to the much simpler expression of the line element:
\begin{equation}
  \label{Met2fois}
  \di s^2 = -\frac{\left( r - \lambda \right) \left( r - \lambda - 1 \right)}{r^2} \di t^2 + \omega \left( \di x +\frac{1}{r} \di t \right)^2 +\frac{1}{4 \left( r -\lambda \right) \left( r - \lambda - 1 \right)} \di r^2 .
\end{equation}
This metric is singular at $r=0, \lambda, \lambda +1$; $r=0$ being a curvature
singularity. On the other hand, the volume form is
$\nicefrac{\sqrt{\omega}}{\left(2 r \right)} \di t \land \di x \land \di r$, which
indicates that the singularities at $r=\lambda$ and $r=\lambda +1 $ may be merely
coordinate singularities, corresponding to horizons. Indeed, this is the case.
If, for instance, we look at the metric around $r=\lambda +1$ (i.e. for $r=\lambda +1 +\epsilon$), we obtain locally:
\begin{equation}
  \di s^2 = \frac {\omega}{\left( 1 + \lambda \right)^2}(\di t + \left( 1 + \lambda
  \right) \di x)^2 -\frac \epsilon {\left( 1 + \lambda \right)^2} \di t \left[\di t + 2
    \frac{\omega}{1+\lambda}\left(\di t+ \left( 1 + \lambda \right)\di x\right) \right]
  +\frac 1 {4\epsilon} \di r^2
\end{equation}
indicating the presence of a horizon. To eliminate the
singularity in the metric, we may introduce Eddington--Finkelstein-like coordinates $u$ and $\xi$:
\begin{subequations}
  \begin{align}
    t &= \left( 1 + \lambda \right) \left (u\ \pm \ \frac{1}{2} \ln \epsilon
    \right) - \omega \xi  ,    \\
    x &= \left( 1 + \frac{\omega}{ 1 + \lambda } \right) \xi - \left( u \pm
      \frac{1}{2} \ln \epsilon \right) .
  \end{align}
\end{subequations}
The same analysis can also be done near the horizon located at
$r=\lambda$. Writing $r=\lambda +\epsilon$, the corresponding
regulating coordinate transformation to use is given by:
\begin{subequations}
  \begin{align}
    t &= \lambda \left( u  \pm  \frac{1}{2} \ln \epsilon \right) + \omega \xi ,\\
    x &= \left( 1 - \frac{\omega}{\lambda} \right) \xi - \left( u\ \pm
      \frac{1}{2} \ln \epsilon \right) .
  \end{align}
\end{subequations}
In order to reach the null Eddington--Finkelstein coordinates, we
must use null rays. The geodesic equations read, in terms of a
function $\Sigma^2[E,P,\varepsilon;r]=\left( E r - P \right)^2 -
\left( P^2/\omega \right) - \varepsilon \left( r - \lambda \right)
\left( r - \lambda - 1 \right)$:
\begin{subequations}
  \begin{align}
    \sigma &= \int \frac 1{4\, \Sigma[E,P,\varepsilon;r]}\di r\\
    t &=\int \frac { \left(E r - P \right) r }{2 \left( r - \lambda \right)
      \left( r - \lambda - 1 \right)\, \Sigma[E,P,\varepsilon;r]} \di r    \\
    x &= -\int \frac{\left( E r - P \right) + P / \omega}{2 \left( r -
        \lambda \right) \left( r - \lambda - 1 \right)\,
      \Sigma[E,P,\varepsilon;r]} \di r
  \end{align}
\end{subequations}
where $E$ and $P$ are the constants of motion associated with $\partial_t$ and
$\partial_x$, $\sigma$ is an affine parameter and $\varepsilon$, equal to $1,0,-1$,
characterizes the time-like, null or space-like nature of the geodesic.
Comparing these equations (with $\varepsilon =0$ and $P=0$) with the coordinates
introduced near the horizons, we see that regular coordinates in their
neighborhoods are given by
\begin{subequations}
  \begin{align}
    t &= T ± \frac{1}{2} \left( \left( 1 + \lambda \right) \ln\abs{ r -
        \lambda - 1 } - \lambda  \ln \abs{ r - \lambda } \right) ,\\
    x &= X \mp \frac{1}{2} \left( \ln\abs{ r - \lambda - 1 } - \ln \abs{ r - \lambda}
    \right) ,
  \end{align}
\end{subequations}
which leads to the metric
\begin{equation}
  \label{EFmet}
  \di s^2 = \left(-1+\frac{ 1 + 2 \lambda }{r} -\frac{\lambda \left( 1 + \lambda \right) -
      \omega} {r^2}\right) \di T^2 + 2 \frac \omega r \di   X \di T + \omega
  \di X^2 \mp \frac 1 r \di T  \di r
\end{equation}

According to the sign, we obtain incoming or outgoing null
coordinates; to build a Kruskal coordinate system we have still to
exponentiate them.

Obviously, we may choose the $X$ coordinate in the metric~(\ref{EFmet}) to
be periodic without introducing closed causal curves.  The question of
performing more general identifications in these spaces will be discussed
now.

We end this section by computing the conserved charges associated
with the asymptotic symmetries of our field configurations
(\ref{SolDoub}). Their expressions provide
solutions of the equations of motion derived from the low-energy
effective action \cite{Kiritsis:1994ta,Callan:1985ia}
\begin{equation}
S = \int d^d x \sqrt{-g} \, \mathrm{e}^{- 2\Phi } \left[R +
4(\nabla\Phi)^2 -\frac{1}{12} H^2 -\frac{k_g}{8} F^2 +
\frac{\delta c}{3} \right]  ,
\end{equation}
in which we have chosen the units such that ${\delta c}=12$.

Expression (\ref{Met2fois}) for the metric is particularly
appropriate for describing the asymptotic properties of the solution.
In these coordinates, the various non-gravitational fields read as
\begin{eqnarray}
F&=&\pm \frac {\sqrt{2} \textsc{h} (1+\lambda )}{r^2\sqrt{k_g}}\di t
\wedge \di
r ,\\
H&=& \mp \frac{\omega}{r^2}\di t\wedge \di x \wedge \di r  ,\\
\Phi&=&\Phi_\star -\frac 12 \ln r .
\end{eqnarray}
By setting $\sqrt{\omega}x=\bar x$ and $r=\rm{e}^{2\bar \rho}$,
near infinity ($\bar \rho\rightarrow \infty$), the metric
asymptotes the standard flat metric: $ds^2 = -dt^2 + d\bar{x}^2 +
d\bar \rho^2 $, while the fields $F$ and $H$ vanish and the
dilaton reads $\Phi=\Phi_\star - \bar \rho$. This allows us to
interpret the asymptotic behavior of our solution (\ref{SolDoub})
as a perturbation around the solution given by $F=0$, $H=0$, the
flat metric and a linear dilaton: $\bar\Phi=\Phi_\star + f_\alpha
X^\alpha$ (here $f_\alpha=(0,0,-1)$). Accordingly, we may define
asymptotic charges associated with each asymptotic reductibility
parameter (see \cite{Glenn:2001}).

For the gauge symmetries we obtain as charges, associated with the
$H$ field
\begin{equation}
Q_H=\pm 2\rm{e}^{-2\Phi_\star}\sqrt{\omega}
\end{equation}
and to the $F$ field
\begin{equation}
Q_F=\pm \frac{2\sqrt{2}{\rm{e}}^{-2\Phi_\star}\textsc{h}
(1+\lambda)}{\sqrt{k_g}} .
\end{equation}
The first one reduces (up to normalization) for $\textsc{h} =0$ to the
result given in \cite{Horne:1991gn}, while the second one provides
an interpretation of the deformation parameter $\textsc{h}$.

Moreover, all the Killing vectors of the flat metric defining
isometries that preserve the dilaton field allow us to define
asymptotic charges. These charges are obtained by integrating the
antisymmetric tensor on the surface at infinity:
\begin{equation}
k^{[\mu \nu]}_{\xi}= {\rm{e}}^{-2\bar \Phi}\left(\xi_\sigma\
\partial_\lambda
{\cal H}^{\sigma \lambda\mu \nu} +\frac 12
\partial_\lambda\xi_\sigma\ {\cal H}^{\sigma \lambda\mu
\nu}+2\left(\xi^\mu h^\nu_\lambda f^ \lambda - \xi^\nu
h^\mu_\lambda f^\lambda\right)\right),\label{kasymp}
\end{equation}
where
\begin{equation}
{\cal H}^{\sigma \lambda\mu\nu}=\bar h^{\sigma\nu}\eta^{\lambda
\mu}+\bar h^{\lambda \mu}\eta^{\sigma\nu}-\bar
h^{\sigma\mu}\eta^{\lambda \nu}-\bar
h^{\lambda \nu}\eta^{\sigma\mu}\\
\end{equation}
is the well known tensor sharing the symmetries of the Riemann
tensor and $ \bar h^{\mu\nu}=h^{\mu\nu}-\frac 12 \eta
^{\mu\nu}\eta^{\alpha\beta}h_{\alpha\beta} $, while the Killing
vector $\xi$ has to fulfill the invariance condition $\xi_\alpha
f^\alpha = 0$. The expression of the tensor $k^{[\mu \nu]}_{\xi}$
depends only on the perturbation $h_{\mu\nu}$ of the metric tensor
because, on the one hand, the $F$ and $H$ fields appear
quadratically in the Lagrangian, and their background values are
zero, while, on the other hand, the perturbation field for the
dilaton vanishes: $\Phi=\bar \Phi$.

Restricting ourselves to constant Killing vectors, we obtain the
momenta (defined for the index $\sigma=t$ and $\bar{x}$)
\begin{equation}\label{momenta}
P^\sigma=\int \di \bar x\ { \rm {e}}^{-2\bar
\Phi}\left(\partial_\lambda {\cal H}^{\sigma \lambda t \bar
\rho}-2 \eta ^{\sigma t} h^\nu_{\bar\rho}\right)
\end{equation}
\emph{i.e.} the density of mass ($\mu$) and momentum ($\varpi $)
per unit length:
\begin{equation}
\mu= 2{\rm{e}}^{-2\Phi_\star}(1+2\lambda)\qquad {\rm and}\qquad
\varpi=-2{\rm{e}}^{-2\Phi_\star}\sqrt{\omega}.
\end{equation}
Of course, if we perform identifications such that the string
acquires a finite length, the momenta (\ref{momenta}) also become
finite.

To reach an ending, let us note that the expressions of $\mu$ and
$\varpi$ which we obtain differ from those given in
\cite{Horne:1991gn} by a normalization factor but also in their
dependance with respect to $\lambda$, even at the limit $\textsc{h} =0$;
indeed, the asymptotic Minkowskian frames used differ from each
other by a boost.

\subsection{Discrete identifications}
\label{sec:btz}

In the same spirit as the original BTZ construction
which we discussed in the first part of this thesis,
 we would like to investigate to what extent discrete
identifications could be performed in the deformed background.
The necessary conditions for a solution~\eqref{EFmet} to remain a ``viable" black
hole can be stated as follows:
\begin{itemize}
\item the identifications must be performed along the orbits of some
  Killing vector $\xi$ of the deformed metric
\item there must be  causally safe asymptotic regions (at spatial infinity)
\item the norm of $\xi$ has to be positive in some region of space-time, and
  chronological pathologies have to be hidden with respect to an asymptotic
  safe region by a horizon

\end{itemize}

The resulting quotient space will exhibit a black hole structure
if, once the regions where $\norm{\xi}<0$ have been removed, we
are left with an almost geodesically complete space, the only incomplete
geodesics being those ending on the locus $\norm{\xi}=0$. It is
nevertheless worth emphasizing an important difference from the
BTZ construction. In our situation, unlike the undeformed
$\mathrm{AdS}_3$ space, the initial space-time where we are to
perform identifications does exhibit curvature singularities.

\subsubsection{Discrete identifications in asymmetric deformations}
\label{sec:btzas}

Our analysis of the residual isometries in purely asymmetric deformations
(Sec.~\ref{sec:backgr-fields-asymm}) shows that the vector $\xi$
used to make up the non-extremal black holes (expressed in terms of the generators of App. \ref{antids})
\begin{equation}\label{next}
 \xi = (r_+ + r_-) R_2 - (r_+ - r_-) L_2
\end{equation}
 survives only in the hyperbolic deformation, whereas the one corresponding to extremal black holes
\begin{equation}\label{ext}
 \xi = 2 r_+  R_2 - (R_1 - R_3) - (L_1 + L_3)
\end{equation}
 is present in the parabolic one. Put another way,
non-extremal BTZ black holes allow for electric deformation, while
in the extremal ones, the deformation can only be induced by an
electro-magnetic wave.  Elliptic deformation is not compatible with
BTZ identifications.

The question that we would like to address is the following: how much of the
original black hole structure survives the deformation? The answer is
simple: a new chronological singularity appears in the asymptotic region of
the black hole. Evaluating the norm of the Killing vector shows that a naked
singularity appears. Thus the deformed black hole is no longer a viable
gravitational background. Actually, whatever Killing vector we consider
to perform the identifications, we are always confronted by such
pathologies.

The fate of the \emph{asymmetric parabolic} deformation of $\mathrm{AdS}_3$
is similar: there is no region at infinity free of closed time-like curves
after performing the identifications.

\subsubsection{Discrete identifications in symmetric deformations}

Let us consider the \emph{symmetric hyperbolic} deformation, whose metric is
given by~\eqref{Met2fois} with $\textsc{h} = 0$, i.e. $\omega = \lambda \left( 1 + \lambda
\right)$. This metric has two residual Killing vectors, manifestly given by
$\partial_t$ and $\partial_x$. We may thus, in general, consider identifications along
integral lines of
\begin{equation}
  \label{KillId}
  \xi=a\,\partial_t + \partial_x .
\end{equation}
This vector has a squared norm:
\begin{equation}
  \norm{\xi}^2=\left( \lambda \left( 1 + \lambda \right) - a^2 \right) + \frac {a \lambda
    \left( 1 + \lambda \right) + a^2 \left( 1 + 2 \lambda \right) }{r} .
\end{equation}
To be space-like at infinity the vector $\xi$ must verify the inequality $\lambda
\left( 1 + \lambda \right) > a^2$. If $a > 0 $, or $-\sqrt{ \lambda \left( 1 + \lambda
  \right) } < a < -2\lambda \left( 1 + \lambda \right) / \left( 1 + 2 \lambda \right)$,
$\xi$ is space-like everywhere. Otherwise, it becomes time-like behind the
inner horizon ($r = \lambda$), or on this horizon if $a=-\lambda$. In this last
situation, the quotient space will exhibit a structure similar to that of
the black string, with a time-like singularity (becoming light-like for
$a=-\lambda$) and two horizons.

\subsubsection{Discrete identifications in double deformations}
\label{sec:btz-ident-comb} 

The norm squared of the identification vector (\ref{KillId}) in the
metric~\eqref{Met2fois} is
\begin{equation}
  \norm{\xi}^2= \left( \omega - a^2 \right) + 2 \frac{a \omega +
    a^2 \left( 1 + 2 \lambda \right) }{r} - \frac{a^2 \left( \lambda \left( 1+\lambda
      \right) - \omega \right)}{r^2} .
\end{equation}
Between $r=0$ and $r=\infty$, this scalar product vanishes once
and only once (if $a\neq 0$).  To be space-like at infinity, we
have to restrict the time component of $\xi$ to $\abs{a} < \omega
$. Near $r=0$ it is negative, while near the inner horizon
($r=\lambda$) it takes the non-negative value $\omega \left(
\lambda + a \right)^2/\lambda^2$. Accordingly, by performing
identifications using this Killing vector, we will encounter a
chronological singularity, located at $r=r^*$, with
$0<r^*\leq\lambda$,  the singularity being of the same type as the
one in the symmetric case (see  Fig. 6.1).

\begin{figure}[ht]
\begin{center}
\includegraphics*[scale=0.5]{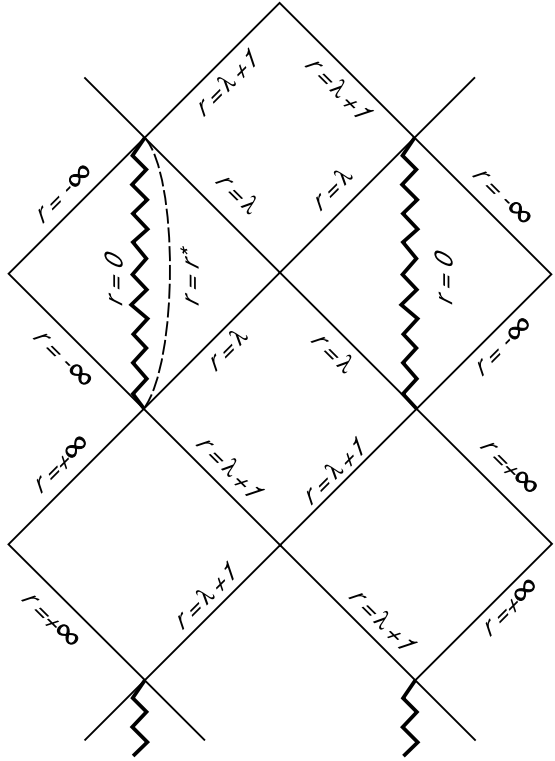}

  \caption{Penrose diagram exhibiting the global structure of the
    double hyperbolic deformation. The time-like curvature singularities
    $r=0$ are represented, as well as the horizons, located at $r=\lambda$ and
    $r=\lambda +1$. When performing identifications along orbits of a Killing
    vector allowing for a causally safe region at infinity,
    chronological singularities appear, which can be time-like and hidden behind an
    outer and an inner horizon ($r=r^*_1$), or space-like and hidden behind
    a single horizon ($r=r^*_2$), while the regions where $r<r^*$ have to be
    removed.}
\end{center}
\label{Penrose}
\end{figure}

\section{D-branes in $\SL$ black holes}\label{DBrBTZ}

We have seen in Sect. \ref{BWZW} that a particular class of D-brane configurations in WZW models is obtained as solutions to
the familiar gluing conditions on the chiral currents         
\begin{equation}
 J(z) = r(\bar{J}(\bar{z})) \quad \mbox{at}\quad z=\zb,
\end{equation}
 where $r$ is a metric preserving Lie algebra automorphism. These gluing conditions describe symmetric D-branes, that is, configurations which preserve conformal invariance and the infinite-dimensional symmetry of the current algebra of the bulk
theory. The geometry of the associated branes is encoded in these gluing conditions: their worldvolumes are shown to lie on (twisted) conjugacy classes in the group manifold. Symmetric D-branes of the $\SL$ WZW model were shown in \cite{BachPetr} to be of three types: two-dimensional hyperbolic planes ($H_2$), de Sitter branes ($dS_2$) and anti-de Sitter branes ($AdS_2$). It was emphasized that the $AdS_2$ worldvolumes, corresponding to twisted conjugacy classes, are the only physically relevant classical configurations, solutions to the DBI equations. Indeed, the worldvolume electric field on the $dS_2$ branes is supercritical (i.e. becomes imaginary), while $H_2$ branes have Euclidean signature and
must therefore be interpreted as instantons.

\subsection{D-branes in BTZ}\label{DBrBTZ1}
Let us first focus on the non-rotating BTZ black hole, which we studied in detail in Sect. \ref{NRBTZ}. We saw in this section that the spinless BTZ
black hole admits a foliation by leaves, the $\rho$ = constant surfaces, which are stable under the action of the BHTZ subgroup and constitute twisted conjugacy classes in $\SL$ ($AdS_2$ spaces). From our previous discussion, each of these leaves constitutes a D1-brane that is, in particular, a solution to the equations of motion derived from the DBI action. We consider as WZW background the extended BTZ metric
\begin{equation}
         \label{metmaxBis} ds^2=L^2 \left ( d\rho^2+\cosh(\rho)^2
         (d\phi^2-(w\,d\phi+dw)^2) \right ) \qquad \rho,w \in \R, \; \phi\in [0,2 \pi \sqrt{M}], 
         \end{equation}
and the B-field
\begin{equation}
 B = (\rho + \sinh \rho \cosh \rho) dw \wedge d\phi \quad ,
\end{equation}
providing a gauge choice for $H=dB$, with $H_{\mu \nu \lambda} = 2 \: \sqrt{- g} \: \varepsilon_{\mu \nu \lambda}$.
Solutions to the DBI equations of motion are:
\begin{eqnarray}\label{Solext}
\rho (x_0,x_1) &=& \rho_0, \quad w (x_0,x_1) = x_0, \quad \phi
(x_0,x_1) = x_1 \nonumber\\ F_{01}(x_0,x_1) &=& -\rho_0 \quad ,
\end{eqnarray}
where $x_0$ and $x_1$ are the worldvolume coordinates on the brane. This yields the expression for the worldvolume electric field, which could also be deduced from \re{omegaomega} (with $\omega = \hat{B} + 2\pi \a' F$).
 These solutions correspond to projections of twisted $AdS_3$
         conjugacy classes that wrap around BTZ space, and are thus
         compatible with the restriction of the $\phi$-variable to its
         range $[0,2\pi\,\sqrt{M}]$. They may be interpreted as closed DBI
         1-branes in BTZ space. In general, the $AdS_3$ branes obtained
         from (\ref{Solext}) by action of isometries do not project into
         closed branes in BTZ, but into infinite branes that extend from
         $\rho=-\infty$ to $\rho=+\infty$. Only the $AdS_3$ branes obtained
         from isometries compatible with the identifications generated by
         the BHTZ subgroup lead to closed DBI branes in extended $BTZ$ space.
         These isometries correspond to the left and right action of the
         subgroup $\exp(\R \h)$ on $\SL$. In terms of the coordinates
         \re{gExt2} or \re{metricExt2}, their Killing vectors read as
\begin{equation} \label{KillVec}
\Xi_- = \partial_{\phi} \quad \mbox{and} \quad \Xi_+ = -w \,
         \partial_{\rho} - w \tanh \rho \, \partial_{\phi} + (w^2 -1)
         \tanh \rho \, \partial_{w} \,.
\end{equation}
\footpourtoi{ce sont toujours des classes de conj.?}
         The action of the corresponding isometries on the brane
         (\ref{Solext}) yields
         \begin{eqnarray} \label{Solextiso}
         \sinh \rho(x_0) &=& + x_0 \cosh \rho_0
         \sinh f_{\pm} +\sinh \rho_0 \cosh f_{\pm} \nonumber \\
         w^2 (x_0) &=& 1 + \frac{\cosh^2 \rho_0}{\cosh^2 \rho (x_0)} (x_0^2
         -1) \nonumber \\
         e^{2 \phi(x_0,x_1)} &=& \frac{\cosh^2 \rho (x_0)}{\cosh^2 \rho_0}
         \, e^{2 (g_{\pm} + x_1)} \, ,
         \end{eqnarray} where $f_{\pm}$ and $g_{\pm}$ are constants related to the
         isometries used to perform the transformations, $f_+ = g_- = 2\pi
         \, \sqrt{M}$ and $f_- = g_+ =0$. In the causally-safe region,
         these solutions are more easily expressed, using the coordinates
         of \re{gReg2} or \re{Metric}, as
         \begin{eqnarray} \label{Solregiso}
         \sinh \rho(x_0) &=& \cos x_0 \cosh \rho_0 \sinh f_{\pm} +
         \sinh \rho_0 \cosh f_{\pm} \nonumber \\
         \sin \tau(x_0) &=& \frac{\cosh \rho_0 \sin x_0}{\cosh \rho
         (x_0)} \nonumber\\
         \theta(x_1) &=& x_1 + g_{\pm} \quad ,
         \end{eqnarray} The corresponding expressions for the worldvolume electric
         field $F_{01}$ can then be deduced from (\ref{eqA}).
The trajectory of a $\rho=cst.$ winding D1-brane is represented in the Penrose diagram of the non-rotating massive BTZ black hole, see Fig.~6.2.
\begin{figure}[ht]
\begin{center}
\includegraphics*[scale=0.5]{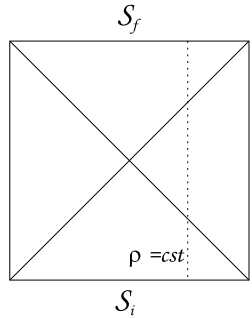}

  \caption{Trajectory of a $\rho=cst$ D1-brane in the BTZ black hole space-time.}
\end{center}
\label{D1brane}
\end{figure}
\footpourtoi{Qd $\rho \ra \infty$, $E\ra \infty$??} 
It is interesting to note that these D-branes do not seem to worry about the presence of the singularity, nor about the fact that they can evolve in a region with closed time-like curves (behind the singularity). It can indeed be checked that the Dirac-Born-Infeld (DBI) action evaluated in the solution \re{Solext}, i.e. the energy of the configuration as measured by an observer sitting at the center of $AdS_3$ (see \cite{BachPetr}, p4),
 is proportional to $\cosh \rho_0$, and so is constant everywhere for a given D1-brane $\rho=\rho_0$.
\footpourtoi{fct hypergeom sont "stables"?? Spectre des fluct quadratiques??}
The stability of the configuration can also be checked by first-order perturbation of solution \re{Solext}.
££

We have thus observed that the foliation of $AdS_3$ into stable (under BTZ identifications) two-dimensional leaves naturally led us to 
a class of winding D1-brane configurations in the non-rotating massive BTZ black hole background.
It is of course tempting to check whether this analysis can be extended to the non-rotating black hole. In this case, the two-dimensional surfaces are not conjugacy classes, and would thus provide an example of non-symmetric branes, with the additional feature that they wind in the compact direction of the black hole. However, the $\tau=cst.$ surfaces do not solve the DBI equations of motion in the background of the non-rotating BTZ black hole.

\subsection{Relation to D-branes in the 2D black hole}\label{DBrBTZ2}


In this section, we would like to relate the D1-branes in the
non-rotating BTZ background to the D0-branes of the 2D Lorentzian
black holes, pointed out in \cite{Yogendran}. We have seen in Sect.\ref{SectSymm} that there is
 a natural way of linking the $\SL$ WZW model to a gauged
version thereof. As shown in \cite{Giveon:1993ph}, the deformation by a
truly marginal operator of a WZW model based on a group $G$
yields, at one end of the deformation line, the gauged model
$G/\mbox{U}(1)$ times a non-compact $\mbox{U}(1)$, while, at the
other end, one finds the dual of the gauged model. In the $\SL$-case, as seen in Sect. \re{sec:backgr-fields-symm}, three
types of marginal symmetric deformations are possible, owing to
the presence of space-like ($\h =\sigma_3$), time-like ($\t = i \sigma_2$) and
light-like ($\e = \frac{1}{2} (\sigma_1 + i \sigma_2)$ and $\f = \frac{1}{2} (\sigma_1 - i \sigma_2)$) generators. Deforming the model in the
$J^{\h}\bar{J}^{\h}$ direction (\emph{hyperbolic} deformation),
one gets, at the end of the deformation line, the 2D Lorentzian
black hole times non-compact $\mbox{U}(1)$, whereas a deformation
in the $J^{\t}\bar{J}^{\t}$ direction (called \emph{elliptic})
leads to the 2D Euclidean black hole times non-compact
$\mbox{U}(1)$. Finally, the deformation in a null direction
(\emph{parabolic}) leads to a linear dilaton background times two
free bosons (see \cite{Israel:2004vv}, Sect.~3 for a review of
references).

We will hereafter naturally focus on the hyperbolic symmetric
deformation. A convenient parametrization of $g \in AdS_3$
is given by the coordinates \re{sphan}:
\begin{equation}
  g = {\rm e}^{{\frac{\psi - \varphi}{2}} \h} {\rm e}^{\beta \t}
  {\rm e}^{\frac{\psi + \varphi }{2} \h},
\end{equation}
where $\psi $ and $\varphi$ \emph{are not} compact coordinates in
$AdS_3$. In performing the BTZ identifications (\ref{Identif}) with
$(g_l,g_r) = (\mathrm{e}^{a n \h}, \mathrm{e}^{-a n
\h})$, the $\varphi-$coordinate becomes periodic.

 The hyperbolic deformation results in the 
background \re{eq:J2J2-deform}, consisting of a metric,
B-field and dilaton. In the undeformed case $\kappa=1$, the coordinate patch
$(\beta,\psi,\varphi)$ does not cover the full $AdS_3$. When
$\kappa < 1$, the background with
metric (\ref{eq:J2J2-deform}) corresponds to a non-extremal black
string~\cite{Horne:1991gn,Horne:1991cn,Horowitz:1993jc}, the
parametrization $(\beta,\psi,\varphi)$ covering the region between
the inner and the outer horizon. When $\kappa \rightarrow \infty$,
 the $\varphi-$direction corresponding to the direction compactified
  by the BTZ identifications decouples and we are left with the
  two-dimensional Lorentzian black-hole\footnote{In the other
  limit, $\kappa \rightarrow 0$, the remaining two-dimensional
  space has a compact space-like direction, so we won't consider
  this limit}. As we noticed, 
the original $\SL \times \SL$ affine symmetry will be broken into
$\mbox{U}(1) \times \mbox{U}(1)$ due to the deformation. The choice of hyperbolic
deformation ensures that the Killing vector used to perform the
BTZ identifications remains a Killing vector of the deformed
background (see previous Section).

We will now analyze how the D-brane configurations of the
preceding section follow the deformation. D-branes in a marginally
deformed WZW model were studied in \cite{Forste:2001gn,Forste:2002uv} in the
$\mbox{SU}(2)$-case, and we expect a similar behavior in the
$\SL$-case.
 First, we notice that the boundary
conditions (\ref{GluingWZW}) are generally not derived from the usual
WZW action, with the integration region being the upper half plane
(an \emph{open} string worldsheet) . Therefore, one has to add a
boundary term to obtain the WZW open string action in the presence
of D-branes (see \cite{BDS}). This additional term can be written
as an integral over the real line, so that it can be identified to
an F-flux on the worldvolume of the D-brane. This F-field, living
only on the brane, can be determined by analyzing the
Dirac-Born-Infeld action for this classical configuration.

In the undeformed case, the worldvolumes of the D-branes described
in Sect.~\ref{DBrBTZ} are given by
\begin{equation}\label{worldvolume}
 \mbox{Tr}(g \h) = 2 \sinh \rho = 2 \cos\beta \sinh\psi = cst. \quad .
 \end{equation}
Furthermore, one checks that with the gauge choice
\begin{equation}\label{BBis2}
B = \cos^2\beta d\phi \wedge d\psi \quad ,
\end{equation}
the DBI equations of motion are solved with a vanishing
worldvolume electric field. Stated another way, the boundary term
obtained from the WZW action with expression (\ref{BBis2}) for the
B-field coincides with the boundary condition (\ref{GluingWZW}). Thus,
no F-flux has to be added to the boundary.

In the deformed case, $\kappa \ne 1$, the same remains true, as
may be verified by a direct computation from the DBI action. Thus,
the branes $\rho=cst.$ survive along the whole deformation line.
These branes correspond to configurations preserving a residual
U(1) symmetry of the U(1) $\times$ U(1) symmetry of the deformed
model, and may be seen as resulting from gluing conditions of the
type (\ref{GluingWZW}) between the remaining chiral and anti-chiral
currents :
\begin{equation}
J^{\h} =\frac{1}{\Delta_{\kappa} (\beta)} (\cos^2 \beta \partial
\psi + \sin^2 \beta
\partial \phi) \quad \mbox{and} \quad  \bar{J}^{\h} = \frac{1}{\Delta_{\kappa} (\beta)}(\cos^2 \beta
 \bar{\partial} \psi -
\sin^2 \beta \bar{\partial} \phi) \quad .
\end{equation}

Notice that, at the end of the deformation line $\kappa
\rightarrow \infty$, the coordinates $(\beta,\psi,\varphi)$ cover
only part of the Lorentzian 2D black hole space-time, namely the
region inside the horizon. To get a better picture, we will turn
to coordinates describing the whole two-dimensional space-time.
This is done by using an equivalent description of the symmetric
hyperbolic deformation, which can be identified with an $(\SL
\times \R)/\R$ coset model, in which the embedding of the dividing
group has a component in both factors \cite{Giveon:1994ph}. In the
case at hand, we will gauge the one dimensional subgroup generated
by $\h$ together with a translation in the $\R$ factor
parameterized by the $x$-variable. We parameterize the elements of
$\SL$ as
\begin{equation}
 g = \left(
         \begin{array}{cc} a & u\\ -v & b
         \end{array}
         \right) \quad \mbox{with} \quad ab+uv = 1 \quad.
\end{equation}
Furthermore, an axial gauging ($g\rightarrow h g h$) allows us to reach the symmetrically deformed model for
$\kappa < 1$, while a vectorial gauging ($g\rightarrow
hgh^{-1}$) covers the $\kappa > 1$ branch of the
deformed geometry. We will focus on the latter. The computation is
analogous to the one in Sect.~2 of \cite{Horne:1991gn}, with the
difference being that the gauging this time allows us to set $u=\pm v$
according to the sign of $1- ab$. In the $\kappa\rightarrow\infty$
limit, the metric reads
\begin{equation}
ds^2 = -\frac{da db}{1-a b} + dx^2 \quad.
\end{equation}
The $(a,b)$-part of the metric corresponds to the 2D Lorentzian
black hole, with horizons located at $ab=0$ and singularities at
$ab=1$. The relation to the other coordinate system is $a={\rm
e}^\psi \cos\beta$, $b={\rm e}^{-\psi}\cos\beta$, and
$x=\frac{\varphi}{\sqrt{\kappa^2 -1}}$, so that with the BTZ
identifications, $x$ becomes periodic. The D-branes
(\ref{worldvolume}) are thus given by
\begin{equation}
 a - b = cst \quad,
 \end{equation}
 which correspond to the unboosted D0-branes in the 2D Lorentzian
 black hole (see \cite{Yogendran}, Sect.~4.2). Finally, let us
 note that in between the extreme points of the deformation,
 there are regions where the BTZ identifications are time-like.
 Indeed, the norm of the identification vector is given by
 \begin{equation}
  ||\partial_x||^2 = \frac{(\kappa^2 -1)(a b -1)}{a b(\kappa^2 -1) -
  \kappa^2} \quad,
\end{equation}
and consequently regions where $1 < a b <
\frac{\kappa^2}{\kappa^2-1}$ will contain closed time-like curves.
The fact that the metric (\ref{eq:J2J2-metric}) describes a part
of a black string space-time appears clearly if one
performs the further change of coordinates $r=-(\lambda+1) +
\sin^2 \beta$, $\hat{t} = \sqrt{-\lambda} \psi$ and
$\hat{x}=\sqrt{-(\lambda+1)} \varphi$, with $\kappa^2 =
\frac{\lambda}{\lambda+1} > 1$, thus $\lambda<-1$, in which the
metric becomes
\begin{equation}
ds^2 = -(1+\frac{\lambda}{r}) \, d\hat{t}^2 + (1 +
\frac{\lambda+1}{r}) \, d\hat{x}^2 +
\frac{dr^2}{4(r+\lambda)(r+\lambda+1)} \quad ,
\end{equation}
which coincides with eq. (12) of \cite{Horne:1991gn}. The metric
has a curvature singularity at $r=0$, corresponding to $a b =
-\lambda = \frac{\kappa^2}{\kappa^2 -1}$, while the horizons are
located at $r=-\lambda-1$ ($a b = 1$) and $r=-\lambda$ ($a b =
0$). Thus, the chronological singularity is located on the inner
horizon $a b=1$, with regions $a b<1$ being causally
safe\footnote{this situation appears as the limiting case for the
identifications in the black string background discussed in
Sect.~\ref{sec:btz}}. When $\kappa \rightarrow \infty$, the
inner horizon collapses on the curvature singularity, located at
$a b =1$ as it should in the two-dimensional black hole.



\cleardoublepage \part{Outlook and discussion}

It is now time to assemble all the pieces of the puzzle. We briefly review in no particular order the results obtained, and speculate on some directions
that could deserve further attention as natural continuations of this work. 

As a first step, we intensively used the underlying group-theoretical structure of BTZ black holes to describe them. Based on foliations of $AdS_3$ in two-dimensional leaves, adapted to the identifications leading to the black hole, we were able to derive global expressions for the metric of  (maximally extended) non-rotating and generic BTZ black holes. We also noticed the prominent role played by two-parameter solvable subgroups of $\SL$, which appeared to be closely related to the black hole's singularities and horizons. We essentially focused on the maximally extended non-rotating BTZ black hole, for two reasons. The first is that in this space-time, the leaves of the foliation are twisted conjugacy classes, and hence represent closed symmetric WZW D1-branes, winding around the compact direction of the black hole. The second lies in the fact that each leaf of the foliation admits an action of a solvable subgroup of $\SL$, which we denoted $AN$. This last observation allowed us to deform the algebra of functions on the brane, in the direction of the (left-)invariant Poisson brackets on the leaf (associated with the WZW $B$-field\footpourmoi{C'est vrai ca? En fait $\theta$ est different de $B$, mais dans notre cas $\theta$ aussi ne depend que de $\rho$, donc est constant sur une feuille, donc c'est le truc invariant a gauche, voir \re{VolumeAN}}). This was partly motivated by the observation that, in flat space-time, the worldvolume of a D-brane, on which open strings end, is deformed into a noncommutative manifold in the presence of a $B$-field. In particular, in the limit $\a' \ra 0$ in which the massive open string modes decouple, the operator product expansion of open string tachyon vertex operators is governed by the Moyal-Weyl product\cite{SeibWitt}\footpourmoi{lien avec la deformation de l'algebre des fonctions : AleksReckSchom!}, and are of the form\footpourmoi{Philippe: preciser!}:
\beq\label{bientotfini}
 \qquad\qquad\qquad\qquad\qquad {\rm e}^{i P.X}(\tau)  {\rm e}^{i Q.X}(\tau') \sim ({\rm e}^{i P.X} \overset{M}{\star} {\rm e}^{i Q.X})(\tau') , \qquad\qquad\qquad\qquad\qquad (\mbox{III}.1) \nonumber
\eeq
where $X$ represents the string coordinates on the brane, $P$ and $Q$ the momenta of the corresponding string states, $\tau$ and $\tau'$ parameterize the worldsheet boundary, and where the classical limit $f \overset{M}{\star} g \ra f.g$ is recovered for $B \ra 0$\footpourmoi{preciser: $\theta$ reprend $\Omega$ ET le $\hbar$, voir SeibWitt p8, PierreStrictp275} \footpourmoi{attention en cas plat, ca a un sens de faire tendre $B$ vers 0, pas dans WZW. Mais on prend la limite $\a'\ra 0$ et $\a' k \ra \infty$, donc dans cette limite $H$ tend vers 0! Voir \cite{AleksReckSchom2}p5-10 et \cite{Schomerus:2002dc,AleksReckSchom1,AleksReckSchom3}}. It is not clear how this kind of relationship could be generalized in curved situations, in particular in WZW models. It was argued in \cite{AleksReckSchom2,Schomerus:2002dc,AleksReckSchom1,AleksReckSchom3} that the relevant limit to take should be $\a' \ra 0$ and $\a' k \ra 0$, where $k$ is the level of the model. There it was observed that a structure similar to (III.1) emerges for the $SU(2)$ WZW model, describing strings on the compact space $S^3$. The appearance of noncommutative structures has also be pointed out in relation with other backgrounds, like the Melvin Universe \cite{Hashimoto:2005hy,Hashimoto:2002nr,Halliday:2006qc,Behr:2003qc,Cai:2002sv} or the Nappi-Witten plane wave \cite{Halliday:2006qc,Halliday:2005zt}, see e.g. \cite{Szabo:2006wx} for a recent review.
In the case of interest here, namely that of open strings on D1-branes in BTZ spaces, one might wonder what would replace (III.1). A natural guess would be that $\overset{M}{\star}$ would be replaced by an operation preserving the (symplectic) structure of the brane\footpourmoi{Pq, si ce n'est que Moyal preserve le $\theta$?}\footpourmoi{le fait qu'on ait un star produit sur BTZ etendu et pas Regular signifie-t-il quelque chose??}\footpourmoi{pq a-t-on cherche des deformations strictes WKB? Dans Seib-Witt, eq(2.9), ce n'est pas Moyal integral qui apparait?}, i.e. the products we obtained and discussed\footpourmoi{Pq dans cette limite particuliere s'attend-on a avoir un truc ASSOCIATIF?(ce qui justifie l'apparition du produit star!!)}.
 To go further, one would have to identify the analogs of the primary fields ${\rm e}^{i P.X}$. This requires the determination of the spectrum of open strings in BTZ space. As strings on BTZ are described by an orbifold (or quotient) of the $\SL$ WZW model, the first step is to determine the spectrum of the original $\SL$ WZW model, and then add the sectors corresponding to strings winding around the compact direction. The first step has actually been a long-standing problem in string theory (see e.g. \cite{Petropoulos:1999nc} for a review and references), and a lot of effort has been devoted to to making this model sensible. It is only recently that Maldacena and Ooguri solved the problem in a series of papers \cite{MO1,MO2,MO3}.
The structure of the Hilbert space is much more complicated than in the flat space, reflecting in part the fact that $\SL$ representation theory  is much more involved than that of flat space. Another reason is that it seems mandatory, for consistency conditions (in particular, to recover the infinite tower of massive string states), to include so-called spectral-flowed states in the spectrum, associated semi-classically with winding strings in $AdS_3$\footpourmoi{winding autour de quoi?? Force due a la cstte cosmologique compensee par le champ B}. With this requirement, the spectrum was shown to be free of ghosts, and the partition function modular invariant. The spectrum for open strings ending on $AdS_2$ has been subsequently studied, namely in \cite{Lee:2001xe,Lee:2001gh,Rajaraman:2001cr,Parnachev:2001gw,Hikida:2001yi,Petropoulos:2001qu}, where it was roughly shown to be the ``holomorphic square root" of the closed string spectrum. Nevertheless, in the BTZ case, there seems to be no convincing determination of the string spectra, neither closed nor open. There have been propositions in \cite{HK,HKExt,Hemming,troostwinding,Satoh:2002nj,Satoh,NatsuumeSatoh,Satoh:1997xf,Hemming:2004gd}, but a number of twilight zones remain: it is not clear how to include the additional winding sectors, nor to answer the question whether the spectrum is ghost-free and the partition function modular invariant. 

After having noticed the existence of the closed winding D1-branes in the non-rotating BTZ black hole, and the possibility of deforming them, we remarked that the behavior of these branes was very similar to existing D0-branes' configurations in the two-dimensional black hole: they cross  horizons and singularities, and do not seem to experience anything particular at the singularities. These objects might perhaps be able to probe the geometry of black holes near singularities\footpourmoi{preciser: Maldacena, Johnson; VOIR p82, reference 33 de Uranga pour le role des branes comme test de singularites, 9911161!!Voir aussi 0605139 (Strominger)}. A reliable analysis would actually have to go beyond a Dirac-Born-Infeld analysis, and should be based on an exact CFT description of these branes in terms of boundary states\footpourmoi{savoir exactement ce que c'est !!}. Again, this would require a detailed understanding of the orbifolded $\SL$ WZW model (for such an analysis of the D0-branes in the two-dimensional black hole, see \cite{Sugawara}).

In the context of string theory, we explored the moduli space of the $\SL$ WZW model, by performing a rather systematic analysis of the deformations it could undergo. Inspired by the BTZ construction, we looked at new solutions that could be obtained by performing discrete identifications in the deformed background. We obtained a large class of new solutions describing black holes and black strings, and computed their charges. We have essentially focused on the geometrical interpretation of the resulting backgrounds. The string spectrum on these backgrounds could, in principle, be determined from that of $AdS_3$, but its derivation will present the same difficulties as for the BTZ black hole. 

We have seen that, in the case of non-rotating BTZ black holes and their higher-dimensional generalizations, the black hole singularities are tighly knit to solvable subgroups. In the three-dimensional case, the relevant groups are the $AN$ and $A\bar{N}$ subgroups of $\SL$; in the $l$-dimensional case, we showed that the closed orbits of the Iwasawa subgroups of $SO(2,l-1)$ define the singularities.
This actually suggests that we tackle the questions of black hole singularities in the context of noncommutative geometry for the following reason.
Noncommutative geometry's basic principle relies on the assumption that space-time may have a noncommutative structure at some scale
\footpourmoi{ (mais nous on deforme partout)}. We stressed in the introduction that the notion of a noncommutative manifold can be algebraically captured in an object, called spectral triple, consisting of the Hilbert space formed by all the spinor fields on the (pseudo-)Riemannian manifold, the Dirac operator associated with its metric, and a noncommutative algebra. In this work, we precisely determined a noncommutative algebra associated with the three-dimensional case (see \cite{Laurent} for the higher-dimensional cases). Suppose now that we have a description of an $AdS$ black hole in terms of a spectral triple. An interesting question to raise could be: how is the singularity encoded in the triple? And most importantly: what happens if the triple is deformed?\footpourmoi{le truc de Connes, c'est que toute l'info sur un espace, on peut le lire sur les fonctions! LIRE le truc de Martinetti!!} Is the singularity ``smeared out" by the deformation? \footpourmoi{ref Connes-Marcolli: triplets spectraux singularites: Lescure}

An important part of this work, somewhat outside its main lines, has been to find $\SL$-invariant associative composition laws on the hyperbolic plane, and to show that these are essentially unique, although constituting a large family of solutions. Incidentally, it would be interesting to see whether or not our formulae can be used to define strict deformations of the hyperbolic plane $H_2$, that is, if an appropriate classical limit can be taken to deform its invariant Poisson bracket. Also, an interesting mathematical problem would be to find the analog of the Schwartz space for the Moyal product, i.e. a space stable for the ordinary pointwise multiplication and for the deformed $\SL$-invariant product. We may finally remark that the hyperbolic plane constitutes a distinguished class of D-branes, in the Euclidean version of $AdS_3$ \cite{Ponsot:2001gt}. Therefore one could also ask if a relation similar to (III.1) holds for open string vertex operators ending on $H_2$ branes, and if the class of $\SL$-invariant products we obtained could be relevant.


As one usually says,
``We hope to return to these questions in future works"!



\cleardoublepage \part{Appendices} 
 
\appendix 

\cleardoublepage \chapter{$AdS_3$ and BTZ coordinate systems}\label{antids}

In this Appendix, we assemble the various coordinate systems mentioned throughout the text.


\section{$\mathrm{AdS}_3$-$SL(2, \mathbb{R})$}
\label{sl2}


The three-dimensional anti-de Sitter space is the universal
covering of the $SL(2,\mathbb{R})$ group manifold. The latter can
be embedded in a Lorentzian flat space with signature $(-,+,+,-)$
and coordinates $(x^0,x^1,x^2,x^3)$:
\begin{equation}
  g  =  L^{-1}\
  \begin{pmatrix}
    x^0 + x^2 & x^1 + x^3 \\ x^1 -
      x^3 & x^0 - x^2
  \end{pmatrix}, \label{4emb}
\end{equation}
where $L$ is the radius of $\mathrm{AdS}_3$.

The isometry group of the $SL(2,\mathbb{R})$ group manifold is
generated by left or right actions on $g$: $g\to hg$ or $g\to gh$
$\forall h \in SL(2,\mathbb{R})$. From the four-dimensional point
of view, it is generated by the Lorentz boosts or rotations
$\zeta_{ab}= i\left( x_a\partial_b - x_b
  \partial_a\right)$ with $x_a=\eta_{ab}x^b$. We explicitly list here the
six Killing vectors, as well as the group action they correspond
to:
\begin{subequations}
  \label{eq:Killing-SL2R}
  \begin{align}
    L_1 &= \frac{i }{2}\left(\zeta_{32} - \zeta_{01}\right), & g &\to
    \mathrm{e}^{-\frac{\lambda}{2}\sigma^1}g,
    \label{L1} \\
    L_2 &= \frac{i }{2}\left(-\zeta_{31}-\zeta_{02} \right), &  g &\to
    \mathrm{e}^{-\frac{\lambda}{2}\sigma^3}g,
    \label{L2} \\
    L_3 &= \frac{i }{2}\left(\zeta_{03} - \zeta_{12}\right), & g&\to
    \mathrm{e}^{i\frac{\lambda}{2}\sigma^2}g,
    \label{L3} \\
    R_1 &= \frac{i }{2}\left( \zeta_{01} + \zeta_{32}\right), & g&\to
    g\mathrm{e}^{\frac{\lambda}{2}\sigma^1},
    \label{R1} \\
    R_2 &= \frac{i }{2}\left(\zeta_{31} - \zeta_{02}\right), &  g&\to
    g\mathrm{e}^{-\frac{\lambda}{2}\sigma^3},
    \label{R2} \\
    R_3 &= \frac{i }{2}\left(\zeta_{03} + \zeta_{12}\right), & g&\to
    g\mathrm{e}^{i\frac{\lambda}{2}\sigma^2}.
    \label{R3}
  \end{align}
\end{subequations}

Both sets satisfy the $\ls$ algebra.
 The norms of the Killing vectors are the following:
\begin{equation}
  \norm{L_1}^2 = \norm{R_1}^2 = \norm{L_2}^2 =
  \norm{R_2}^2 =- \norm{L_3}^2=-\norm{R_3}^2 = \frac{L^2}{4}.
\end{equation}
Moreover $L_i \cdot L_j = 0$ for $i\neq j$ and similarly for the
right set. Left vectors are not orthogonal to right ones.

The isometries of the $SL(2,\mathbb{R})$ group manifold turn into
symmetries of the $SL(2,\mathbb{R})_k$ WZW model, where
they are realized in terms of conserved currents.
The reader will find details on those issues in the Appendices of
\cite{Israel:2004vv}.

\section{``Symmetric" coordinates} \label{antidss}

One introduces Euler-like angles by
\begin{equation}
g = \mathrm{e}^{i {\tau+\phi \over 2} \sigma^2}
\mathrm{e}^{\rho \sigma^1} \mathrm{e}^{i{\tau-\phi \over 2}
\sigma^2} , \label{euler}
\end{equation}
which provide good global coordinates for $\mathrm{AdS}_3$ when
$\tau\in ]-\infty,+\infty[$, $\rho\in [0,\infty[$, and $\phi\in
[0,2\pi]$. In Euler angles, the invariant metric reads:
\begin{equation}
d s^2= L^2\left[- \cosh ^2 \rho  \, d \tau ^2 +d \rho^2 +
\sinh^2 \rho \, d \phi^2\right]. \label{dseul}
\end{equation}
The Ricci scalar of the corresponding Levi--Civita connection is
$\mathcal{R}=-6/L^2$. The volume form reads:
\begin{equation}
  \label{vfeul}
  \omega_{[3]} = \frac{L^3}{2}\sinh 2\rho
     d \rho \land d \phi  \land d \tau,
\end{equation}
whereas $L_3 = \frac{1}{2}\left( \p_\tau + \p_\phi\right)$ and
$R_3 = \frac{1}{2}\left( \p_\tau - \p_\phi\right)$.

Another useful, although not global, set of coordinates is defined
by
\begin{equation}
  g = {\rm e}^{{\frac{\psi - \varphi}{2}} \sigma^3} {\rm e}^{i t \sigma^1}
  {\rm e}^{\frac{\psi + \varphi }{2} \sigma^3},
\end{equation}
($\psi $ and $\varphi$ \emph{are not} compact coordinates).  The metric reads:
\begin{equation}
  d s^2 = L^2\left[ \cos ^2 t  d \psi^2 -d t^2 + \sin^2 t\,
    d \varphi^2\right], \label{dssphan}
\end{equation}
with volume form
\begin{equation}
  \label{vfsphan}
  \omega_{[3]} = \frac{L^3}{2}\sin 2t d t \land d \psi  \land d \varphi.
\end{equation}
Now $L_2 = \frac{1}{2}\left( \p_\psi - \p_\varphi\right)$ and $R_2 =
\frac{1}{2}\left( \p_\psi + \p_\varphi \right)$.

Finally, the Poincar\'e coordinate system is defined by
  \begin{equation} \label{eq:ads-poinc-tr}
    \begin{cases}
    x^0 + x^2 &=\frac{L}{u},\\
    x^0 - x^2 &= Lu + \frac{L x^+ x^-}{u},\\
    x^1 \pm x^3 &= \frac{L x^\pm}{u} .
    \end{cases}
  \end{equation}
  For $\{u,x^+,x^-\} \in \mathbb{R}^3$, the Poincar\'e
  coordinates cover the $SL(2\mathbb{R})$ group manifold once. Its
  universal covering, $\mathrm{AdS}_3$, requires an infinite
  number of such patches. Moreover, these coordinates exhibit
  a Rindler horizon at $\vert u \vert \to \infty$; the conformal
  boundary is at $\vert u \vert \to 0$.
  Now the metric reads:
  \begin{equation}
    d s^2 = \frac{L^2}{u^2} \left( d u^2 + d x^+ d x^-
    \right),
  \end{equation}
  and the volume form:
  \begin{equation}
    \label{eq:ads-poinc-volume}
    \omega_{[3]} = \frac{L^3}{2u^3} d u \land d x^+ \land d
    x^-.
  \end{equation}
We also have $L_1+L_3 = -\p_-$ and $R_1+R_3 = \p_+$.

\section{``Asymmetric" coordinates} \label{antidsas}

The above three sets of AdS$_3$ coordinates are suitable for
implementing symmetric parabolic, elliptic or hyperbolic
deformations, driven by $\left( J^1 +J^3
\right)\left(\bar J^1 + \bar J^3 \right)$, $J^3\bar J^3$ or  $J^2
\bar J^2$ respectively. For asymmetric elliptic or hyperbolic deformations, we
must use different coordinate systems, where the structure of
$\mathrm{AdS}_3 $ as a Hopf fibration is more transparent. They
are explicitly described in the following.
\begin{itemize}
\item The coordinate system used to describe
  the elliptic asymmetric deformation is defined as follows:
  \newcommand{\CR}[0]{\cosh \frac{\rho}{2}}
  \newcommand{\SR}[0]{\sinh \frac{\rho}{2}}
  \newcommand{\CPHI}[0]{\cosh \frac{\phi}{2}}
  \newcommand{\SPHI}[0]{\sinh \frac{\phi}{2}}
  \newcommand{\CT}[0]{\cos \frac{t}{2}}
  \newcommand{\ST}[0]{\sin \frac{t}{2}}
  \begin{equation}
    \begin{cases}
      \frac{x_0}{L} &= \CR \CPHI \CT - \SR \SPHI \ST, \\
      \frac{x_1}{L} &= -\SR \SPHI \CT - \CR \SPHI \ST,\\
      \frac{x_2}{L} &= -\CR \SPHI \CT + \SR \CPHI \ST, \\
      \frac{x_3}{L} &= -\SR \SPHI \CT - \CR \CPHI \ST.
    \end{cases}
  \end{equation}
  The metric now reads:
   \begin{equation}
     \label{eq:ads-rhotphi-metric}
     d s^2 = \frac{L^2}{4} \left( d \rho^2 + d \phi^2 - d t^2 -
       2 \sinh \rho d t d \phi\right),
   \end{equation}
   and the corresponding volume form is
   \begin{equation}
     \label{eq:ads-rhotphi-vf}
     \omega_{[3]} = \frac{L^3}{8}\cosh \rho
     d \rho \land d \phi  \land d t.
  \end{equation}
 This coordinate
  system is such that the $t$-coordinate lines coincide with the integral
  curves of the Killing vector $L_3 = - \p_t$, whereas the $\phi$-lines are
  the curves of $R_2 = \p_\phi$.
\item The coordinate system used to describe the asymmetric
  hyperbolic deformation is defined as follows:
  \renewcommand{\CR}[0]{\cosh \frac{r}{2}}
  \renewcommand{\SR}[0]{\sinh \frac{r}{2}}
  \newcommand{\CX}[0]{\cosh \frac{x}{2}}
  \newcommand{\SX}[0]{\sinh \frac{x}{2}}
  \renewcommand{\CT}[0]{\cos \frac{\tau}{2}}
  \renewcommand{\ST}[0]{\sin \frac{\tau}{2}}
  \begin{equation}
    \label{eq:ads-rxt-coo}
    \begin{cases}
      \frac{x_0}{L} &= \CR \CX \CT + \SR \SX \ST, \\
      \frac{x_1}{L} &= - \SR \CX \CT + \CR \SX \ST, \\
      \frac{x_2}{L} &= - \CR \SX \CT - \SR \CX \ST, \\
      \frac{x_3}{L} &= \SR \SX \CT - \CR \CX \ST.
    \end{cases}
  \end{equation}
  For $\{r,x,\tau\} \in \mathbb{R}^3$, this patch covers the
  whole $\mathrm{AdS}_3$ exactly once, and is regular
  everywhere~\cite{Coussaert:1994tu}.  The metric is then given by
  \begin{equation}
     \label{eq:ads-rxt-met}
    d s^2 = \frac{L^2}{4} \left( d r^2 + d x^2 - d \tau^2 +
      2 \sinh r d x d \tau \right),
  \end{equation}
  and correspondingly the volume form is
  \begin{equation}
    \label{eq:ads-rxt-vf}
    \omega_{[3]} = \frac{L^3}{8} \cosh r d r \land d x \land d \tau .
  \end{equation}
  In this case, the $x$-coordinate lines
  coincide with the integral curves of the Killing vector $L_2 = \p_x$,
  whereas the $\tau$-lines are the curves of $R_3=-\p_\tau$.
\end{itemize}

\section{Original BTZ coordinates}
\label{btzcooor}

The BTZ coordinates are defined as follows ($t, \varphi$
and $r$ should not be confused with other coordinates introduced
elsewhere in this Appendix):
\begin{equation}
  x^0 \pm x^2 =
  \begin{cases}
    \pm \epsilon_1 L \left(r^2_{\vphantom+}- r^2_+ \over r^2_+ -
      r^2_-\right)^{1/2} \exp \pm \left(r_+ {t} - r_- \varphi \right) &
    \text{for $r > r_+$,} \\
    \phantom{\pm} \epsilon_2 L \left(r^2_+ - r^2_{\vphantom+} \over r^2_+ -
      r^2_-\right)^{1/2} \exp \pm \left(r_+ {t} - r_- \varphi \right) &
    \text{for $r < r_+$}
  \end{cases}
\label{tr0}
\end{equation}
and
\begin{equation}
  x^1 \pm x^3 =
  \begin{cases}
    \pm \epsilon_3 L \left(r^2_{\vphantom+}- r^2_- \over r^2_+ -
      r^2_-\right)^{1/2} \exp \pm \left(r_+ \varphi - r_- {t} \right) &
    \text{for $r > r_-$,} \\
    \phantom{\pm}\epsilon_4 L \left(r^2_- - r^2_{\vphantom+} \over r^2_+ -
      r^2_-\right)^{1/2} \exp \pm \left(r_+ \varphi - r_- {t} \right) &
    \text{for $r < r_-$.}
  \end{cases}
 \label{tr1}
\end{equation}
Here the $\epsilon$s are pure signs -- several different choices of them
are necessary in order to cover entirely the part of a hyperboloid where
$\xi\equiv \partial_\varphi$ is space-like.  Note also that the above
transformations degenerate in the extremal case $r_+=r_-$, where they no
longer apply (the precise coordinate transformations are more complicated
and we will not show them here). The AdS$_3$ metric reads:
\begin{equation}
  d s^2 =L^2 \left[ -f^2(r) \, d t^2 + f^{-2}(r) \, d r^2 +
    r^2 \left( d \varphi-{r_+ \, r_- \over   r^2}\, d t
    \right)^2\right] , \label{btz1}
\end{equation}
with
\begin{equation}
  f(r) ={1\over r}\sqrt{\left( r^2_{\vphantom +}-r^2_{+} \right)
    \left(r^2_{\vphantom -}-r^2_{-}\right)} . \label{btz2}
\end{equation}

The genuine BTZ geometry is obtained by setting $\varphi
\cong \varphi+2\pi$. Regions of AdS$_3$ where
$\norm{\partial_\varphi}$ is negative must be excised, in order to
avoid closed time-like curves. The boundary of such regions is the
black-hole singularity. Note that the BTZ coordinates
are not global, neither for the AdS$_3$ nor for the black hole
itself: they do not entirely cover even the remaining part of the
hyperboloid, where $\partial_\varphi$ is space-like.

\section{Global BTZ coordinates}
\label{btzcooss}

Particular coordinate systems exist covering (part of)
$\mathrm{AdS}_3$ that are well-adapted to the description of
BTZ black holes. In the non-rotating case, the following
parametrization covers the black hole region (where the identifications are
space-like) exactly once :
\begin{equation}
  \label{gReg}
  g(\rho,\theta_,\tau) =
  \begin{pmatrix}
    \cos\tau \cosh\rho + \sinh \rho &   e^\theta \cosh \rho \sin \tau \\
    -e^{-\theta} \cosh \rho \sin \tau &  \cos\tau \cosh\rho - \sinh \rho
  \end{pmatrix}
 \quad,
\end{equation}
leading to the metric
\begin{equation}
  \label{metricReg}
  d s^2 = d \rho^2 + \cosh^2 \rho \left( -d \tau^2 + \sin^2 \tau d \theta^2 \right)
  \quad,
\end{equation}
where $-\infty < \rho < +\infty$, $0<\tau<\pi$, $\theta$ becoming periodic
after the identifications. The chronological singularities are located at
$\tau = 0$ and $\tau = \pi$.  There is also another coordinate system going
behind the chronological singularities :
\begin{equation}
  \label{gExt}
  g(\rho,\phi_,w) =
  \begin{pmatrix}
    w \cosh\rho + \sinh \rho &   e^\phi \cosh \rho \left( w^2 - 1 \right)\\
    e^{-\phi} \cosh \rho  &  w \cosh\rho - \sinh \rho
  \end{pmatrix} \quad,
\end{equation}
with the corresponding metric
\begin{equation}
  \label{metricExt}
  d s^2  = d \rho^2 + \cosh^2 \rho \left( d \phi^2 - \left( w d \phi +
      d w \right)^2\right) \quad.
\end{equation}
The black hole singularities are located at $w=\pm 1$, while the horizons
correspond to the surfaces
\begin{equation}
  w = \pm \tanh \rho \quad.
\end{equation}
The two coordinate systems are related by\footnote{We correct a misprint in
  Eq.~(2.22) of~\cite{Bieliavsky:2004yp}}
\begin{equation}
  w = \cos \tau \quad \text{and} \quad e^{\phi} = \frac{e^{\theta}}{\sin \tau} \, .
\end{equation}
The parameterizations~(\ref{sphan}) and~(\ref{gExt}) are related by
\begin{subequations}
  \begin{align}
    e^{2 \psi} &= \frac{ w \cosh \rho + \sinh \rho}{w \cosh \rho -
      \sinh \rho} \quad , \\
    e^{2\varphi} &= e^{2\phi} \left( 1 - w^2 \right) \quad, \\
    \cos 2 t &= \cosh^2 \rho \left( 2 w^2 -1 \right) - \sinh^2 \rho \quad.
  \end{align}
\end{subequations}
In the case of the rotating BTZ black hole, the following parametrization of
the group elements has been used in \cite{BDHRS}:
\begin{equation}
  \label{CoordRot}
  g = \mathrm{e}^{\vartheta \sigma_3} \mathrm{e}^{u \left(\imath \sigma_1 -
  \sigma_2 \right)} \mathrm{e}^{\imath \tau \sigma_1} \mathrm{e}^{\alpha \vartheta
  \sigma_3}\quad,
\end{equation}
where the constant $\alpha$ was related to the mass and angular momentum of
the black hole through $M=\frac{1 + \alpha^2}{2}$ and
$J=\frac{1-\alpha^2}{2}$. Notice that, for $|\alpha| < 1$, this defines a
global coordinate system on $AdS_3$, with $\alpha = 0$ corresponding to the
usual Iwasawa decomposition of $\SL$.  With this choice of coordinates, the
identification Killing vector of the generic (non-extremal) rotating black
hole, $\xi = \alpha R_2 - L_2$, simply reads as $\xi = \partial_\vartheta$.
The coordinate transformation between this parametrization and (\ref{sphan})
is
\begin{subequations}
  \begin{align}
    e^{2 \psi} &= \mathrm{e}^{2 \left( \alpha +1 \right)\vartheta} \left( 1
      - u \tan \tau\right) \quad , \\
    e^{2\varphi} &= \mathrm{e}^{2 \left( 1 - \alpha \right)\vartheta} \left(
      1 + u \cot \tau \right) \quad,\\
    \cos^2 t &= \cos^2 \tau \left(1- u \tan \tau \right)\quad.
  \end{align}
\end{subequations}

\cleardoublepage \chapter{Symmetric spaces}\label{AppSymm}

The terminology \emph{symmetric space}\index{symmetric space} is used in various contexts, and embodies a large class of spaces. We will here try to collect different definitions encountered in the literature and analyze to what extent they overlap.
The theorems and results will mostly be stated without proof.
For a general account of symmetric spaces, the patient reader is referred to \cite{Helgason,Loos,Kobayashi:1969}. Affine and Riemannian symmetric spaces are treated in \cite{Helgason}. Symplectic symmetric spaces are presented in \cite{SSSS, ThesePierre}. Useful shortcuts include \cite{Laurent,LocSymmSp,Caselle:2003qa}.

Let us start with a general definition \cite{Loos}. A \emph{symmetric space} is a manifold $M$ and an analytic ``multiplication" $\mu : M \times M \ra M$, written $s_x(y) \eqdef \mu(x,y)$, such that
\begin{enumerate}
{\item $\forall x\in M$, $s_x$ is an involutive diffeomorphism\footnote{i.e. $(s_x)^2 = Id$} of $M$ called ``the \emph{symmetry}\index{symmetry} at $x$",}

{\item $\forall x\in M$, $x$ is an isolated fixed point of $s_x$,}
{\item $\forall x,y\in M$, $s_x \circ s_y \circ s_x = s_{s_x(y)}$.}
\end{enumerate} 

A homomorphism of symmetric spaces $(M,s)$ and $(M',s')$ is an analytic map $\varphi : M\ra M'$ satisfying
\begin{equation}
 \varphi (s_x(y)) = s'_{\varphi(x)}(\varphi(y)).
\end{equation}
\footpourtoi{Why automorphism? Why normal subgroup; Laurent p183}
\footpourtoi{transvection = displacement?}
For all $x\in M$, the symmetry $s_x$ is an automorphism of $M$ (as symmetric space).The group generated by all the $s_x \circ s_y$, $x,y\in M$, is the \emph{displacement group}\index{displacement group} or \emph{transvection group}\index{transvection group}, denoted $G(M)$. The displacement group is a normal subgroup of $Aut (M)$.

As a simple example, let $L$ be a Lie group endowed with the structure
\begin{equation}
s_x y = x y^{-1} x.
\end{equation}
One may check that $\forall x,y \in L$, $(s_x \circ s_x)(y) = y$ and $s_x \circ s_y \circ s_x = s_{s_x(y)}$. The last condition may be verified as follows. First note that $x (s_y(z)) = (xy) (xz)^{-1} (xy) = s_{xy}(xz)$, so that it is sufficient to check the property on $s_e$ because the left translation is analytic. Since $s_e(y) = y^{-1}$, the property stems from the fact that $e$ is an isolated fixed point for the inversion in a topological group.

\section{Affine and Riemannian symmetric spaces}\label{AppAffRiem}
We first recall some elementary definitions of differential geometry. 

An \emph{affine connection}\index{affine connection} on a manifold $M$ is a rule $\nabla$ which assigns a linear mapping $\nabla_X : \varkappa (M) \ra \varkappa (M)$ to each vector field $X \in \varkappa (M)$ satisfying the following two conditions: 
\begin{itemize}
\item $\nabla_{f X+ g Y} = f \nabla_X + g \nabla_Y$ ;
\item $\nabla_X(f Y) = f \nabla_X Y + (X f)Y$, 
\end{itemize}
$\forall f,g \in C^\infty(M)$, $X,Y \in \varkappa(M)$. The operator $\nabla_X$ is called the \emph{covariant derivative}\index{covariant derivative} in the direction of (or with respect to) $X$.

Let $c_V : I \subset \R \ra M$ be a curve in $M$, such that $c_V(0) = x \in M$. The tangent vector $V$ of $c_V(t)$ at $x$ can be seen as the first order differential operator acting on functions defined in a neighborhood of $x$ as 
\begin{equation}\label{Vect}
 \dot{c_V}(0)f \eqdef V(0)f= (f \circ c_V)'(0) = \ddto [f(c_V(t))]. 
\end{equation}
If $(U,\phi)$ is a local chart around $x$, we may rewrite \re{Vect} as
\begin{equation}\label{Vect2}
 V(0)f = \ddto [\tilde{f} \circ \widetilde{c_V}] = \frac{d \widetilde{c_V}^\a}{dt}_{|t=0} \frac{\p \tilde{f}}{\p x^\a}_{|x^\a = \widetilde{c_V}^\a (0)} ,
\end{equation}
where $\widetilde{c_V} \eqdef \phi \circ c_V$ is a curve in $\R^n$ ($dim M=n$), represented by $\widetilde{c_V}^\a (t)$, $\tilde{f} = f \circ \phi^{-1} \in Fun (\R^n)$, and $\phi(x) = (x^\a) \in \R^n$. Eq. \re{Vect2} is the precise meaning of the notation
\begin{equation}
 V_x = V^\a (x) \p_\a
\end{equation}
for the tangent vector of $c_V$ at $x$. 

Let $X$ now be a vector field defined (at least) along the curve $c_V (t)$:
\begin{equation}
 X_{|c_V(t)} = X^\mu (c_V (t)) \p_\mu.
\end{equation}
If $X$ satisfies the condition 
\begin{equation}
 \nabla_V X =0 \quad , \quad \forall t\in \R,
\end{equation}
then $X$ is said to be \emph{parallel transported}\index{parallel transport} along $c_V (t)$.

If the tangent vector $V$ itself is parallel transported along $c_V (t)$, namely if
\begin{equation}
 \nabla_V V = 0 ,
\end{equation}
then the curve $c_V(t)$ is called a \emph{geodesic}\index{geodesic}. 

Let us state the following proposition \cite{Helgason}. Let $M$ be a differentiable manifold with an affine connection. Let $x$ be any point in $M$ and let $V \in T_x M$, with $V \ne 0$. There then exists a unique maximal geodesic\footnote{A geodesic is called \emph{maximal}\index{maximal geodesic} if it is not a proper restriction of any geodesic.} $t \ra c(t)$ in $M$ such that  
\begin{equation}
 c(0)= x \quad , \quad \dot{c}(0) = V.
\end{equation}
This geodesic is obviously denoted by $c_V$. If $V=0$, we put $c_V (t) = x$, $\forall t\in \R$. We then have the following theorem \cite{Helgason}. For all $x \in M$, there exists an open neighborhood $N_0$ of $O$ in $T_x M$ and an open neighborhood $N_x$ of $x$ in $M$ such that the mapping 
\begin{equation}
 N_0 \ra N_x : V \ra c_V(1)
\end{equation}
is a diffeomorphism of $N_0$ onto $N_x$.
The mapping $V \ra c_V(1) \eqdef \mbox{Exp}_x (V)$ is called the \emph{exponential mapping}\index{exponential mapping} at $x$.
In general, the exponential mapping is only locally defined, that is, it only takes a small neighborhood of the origin at $T_x M$, to a neighborhood of $x$ in the manifold.

The exponential mapping is maybe more familiar from Lie theory, where it represents a map from the Lie algebra of a Lie group to the group, completely allowing us to recapture the local group structure from the Lie algebra. The existence of the exponential map is one of the primary justifications for the study of Lie groups at the level of Lie algebras. Explicitly, if $G$ is a Lie group and $\mathfrak{g}$ its Lie algebra, identified with $T_e G$, the exponential mapping is

\begin{equation}
 \exp : \mathfrak{g}\ra G ,
\end{equation}
with $\exp(X) = \gamma(1)$, where $\gamma : \R \ra G$ is the unique one-parameter subgroup of $G$ whose tangent vector at the identity is equal to $X$. It follows easily from the chain rule that $\exp(tX) = \gamma(t)$. The map $\gamma$ may be constructed as the integral curve of either the right- or left-invariant vector field associated with X. That the integral curve exists for all real parameters follows by right- or left-translating the solution near zero. If G is a matrix Lie group, then the exponential map coincides with the matrix exponential and is given by the ordinary series expansion.

The two notions of the exponential mapping coincide in the case of Lie groups equipped with bi-invariant metrics (i.e. Riemannian metrics invariant under left and right translation). In this case the geodesics through the identity are precisely the one-parameter subgroups of G.

An open neighborhood $N_0$ of the origin in $T_x M$ is said to be \emph{normal}\index{normal neighborhood} if : (1) the mapping $\mbox{Exp}$ is a diffeomorphism of $N_0$ onto an open neighborhood $N_x$ of $x$ in M; (2) if $X\in N_0$ and $0\leq t \leq 1$, then $t X\in N_0$.
The last condition means that $N_0$ is star-shaped. A neighborhood $N_x$ of $x$ in $M$ is called a \emph{normal neighborhood} of $x$ in $M$ if $N_x =\mbox{Exp} N_0$ where $N_0$ is a normal neighborhood of $0$ in $T_x M$. Assuming this to be the case, and letting $X_1, \cdots, X_n$ denote some basis of $T_xM$, the inverse mapping
\begin{equation}
 \mbox{Exp}_x (a_1 X_1+\cdots + a_n X_n) \ra (a_1,\cdots,a_n)
\end{equation}
of $N_x$ into $\R^n$ is called a system of \emph{normal coordinates}\index{normal coordinates} at $x$.

We now come to the definition of an \emph{affine symmetric space}\index{affine symmetric space}. Let $M$ be a smooth manifold with an affine connection $\nabla$. Let $x\in M$ and $N_0$ be a normal neighborhood of the origin $O\in T_x M$, symmetric with respect to $0$. As usual, we put $N_x = \mbox{Exp}_x N_0$. For each $y \in N_x$, consider the geodesic $t\ra c(t)$ within $N_x$ passing through $x$ and $y$ and such that $c(0)=x$, $c(1)=y$. 
\footpourtoi{$s_x$ satisfait toutes les conditions pour espace sym.? voir Laurent p188}
We put $y'=c(-1) \eqdef s_x(y)$. The mapping $s_x : N_x \ra N_x : y \ra s_x(y) = y'$ is called the \emph{geodesic symmetry}\index{geodesic symmetry} with respect to $x$. In normal coordinates $\{x_1,\cdots,x_n\}$ at $x$, $s_x$ has the expression $\{x_1,\cdots,x_n\} \ra \{-x_1,\cdots,-x_n\}$. In particular, $s_x$ is a diffeomorphism of $N_x$ onto itself and $(d s_x)_x = -Id_x$.

$M$ is called \emph{affine locally symmetric}\index{affine locally symmetric} if each point $x \in M$ has an open neighborhood $N_x$ on which the geodesic symmetry is an affine transformation\footnote{Suppose that $\nabla$ is an affine connection on $M$ and that $\phi$ is a diffeomorphism of $M$. A new affine connection $\nabla'$ can be defined on $M$ by $\left( \nabla'_X Y \right)_x \eqdef \phi^{-1}_{x * \phi(x)} (\nabla_{\phi_{\phi(x)*x}X}\phi_{\phi(x)*x}Y)$, or in short $\left( \nabla'_X Y \right)_x \eqdef \phi^{-1}_* (\nabla_{\phi_* X}\phi_*Y)$, $X,Y\in \varkappa(M)$. $\nabla'$ indeed defines an affine connection \cite{Helgason}. The affine connection $\nabla$ is called \emph{invariant}\index{invariant connection} under $\phi$ if $\nabla' = \nabla$. In this case $\phi$ is called an \emph{affine transformation}\index{affine transformation} of $M$.}.

We have the following theorem \cite{Helgason}. A manifold $M$ is affine locally symmetric iff $T=0$ and $\nabla_Z R=0$ for all $Z\in \varkappa(M)$, where $T$ and $R$ represent the torsion and curvature tensors of the affine connection.
A Riemannian manifold $M$ is called a \emph{Riemannian locally symmetric space}\index{Riemannian locally symmetric space} if for each $x \in M$, there exists a normal neighborhood of $x$ on which the geodesic symmetry with respect to $x$ is an isometry.
\footpourtoi{je vois pas trop extension globale..?}
A space is said to be an \emph{affine symmetric space}\index{affine symmetric space} if it is affine locally symmetric and if for all $x \in M$, the symmetry can be globally extended to an affine transformation of $M$.
For the definition of \emph{Riemannian symmetric space}\index{Riemannian symmetric space}, one has to replace affine by Riemannian and affine transformation by isometry in the preceding definition \cite{LocSymmSp}.
One can show that an affine (Riemannian) symmetric space is complete\footnote{The completeness of an affine (Riemannian) manifold is equivalent to the fact that each maximal geodesic in $M$ has the form $c_V (t)$, $-\infty < t < +\infty$, i.e. "has infinite length", \cite{Helgason}, p56.}.

An important property is that the group $\mbox{Aff}(M)$ of affine transformations of an affine symmetric space is transitive on $M$\cite{Kobayashi:1969}. One can moreover prove that $\mbox{Aff}(M)$ is a Lie group. We denote its identity component by $G$. Since $G$ clearly acts transitively on $M$, $M$ can be represented is a homogeneous space $G/H$ (see definition Sect.~\ref{SectLTG}).
\footpourtoi{lien entre cette def et Th.2.4PierreStrict: l\`a, $M=G/K$, mais $G$= {\bf smallest} subset of $\mbox{Aff}(M)$.?? {\bf check this theorem!} }
More precisely, we have the following theorem. Let $G$ be the largest connected group of affine transformations of an affine symmetric space $M$, and $H$ the isotropy group of a fixed point $o \in  M$, so that $M=G/H$ as a homogeneous space. Let $s_o$ be the symmetry of $M$ at $o$, and $\tilde{\sigma}$ the involutive automorphism of $G$ defined by
\begin{equation}
  \tilde{\sigma}(g) = s_o \circ g \circ s_o.
\end{equation}
Let $G_{\tilde{\sigma}}$ be the closed subgroup of $G$ which fixes $\tilde{\sigma}$. Then $G_{\tilde{\sigma}}^0 \subset H \subset G_{\tilde{\sigma}}$, where $G_{\tilde{\sigma}}^0$ is the neutral connected component of $G_{\tilde{\sigma}}$.

\section{Symmetric spaces as special homogeneous spaces}\label{SectD2}
The last theorem of the previous section leads to an interesting statement: it establishes that every affine symmetric space can be identified with a homogeneous space $G/H = \left\{[g] = \{g h \, | \, h \in H\}, g \in G \right\}$ \cite{Barut}, endowed with a particular involution $\tilde{\sigma}$ of $G$. 
An affine symmetric space can therefore be described by a triple $(G,H,\tilde{\sigma})$, where
\begin{itemize}
\item $G$ is a connected Lie group;
\item $H$ is a closed subgroup of $G$; 
\item $\tilde{\sigma}$ is an involutive automorphism of $G$ such that $G_{\tilde{\sigma}}^0 \subset H \subset G_{\tilde{\sigma}}$,
\end{itemize}
where $G_{\tilde{\sigma}} = \{ g\in G | \tilde{\sigma}(g)=g \}$. Symmetric spaces can be viewed as homogeneous spaces, thus of the form $G/H$, where the subgroup $H$ is fixed by an involutive automorphism.

A \emph{symmetric Lie algebra}\index{symmetric Lie algebra} is a triple $(\mathfrak{g},\mathfrak{h},\sigma)$ with 
\begin{itemize}
\item $\mathfrak{g}$ a Lie algebra;
\item $\mathfrak{h}$ a Lie subalgebra of $\mathfrak{g}$; 
\item $\sigma$ an involutive automorphism of $\mathfrak{g}$ whose set of fixed points is $\mathfrak{h}$.
\end{itemize}
All symmetric spaces $(G,H,\tilde{\sigma})$ give rise to a symmetric Lie algebra $(\mathfrak{g},\mathfrak{h},\sigma)$, where $(\mathfrak{g}$ and $\mathfrak{h}$ are the Lie algebras of $G$ and $H$, while $\sigma = d\tilde{\sigma}_e$). The converse is true under certain analyticity hypothesis. 

Let $(\mathfrak{g},\mathfrak{h},\sigma)$ be a symmetric algebra. As linear transformation of $\mathfrak{g}$, $\sigma$ has eigenvalues $+1$ and $-1$ (because it is involutive), so that it induces a decomposition
\begin{equation}
   \mathfrak{g} = \mathfrak{h} \oplus \mathfrak{q},
\end{equation}
where $\mathfrak{h}$ is the $+1$ eigenspace and $\mathfrak{q}$ the $-1$ eigenspace. 
This is the \emph{canonical decomposition}\index{canonical decomposition}. 
\footpourtoi{donc sym\'etrique est plus fort que r\'eductif? NON (et que goldorak? P-E)}
This decomposition fulfils
\begin{equation}\label{reduct}
[\mathfrak{h},\mathfrak{h}] \subset \mathfrak{h}, \quad  [\mathfrak{h},\mathfrak{q}] \subset \mathfrak{q}, \quad [\mathfrak{q},\mathfrak{q}] \subset \mathfrak{h}.
\end{equation}

The associated homogeneous space, if it exists, is then automatically reductive\footnote{The homogeneous space $G/H$ is said to be \emph{reductive}\index{reductive} if we can find a subspace $\mathfrak{q}$ of $\mathfrak{g}$ such that (1)$\mathfrak{g}= \mathfrak{h} \oplus \mathfrak{q}$ and (2) $[\mathfrak{h},\mathfrak{q}] \subset \mathfrak{q}$. Because of the second condition, $\mathfrak{q}$ is said $H-$invariant. }. On the other hand, if we have a Lie algebra $\mathfrak{g}$ and a direct decomposition $\mathfrak{g}= \mathfrak{h} \oplus \mathfrak{q}$ fulfilling \re{reduct}, then the definition $\sigma = Id_{\mathfrak{h}} \oplus -Id_{\mathfrak{q}}$ gives rise to a symmetric Lie algebra $(\mathfrak{g},\mathfrak{h},\sigma)$ (indeed condition \re{reduct} ensures that $\sigma$ be an automorphism, see \cite{Laurent}, p184).

An homogeneous space is symmetric iff it is reductive.

We will use the following properties (see \cite{Kobayashi:1969}, Vol.2, pp230-233), especially the third one, in Chapter~\ref{ChapBHAdSl}.

With respect to the canonical connection of a symmetric space
$(G,H,\sigma)$, the homogeneous space $M=G/H$ is a (complete) affine
symmetric space with symmetries $s_x$ and possesses the following
properties :
\begin{enumerate}
\item $T=0$, $\nabla R=0$ and $R(X,Y)Z=[ [X,Y],Z]$ for $X,Y,Z\in
\mathfrak{q}$ where $\mathfrak{q}$ is identified with $T_oM$, $o$ being
the origin of $M$;
\item for each $X\in\mathfrak{q}$, the parallel displacement along
$\pi(\exp tX)$ coincides with the differential of the transformation
$\exp tX$ on $M$;
\item For each $X\in\mathfrak{q}$, $\pi(\exp tX)=(\exp tX)\cdot o$ is a
geodesic starting from $o$ and conversely, every geodesic from $o$ is of
this form.
\end{enumerate}

Let us end by introducing another commonly used decomposition. Let $B$ be the Killing form of $\mathfrak{g}$. An involution $\theta : \mathfrak{g} \ra \mathfrak{g}$ such that the quadratic form $B_\theta$ defined by 
\begin{equation}
 B_\theta (X,Y) := -B(X,\theta Y)
\end{equation}
is positive definite is called a \emph{Cartan involution}\index{Cartan involution}.

Let $\mathfrak{g}$ be a real semi-simple Lie algebra, $\theta$ a Cartan involution and $\sigma$ another involution. There then exists an inner automorphism of $\mathfrak{g}$, $\varphi \in Int(\mathfrak{g}) \subset Aut(\mathfrak{g})$ such that $[\varphi \theta \varphi^{-1}, \sigma]=0$ (see \cite{Laurent},p91). Also, any two Cartan involutions of a real semi-simple Lie algebra are conjugate by an inner automorphism.

A vector space decomposition 
\begin{equation}
 \mathfrak{g} = \mathfrak{k} \oplus \mathfrak{p}
\end{equation}
is a \emph{Cartan decomposition}\index{Cartan decomposition} if the Killing form is negative definite on $\mathfrak{k}$ and positive definite on $\mathfrak{p}$, and if the following relationships hold:
\begin{equation}
 [\mathfrak{k},\mathfrak{k}]\subseteq \mathfrak{k} , \quad   [\mathfrak{k},\mathfrak{p}]\subseteq \mathfrak{p}, \quad [\mathfrak{p},\mathfrak{p}] \subseteq \mathfrak{k}    .
\end{equation}
If $\theta$ is a Cartan involution, then $\mathfrak{k}$ ($\mathfrak{p}$) is its $+1$ ($-1$) eigenspace. Let $\mathfrak{a}$ be a maximal Abelian subalgebra of $\mathfrak{p}$. The dimension of $\mathfrak{a}$ is called the \emph{real rank}\index{real rank} of $\mathfrak{g}$. For $\lambda \in \mathfrak{a}^*$, we define the space 
\begin{equation}\label{RootGen}
 \mathfrak{g}_\lambda = \{ X \in \mathfrak{g} | \forall H \in \mathfrak{a}, [H,X]=\lambda(H) X \}.
\end{equation}
If $\mathfrak{g}_\lambda \ne 0 $ and $\lambda \ne 0$, $\lambda$ is called a \emph{restricted root}\index{restricted root} of $\mathfrak{g}$. We denote by 
$\Phi$ the set of all restricted roots of $\mathfrak{g}$. Let us consider a notion of positivity\footnote{Let $V$ be a vector space. A \emph{positivity notion}\index{positivity notion} is the data of a subset $V^+ \subset V$ such that (i) $\forall v \ne 0 \in V, v \in V^+$ or $-v \in V^+$; (ii) $\forall v,w \in V^+$ and $\forall \mu \in \R^+_0$, $v+w \in V^+$ and $\mu v \in V^+$.  If $v\in V^+$, $v$ is said positive.} on $\mathfrak{a}^*$,  and denote by $\Phi^+$ the set of positive roots. Let us define
\begin{equation}
 \mathfrak{n} = \oplus_{\lambda \in \Phi^+} \mathfrak{g}_\lambda ,
\end{equation}
and 
\begin{equation} \label{nbarre}
 \bar{\mathfrak{n}} = \oplus_{\lambda \in \Phi^-} \mathfrak{g}_\lambda .
\end{equation}
We then have the \emph{Iwasawa decomposition}\index{Iwasawa decomposition}\cite{Helgason}
\begin{equation}
 \mathfrak{g} = \mathfrak{a} \oplus \mathfrak{n} \oplus \mathfrak{k} ,
\end{equation}
which induces a global diffeomorphism
\begin{equation}
 \phi : \ca \times \cn \times \ck \ra G :(a,n,k)\ra \phi(a,n,k) = a n k,
\end{equation}
where $\ca$, $\cn$ and $\ck$ are the connected analytic subgroups of $G$ whose Lie algebras are $\mathfrak{a}$, $\mathfrak{n}$ and $\mathfrak{k}$ respectively.
We have the property that the Lie algebra $\mathfrak{n}$ is nilpotent, while $\mathfrak{a} \oplus \mathfrak{n}$ is a solvable\footnote{Let $\mathfrak{g}$ be a Lie algebra. The descending central series $C^k(\mathfrak{g})$ is defined recursively by $C^1(\mathfrak{g})=\mathfrak{g}$, $C^k(\mathfrak{g})=[C^{k-1}(\mathfrak{g}),\mathfrak{g}]$. We also define the derived series $D^k(\mathfrak{g})$ through $D^1(\mathfrak{g})=[\mathfrak{g},\mathfrak{g}]$, $D^k(\mathfrak{g})=[D^{k-1}(\mathfrak{g}),D^{k-1}(\mathfrak{g})]$. The Lie algebra $\mathfrak{g}$ is said \emph{nilpotent}\index{nilpotent} if there exists $n\geq 1$ such that $C^n(\mathfrak{g})=\{0\}$. The Lie algebra $\mathfrak{g}$ is said \emph{solvable}\index{solvable} if there exists $n\geq 1$ such that $D^n(\mathfrak{g})=\{0\}$.} Lie subalgebra of $\mathfrak{g}$. $AN\,$ is a \emph{solvable subgroup}\index{solvable subgroup}, or \emph{minimal parabolic subgroup}\index{minimal parabolic subgroup} of $G$, and is sometimes called the \emph{Iwasawa group}\index{Iwasawa group}.

The Iwasawa decomposition induces a global diffeomorphism between the group manifold $R = AN\,$ and the symmetric space of the non-compact type\footnote{A symmetric space $G/H$ is said of the \emph{non-compact type}\index{symmetric space of the non-compact type} when $G$ is non-compact, and $H=\{g \in G |\theta(g) = g\}$, where $\theta$ is a Cartan involution. In this situation, $H$ is a maximal compact subgroup $K$ of $G$.} $G/K$:
\begin{equation}\label{RGsurK}
 R \ra G/K : (a,n) \ra an K.
\end{equation}
The latter map allows us naturally to identify $\mathfrak{p}$ with the tangent space
$T_{[e]}M = T_K M$.
\begin{equation}
\end{equation}



\section{Symplectic symmetric spaces}
This section is entirely taken from \cite{PierreStrict}, see also \cite{ThesePierre}. Some definitions are redundant with the previous sections, making it self-contained.

Let $(M,\omega,\nabla)$ be a $2n$-dimensional \emph{affine symplectic 
manifold}\index{affine symplectic 
manifold}, that is, $(M, \omega)$ is a smooth connected symplectic manifold 
and $\nabla$ is a torsion-free affine  connection on $M$ such that 
$\nabla\omega=0$. Its {\bf automorphism group} $\mbox{Aut}(M,\omega,\nabla)$ 
is defined as  
$$
\mbox{Aut}(M,\omega,\nabla)=\mbox{Aff}(\nabla)\cap\mbox{Symp}(\omega)
$$
where $\mbox{Aff}(\nabla)$ is the group of affine transformations of 
the affine manifold $(M,\nabla)$ and where $\mbox{Symp}(\omega) $
denotes the group of symplectomorphisms of $(M,\omega)$. Note that, 
since $\mbox{Aff}(\nabla)$ is a Lie group of transformations of $M$ 
(cf.~\cite{Kobayashi:1969}), so is $\mbox{Aut}(M,\omega,\nabla)$.\\

A \emph{symplectic symmetric space}\index{symplectic symmetric space} 
is a triple $(M, \omega,s)$, where $(M,\omega)$ is a smooth connected 
symplectic manifold, and where $s : M \times M \to M$ is a smooth map such 
that 
  
\begin{enumerate} 
\item[(i)] for all $x$ in $M$, the partial map $s_x : M \to M : y \mapsto 
s_x (y) := s(x,y)$ is an involutive symplectic diffeomorphism of $(M,\omega)$ 
called the {\bf symmetry} at $x$. 
\item[(ii)] For all $x$ in $M$, $x$ is an isolated fixed point of $s_x$. 
\item[(iii)] For all $x$ and $y$ in $M$, one has $s_x \circ s_y \circ s_x=s_{s_x (y)}$. 
\end{enumerate}

Two symplectic symmetric spaces  $(M,\omega,s)$ 
and $(M',\omega ',s')$ are isomorphic if there exists a symplectic 
diffeomorphism $\varphi: (M,\omega) \rightarrow (M',\omega')$ such that 
$\varphi s_x=s'_{\varphi (x)} \varphi$. Such a $\varphi $ is called an {\bf
isomorphism} of $(M,\omega,s)$ onto $(M',\omega',s')$. When 
$(M,\omega,s) = (M',\omega',s')$, one talks about \emph{automorphisms}\index{automorphism}. 
The group of all automorphisms of the symplectic symmetric space 
$(M,\omega,s)$ is denoted by $Aut(M,\omega,s)$.

On a symplectic symmetric space $(M,\omega,s)$, there exists 
one and only one affine connection $\nabla$ which is invariant under the 
symmetries. Moreover, this connection satisfies the following properties: 

\begin{enumerate} 
\item[(i)] For all smooth tangent vector fields $X,Y,Z$  on $M$ and all 
points $x$ in $M$, one has $$ \omega_{x} (\nabla _X Y,Z)=\frac{1}{2}
X_x.\omega (Y+s_{x_{\star}}Y,Z).$$ \item[(ii)] $(M,\nabla )$ is an affine 
symmetric space. In particular $\nabla$ is torsion-free and its curvature 
tensor is parallel.  
\item[(iii)] The symplectic form $\omega$ is parallel; $\nabla$ is therefore 
a symplectic connection.   
\item[(iv)] One has
$$\mbox{Aut}(M,\omega,s)=\mbox{Aut}(M,\omega,\nabla)=\mbox{Aff}(\nabla)\cap\mbox{Symp}(\omega).$$
\end {enumerate} 

The connection $\nabla$ on the symmetric space $(M,s)$ is called the \emph{Loos connection}. The following facts are taken from \cite{Loos},V.1,\cite{Kobayashi:1969},v.~II, Chapters X and XI. 

Let $(M,\omega,s)$ be a symplectic symmetric space and $\nabla$ its 
Loos connection. Fix $o$ in $M$ and denote by $H$ the stabilizer of 
$o$ in $Aut(M,\omega,s)$. Denote by $G$ the {\bf transvection group} of 
$(M,s)$ (i.e.~the subgroup of $Aut(M,\omega,s)$ generated by $\{ s_x \circ 
s_y \, ; \, x,y \in M \}$)  and set $K=G \cap H$. Then, 

\footpourtoi{Trucs \`a piger ou eliminer dans les pts (iii) et (iv) !!}
\begin{enumerate} 
\item[(i)] the transvection group $G$ turns out to be a connected Lie transformation 
group of $M$. It is the smallest subgroup of $Aut(M,\omega,s)$ which is transitive 
on $M$ and stabilized by the conjugation $\tilde{\sigma} : 
Aut(M,\omega,s) \to Aut(M,\omega,s)$ defined by $\tilde{\sigma}(g)=s_o \circ g \circ s_o$.
\item[(ii)] The homogeneous space $M=G/_{\textstyle{K}}$ is reductive. The 
Loos connection $\nabla$ coincides with the canonical connection induced 
by the structure of reductive homogeneous space. 
\item[(iii)] Denoting by $G^{\tilde{\sigma}}$ the set of $\tilde{\sigma}$-fixed 
points in $G$ and by $G_0^{\tilde{\sigma}}$ its neutral connected component, 
one has $$ G_0^{\tilde{\sigma}} \subset K \subset G^{\tilde{\sigma}}.$$ 
Therefore, the Lie algebra $\mathfrak{k}$ of $K$ is isomorphic to the holonomy algebra 
with respect to the canonical connection $\nabla$.
\item[(iv)] Denote by $\sigma$ the involutive automorphism of the Lie 
algebra $\mathfrak{g}$ of $G$ induced by the automorphism $\tilde{\sigma}$. Denote 
by $\mathfrak{g} = \mathfrak{k} \oplus \mathfrak{p}$ the decomposition in $\pm 1$-eigenspaces for $\sigma$. 
Then, identifying $\mathfrak{p}$ with $T_o(M)$, one has $$ exp (X) = 
s_{Exp_o(\frac{1}{2} X)} \circ s_o $$ for all $X$ in a neighborhood of 
$0$ in $\mathfrak{p}$. Here $exp$ is the exponential map $\exp : \mathfrak{g} \to G$ and 
$\mbox{Exp}_o$ is the exponential map at point $o$ with respect to the connection 
$\nabla$.   

\end{enumerate}

\cleardoublepage \chapter{Symplectic and Poisson geometry}\label{PoissSympl}

This Appendix is essentially based on \cite{CannasWein,WeinsteinPoisson,WeinsteinMorita,CoursGutt,marle}\footnote{see also various notes at \url{http://math.berkeley.edu/~alanw/}}.
\footpourtoi{Lie bracket implique antisymetrie et Jacobi!}

\section{Definitions}

Let $M$ be a smooth manifold. A \emph{Poisson structure}\index{Poisson structure}, or \emph{Poisson bracket}\index{Poisson bracket} on $M$, is a $\R-$ bilinear Lie bracket $\{.,.\}$ satisfying the Leibniz rule
\begin{equation}
 \{f,gh\} = \{f,g\}h + g\{h,f\} \quad , \quad \forall f,g,h \in C^\infty(M).
\end{equation}
A \emph{Poisson algebra}\index{Poisson algebra} is a vector space $V$ endowed with two products: a commutative product 
\begin{equation}
 .\quad : (f,g) \ra f.\,\,g,
\end{equation}
that makes $V$ a commutative algebra, and a Poisson bracket
\begin{equation}
 \{.,.\} : (f,g) \ra \{f,g\}.
\end{equation}
A Poisson algebra is thus in particular a Lie algebra.
A \emph{Poisson manifold}\index{Poisson manifold} is a manifold $M$, whose algebra of smooth functions $C^\infty(M)$ is a Poisson algebra with the pointwise multiplication as commutative product. 

For a function $f\in C^\infty(M)$, we may define the \emph{Hamiltonian vector field}\index{Hamiltonian vector field} $X_f \in \varkappa(M)$ of $f$ by
\begin{equation}
 X_f =\{f,.\} \qquad .
\end{equation}
$\varkappa(M)$ denotes the set of vector fields on $M$ \index{$\varkappa(M)$}.
Since the Poisson bracket satisfies the Leibniz rule in every component, it has to come from a bivector field $\Pi \in \Lambda^2 TM$ such that 
\begin{equation}\label{PoissonBiVector}
 \{f,g\} = \Pi(df,dg).
\end{equation}
In local coordinates $\{x^i\}$, a bivector field $\Pi$ is represented by functions $\Pi^{ij}(x)$, called the \emph{structure functions}\index{structure functions}, given by 
\begin{equation}
\Pi^{ij}(x) =  \{x^i,x^j\} ,
\end{equation}
such that the Poisson bracket of two functions in coordinate notation is
\begin{equation}
 \{f,g\} = \Pi^{ij}(x) \p_i f \p_j g.
\end{equation}
The Jacobi identity for the Poisson bracket can be expressed in terms of $\Pi$ as
\begin{equation}\label{F7}
\sum \Pi^{ir}\p_r \Pi^{jk} = 0,
\end{equation}
 where the sum is taken over the cyclic permutations of $i,j,k$. 
If the matrix $\Pi^{ij}$ is invertible at each $x\in M$, then $\Pi$ is called nondegenerate or \emph{symplectic}\index{symplectic bivector field}. In this case, the local matrices $\omega_{ij}$ such that $\omega_{ij} \Pi^{jk} = -\delta^k_j$ define a global 2-form $\omega \in \Omega^2 (M) = \Gamma (\Lambda^2 T^*M)$. This form is such that $\omega_x$ is nondegenerate for each $x\in M$, i.e. if $x\in M$ and $\omega_x (v,z) =0$ for all $z\in T_x M$, then $v=0$. Furthermore, the condition \re{F7} is equivalent to $d\omega = 0$.

This leads us to the following definition. A \emph{symplectic manifold}\index{symplectic manifold} is a pair $(M,\omega)$ consisting in a smooth manifold $M$ and a closed two-form $\omega$ on $M$ such that $\omega_x$ is nondegenerate for each each $x\in M$. The two-form is called a \emph{symplectic structure}\index{symplectic structure} on $M$. A symplectic manifold is thus a Poisson manifold whose bivector field is symplectic. In the sequel, we will mostly restrict ourselves to symplectic manifolds. We would like therefore re-express the above quantities in terms of $\omega$ instead of $\Pi$. If $(M,\omega)$ is a symplectic manifold, the Hamiltonian vector field $X_f \in \varkappa(M)$ \index{Hamiltonian vector field} associated with a function $f \in C^\infty (M)$ can be expressed as
\begin{equation}
 i(X_f) \omega = -df \quad,\quad \mbox{i.e.}\; \omega_x(X_f,v)=-v(f)= -df(v) \; \forall v\in T_xM. 
\end{equation}
This relation defines $X_f$ since $\omega_x$ is nondegenerate for all $x\in M$. If $\omega$ is written in local coordinates $\{p_i,q^i\}$, $i=1,\cdots,n$, $dim(M)=2n$ as
\begin{equation}
 \omega = \sum_{i=1}^n dp_i \wedge dq^i ,
\end{equation}
then 
\begin{equation}
 X_f = \sum_{i=1}^n (\frac{\p f}{\p p_i}\frac{\p}{\p q^i} - \frac{\p f}{\p q^i}\frac{\p}{\p p_i}).
\end{equation}
The Poisson bracket on $M$ is then given by
\begin{equation}\label{LieOmega}
 \{f,g\} = \omega(X_f,X_g) = X_f(g) = -X_g(f).
\end{equation}
We have the following isomorphism of Lie structures:
\begin{equation}
 X_{\{f,g\}} = [X_f,X_g],
\end{equation}
which follows from the fact that $i([X,Y])\a = {\cal L}_X i(Y)\a -  i(Y) {\cal L}_X \a$ for all p-forms $\a$ and for all vector fields $X$ and $Y$, where ${\cal L}$ denotes the Lie derivative.

Let us mention the \emph{Darboux theorem}\index{Darboux theorem}. Let $(M,\omega)$ be a symplectic manifold, and let $x \in M$. There then exists an open neighborhood $U$ of $x$ in $M$ and local coordinates $q^1, \cdots,q^k,p_1,\cdots,p_k$ on $U$ so that 
\begin{equation}
\omega_{|U} = \sum_{i=1}^k dp_i \wedge dq^i.
\end{equation} 
In particular, $dim M = 2k$, so the manifold is even-dimensional. The coordinates $\{q^1, \cdots,q^k,p_1,\cdots,p_k\}$ are called \emph{canonical coordinates}\index{canonical coordinates}.

\section{Lie transformation groups}\label{SectLTG}
Let $M$ be a smooth manifold and $G$ be a Lie group. The group $G$ is called a \emph{Lie transformation group}\index{Lie transformation group} of $M$ if to each $g\in G$ is associated a diffeomorphism of $M$ onto itself such that
\begin{enumerate}
\item $g_1 g_2. x = g_1. (g_2. x)$ for all $x\in M$ and for all $g_1, g_2 \in G$
\item the map $G\times M \ra M : (g,x) \ra g. x$ is $C^\infty$.
\end{enumerate}
One says that $G$ \emph{acts}\index{group action} on $M$. The \emph{action} of $G$ is called \emph{effective}\index{effective} if $e\in G$ is the only element of $G$ which leaves each $x \in M$ fixed. The action is said to be \emph{transitive}\index{transitive} if, for any $x \in M$, $G. x = \{g. x| g\in G\}$ is equal to $M$.

We mention the following theorem. Let $G$ be a transitive Lie transformation group of a smooth manifold $M$, let $x_0$ be a point in $M$ and $G_{x_0}$ its stabilizer ( $G_{x_0}= \mbox{Stab}(x_0) = \{g \in G| g. x_0 = x_0\}$). Then $M$ is diffeomorphic to the coset space $G/G_{x_0}$, whose elements are $\{[g] = g h|h\in G_{x_0}\}$  (under reasonable conditions, see also \cite{Barut},p 123). The space $M$ is called a \emph{$G$-homogeneous space}\index{$G$-homogeneous space}\index{homogeneous space}. One also denotes by $\pi : G \ra M=G/G_{x_0}$ the \ind{canonical projection}. If $[e]:=\vartheta$, one notices that $d \pi_e : \mathfrak{g} \ra T_\vartheta M$ is surjective. \footpourmoi{Pq?}

\section{Symplectomorphisms}
Let $\phi$ be a diffeomorphism of $M$. It is called a \emph{symplectomorphism}\index{symplectomorphism} or \emph{canonical transformation}\index{canonical transformation} if it preserves $\omega$, in the sense \footnote{Recall that the pull-back\index{pull-back} of a p-form $\omega$ is defined as $(\phi^* \omega)_{\phi^{-1}(y)} (X_1,\cdots,X_p) \eqdef \omega_y (\phi_{y*\phi^{-1}(y)}X_1,\cdots,\phi_{y*\phi^{-1}(y)}X_p)$, for $X_i \in T_{\phi^{-1}(y)}M$. The pull-back of a function is $\phi^* f = f \circ \phi$. The push-forward\index{push-forward} $\phi_* X$ of a vector $X\in T_x M$ is a vector belonging to $T_{\phi(x)}M$, defined by $(\phi_* X)_{\phi(x)}f = X_x (\phi^* f)$, where $f$ is defined in a neighborhood of $\phi(x)$.}
 \begin{equation}\label{symplecto}
  \phi^* \omega = \omega
\end{equation}
This expression obviously means that $\left[\phi^*\omega_{|\phi(x)}\right]_{|x} = \omega_x$. It can be shown that \re{symplecto} is the same condition as
\begin{equation}
 \{ \phi^* f,\phi^* g\} = \phi^* \{f,g\} .
\end{equation}
Let us show one side of this statement in some detail. Let us first compute, for $X\in T_x M$, $Y\in T_{\phi^{-1}(x)}M$,
\begin{eqnarray}
\left[ i(\phi^{-1}_* X_f) \omega\right]_{\phi^{-1}(x)} (Y_{\phi^{-1}(x)}) &=& \omega_{\phi^{-1}(x)} (\phi^{-1}_*(X_f),Y_{\phi^{-1}(x)}) \nonumber\\
 																							&=& [\phi^*\omega_x]_{\phi^{-1}(x)} (\phi^{-1}(X_f),Y_{\phi^{-1}(x)}) \nonumber\\
 																							&=& \omega_x (\phi_{x*\phi^{-1}(x)}\phi^{-1}_*(X_f),\phi_{x*\phi^{-1}(x)}Y_{\phi^{-1}(x)}) \nonumber\\
 																							&=& \omega_x (X_f, \phi_*Y) \nonumber\\
 																							&=& -df_x (\phi_* Y) \nonumber\\
 																							&=& -\phi^*(df)_{\phi^{-1}(x)}(Y_{\phi^{-1}(x)}) \nonumber\\
 																							&=& -d(\phi^*f)_{\phi^{-1}(x)}(Y_{\phi^{-1}(x)}) \nonumber\\
 																							&\eqdef& \left[i(X_{\phi^* f})\omega\right]_{\phi^{-1}(x)}(Y_{\phi^{-1}(x)})
 \end{eqnarray}
 We have used the definition of the pull-back, the invariance \re{symplecto} and the fact that the exterior derivative commutes with the pull-back.
 From this we may conclude that
 \begin{equation}
  \phi^{-1}_* X_f = X_{\phi^* f}
\end{equation}
for any symplectomorphism. We may now write with \re{LieOmega}
\begin{eqnarray}
\{ \phi^* f,\phi^* g\} (x) &=& \omega_x (X_{\phi^* f},X_{\phi^* g}) \nonumber\\
														&=& \omega_x(\phi^{-1}_* X_f , \phi^{-1}_* X_g) \nonumber\\
														&=& [\phi^* \omega_{\phi(x)}]_x (\phi^{-1}_* X_f , \phi^{-1}_* X_g) \nonumber\\
														&=& \omega_{\phi(x)} (X_f,X_g) \nonumber\\
														&=& \{f,g\}_{\phi(x)} = \phi^* \{f,g\}.
 \end{eqnarray}
Let us state a proposition, without a proof : a vector field $X$ on $(M,\omega)$ generates one-parameter local symplectomorphisms of $M$ if and only if $i(X)\omega$ is a closed one-form. One defines 
\begin{equation}
\mbox{aut}(M,\omega) = \{X \in \varkappa(M)\quad s.t.\quad i(X)\omega \; \mbox{is \;closed}\}.   
\end{equation}
and
\begin{equation}
 \mbox{Ham}(M,\omega) = \{X_f, f\in C^\infty (M) \}.
\end{equation}

\section{Lie groups of symplectomorphisms}\label{Hamilton}
Let $M$ be a manifold and $G$ a Lie group. Let $\sigma$ be an action of $G$ on $M$:
\begin{equation}
 \sigma : G \times M \ra M : (g,x)\ra \sigma (g).x \quad , \quad \sigma(g.g') =\sigma (g) \circ \sigma (g') \; \forall g,g'\in G.
\end{equation}
Each $X \in \mathfrak{g}$ defines a one-parameter group of diffeomorphisms $\Psi_t^\xi$:
\begin{equation}
  \Psi_t^X (x) = \sigma(\exp t X).x
\end{equation}
The \emph{fundamental vector field}\index{fundamental vector field} $X^*$ on $M$ associated with $X\in \mathfrak{g}$ is, by definition,
\begin{equation}
 X^*_x = \ddto \sigma (\exp -t X).x  \quad.
\end{equation}
One may show that the map $\mathfrak{g} \ra \varkappa (M) : X \ra X^*$ is a homomorphism of Lie algebras, i.e. $[X,Y]^* = [X^*,Y^*]$.

The action of $G$ on $M$ is said to be \emph{almost Hamiltonian}\index{almost Hamiltonian action} or \emph{weakly Hamiltonian}\index{weakly Hamiltonian action}  if for each $Y\in \mathfrak{g}$, the fundamental vector field $Y^*$ is Hamiltonian, i.e. there exists a function $\mu_Y \in C^\infty(M)$ such that $Y^* = X_{\mu_Y}$, $i(Y^*)\omega = -d \mu_Y$, or $Y^* (g) = \{\mu_Y ,g\}$ for all $g\in C^\infty(M)$, see \re{LieOmega}.  The action is \emph{Hamiltonian}\index{Hamiltonian action} if for each $Y \in\mathfrak{g}$, there exists a function $\lambda_Y$ such that $Y^* = X_{\lambda_Y}$ \emph{and} the correspondence $\lambda : \mathfrak{g} \ra C^\infty (M) : Y\ra \lambda_Y$ is a Lie algebra homomorphism, i.e. $\lambda_{Y+Z} = \lambda_Y + \lambda_Z$, $\lambda_{k Y} = k \lambda_Y$, $\forall k \in \R$, and $\lambda_{[Y,Z]} = \{\lambda_Y , \lambda _Z\}$.

\section{Orbits of a Lie group in the dual of its Lie algebra}

We are going to illustrate the notions of the previous sections with an example that will be very useful for our purposes. We will consider the class of symplectic manifolds consisting of orbits of a Lie group $G$ in the dual of its Lie algebra.  Let $G$ be a connected Lie group with Lie algebra $\mathfrak{g}$. Let $\mathfrak{g}^*$ be the dual of $\mathfrak{g}$ (the space of all real linear forms on $\mathfrak{g}$). 
The group $G$ acts on $\mathfrak{g}^*$ by the so-called \emph{co-adjoint action}\index{co-adjoint action}: 
\begin{equation}
 G \times \mathfrak{g}^* \ra \mathfrak{g}^* : (g,f) \ra g. f = f \circ Ad_{g^{-1}} ,
\end{equation}
i.e.
\begin{equation}
 \langle g. f,X \rangle = \langle f, Ad_{g^{-1}}X \rangle \quad, \quad g\in G, f\in \mathfrak{g}^*, X\in \mathfrak{g}.  
\end{equation}
Remark that $g. f$ is sometimes also denoted by $Ad^* g f$ or $\sigma(g). f$.
If $f\in \mathfrak{g}^*$, let $\theta_f = G\circ f = \{g.  f | g\in G \}$ be the orbit of $f$ under this action of $G$. If $x\in \theta_f \subset \mathfrak{g}^*$, the tangent space $T_x \theta_f$ is spanned by the vectors $X^*_x$ for any $X\in \mathfrak{g}$, where
\begin{equation}\label{fvf}
 X^*_x = \ddto (\exp -tX).x = \ddto x \circ Ad_{\exp tX} = x \circ ad_X,
\end{equation}
where the exponential map $\exp : \mathfrak{g} \ra G$ is defined by
\begin{equation}
\ddto \exp tX = X \quad , \quad \mathfrak{g} \simeq T_e G.
\end{equation}
Then one defines 
\begin{equation}
 \omega_x (X^*_x, Y^*_x) = \langle x,[X,Y] \rangle \quad , \quad x\in \theta_f\subset \mathfrak{g}^*, \; X,Y\in \mathfrak{g}^*,\; X^*_x, Y^*_x \in T_x\theta_f, 
\end{equation}
where $[.,.]$ is the Lie bracket in $\mathfrak{g}$.
One may then show that \cite{CoursGutt}:
\begin{enumerate}
{\item $\omega_x$ is a well defined two-form on $T_x\theta_f$,}
{\item $\omega_x$ is nondegenerate,}
{\item $\omega$ is closed.}
\end{enumerate}
Thus every orbit of a connected Lie group $G$ in the dual of its Lie algebra has a natural symplectic structure.

Let us now take $M= \theta_f$. Being the orbit in $\mathfrak{g}^*$ of $f$ under the group $G$, $M$ obviously admits an action of $G$:
\begin{equation}\label{ActionSurTheta}
 \sigma : G \times M \ra M : (g,x) \ra \sigma(g). x \quad,
\end{equation}
 whose fundamental vector fields are given by \re{fvf}. One can show that $G$ actually acts by symplectomorphisms, that is 
 \begin{equation}
 \sigma (g)^* \omega = \omega.
\end{equation}
 Next, consider the action of a one-parameter subgroup of $G$, $\sigma(\exp (t Z))$, with $Z\in \mathfrak{g}$. This gives a one-parameter group of symplectomorphisms of the orbit $M$. It can be checked that its associated fundamental vector field $Z^*$ is Hamiltonian and
\begin{equation}
 i(Z^*)\omega = -d\lambda_Z \quad , \quad \mbox{where} \; \lambda_Z (x) = \langle x , Z \rangle .
\end{equation}
Furthermore, the correspondence $\mathfrak{g} \ra C^\infty(M)$ can be shown to be a Lie algebra homomorphism. Therefore, the action of $G$ on $M$ given by \re{ActionSurTheta} is Hamiltonian.


\section{Symplectic foliation of Poisson manifolds}
\footpourmoi{je ne me sers pas du tout de cette section...}

Let us first roughly state some results on foliations. Let $M$ be a smooth manifold, $dim M =n$. Let us consider a set of $k$ independent one-forms $\theta^\a$, $\a=1,\cdots,k<n$. The system of differential forms
\begin{equation}
\theta^\a = 0 \quad , \quad \theta^\a = \theta^\a_a dx^a \quad , \quad a=1,\cdots,n
\end{equation}
is said \emph{integrable}\index{integrable} if and only if there exist $k$ functions $f^\beta$ and an invertible matrix $S^\a_\beta$ such that
\begin{equation}
 \theta^\a = S^\a_\beta \, df^\beta .
\end{equation}
The \emph{Frobenius theorem}\index{Frobenius theorem} states that $\theta^\a = 0$ is integrable iff 
\begin{equation}
 \theta^1 \wedge \theta^2 \wedge \cdots \wedge \theta^k \wedge d\theta^\a = 0 \quad, \quad \forall \a = 1,\cdots,k.
\end{equation}
At each point $x\in M$, the $k$ one-forms $\theta^\a$ define a sub-vector space of $T_x M$ via the set of equations
\begin{equation}
 \theta^\a (X) = 0 \quad , \quad X\in T_x M. 
\end{equation}
Vectors of $T_x M$ satisfying this relation form a vector space of dimension $(n-k)$. The data of a system of differential forms $\theta^\a = 0$ is thus equivalent to the data, at each point of $M$, of a sub-vector space of dimension $(n-k)$ of the tangent space at that point. This field of sub-vector spaces is called a \emph{distribution}\index{distribution} of rank $n-k$, and is denoted by $\Delta$.
The dual form of Frobenius' theorem states that the system of differential forms $\theta^\a = 0$ is integrable iff $\forall X,Y \in \Delta \subset \varkappa(M)$, we have $[X,Y] \in \Delta$. $\Delta_x$ is sometimes called the \emph{fiber}\index{fiber} at $x$ \cite{marle}.
A rank-$k$ distribution is said to be differentiable if every point $x\in M$ has a neighborhood $U$ on which there exist $k$ differentiable vector fields $X_i$, $1 \leq i \leq k$, whose values at each point $y\in U$ form a basis of $\Delta_y$. The \emph{integral}\index{integral of a distribution} of the differentiable distribution $\Delta$ on $M$ is an immersion\footnote{Let $M$ and $N$ be smooth manifolds and $f: M\ra N$ be a smooth map. It is called an \emph{immersion}\index{immersion} if for any point $x\in  M$, the differential of the map $(df)_x : T_x M \ra T_{f(x)}N$ is injective.} $f : S \ra M$ of a connected manifold $S$ into $M$, such that for all $x\in S$, $(df)_x (T_x S)$ is a sub-vector space of $\Delta_{f(x)}$. When this is the case, the dimension of $S$ is necessarily less or equal to $k$, because $(df)_x$ is injective.
An \emph{integral submanifold}\index{integral submanifold} of $\Delta$ is a connected submanifold $S$ of $M$ such that, for all $x\in S$, $T_x S \subset \Delta_x$. It said to be of maximal dimension if for all $x\in S$, $T_x S = \Delta_x$.   
An integral submanifold of $\Delta$ is said to be \emph{maximal}\index{maximal integral submanifold} if every integral submanifold of $\Delta$ containing it is necessarily equal to it.

We may now return to the concept of integrability.
\footpourtoi{feuilletage = foliation?}
 A differentiable distribution $\Delta$ of rank $k$ on $M$ is said to be (completely) \emph{integrable}\index{integrable} if, for all $x\in M$, there exists an integral submanifold of $\Delta$ of dimension $k$ passing through the point $x$. When this is the case, one can show that there is a unique maximal integral submanifold of $\Delta$ through each point of $M$. $M$ is then a disjoint union of integral submanifolds of $\Delta$, each of them being maximal and of maximal dimension. One says that $\Delta$ defines a \emph{foliation}\index{foliation} of $M$. The maximal integral submanifolds of $\Delta$ are called \emph{leaves}\index{leaf} of this foliation. Frobenius' theorem gives necessary and sufficient conditions for a rank-$k$ distribution to be (completely) integrable. The notions of integrability and foliations can be extended to differentiable distributions whose rank is not necessarily constant (for more details, see \cite{marle}), which naturally appear when studying Poisson manifolds.
 
Let us return to the Poisson bi-vector field $\Pi$, defined in \re{PoissonBiVector}. We denote by $\Pi^\sharp : T^*M \ra TM$ the morphism of vector bundles defined in the following way. For all $x\in M$, $\a \in T^*_x M$, $\Pi^\sharp(\a)_x$ is the unique element of $T_x M$ such that
\begin{equation}
 \beta \left( \Pi^\sharp (x) (\a)\right) = \Pi_x (\a, \b) \quad , \quad \forall \b \in T^*_x M. 
\end{equation} 
This way, the Hamiltonian vector field can be rewritten as 
\begin{equation}
 X_f =  \Pi^\sharp (df).
\end{equation} 

Let $(M,\Pi)$ be a Poisson manifold. The \emph{characteristic field} of this Poisson manifold is $C =  \Pi^\sharp (T^* M)$. For all $x\in M$, the fiber $C_x = \Pi^\sharp (T_x^* M)$ of the characteristic field at $x$ in called \emph{characteristic space} at $x$. At any point $x \in M$, $C_x$ is a sub-vector space of even dimension of $T_x M$. This stems from the fact that the rank of a bilinear antisymmetric form on a finite-dimensional vector space is always even. The characteristic field $C$ of a Poisson manifold is not, in general, a sub-vector bundle of the tangent bundle, since the dimension of $C_x$ may change from point to point \cite{marle}. Finally, each characteristic space $C_x$ may be endowed with a symplectic form, which makes $C_x$ a symplectic vector space (i.e. a vector space endowed with a bilinear antisymmetric form whose rank equals the dimension of the space). Now, $C$ defines a distribution in the above sense, with the only subtlety that its rank may vary from point to point.

When the distribution $\Delta$ is differentiable and of constant rank, i.e. when all fibers $\Delta_x$ of $\Delta$ have the same dimensions, the (dual form of) Frobenius' theorem may be applied. This may be generalized to the case where $\Delta$ is differentiable, but not necessarily of constant rank. $\Delta$ is then (completely) integrable iff $\forall X,Y \in \Delta \subset \varkappa (M)$, $(\Phi_t)_* Y \in \Delta$ for all $t\in \R$, where $X = \ddt \Phi_t$ \cite{Sussmann,Stefan}.

One may then show that the characteristic field of a Poisson manifold $(M,\Pi)$ is completely integrable, and defines a foliation (in a generalized sense since all leaves do not necessarily have same dimensions), whose leaves are symplectic manifolds.

This can be made more precise through the \emph{local splitting theorem}\index{local splitting theorem} \cite{WeinLST} : on a Poisson manifold $(M,\Pi)$, any point $x\in M$ has a coordinate neighborhood with coordinates $(q_i,p^j,y^m)$, $i,j=1,\cdots,k$, $m=1,\cdots,l$ centered at $x$, such that
\begin{equation}\label{splitting}
\Pi = \sum_{i=1}^k \frac{\p}{\p q_i} \wedge \frac{\p}{\p p^i} + \frac{1}{2} \sum_{i,j=1}^l \varphi_{ij}(y) \frac{\p}{\p y^i}\wedge \frac{\p}{\p y^j} \quad \mbox{and} \quad \varphi_{ij}(0) =0.
\end{equation}
When $l=0$, the Poisson structure is called symplectic, and the theorem
is Darboux's theorem. In this situation, the rank of $\Pi$ is equal to $dim M$ everywhere, and we are dealing with a symplectic manifold.  Hence, a general Poisson manifold is isomorphic, near each point, to the product of an open subset of a standard symplectic
manifold $\R^{2k}$ and a Poisson manifold for which the Poisson tensor vanishes at the point in question.

In the general case, points where $\Pi_{ij}$ has locally constant rank are called \emph{regular}\index{regular points}. If all points
of $M$ are regular, M is called a \emph{regular Poisson manifold}\index{regular Poisson manifold}, endowed with a \emph{regular Poisson structure}\index{regular Poisson structure}.
\footpourtoi{If Poisson str. not regular, then foliation is singular? signifie quoi?}
When the Poisson structure is regular, there is a natural foliation of $M$ by symplectic manifolds whose dimension is the rank of $\Pi$.
This is called \emph{symplectic foliation}\index{symplectic foliation}, the leaves being the \emph{symplectic leaves}\index{symplectic leaf}.

But as we stated above, symplectic leaves exist through every point, even
on Poisson manifolds where the Poisson structure is not regular. The symplectic leaves are determined locally by the splitting theorem \re{splitting}. For any point $x$ of the Poisson manifold, if $(q_i, p^j, y^k)$ are the normal coordinates as in \re{splitting}, then the symplectic leaf through $x$ is given locally by the equation y = 0.








\cleardoublepage \chapter{Technical Appendices}

\section{Determination of $AN\,$-invariant kernels}\label{AppendixAN}

In this section, we show how to successively implement conditions $(i)$ to $(vi)$ of Sect. \ref{ANsurAN} in order to get an $AN\,$-invariant associative product on the $AN\,$ group manifold. The group manifold variables are denoted by $a$ and $n$. The
         infinitesimal left translations are generated by the vector fields
         $\partial_a$ and $\exp[-\,a]\partial_n$. The star product is
         written as:
         \begin{equation} \label{SP}
         (u*v)({\bf x)}= \int {\rm K}[{\bf x},{\bf y},{\bf z}]\,u({\bf
         y})\,v({\bf z})\,d\mu_{\bf y}\,d\mu_{\bf z}\,
         \end{equation} where the (left-invariant) measure used is simply
         $d\mu_{\bf x}= da_{\bf x}\ dn_{\bf x}$. To be left-invariant, the
         kernel ${\rm K}[{\bf x},{\bf y},{\bf z}]$ has to verify the
         equations:
         \begin{eqnarray} &&\left ( \partial_{a_{\bf x}}+\partial_{a_{\bf
         y}}+\partial_{a_{\bf z}}\right){\rm K}[{\bf x},{\bf y},{\bf z}]=0\qquad ,\\
         &&\left (\exp[-\,a_{\bf x}]\,
         \partial_{n_{\bf x}}+\exp[-\,a_{\bf y}]\,\partial_{n_{\bf y}}+\exp[-\,a_{\bf
         z}]\,\partial_{n_{\bf z}}\right){\rm K}[{\bf x},{\bf y},{\bf
         z}]=0\qquad .
         \end{eqnarray} Hence it depends on four variables instead of six:
         \begin{equation} {\rm K}[{\bf x},{\bf y},{\bf
         z}]=K^L[\al{x}{y},\al{x}{z};\un{y}{x},\un{z}{x}] \qquad ,\label{KL}
         \end{equation} where we have set
         \begin{equation}
         \al{x}{y}:=a_{\bf x}-a_{\bf y}\quad , \quad \un{x}{y}:={n_{\bf
         x}-\exp[-(a_{\bf x}-a_{\bf y})]\,n_{\bf y}} \qquad .
         \end{equation} This condition ensures the left invariance of the star product
         under the $AN\,$ group. We now proceed to impose four additional
         conditions: the two that define a star product, i.e. the
         associativity and the existence of right and left units, as well as
         a condition on the trace and the hermiticity.

First of all, the existence of a unit element, $u\st LG 1=u$ and
         $1\st LG u=u$, imposes the conditions:
         \begin{eqnarray}
         \int K^L[\al{x}{y},\al{x}{z};\un{y}{x},\un{z}{x}]\, d\mu_{\bf
         z}=\delta ^2[\bf{
         x- y}] &&\label{unitr} \qquad ,\\
         \int K^L[\al{x}{y},\al{x}{z};\un{y}{x},\un{z}{x}]\, d\mu_{\bf
         y}=\delta ^2[\bf{ x- z}] &&\label{unitl} \qquad .
         \end{eqnarray} To fulfill these conditions, we assume that:
         \begin{equation} K^L(\al{x}{y},\al{x}{z};\un{y}{x},\un{z}{x})= \frac
         1{(2\,\pi\,\lambdabar)^2}B(\al{x}{y},\al{x}{z})\,\exp\left \{ i
         \Psi(\al{x}{y},\al{x}{z}; \un{y}{x},
         \un{z}{x})\right\}\label{ampliphas}
         \end{equation} where:
         \begin{equation}\Psi(\al{x}{y},\al{x}{z}; \un{y}{x}, \un{z}{x}):=
         Y(\al{x}{y},\al{x}{z})\,\un{y}{x}+Z(\al{x}{y},\al{x}{z})\,\un{z}{x}
         \quad ,\label{defpsi}
         \end{equation} where $Y(a,\, b)$ and $Z(a,\, b)$ are real functions, and $B(a,\, b)$
         complex. This special choice of the phase $\Psi$, linear in the
         $\un yx$ and $\un zx$ variables, as well as the independence of
         the function $B$ in these variables, is dictated by the structure
         of the Fourier transform of the Dirac delta distribution. Eq.
         \ref{unitr} now reads :
         \begin{equation}
         \frac 1{2\,\pi\,\lambdabar\,^2}\int
         B(\al{x}{y},\al{x}{z})\delta[Z(\al{x}{y},\al{x}{z})]\exp\left \{
         iY(\al{x}{y},\al{x}{z})\un{y}{x}\right\} d a_{\bf z} = \delta ^2[\bf
         x-\bf y]\label{unitr2}\qquad .
         \end{equation} To reproduce the distribution of the right-hand side of eq.
         (\ref{unitr2}), we have to assume that:
         \begin{equation}
         \delta[Z(\al xy,\al xz)]=\delta[\al xy]/\zeta(\al xz)\quad{\rm i.\
         e.}\quad Z(a,\,b)=0 \quad {\rm iff}\quad a=0\qquad .\label{dZ}
         \end{equation} Hence, to verify eq.(\ref{unitr}), the
         function $B(a,\, b)$ has to satisfy the condition:
         \begin{equation}
         \lambdabar^2 \, \zeta(b)\,\partial_b Y(0,b)= B(0,b)\qquad ,
         \label{unitr3}
         \end{equation} with the sign being fixed by the requirement that $Y(0,b)$ runs
         from $-\infty$ to $+\infty$ when $b$ goes from $-\infty$ to
         $+\infty$. An analogous calculation shows that eq. (\ref{unitl})
         implies:
         \begin{equation}
         \lambdabar^2\,\eta(a)\,\partial_a Z(a,0)= B(a,0)\qquad ,
         \label{unitl3}
         \end{equation} and leads to a relationship similar to eq. (\ref{dZ}):
         \begin{equation}
         \delta[Y(\al xy,\al xz)]=\delta[\al xz]/\eta(\al xy)\quad{\rm i.\
         e.}\quad Y(a,\,b)=0 \quad {\rm iff}\quad b=0 \qquad ,\label{dY}
         \end{equation} with $Z(a,0)$ running from $-\infty$ to $+\infty$ when $a$
         varies from $-\infty$ to $+\infty$.

         Let us now impose the following trace condition on the star
         product :
         \begin{equation}
         \int (u \st LG v)({\bf x})d\mu_{\bf x}:=\int k[\bbf y,\,\bbf z]u({\bf
         y})\,v({\bf z})d\mu_{\bf y}d\mu_{\bf z}=\int k[\bbf z,\,\bbf y]u({\bf
         y})\,v({\bf z})d\mu_{\bf y}d\mu_{\bf z}\qquad ,\label{contr}
         \end{equation} which implies that the two-point kernel:
         \begin{equation}\label{kernel}
         \int K^L[\al{x}{y},\al{x}{z};\un{y}{x},\un{z}{x}]\, d\mu_{\bf
         x}:=k[{\bf y},\,{\bf z}]=k[{\bf z},\,{\bf y}]
         \end{equation}
         is symmetric. We will not discuss this condition in all its
         generality here, but initially restrict ourselves to the special case:
         \begin{equation}\label{kdelta}
         k[\bbf y,\,\bbf z]=\delta^2[\bbf y,\,\bbf z]\qquad ,
         \end{equation}
         and discuss a slightly more general situation later. After
         integration over $n_{\bbf x}$ in (\ref{kernel}), this condition may
         be rewritten as:
         \begin{eqnarray} &\frac 1{2\,\pi\,\lambdabar^2}\int
         B(\al{x}{y},\al{x}{z})\delta[Y(\al{x}{y},\al{x}{z})\exp(\,\al xy)+
         Z(\al{x}{y},\al{x}{z})\exp(\,\al xz)]\qquad &\nonumber\\ & \exp\left
         \{ iY(\al{x}{y},\al{x}{z})n_{\bf y}+Z(\al{x}{y},\al{x}{z})n_{\bf
         z})\right\} d a_{\bf x} = \delta(a_{\bbf y}-a_{\bbf z}) \,
         \delta(n_{\bbf y}-n_{\bbf z})\quad .&\label{tracex}
         \end{eqnarray} To satisfy this relationship, we assume that the delta
         distribution appearing in the integrand is equivalent to $\delta
         [\al yx -\al zx]$. In other words, we assume that:
         \begin{equation} Y(a,\,b)\,{\rm e}^{\,a}+Z(a,\, b)\, {\rm e}^{\,b}=0
         \Leftrightarrow a=b\qquad .\label{halfdelta}
         \end{equation} This condition is sufficient to pursue the construction of
         the star product; we shall return to it later.

         We now analyze the conditions implied by the associativity. The
         star product will be associative if and only if $I_R=I_L$, where:
         \begin{equation} I_R = \int K^L(\bbf x,\bbf p,\bbf y)\,K^L(\bbf y,\bbf q,\bbf r)\, d
         \mu_\bbf y \qquad {\rm and } \qquad I_L = \int K^L(\bbf x,\bbf y',\b
         r)\,K^L(\bbf y',\bbf p,\bbf q)\, d \mu_\bbf {y'} \qquad .
         \end{equation} After integrations on $n_{\bbf y}$ and $n_{\bbf {y'}}$ we obtain:
         \begin{eqnarray} I_R&=&\frac 1{8\,\pi^3\,\lambdabar^4}\int B(\al xp,\al xy)\,B(\al yq,\al yr){\rm e}^{-i\left\{\left(Y(\al xp,\al xy)\,{\rm e}^{\,\al xp} +Z(\a xp,\al xy)\,{\rm e}^{\,\al xy}\right)n_{\bbf x}\right\}}\nonumber\\
         &&\quad {\rm e}^{i\left\{Y(\al xp,\al xy)\,n_{\bbf p}+Y(\al yq,\al yr)\,n_{\bbf q} +Z(\al yq,\al yr)\,n_{\bbf r}\right\}} \nonumber\\
         &&\quad \delta[Z(\al xp,\,\al xy)-Y(\al yq,\,\al yr){\rm e}^{\,\al yq}-Z(\al yq,\,\al yr){\rm e}^{\,\al yr}] \ d a_\bbf y\label{IR}
         \\ &&\nonumber\\ I_L&=&\frac 1{8\,\pi^3\,\lambdabar^4}\int B(\al x{y'},\al xr)\,B(\al {y'}p,\al {y'}q){\rm e}^{-i\left\{\left(Y(\al x{y'},\al xr)\,{\rm e}^{\,\al x{y'}} +Z(\al x{y'},\al xr)\,{\rm e}^{\,\al xr}\right)n_{\bbf x}\right\}}\nonumber\\ &&\quad {\rm e}^{i\left\{Y(\al {y'}p,\al {y'}q)\,n_{\bbf p}+Z(\al {y'}p,\al {y'}q)\,n_{\bbf q} +Z(\al x{y'},\a xr)\,n_{\bbf r}\right\}} \nonumber\\ &&\quad \delta[Y(\al x{y'},\,\a xr)-Y(\al {y'}p,\,\al {y'}q){\rm e}^{\,\al {y'}p}-Z(\al {y'}p,\,\al {y'}q){\rm e}^{\,\al {y'}q}] \ d a_\bbf {y'}\label{IL}
         \end{eqnarray} Using eq. (\ref{halfdelta}), we see that points such
         that $a_\bbf q =a_\bbf r$ and $a_\bbf p=a_\bbf x$ belong to the support of the delta
         distribution appearing in $I_R$, while those such that $a_\bbf p
         =a_\bbf q$ and $a_\bbf r=a_\bbf x$ belong to the support of the delta
         distribution appearing in $I_L$. To obtain $I_R=I_L$, the supports
         of these delta distributions must coincide (possibly after
         redefinition of the $a_\bbf {y'}$ variable). To continue, we assume
         that the support of both delta distributions are located on the
         subset defined by:
         \begin{equation} a_{\bbf p} - a_{\bbf q} + a_{\bbf r} - a_{\bbf x} =0 \qquad
         .\label{supdel}
         \end{equation} This condition and the requirement for the equality of the
         phases of the two integrals (\ref{IR} and \ref{IL}), lead to the
         set of four equations:
         \begin{eqnarray} Y(\al rq,\al xy)&=&Y(\al {y'}p,\al {y'}q)\label{np}\\ Y(\al yq,\a yr)&=&Z(\al {y'}p,\al {y'}q)\label{nq}\\ Z(\al yq,\al yr)&=&Z(\al x{y'},\al pq)\label{nr}\\ Y(\al rq,\al xy){\rm e}^{\al rq}+Z(\al rq,\al xy){\rm e}^{\al xy}&=& Y(\al x{y'},\al pq){\rm e}^{\al x{y'}}+Z(\al x{y'},\al pq){\rm e}^{\al pq}\qquad .\label{nx} \end{eqnarray}
          To
         ensure the compatibility between the functional relationships
         (\ref{np}-\ref{nr}) and eqs (\ref{dZ}, \ref{dY}) we must impose
         the conditions:
         \begin{eqnarray} a_\bbf y =a_\bbf x =a_\bbf p - a_\bbf q+a_\bbf r &\Leftrightarrow& a_\bbf {y'}=a_\bbf q \qquad \label{a1bis},\\ a_\bbf y =a_\bbf r &\Leftrightarrow&
         a_\bbf {y'}=a_\bbf p \qquad ,\\ a_\bbf y =a_\bbf q &\Leftrightarrow& a_\bbf {y'}=a_\bbf x =a_\bbf p+a_\bbf r - a_\bbf q \qquad \label{a3} .
         \end{eqnarray}
         
          The only linear relationship between $a_\bbf y$ and
         $a_{\bbf {y'}}$ satisfying relationships (\ref{a1})-(\ref{a3})is
         \begin{equation} a_\bbf {y'}=-a_\bbf y +a_\bbf p+ a_\bbf r\qquad ;\label{ayp}
         \end{equation} the linearity of the relationship is imposed by the equality of the
         integrals $I_R$ and $I_L$. Furthermore, by inserting the relationships
         (\ref{supdel}, \ref{ayp}) in eqs (\ref{np}, \ref{nr}) we deduce
         that the functions $Y$ and $Z$ depend on one variable only:
         \begin{eqnarray} Y(a,\, b)&=&\tilde Y(b)\quad {\rm and}\quad \tilde
         Y(0)=0\qquad ,\\ Z(a,\, b)&=&\tilde Z(a)\quad {\rm and}\quad
         \tilde Z(0)=0\qquad.
         \end{eqnarray} Moreover eq. (\ref{nq}) implies:
         \begin{equation}
         \tilde Y(a)=\tilde Z(-a)\qquad .
         \end{equation} Condition (\ref{nx}) now becomes:
         \begin{eqnarray}
         \tilde Y(\al pq +\al ry)\ {\rm e}^{\, \al rq}+\tilde Y(\al qr)\ {\rm
         e}^{\,(
         \al pq +\al ry) }&=&\nonumber\\
         \tilde Y(\al pq)\ {\rm e}^{\, \al yq}+\tilde Y(\al qy)\ {\rm e}^{\,
         \al pq}\qquad .\label{LC}&&
         \end{eqnarray} If in this last condition we set $a _\bbf p= a_\bbf y$ and $a _\b
         q= a_\bbf r$, we deduce that $\tilde Y$ and $\tilde Z$ are odd
         functions satisfying:
         \begin{equation}
         \tilde Y(a)=-\tilde Z(a)=\tilde Z(-a)=-\tilde Y(-a)\qquad .
         \end{equation} Finally, considering the condition (\ref{LC}) for $a _\bbf q=
         a_\bbf y$ and $a _\bbf p= a_\bbf r$, we obtain:
         \begin{equation}
         \tilde Y(2\,a)=2\,\tilde Y(a)\, \cosh (\, a) \qquad ,
         \end{equation} whose only continuous solution, with the condition that $\tilde
         Y(a)$ runs from $-\infty$ to $+\infty$ when $a$ starts from
         $-\infty$, is:
         \begin{equation}
         \tilde Y(a)= \lambdabar^{-1} \sinh( \, a)\qquad ,
         \end{equation} where $\lambdabar$ is a positive constant. These considerations
         allow us to fix the $\Psi$-phase of the kernel defining the star
         product as:
         \begin{equation}
         \Psi=\lambdabar^{-1} \left\{ \sinh\left[({a_{\bbf y}}-{a_{\bbf x}})\right]\, {n_{\bbf z} }+ \sinh\left[({a_{\bbf z}}-{a_{\bbf y}})\right]\, {n_{\bbf x} }+ \sinh\left[({a_{\bbf x}}-{a_{\bbf z}})\right]\, {n_{\bbf y} } \right\}\label{phase}\qquad .
         \end{equation} Indeed, it is straightforward to check that this
         expression of $\Psi$
         satisfies the assumption (\ref{supdel}) made about the supports of
         the delta distributions appearing in eqs (\ref{IR} and \ref{IL}).

         Having obtained $\Psi$, we still have to determine the
         $B$-function of the kernel introduced in eq. (\ref{ampliphas}).
         From the existence of right and left units we obtain, using eqs
         (\ref{dZ}, \ref{dY}):
         \begin{equation} \label{BO} B(0,\al xz)=\,\cosh(\,\al xz)\qquad {\rm
         and}\qquad B(\al xy,\, 0)=\,\cosh(\,\al xy)\qquad ,
         \end{equation} while the trace condition (\ref{tracex}) implies the diagonal
         relationship:
         \begin{equation} B(\al xy,\,\al xy)=\, \cosh(\,\al xy)\qquad
         .\label{Adiag}
         \end{equation} In addition, the associativity condition requires that the
         $B$-function obeys the quadratic functional equation:
         \begin{equation}
         \frac{B(\al rq,\,\al rq +\al py)\,B(\al yq,\,\al yr)}{\cosh(\al qr)}=\frac {B(\al yq,\,\al pq )\,B(\al ry,\,\al rq + \al py)}{\cosh(\al qp)}\label{quadasscon}
         \end{equation} from which we easily deduce (by considering the two special
         cases $a_\bbf q =a_\bbf r$ and $a_\bbf p =a_\bbf y$) that:
         \begin{equation} B(a,b)\,B(b,a)= \cosh(\,a)\cosh(\,b)\cosh(a-b)\qquad .
         \end{equation}

         The last condition that we impose on the star product is the
         hermiticity condition:
         \begin{equation}
         \overline{\left( u\st LG v\right)}=\left( \overline{v}\st LG
         \overline{u}\right)\qquad .\label{hermcon}
         \end{equation} It implies:
         \begin{equation} B(a,\,b)=\overline {B(b,\, a)} \label{amplcon}
         \end{equation} and the antisymmetry of $\Psi$ with respect to this exchange, a
         condition that is already satisfied. So, if we assume the
         hermiticity condition, we obtain:
         \begin{equation} \left | B(\al xy ,\,\al xz) \right |=
         \sqrt{\cosh(\,\al xy)\cosh(\,\al xz)\cosh(\,\al yz)}\qquad
         .\label{ampl}
         \end{equation} The phase $\psi(\al xy,\al xz)$ of the complex function $B$
         (not to be confused with $\Psi$) is an arbitrary odd function of
         two variables, vanishing when one of the variables is an integer
         multiple of $\pi$. Indeed, eqs (\ref{unitr3}, \ref{unitl3}) imply
         that $B(a,0)$ and $B(0,a)$ are real.
         \begin{equation}
         \psi(a,0)=0\qquad.\label{Bphasecon}
         \end{equation} It is a matter of trivial calculation to check that the kernel
         built with the $\Psi$-phase (\ref{phase}) and the function $B$ so
         defined provides a hermitian star product, satisfying the trace
         condition (\ref{contr}) and admitting a left and right unit. Let us
         also note that both the phase (\ref{phase}) and
         the amplitude (\ref{ampl}) admit a geometrical interpretation in
         terms of geodesic triangles built on the
         three points $\bbf x$, $\bbf y$ and $\bbf z$ (see refs \cite{PierreStrict, ZQian}).

         Another left-invariant star product under the $\ca \, \cn$ group
         was obtained previously starting from completely
         different considerations \cite{PierreStrict}. Its phase is also given
         by eq. (\ref{phase}) but $B(\al xy ,\,\al xz)$ is real and given by
         :
         \begin{equation} B(\al xy ,\,\al xz) = \cosh(a_\bbf y-a_\bbf z)\qquad .
         \end{equation} This star product does not satisfy the trace condition
         (\ref{contr}) with (\ref{kdelta}) but instead a twisted trace
         condition involving a $K_0$ Bessel function:
         \begin{equation}
         \int (u \st LG v)({\bf x})d\mu_{\bf x}=\frac 1{\pi\,\lambdabar}
         \int K_0\left[ {\lambdabar}^{-1}(n_{\bbf y}-n_{\bbf z})\right]
         \delta(a_{\bbf y}-a_{\bbf z}) u({\bf y})\,v({\bf z})d\mu_{\bf
         y}d\mu_{\bf z}\quad .
         \end{equation}

         This star product may be reobtained and generalized as follows.
         The phase $\Psi$ given by eq. (\ref{phase}) was built without any
         use of the hermiticity condition. Modulo the computational
         assumptions introduced, its expression results essentially from
         the invariance conditions and one ``half" of the trace condition
         (\ref{halfdelta}) : the condition that leads to the factor
         $\delta(a_\bbf y -a_\bbf z)$ on the right hand side of eq.
         (\ref{tracex}). A similar computation as the one described here
         above, but ignoring the other ``half" of the trace condition (eq.
         (\ref{Adiag})), leads to:
         \begin{equation} B(a,\,b)\,B(b,\,a)= \frac{B(a,\,a)\,B(b,\,b)}{B(a-b,\,a-b)}
         \cosh^2[(a-b)]\label{Bgen}\qquad .
         \end{equation} Note that by interchanging $a$ and $b$ in this equation, we
         find that $B(a,\,a)=B(-a,\,-a)$. As a consequence, $B(a,a)$ is
         an even function and the star product satisfies a
         twisted trace condition in general, {\it i.e.} a trace condition that instead
         of $\delta^2[\bbf y-\bbf z]$ in eqs (\ref{contr}, \ref{tracex}),
         involves a slightly more general, but nevertheless always
         invariant, symmetric kernel $F(n_{\bbf y}-n_{\bbf z})\ \delta(a_\b
         y-a_\bbf z)$ in its right hand side, with:
         \begin{equation} F(n_{\bbf y}-n_{\bbf z})=\frac{1}{2\,\pi\,\lambdabar} \int
         B(a,a) {\rm e}^{\frac i{\lambdabar}(n_{\bbf y}-n_{\bbf z})\sinh
         a}da\qquad .
         \end{equation} Conversely, if we fix the distribution $F(n_{\bbf y}-n_{\bbf z})$,
         $B(a, a)$ must be equal to:
         \begin{equation} B(a, a)=\tilde\cf(a)=\,\cosh(a)\, \hat {\rm
         F}(\lambdabar^{-1}\sinh(a)) \qquad ,\label{Ftilde}
         \end{equation} where $\hat {\rm F}(k)=\int F(n)\exp(-i\,k\,n)\, dn $ is the
         Fourier transform of $F$.

         The most general solution of eq. (\ref{Bgen}) can easily be
         expressed by decomposing the function $B(a,b)$ into its symmetric
         $B_s$ and antisymmetric $B_a$ parts:
         \begin{equation} B_s(a,b)=\frac12\left(B(a,b)+B(b,a)\right)\qquad ,\qquad
         B_a(a,b)=\frac12 \left(B(a,b)-B(b,a)\right)\qquad .
         \end{equation} Equations (\ref{BO}) imply that
         the function $B_a$ vanishes when one of its arguments is zero
         ($B_a(a,0 )=0$), but otherwise is arbitrary. The symmetric part of
         the $B$-function depends on the function $\tilde\cf(a)$ defined in
         eq. (\ref{Ftilde}), with the additional condition that $\tilde\cf
         (0)=1$ ensuring that $\int_{-\infty}^\infty F(n) dn=1$. It is
         given by:
         \begin{equation}
         B_s^2(a,b)=\frac{{\tilde\cf}(a){\tilde\cf}(b)} {{\tilde\cf}
         (a-b)}\cosh^2(a-b)+B_a^2(a,b)\label{amplsymgen}
         \end{equation}
          Choosing $B$ as symmetric and real with
         $\tilde\cf(\alpha)= 1$ corresponds to the star product presented
         in \cite{PierreStrict}, while $\tilde\cf(\alpha)=\,\cosh \alpha $ corresponds to a
         hermitian star product such that the trace condition is given by
         eqs. (\ref{contr}) and (\ref{kdelta}).

Let us also mention that some of the star products we have
         discussed above are easily related to the Moyal-Weyl star
         product (denoted $\stm$)\footnote{Formally, all star products are
         equivalent to a Moyal-Weyl star product; the transformation $T$
         that makes the correspondence can always be constructed, step by
         step, as a formal series. The point here is that we have explicit
         transformations that allow to define the functional space on which
         the star product constitutes an internal composition law.}:
         \begin{equation} (u \stm v)(\bbf x):=\frac 1{(2\pi\,\lambdabar)^2} \int {\rm
         e}^{\frac{i}{\lambdabar} (\bbf x-\bbf y)\wedge (\bbf x -\bbf z)}u(\bbf y)\,
         v(\bbf z)\, d\mu_\bbf y d\mu_\bbf z\label{stM} \qquad ,
         \end{equation} via the sequence of transformations:
         \begin{equation} (u\st LG v)=T^{-1}\left[T[u]* T[v]\right]\label{stTM}
         \qquad ,
         \end{equation} where:
         \begin{equation} T[u](a,\, n):=\frac {1}{2\,\pi\,\lambdabar}\int {\rm
         e}^{-\frac i{\lambdabar}\xi\,n}\cp(\xi){\rm e}^{\frac
         i{\lambdabar} \sinh(\xi)\nu}u(a,\, \nu)\, d\nu\, d\xi \qquad ,
         \label{TtransfAPP}
         \end{equation} slightly generalizing, by an extra multiplication by the
         (non-vanishing) complex function $\cp$, a similar transformation
         first obtained in ref. \cite{PierreStrict}. The kernel of the star
         product so defined is given by:
         \begin{equation} \label{noyauP}
         K^L[\al{x}{y},\al{x}{z};\un{y}{x},\un{z}{x}]=\frac{1}{(2\,\pi\,\lambdabar)^2 }
         \frac{\cp(\al yx)\cp(\al xz)}{\cp(\al yz)}\cosh(\al yz)\, \exp(i\Psi)\qquad .
         \end{equation}

         formulae (\ref{stTM}, \ref{TtransfAPP}) clarify some of the
         constraints imposed to the function $\cp$: $\cp(0)=1$ is necessary
         to obtain $u\st LG 1=1\st LG u=u$, as otherwise we would have
         $u\st LG 1=1\st LG u= \cp(0)\ u$; its positivity on the real axis
         is necessary for the existence of $T^{-1}$, and it is only if
         $\cp(a)=\overline{\cp(-a)}$ that the hermiticity condition
         (\ref{hermcon}) can be fulfilled.

         A tedious but elementary calculation shows that all the star
         products considered here may be seen as deformations of the
         canonical symplectic structure defined by the surface element
         under the $\ca \cn$ group (if $\cp(0)$ is properly normalized).
         Indeed, using the kernel (\ref{ampliphas}), one finds at first
         order in $\lambdabar$ :
         \begin{eqnarray} \label{firstord} (u\st LG v)(\bbf x)&=&u(\bbf x)\,v(\b
         x)-i \,\lambdabar
         \, B(0,0)\, u(\bbf x)(\,\lpartial_{a_{\bbf x}}\rpartial_{n_{\b
         x}}-\lpartial_{n_{\bbf x}}\rpartial_{a_{\bbf x}})v(\bbf x) \nonumber\\
         && + i \,\lambdabar
         \,(v(\bbf x) \partial_{n_{\bbf x}}u(\bbf x) B^{(0,1)}(0,0) -
         u(\bbf x) \partial_{n_{\bbf x}}v(\bbf x) B^{(1,0)}(0,0)) + O(\lambdabar ^2) \, .
         \end{eqnarray}
         In order to define an invariant star product, the amplitude
         $B(a,b)$ has in particular to satisfy the relationships (\ref{BO}),
         which force $B(0,0) = 1$ and $(\partial_a B)(0,0)=(\partial_b
         B)(0,0)=0$. Equivalently, this implies that $\cp(0)=1$ in the
         kernel (\ref{noyauP}).

\section{Solving the $\Box$ equations for the kernels $u$ and $v$} \label{App-Box}

We have seen in Sect. \ref{SectSL} that the kernel of the operator $U$ is given as solution to the equation
\begin{equation}
- \rho_\nu^1(F)_{|{\bf z}} u({\bf z}^{-1}{\bf x}) = F^*_{|{\bf x}}
u({\bf z}^{-1}{\bf x}) \quad .
\end{equation}
This equation could be rewritten as
\begin{equation}\label{BoxBrutApp}
- \Box_{|\bar{{\bf z}}} W({\bf x},\bar{{\bf z}}) = F^*_{|{\bf x}}
W({\bf x},\bar{{\bf z}}) \quad ,
\end{equation}
with
\begin{equation}\label{DefWApp}
W({\bf x},\bar{{\bf z}}) = \left( (\phi_\nu^*)^{-1} \circ F
\circ T^T \right)_{|{\bf z}} u({\bf z}^{-1}{\bf x}) \quad ,
\end{equation}
by introducing the operator $\Box$ defined
as
\begin{equation}\label{BoxSympaApp}
\left( (\phi_\nu^*)^{-1} F \rho_\nu(F)f \right)(a,b) =
\Box((\phi_\nu^*)^{-1} F f)(a,b) \quad ,
\end{equation} where
\begin{equation}
\phi_\nu (a,l) = (a,\frac{1}{2 i \nu}\sinh 2 i \nu l) \quad,
\end{equation}
and
\begin{equation} \label{FApp}
F(g)(a,b) = \int e^{-i b l} g(a, l) dl \quad .
\end{equation}
In what follows, we will set $\l = i \n$.
With 
\begin{equation}
\phi_\lambda^{-1} (a,l) = (a, \frac{1}{2\lambda} \mbox{arcsinh} 2
\lambda l) \quad ,
\end{equation}
as well as eq. \re{lambdaF}, and the relationships
\begin{eqnarray}
F[\sinh (2 \nu \partial_l) \, f(a,l)] &=&
\sinh (2 i \nu b) \, \hat{f}(a,b) \quad , \\
F[ l \cosh (2 \nu \partial_l) \, f(a,l) ] &=& - \frac{1}{i}
\partial_b
(\cosh (2 i \nu b) \, \hat{f}(a,b)) \quad , \\
F[ l^2 \sinh (2 \nu \partial_l) \, f(a,l) ] &=& -
\partial_b^2 (\sinh (2 i \nu b) \, \hat{f}(a,b))
\quad ,
\end{eqnarray}
the operator $\Box$ can be explicitly determined as
\begin{equation}\label{Box0}
\Box_{|{\bf y}} = \Box_{(a,b)} = i e^{2a} \left[ b \lambda^2
\partial_a^2 + b (1+4 \lambda^2 b^2) \partial_b^2 + (1+4 \lambda^2
b^2) \partial_a
\partial_b + 4 \lambda^2 b \partial_a + 2(1 + 6 \lambda^2 b^2)
\partial_b - b(k-4\lambda^2) \right] \quad .
\end{equation}
On the other hand, the relationship (\ref{DefWApp})
constrains the functional form of the function $W({\bf
x},\bar{{\bf z}}) = W(c,p,a,b)$, where ${\bf x}=(c,p)$ and ${\bf
z}=(a,l)$, $b$ being the variable conjugated to $l$ through the
transformations $T^T$ and $F$ which only affect the second
coordinate. It is written as
\begin{equation}\label{FormeW}
W_\lambda (x,p,b) = - (1+ 4 \lambda^2 b^2)^{1/4} \, \frac{x}{4\pi}
\, e^{-\frac{i}{4\pi}b p x} \, F_\lambda (b, x) \quad,
\end{equation}
with $x= 4 \pi e^{2(c-a)}$ and
\begin{equation}\label{uF}
F_\lambda (b, x) = \int\, e^{\frac{i}{4 \pi} b x y} \, u_\lambda
(\frac{1}{2} \ln \frac{x}{4\pi},y) \, dy \quad.
\end{equation}
Equation (\ref{BoxBrutApp}) is explicitly written, using eqs.
(\ref{fundamental}) and (\ref{Box0}), the prescribed form of
$W({\bf x},{\bf z})$ and the change of variable $x= 4 \pi
e^{2(c-a)}$, as
\begin{eqnarray}\label{BoxBox}
i [ \, 4 \lambda^2 b x^2 \partial_\x^2 + b (1 + 4 \lambda^2 b^2)
\partial_b^2 - 2x (1+4 \lambda^2 b^2)
\partial_\x \partial_b - 4 \lambda^2 b x \partial_\x +
2(1+6\lambda^2 b^2)\partial_b - \nonumber\\ b(k - 4 \lambda^2) ]
W(x,p,b) = -\frac{x}{4\pi}\left( 2 x p \partial_\x -
(m+p^2)\partial_p \right) W(x,p,b) \quad ,
\end{eqnarray}
which further reduces to the following equation for $F_\lambda (b,
x)$, plugging (\ref{FormeW}) into (\ref{BoxBox}) :
\begin{eqnarray}\label{BoxComplet}
(-4 b x^2 \lambda^2 (1+4 b^2 \lambda^2) \partial_x^2 - b(1+4 b^2
\lambda^2)^2 \partial_b^2  + 2x(1+4 b^2 \lambda^2)^2 \partial_b
\partial_x \nonumber\\ - (1+4 b^2 \lambda^2) 8 b^2 \lambda^2 \partial_b +
V(b,x) ) F_\lambda (b,x) = 0 \quad,
\end{eqnarray}
with $V(b,x)=b(-2 \lambda^2 (1+2 b^2 \lambda^2) + (1+4 b^2
\lambda^2) (k - m (\frac{x}{4 \pi})^2))$.

The problem thus amounts to solve this equation for $F_\lambda (b,
x)$, from which we could find the ``function" $u_\lambda$ by
inverting (\ref{uF}).

For further purposes, we slightly generalize (\ref{FormeW}) for
\begin{equation}\label{DefWAppGen}
W({\bf x},\bar{{\bf z}}) = \left( (\phi_\lambda^*)^{-1} \circ F
\circ T_{n} \right)_{|{\bf z}} u({\bf z}^{-1}{\bf x}) \quad ,
\end{equation}
where $T^{n}$ denotes the transform (\ref{TTransform}) for a
general $AN-$invariant product, not necessarly $\st T{}$ . By
restricting ourselves to ${\cal P}(x) = (\cosh 2 x)^n$, we get
\begin{equation}\label{FormeWGen}
W_\lambda (x,p,b) = - (1+ 4 \lambda^2 b^2)^{n/2} \, \frac{x}{4\pi}
\, e^{-\frac{i}{4\pi}b p x} \, F_\lambda (b, x) \quad.
\end{equation}
Eq. (\ref{BoxBrutApp}) then imposes the following equation for
$F_\lambda (b, x)$ :
\begin{eqnarray}\label{BoxCompletFGen}
(-4 b x^2 \lambda^2 (1+4 b^2 \lambda^2) \partial_\x^2 - b(1+4 b^2
\lambda^2)^2 \partial_b^2 + 2x(1+4 b^2 \lambda^2)^2 \partial_b
\partial_\x - \nonumber\\
(1+4 b^2 \lambda^2) 4 b^2 \lambda^2 (1+2 n)
\partial_b - 4 (1-2 n) b x \lambda^2 (1+4 b^2 \lambda^2)
\partial_x
+ V(b,x) ) F_\lambda (b,x) = 0 \quad,
\end{eqnarray}
with $V(b,x)=b(-4 n \lambda^2 (1+4 n b^2 \lambda^2) + (1+4 b^2
\lambda^2) (k - m (\frac{x}{4 \pi})^2))$.


Writing (\ref{BoxComplet}) as $\left[g^{\mu \nu} (\nabla_\mu +
A_\mu)(\nabla_\nu + A_\nu) + \overline{Q}\right] F_\lambda(b,x)$,
we find
\begin{equation}
A_x = \frac{1}{2x} \quad , \quad A_b=\frac{2 b \lambda^2}{1 + 4
b^2 \lambda^2} \quad,
\end{equation}
thus $A = df$, with $f(b,x) = \frac{1}{2} \ln (2 x (1+4 b^2
\lambda^2)^{1/2})$. We also find that
\begin{equation}
\overline{Q}(b,x) = V(b,x) + b \lambda^2 \quad.
\end{equation}

To eliminate the gauge field $A_\mu$, we set $G(b,x) = h(b,x)
F(b,x)$, with $h(b,x) = e^{f(b,x)} = \sqrt{2x} (1+ 4 b^2
\lambda^2)^{1/4}$, in terms of which the equation becomes
\begin{equation}
\left[ \Delta_{(b,x)} + \tilde{Q}(b,x) \right] G(b,x) = 0 \quad,
\end{equation}
$\Delta_{(b,x)}$ denoting the Laplacian of the metric
\begin{equation}\label{metr}
ds^2 = \frac{1}{x (1+4 \lambda^2 b^2)} \left( \frac{4 b
x\lambda^2}{1+4 \lambda^2 b^2} db^2 + 2 db dx + \frac{b}{x}
dx^2\right) \quad
\end{equation}
and
\begin{equation}
\tilde{Q}(b,x) = -b (1+4 \lambda^2 b^2) (\lambda^2 - k + m
\frac{x^2}{16 \pi^2}) \quad .
\end{equation}
With the further change of variable $b=\frac{1}{2\lambda}\sinh 2
\lambda \theta$, $x=e^\alpha$ and
\begin{equation}
U = e^{\alpha/2} \sinh \theta \lambda \quad, \quad V =
e^{\alpha/2} \cosh \theta \lambda\quad,
\end{equation}
the metric (\ref{metr}) is rewritten as
\begin{equation}
 ds^2 = -\frac{4}{\lambda} \frac{U^2 - V^2}{(U^2 + V^2)^2} dU dV
 \quad,
 \end{equation}
and we finally get
\begin{equation}
\left[\partial_U \partial_V + \tilde{Q}(U,V)\right] G(U,V) = 0
\quad,
\end{equation}
with
\begin{equation}
\tilde{Q}(U,V) = -\frac{1}{\lambda^2} \frac{UV}{(U^2 - V^2)^2}
(\lambda^2 - k + \frac{m}{16 \pi^2}(U^2 - V^2)^2) \quad.
\end{equation}
 Let $U^2=\frac{1}{2}(t-x)$,
$V^2=\frac{1}{2}(t+x)$ and $G(b,x)=H(t,x)$, and the equation becomes
$$\left\{x^2(\partial^2_t-\partial^2_x)-\frac
1{4\lambda^2}[\lambda^2-k+\frac m{16 \pi^2}x^2]\right\}H(t,x)=0$$

admitting solutions by separating variables: $H(t,x)=k_s(x)
exp(i\,s\,t)$, with
\begin{equation}\label{ks}
k_s(x)=\sqrt{x}\left (A J_{\frac
{\sqrt{k}}{2\,\lambda}}[\sqrt{s^2+\frac m {(8 \pi \lambda)^2}}\,\,
x] + B J_{-\frac{\sqrt{k}}{2\,\lambda}}[\sqrt{s^2+\frac m {(8 \pi
\lambda)^2}}\,\, x]\right),
\end{equation}
where $J_\m$ denote the Bessel function of the first kind\footnote{i.e. $[x^2 \frac{d^2}{dx^2} + x \frac{d}{dx} + (x^2 - \m^2)] J_\m(x) =0$.}.
The general solution is then obtained as a superposition of these
modes.


Putting this all together, we get
\begin{equation}
F_{\lambda,s} (b,x) = \frac{1}{\sqrt{2x}} k_s(x) (1+4 \lambda^2
b^2)^{-1/4} e^{i s x\sqrt{1+4 \lambda^2 b^2} }\quad,
\end{equation}
and from (\ref{uF})
\begin{equation}
u(a,l) = \frac{e^{2a}}{4 \pi}\int\, e^{i b l e^{2a}} F(b,4 \pi
e^{2a}) \, db \quad,
\end{equation}
thus (up to constant factors)
\begin{equation}\label{uGen}
u^T_{\lambda,s}(a,l) =  e^a \, k_s(4\pi e^{2a})\, \int \, {\cal
H}_s(a,b) \, e^{i b l e^{2a}} \, db \quad,
\end{equation}
with
\begin{equation}\label{uGenbis}
{\cal H}_s(a,b) = (1+ 4 \lambda^2 b^2)^{-1/4} e^{4 \pi i s
\sqrt{1+4 \lambda^2 b^2} e^{2a}} \quad,
\end{equation}
where the subscript $^T$ refers to the fact that the operator
$U^T$ whose kernel is $u^T_{\lambda,s}(a,l)$ is such that $f \# g
= U^T\left((U^T)^{-1}f \st{T}{} \;(U^T)^{-1}g \right)$.

 Note that the resolution of (\ref{BoxCompletFGen})
yields
\begin{equation}\label{FLambda}
F_{\lambda,s} (b,x) = \frac{1}{\sqrt{2x}} k_s(x) (1+4 \lambda^2
b^2)^{-n/2} e^{i s x\sqrt{1+4 \lambda^2 b^2} }\quad,
\end{equation}
and
\begin{equation}\label{uGen2}
u^{n}_{\lambda,s}(a,l) =  e^a \, k_s(4\pi e^{2a})\, \int \,
{\cal H}^{n}_s(a,b) \, e^{i b l e^{2a}} \, db \quad,
\end{equation}
with
\begin{equation}\label{uGen2bis}
{\cal H}^{n}_s(a,b) = (1+4 \lambda^2 b^2)^{-n/2} e^{4 \pi i
s \sqrt{1+4 \lambda^2 b^2} e^{2a}} \quad,
\end{equation}
with ${\cal P}(x) = (\cosh 2 x)^n$.

 The kernel (\ref{uGen2}), obtained by solving the Box equation
 with the functional form (\ref{DefWAppGen}), with ${\cal
P}(x) = (\cosh 2x)^n$, actually corresponds to an intertwiner
between an $AN-$invariant product with ${\cal P}'(x) = (\cosh
2x)^{(1-n)}$ (and {\bf not} $(\cosh 2x)^{n}$!)  and $\#$. Indeed,
from (\ref{uGen}), one finds that the kernel of $U^{{\cal P}'} =
U^T \circ (T^T)^{-1} \circ T^{{\cal P}'}$ (such that $f \# g =
U^{{\cal P}'}\left((U^{{\cal P}'})^{-1}f \st{{\cal P}'}{}
(U^{{\cal P}'})^{-1}g \right)$, with ${\cal P}'(x) = (\cosh
2x)^{n'}$) reads
\begin{equation}\label{uGen3}
u^{n'}_{\lambda,s}(a,l) =  e^a \, k_s(4\pi e^{2a})\, \int \, (1+4
\lambda^2 b^2)^{\frac{n' - 1}{2}} e^{4 \pi i s \sqrt{1+4 \lambda^2
b^2} e^{2a}} \, e^{i b l e^{2a}} \, db \quad.
\end{equation}
Comparing (\ref{uGen2}) and (\ref{uGen2bis}) to (\ref{uGen3}), one
indeed has $\frac{n' -1}{2} = -\frac{n}{2}$.

Finally, the corresponding function $W(x,p,b)$ reads, in all
cases,
\begin{equation}\label{ModeW}
 W(x,p,b) = - \frac{\sqrt{x}}{4\pi} \, e^{-\frac{i}{4\pi} b x
p} k_s(x) e^{i s x (1+ 4 \lambda^2 b^2)^{1/2}} \quad ,
\end{equation}
with $s\in\R$.

We may use a similar technique to determine the kernel of a given operator $U^{-1}$. By writing
\begin{equation}
 [U^{-1}f]({\bf x})=\int \, v({\bf z}^{-1}{\bf x}) \, f({\bf z}) d{\bf z} \quad,
 \end{equation}

one finds from (\ref{cond1}) that
\begin{equation}\label{BoxBrutAppInv}
 \Box_{|\bar{{\bf x}}} M(\bar{{\bf x}},{\bf z}) = - F^*_{|{\bf z}} M(\bar{{\bf x}},{\bf z}) \quad
 \end{equation}
with
\begin{equation}\label{M}
M(\bar{{\bf x}},{\bf z}) = \left( (\phi_\lambda^*)^{-1} \circ F
\circ T^{{\cal P}} \right)_{|{\bf x}} v^{{\cal P}}({\bf
z}^{-1}{\bf x}) \quad .
\end{equation}
Note that, contrary to (\ref{BoxBrutApp}), (\ref{BoxBrutAppInv}) can
be written using any $T_{n}$ transform, and not necessarly $T^{T}$.
This comes from the fact that $F^*$ is always a self-adjoint operator, while the operator
$\rho_\nu^1(F)$ is only self-adjoint when it is constructed with an $AN\,$-invariant
star product satisfying the trace condition.

Solutions to (\ref{BoxBrutAppInv}) can be deduced from solutions to
(\ref{BoxBrutApp}), as $v^{n}({\bf z}^{-1}{\bf x})= u^{n}({\bf x}^{-1}{\bf z})$, with $u^{n}$ given by
(\ref{uGen2}) and (\ref{uGen2bis}). Indeed, from (\ref{BoxBrutApp})
and (\ref{BoxBrutAppInv}), we see that $M(\bar{{\bf x}},{\bf z}) =
W({\bf z},\bar{{\bf x}}) $ and hence, if $M({\bf x},{\bf z}) =
\left( (\phi_\lambda^*)^{-1} \circ F \circ T_{n}
\right)_{|{\bf x}} v^{n}({\bf z}^{-1}{\bf x})$ and $W({\bf
x},{\bf z}) = \left( (\phi_\lambda^*)^{-1} \circ F \circ T^{{\cal
P}} \right)_{|{\bf z}} u^{n}({\bf z}^{-1}{\bf x})$, this
implies that $v^{n}({\bf z}^{-1}{\bf x})= u^{n}({\bf
x}^{-1}{\bf z})$. We finally get
\begin{equation}\label{vGen2}
v^{n}_{\lambda,s}(a,l) = e^{-a} \, k_s(4\pi e^{-2a})\, \int
\, (1+ 4 \lambda^2 b^2)^{-n/2} e^{4 \pi i s \sqrt{1+4 \lambda^2
b^2} e^{-2a}} \, e^{-i b l} \, db \quad,
\end{equation}


\section{About the unicity of the $\#$-product}\label{AppUnicity}
In this Appendix, we discuss the unicity of the construction of Sect. \ref{SectSL}, by analyzing to what extent our derivation gives \emph{all} $\SL$-invariant kernels. To this end, we will start from an arbitrary $\SL$-invariant product, related to a given $AN\,$-invariant product by an intertwiner $U$, and
show that the kernel of this operator, defined in \re{UTransfTw}, necessarily satisfies \re{BoxBrutApp}, up to all possible minor modifications and/or redefinitions.

Before proceeding, let us introduce a useful concept, the \ind{Chevalley cohomology}.
 Let $(V,\rho)$
be a representation of a Lie algebra $\mathfrak{g}$. A p-cochain with values
in V is a rule $c$ which assigns to each (p)-uple
$(X_0,\dots,X_{p-1})$ in $\mathfrak{g} \times \cdots \times \mathfrak{g}$ an element
$c(X_0,\cdots,X_{p-1})$ in V, totally skew-symmetric in its
arguments. The (abelian) group of all p-cochains is denoted by
$C^p(\mathfrak{g},V)$. The co-boundary operator $\delta$ of the Chevalley
cohomology associated to the representation $\rho$ is defined as
\begin{equation}\label{DelChev}
(\delta c)(X_0,\dots,X_{p}) = \underset{i}\sum (-1)^i
\rho(X_i)\,c(X_0,-,\hat{X_i},-,X_p) + \underset{i<j}\sum
(-1)^{i+j}\,c([X_i,X_j],X_0,-,\hat{X_i},-,\hat{X_j},-,X_p)
\end{equation}
One checks that $\delta \circ \delta = 0$. The $p^{th}$ Chevalley
cohomology space $H^p(\mathfrak{g},V)$ is the
quotient\footpourmoi{Chevalley-Eilenberg, Bourbaki}
\begin{equation}
H^p(\mathfrak{g},V) = \{ \mbox{p-cocycles, i.e. p-cochains s.t.} \,\delta
c=0\}\, / \, \{\mbox{p-coboundaries, i.e. p-cochains s.t.}
\,c=\delta b\} \quad .
\end{equation}
We have the \ind{Whitehead's lemma}\cite{deAzcarraga:1995jw}:

Let $\mathfrak{g}$ be a finite-dimensional semi-simple Lie algebra and
$\rho$ a non-trivial representation of $\mathfrak{g}$ on $V$. Then
\begin{equation}
 H^q(\mathfrak{g},V) = 0 \quad  q=1,2 \, .
 \end{equation}

Suppose we start with a $G$-invariant product $\#$ on $R=AN\,$.
This is equivalent to the statement that there exists a Lie algebra
homomorphism $\mathfrak{g} \rightarrow \mbox{Der}(\#) : X \rightarrow X^*$.
Furthermore, because all (formal) star products on $R$ are
classified by $H^2_{dRh} (R) = 0$, all products are equivalent to
$\overset{M}{\star}$. This implies that if $D$ is a derivation for
$\overset{M}{\star}$, then $M D M^{-1}$ will be a derivation for $\#$, with
$f \overset{M}{\star} g = M^{-1}(M f \# M g)$. But all the derivations for
$\overset{M}{\star}$ are interior\footpourmoi{R\'ef\'erence? + pr\'eciser le
type de fonction}, that is, for all $D\in \mbox{Der}(\overset{M}{\star})$,
there exists $\lambda_D \in C(R)$, such that $D f =
[\lambda_D,f]_{\overset{M}{\star}}$. Consequently, $\mbox{Der}(\#) =
\mbox{int}(\#)$, and there is an application
\begin{equation}
 \mu : \mathfrak{g} \rightarrow C^\infty(R)[[\nu]] : X \rightarrow \tilde{\mu}_X
 \quad,
 \end{equation}
 with
 \begin{equation}
  X^* = [\tilde{\mu}_X, \, . \,]_\# \overset{\triangle}{=} ad_\# \tilde{\mu}_X \quad.
  \end{equation}
In what follows, we will often set $2 \n =1$ for simplicity. 
Because of the Lie algebra homomorphism $\mathfrak{g} \rightarrow
\mbox{Der}(\#) : X \rightarrow X^*$, we have $[X,Y]^* =
[X^*,Y^*]$. The right-hand side $[ad_\# \tilde{\mu}_X, ad_\#
\tilde{\mu}_Y]$ can be written, using the Jacobi identity, as
$ad_\# [\tilde{\mu}_X,\tilde{\mu}_Y]_\#$, while the left-hand side is by
definition $ad_\# \tilde{\mu}_{[X,Y]}$. It follows that
\begin{equation}
 c (X,Y) := \tilde{\mu}_{[X,Y]} - [\tilde{\mu}_X,\tilde{\mu}_Y]_\#
 \end{equation}
belongs to the center of the algebra, i.e. it is a 2-cochain
\begin{equation}
c : \Lambda^2 (\mathfrak{g}) \rightarrow {\cal C}[[\nu]]
\end{equation}
with values in the (formal) constants. With (\ref{DelChev}), one
also checks that $\delta c = 0$, thus, by virtue of the
Whitehead's lemma\footpourmoi{\bf applicable si representation
triviale?}, $c$ has to be a co-boundary :
\begin{equation}
 c = \delta b, \quad \mbox{with} \quad b : \mathfrak{g}\rightarrow {\cal
 C}[[\nu]] \quad.
 \end{equation}
The application $\mu = \tilde{\mu} + b$ then defines a Lie algebra
homomorphism, because one can check that $\mu_{[X,Y]} - [\mu_X,\mu_Y]_\# = 0$.

Now, consider an operator $U$ such that
\begin{equation}\label{sharp}
f \# g = U ( U^{-1} f \star^1_\n U^{-1} g) ,
\end{equation}
for a given $AN\,-$invariant product $\star^1_\n$. We would like to
check if this operator necessarly satisfies (\ref{BoxBrutApp}). This
operator has to be a convolution in order to preserve the
$AN-$invariance and must commute with the action of $R^*$ :
\begin{equation}
 U X^* U^{-1} f = X^* f \quad , \, \forall X\in \mathfrak{a} \oplus \mathfrak{n} \quad.
 \end{equation}
With (\ref{SolAN}) and (\ref{ad1}), the left-hand side is rewritten as
$U[\Lambda_X,U^{-1}f]_{\st{1}{}}$, while the right-hand side is $[\mu_X
,f]_\# = U[U^{-1}\mu_X,U^{-1}f]_{\st{1}{}}$, from which we
deduce
\begin{equation}\label{MuLambda}
\Lambda_X = U^{-1} \mu_X + c(X)  \quad, \quad \forall X \in \mathfrak{a} \oplus \mathfrak{n} \quad.
 \end{equation}
where $c(X)$ are a priori arbitrary constants. Let us first assume that these constants vanish. We will return to this assumption at the end of the section. The next step is to define
\begin{equation}\label{rhosharp}
 \rho^\# (X) = U \rho^1 (X) U^{-1} \quad, \forall X \in \mathfrak{g}.
 \end{equation}
 By construction, this operator is a derivation for $\#$, and because all
 derivations are interior we have
 \begin{equation}
  \rho^\# (X) = [\Gamma_X,f]_\# \quad, \mbox{with}\quad \Gamma_X =
  U\Lambda_X \quad,
  \end{equation}
because of (\ref{sharp}) and (\ref{rhosharp}). Following the same
argument that led to (\ref{Homo1}), we know that $\rho^\#$ is a
homomorphism, by construction :
\begin{equation}\label{Homosharp}
[\rho_\nu^\#(X),\rho_\nu^\#(Y)]=\rho_\nu^\#([X,Y]) \, \forall
X,Y\in \mathfrak{g} \quad.
\end{equation} Moreover, we note that,
$\forall X \in \mathfrak{a} \oplus \mathfrak{n} \subset \mathfrak{g}$, $\Gamma_X = \mu_X$.  On the other hand, we know that the vector field $F^*$ may be written as
\begin{equation}
F^* = [\mu_F, \,. \,]_\# \quad,
\end{equation}
where the function $\mu_F$ is not known at this stage.

In our approach, to determine the operator $U$, we solved the equation $\rho^\# (F)
= F^*$ by taking $\mu_F = \Gamma_F$. We would like to analyze to
what extent the equation would be modified for a general $\mu_F$.
To this end, we first have to identify $\mu_F$. On the one hand,
 we saw that $\mu$ is a homomorphism :
 \begin{equation}
 [\mu_x,\mu_Y]_\# = \mu_{[X,Y]}  \quad.
 \end{equation}
 This implies that $U^{-1}\mu$ is also a homomorphism, because
 \begin{eqnarray}\label{HomoU}
 U^{-1}[U U^{-1} \mu_X,U U^{-1} \mu_Y]_\# &=& U^{-1}\mu_{[X,Y]}
 \nonumber\\
\hookrightarrow [U^{-1}\mu_X,U^{-1}\mu_Y]_{\st{1}{}} &=&
U^{-1}\mu_{[X,Y]} \quad.
\end{eqnarray}
Thus,
\begin{eqnarray}
[U^{-1}\mu_H,U^{-1}\mu_F]_{\st{1}{}} &=&  -2 U^{-1}\mu_F  \label{a1}\\
                                       &=& [\Lambda_H ,U^{-1}\mu_F]_{\st{1}{}} \label{a2bis}\\
                                       &=& H^* (U^{-1}\mu_F)
                                       \label{a3bis} \quad ,
\end{eqnarray}
where (\ref{a2bis}) holds because of (\ref{MuLambda}) with $c(X)=0$, and where (\ref{a3bis})
is due to (\ref{SolAN}) and  (\ref{ad1}). Thus
\begin{equation}\label{EqH}
H^* (U^{-1}\mu_F) = -2 (U^{-1}\mu_F) \quad.
\end{equation}
Next, we have
\begin{eqnarray}
[U^{-1}\mu_E,U^{-1}\mu_F]_{\st{1}{}} &=&   U^{-1}\mu_H  \label{b1}\\
                                    &=& [\Lambda_E ,U^{-1}\mu_F]_{\st{1}{}} \label{b2}\\
                                       &=& E^* (U^{-1}\mu_F)
                                       \label{b3} \quad ,
\end{eqnarray}
On the other hand, we also have the homomorphism (\ref{Homo1bis})
(with $2\nu =1 $ ):
\begin{equation}
[\Lambda_X,\Lambda_Y]_{\star^1_\nu} = \Lambda_{[X,Y]} \, ,
\, \forall X,Y \in \mathfrak{g} = \ls \quad .
\end{equation}
Thus, (\ref{b1}) reads, with the help of (\ref{MuLambda}) (and $c(X)=0$)
\begin{eqnarray}
U^{-1}\mu_H &=& U^{-1}\Gamma_H \\
            &=&  \Lambda_H \\
            &=& [\Lambda_E,\Lambda_F]_{\st{1}{}} \\
            &=& E^* \Lambda_F \label{c3} \quad.
\end{eqnarray}
From this we have
\begin{equation}\label{EqEcE}
E^* (U^{-1}\mu_F - \Lambda_F) = 0 \quad.
\end{equation}
Finally,
\begin{eqnarray}
 [\Lambda_H, \Lambda_F]_{\st{1}{}} &=& \Lambda_{[H,F]} = -2
 \Lambda_F\\
 &=& H^* \Lambda_F \label{d3} \quad.
 \end{eqnarray}
From (\ref{EqH}) and (\ref{d3}), we get, with $\zeta_F =
U^{-1}\mu_F -\Lambda_F$
\begin{equation}\label{EqEcH}
H^* \zeta_F = -2 \zeta_F \quad.
\end{equation}
Solving (\ref{EqEcE}) and  (\ref{EqEcH}) yields
\begin{equation}
 \zeta_F(a,l)  =  C \, \exp(2a) \quad .
 \end{equation}


The equation to be satisfied by any operator $U$ now reads
\begin{eqnarray}
 F^* = [\mu_F, \, .\,]_\# &=& [U(\Lambda_F + \zeta_F),\, . \,]_\# \\
                          &=& U[\Lambda_F + \zeta_F,U^{-1} \, .\,]_{\st{1}{}} \\
                          &=& U ad_{\star^1}(\Lambda_F + \zeta_F)
                          U^{-1} \quad.
                          \end{eqnarray}
With $\rho_\nu^1 (F) = [\Lambda_F, \, .\,]_{\st{1}{}}$, equation
(\ref{eqverif}) now becomes :
\begin{equation}\label{eqverifbis}
 (\rho_\nu^1(F) + ad_{\st{1}{}}(\zeta_F))_{|{\bf z}} u({\bf
z}^{-1}{\bf x}) = -F^*_{|{\bf x}} u({\bf z}^{-1}{\bf x}) \quad .
\end{equation}
Again, we made use a strongly closed $\star^1$, but this is not restrictive.
By noticing that
\begin{equation}
ad_{\st{1}{}}(\zeta_F) = (T^T)^{-1}  ad_{\overset{M}{\star}}(T^T \zeta_F)
T^T \quad,
\end{equation}
the left-hand side of $\ref{eqverifbis}$ is
\begin{equation}
 \left[(T^T)^{-1} (\rho_\nu^0(F) + ad_{\overset{M}{\star}}(T^T \zeta_F))
 T^T\right]_{\bbf z} u(\bbf z^{-1}\bbf x) \quad.
 \end{equation}
The additional term can be evaluated with the help of
(\ref{MoyalAsympt}) to be :
\begin{equation}
ad_{\overset{M}{\star}}(T^T \zeta_F)) = ad_{\overset{M}{\star}}(\zeta_F)) = C e^{2a}
\sinh(2\nu \partial_l) \quad.
\end{equation}
By defining a new operator
\begin{equation}\label{BoxSympaAppBis}
 \Box' ((\phi_\lambda^*)^{-1} F f)(a,b) =
 \left( (\phi_\lambda^*)^{-1} F (\rho_\nu(F) + ad_{\overset{M}{\star}}(T^T \zeta_F))f
 \right)(a,b)\quad,
 \end{equation}
 we find that this one is related to the previous box operator (\ref{Box0})
 by
 \begin{equation}
 \Box'_{(a,b)} =  \Box_{(a,b)} + i e^{2a} C b \quad.
 \end{equation}
Consequently, the additional term just amounts to redefining the
constant $k$ of (\ref{Box0}), and all the functions $W(\bbf x,\bbf z)$
(i.e. all the convolution kernels $u(\bbf z^{-1}\bbf x)$) have to satisfy
eq.(\ref{BoxBrutApp}).


To end up, we now discuss the status of the constants $c(E)$ and $c(H)$ which we discarded in \re{MuLambda}.
 From  (\ref{Homo1bis}) and
(\ref{HomoU}), we see that $c(E) = 0$, but $c(H)$ remains
arbitrary. We may define
\begin{equation}
 \Lambda'_X = \Lambda_X + k_X \quad, X \in \mathfrak{a} \oplus \mathfrak{n},
 \end{equation}
 with  $k_E = 0$, and $k_H = A =cst.$, and observe that we still
 have ($2 \nu =1$)
\begin{equation}\label{Homo1bisbis}
[\Lambda'_X,\Lambda'_Y]_{\star^1_\nu}  =\Lambda'_{[X,Y]} \, , \,
\forall X,Y \in \mathfrak{a} \oplus \mathfrak{n} \quad .
\end{equation}
This way, we can always redefine $\Lambda \rightarrow \Lambda'$
such that (\ref{HomoU}) is true\footpourmoi{Le fait que ce ne soit
 un homomorphisme que pour $\mathfrak{a} \oplus \mathfrak{n}$ et pas pour $\mathfrak{g}$ entier
joue-t-il ??}. Let us analyze the effects of this redefinition. We
have $\Lambda_H \rightarrow \Lambda_H + A$, with $A$ being constant.
From (\ref{ad0}), (\ref{SolAN}) and (\ref{ad1}), we can see how
this translates on the moment maps (\ref{moments}). First note
that
\begin{equation}
U_1^{-1} \Lambda_X = \lambda_X + d(X) \quad,
\end{equation}
where $d(X)$ again denotes a constant (set to 0 up to now). Thus
the shift $\Lambda_H \rightarrow \Lambda_H + A$ induces the shift
$\lambda_H \rightarrow \lambda_H + \alpha$. Consequently, the net
effect is simply the translation $l \rightarrow l+\alpha$. Now, we
can wonder how equation (\ref{BoxBrutApp}) will be modified by this
change of coordinates. With (\ref{lambdaF}) and (\ref{BoxSympaApp}),
we see that the left-hand side of (\ref{BoxBrutApp}) will remain unchanged if
we redefine the Fourier transform (\ref{F}) by
\begin{equation} \label{Falpha}
F_\alpha(g)(a,b) :=  \int e^{-i b u} g(a, u - \alpha) du \quad .
\end{equation}
On the other hand, the right-hand side will be modified by the substitution
$l \rightarrow l + \alpha$ in $F^*$, see eq. (\ref{fundamental}).
Eq. (\ref{BoxBrutApp}) then becomes, with $\x = (c,p)$, $\bar{\z} =
(a,b)$, and $x = e^{2(c-a)}$:
\begin{eqnarray}\label{BoxBoxBis}
i [ \, 4 \lambda^2 b x^2 \partial_\x^2 + b (1 + 4 \lambda^2 b^2)
\partial_b^2 - 2x (1+4 \lambda^2 b^2)
\partial_\x \partial_b - 4 \lambda^2 b x \partial_\x +
2(1+6\lambda^2 b^2)\partial_b - \nonumber\\ b(k - 4 \lambda^2) ]
W(x,p,b) = -x\left( 2 x p \partial_\x - ((m +
\alpha^2)+p^2)\partial_p + 2 \alpha x \partial_x -2 \alpha p \,
\partial_p \right) W(x,p,b) \quad .
\end{eqnarray}
Plugging $W_\lambda (x,p,b) = - (1+ 4 \lambda^2 b^2)^{1/4} \, x \,
e^{-i b p x} \, F_\lambda (b, x)$ into this,
 we get
\begin{eqnarray}\label{BoxCompletBis}
(-4 b x^2 \lambda^2 (1+4 b^2 \lambda^2) \partial_\x^2 - b(1+4 b^2
\lambda^2)^2 \partial_b^2 + 2x(1+4 b^2 \lambda^2)^2 \partial_b
\partial_\x  - (1+4 b^2 \lambda^2) 8 b^2 \lambda^2 \partial_b \nonumber\\ +
2 i (1+4\lambda^2 b^2) \alpha x^2 \partial_x + V(b,x) ) F_\lambda
(b,x) = 0 \quad,
\end{eqnarray}
with $V(b,x)=b(-2 \lambda^2 (1+2 b^2 \lambda^2) + (1+4 b^2
\lambda^2) (k - (m+\alpha^2) x^2 )) + 2 i x \alpha (1+4 \lambda^2
b^2)$. Comparing this to (\ref{BoxComplet}), we see that the equation
acquires an imaginary part, while the real part remains unchanged,
modulo the redefinition $m \rightarrow m + \alpha^2$. Writing
$F_\lambda (b,x) =  F^R_\lambda (b,x) + i F^I_\lambda (b,x)$, the
real part is still given by (\ref{FLambda}), while $F^I_\lambda
(b,x)$ satisfies
\begin{equation}
 ( 1 + x \partial_x)\,  F^I_\lambda
(b,x) = 0 \quad,
\end{equation}
and is thus given by
\begin{equation}
 F^I_\lambda (b,x) = \frac{C}{x} \quad.
 \end{equation}

In conclusion, we have thus shown that every $\SL$-invariant product on $\Pi$ is of the form discussed in Sect. \re{SectSL}, i.e. characterized by an intertwining kernel $u$ satisfying the second order differential equation \re{BoxBrutApp}.

\printindex
\addcontentsline{toc}{chapter}{\numberline{}Index}
\bibliography{BibliaThese}
\addcontentsline{toc}{chapter}{\numberline{}Bibliography}



\end{document}